\documentclass[longauth]{aa}

\usepackage{graphicx}
\usepackage{natbib}
\usepackage{scalerel}

\usepackage[table]{xcolor}

\usepackage{txfonts}
\usepackage[pdfencoding=auto,psdextra]{hyperref}
\hypersetup{
    colorlinks=true,
    linkcolor=blue,
    filecolor=magenta,      
    urlcolor=blue,
    citecolor=blue
}
\makeatletter
\renewcommand*\aa@pageof{, page \thepage{} of \pageref*{LastPage}}
\makeatother

\usepackage[utf8]{inputenc}

\usepackage{euclid}

\newcommand{\av}{$A_V$}
\newcommand{\rv}{$R_V$}
\newcommand{\ebv}{$E(B-V)$}
\newcommand{\zsun}{$Z_\odot$}

\newcommand{\mulim}{$\mu_\mathrm{lim}$}
\newcommand{\hi}{\ion{H}{i}} 
\newcommand{\hii}{\ion{H}{ii}} 

\newcommand{\mstar}{$M_*$}

\newcommand{\msun}{$M_\odot$}

\newcommand{\hst}{{HST}}
\newcommand{\jwst}{{JWST}}

\newcommand{\hers}{\textit{Herschel}}

\newcommand{\lcdm}{$\Lambda$CDM}
\newcommand{\magarc}{mag\,arcsec$^{-2}$}

\newenvironment{hangingpar}[1]
  {\begin{list}
          {}
          {\setlength{\itemindent}{-#1}
           \setlength{\leftmargin}{#1}
           \setlength{\itemsep}{0pt}
           \setlength{\parsep}{\parskip}
           \setlength{\topsep}{\parskip}
           }
    \setlength{\parindent}{-#1}
    \item[]
  }
  {\end{list}}

\begin{document}
%
%
\title{\Euclid: Early Release Observations -- Deep anatomy of nearby galaxies\thanks{This paper is published on behalf of the Euclid Consortium.}}    

\renewcommand{\orcid}[1]{} 
\author{L.~K.~Hunt\orcid{0000-0001-9162-2371}\thanks{\email{leslie.hunt@inaf.it}}\inst{\ref{aff1}}
\and F.~Annibali\inst{\ref{aff2}}
\and J.-C.~Cuillandre\orcid{0000-0002-3263-8645}\inst{\ref{aff3}}
\and A.~M.~N.~Ferguson\inst{\ref{aff4}}
\and P.~Jablonka\orcid{0000-0002-9655-1063}\inst{\ref{aff5}}
\and S.~S.~Larsen\orcid{0000-0003-0069-1203}\inst{\ref{aff6}}
\and F.~R.~Marleau\orcid{0000-0002-1442-2947}\inst{\ref{aff7}}
\and E.~Schinnerer\orcid{0000-0002-3933-7677}\inst{\ref{aff8}}
\and M.~Schirmer\orcid{0000-0003-2568-9994}\inst{\ref{aff8}}
\and C.~Stone\orcid{0000-0002-9086-6398}\inst{\ref{aff9}}
\and C.~Tortora\orcid{0000-0001-7958-6531}\inst{\ref{aff10}}
\and T.~Saifollahi\orcid{0000-0002-9554-7660}\inst{\ref{aff11},\ref{aff12}}
\and A.~Lan\c{c}on\orcid{0000-0002-7214-8296}\inst{\ref{aff11}}
\and M.~Bolzonella\orcid{0000-0003-3278-4607}\inst{\ref{aff2}}
\and S.~Gwyn\orcid{0000-0001-8221-8406}\inst{\ref{aff13}}
\and M.~Kluge\orcid{0000-0002-9618-2552}\inst{\ref{aff14}}
\and R.~Laureijs\inst{\ref{aff15}}
\and D.~Carollo\orcid{0000-0002-0005-5787}\inst{\ref{aff16}}
\and M.~L.~M.~Collins\orcid{0000-0002-1693-3265}\inst{\ref{aff17}}
\and P.~Dimauro\orcid{0000-0001-7399-2854}\inst{\ref{aff18},\ref{aff19}}
\and P.-A.~Duc\orcid{0000-0003-3343-6284}\inst{\ref{aff20}}
\and D.~Erkal\orcid{0000-0002-8448-5505}\inst{\ref{aff17}}
\and J.~M.~Howell\inst{\ref{aff4}}
\and C.~Nally\orcid{0000-0002-7512-1662}\inst{\ref{aff4}}
\and E.~Saremi\orcid{0000-0002-5075-1764}\inst{\ref{aff21}}
\and R.~Scaramella\orcid{0000-0003-2229-193X}\inst{\ref{aff18},\ref{aff22}}
\and V.~Belokurov\inst{\ref{aff23}}
\and C.~J.~Conselice\orcid{0000-0003-1949-7638}\inst{\ref{aff24}}
\and J.~H.~Knapen\orcid{0000-0003-1643-0024}\inst{\ref{aff25},\ref{aff26}}
\and A.~W.~McConnachie\orcid{0000-0003-4666-6564}\inst{\ref{aff13}}
\and I.~McDonald\orcid{0000-0003-0356-0655}\inst{\ref{aff24}}
\and J.~Miro~Carretero\orcid{0000-0003-1808-1753}\inst{\ref{aff27},\ref{aff28}}
\and J.~Roman\orcid{0000-0002-3849-3467}\inst{\ref{aff26},\ref{aff25}}
\and M.~Sauvage\orcid{0000-0002-0809-2574}\inst{\ref{aff3}}
\and E.~Sola\orcid{0000-0002-2814-3578}\inst{\ref{aff23}}
\and N.~Aghanim\inst{\ref{aff29}}
\and B.~Altieri\orcid{0000-0003-3936-0284}\inst{\ref{aff30}}
\and S.~Andreon\orcid{0000-0002-2041-8784}\inst{\ref{aff31}}
\and N.~Auricchio\orcid{0000-0003-4444-8651}\inst{\ref{aff2}}
\and S.~Awan\inst{\ref{aff32}}
\and R.~Azzollini\orcid{0000-0002-0438-0886}\inst{\ref{aff32}}
\and M.~Baldi\orcid{0000-0003-4145-1943}\inst{\ref{aff33},\ref{aff2},\ref{aff34}}
\and A.~Balestra\orcid{0000-0002-6967-261X}\inst{\ref{aff35}}
\and S.~Bardelli\orcid{0000-0002-8900-0298}\inst{\ref{aff2}}
\and A.~Basset\inst{\ref{aff36}}
\and R.~Bender\orcid{0000-0001-7179-0626}\inst{\ref{aff14},\ref{aff37}}
\and D.~Bonino\orcid{0000-0002-3336-9977}\inst{\ref{aff38}}
\and E.~Branchini\orcid{0000-0002-0808-6908}\inst{\ref{aff39},\ref{aff40},\ref{aff31}}
\and M.~Brescia\orcid{0000-0001-9506-5680}\inst{\ref{aff41},\ref{aff10},\ref{aff42}}
\and J.~Brinchmann\orcid{0000-0003-4359-8797}\inst{\ref{aff43},\ref{aff44}}
\and S.~Camera\orcid{0000-0003-3399-3574}\inst{\ref{aff45},\ref{aff46},\ref{aff38}}
\and G.~P.~Candini\orcid{0000-0001-9481-8206}\inst{\ref{aff32}}
\and V.~Capobianco\orcid{0000-0002-3309-7692}\inst{\ref{aff38}}
\and C.~Carbone\orcid{0000-0003-0125-3563}\inst{\ref{aff47}}
\and J.~Carretero\orcid{0000-0002-3130-0204}\inst{\ref{aff48},\ref{aff49}}
\and S.~Casas\orcid{0000-0002-4751-5138}\inst{\ref{aff50}}
\and M.~Castellano\orcid{0000-0001-9875-8263}\inst{\ref{aff18}}
\and S.~Cavuoti\orcid{0000-0002-3787-4196}\inst{\ref{aff10},\ref{aff42}}
\and A.~Cimatti\inst{\ref{aff51}}
\and G.~Congedo\orcid{0000-0003-2508-0046}\inst{\ref{aff4}}
\and L.~Conversi\orcid{0000-0002-6710-8476}\inst{\ref{aff52},\ref{aff30}}
\and Y.~Copin\orcid{0000-0002-5317-7518}\inst{\ref{aff53}}
\and L.~Corcione\orcid{0000-0002-6497-5881}\inst{\ref{aff38}}
\and F.~Courbin\orcid{0000-0003-0758-6510}\inst{\ref{aff5}}
\and H.~M.~Courtois\orcid{0000-0003-0509-1776}\inst{\ref{aff54}}
\and M.~Cropper\orcid{0000-0003-4571-9468}\inst{\ref{aff32}}
\and A.~Da~Silva\orcid{0000-0002-6385-1609}\inst{\ref{aff55},\ref{aff56}}
\and H.~Degaudenzi\orcid{0000-0002-5887-6799}\inst{\ref{aff57}}
\and A.~M.~Di~Giorgio\orcid{0000-0002-4767-2360}\inst{\ref{aff58}}
\and J.~Dinis\inst{\ref{aff55},\ref{aff56}}
\and F.~Dubath\orcid{0000-0002-6533-2810}\inst{\ref{aff57}}
\and X.~Dupac\inst{\ref{aff30}}
\and S.~Dusini\orcid{0000-0002-1128-0664}\inst{\ref{aff59}}
\and M.~Farina\orcid{0000-0002-3089-7846}\inst{\ref{aff58}}
\and S.~Farrens\orcid{0000-0002-9594-9387}\inst{\ref{aff3}}
\and S.~Ferriol\inst{\ref{aff53}}
\and P.~Fosalba\orcid{0000-0002-1510-5214}\inst{\ref{aff60},\ref{aff61}}
\and M.~Frailis\orcid{0000-0002-7400-2135}\inst{\ref{aff16}}
\and E.~Franceschi\orcid{0000-0002-0585-6591}\inst{\ref{aff2}}
\and M.~Fumana\orcid{0000-0001-6787-5950}\inst{\ref{aff47}}
\and S.~Galeotta\orcid{0000-0002-3748-5115}\inst{\ref{aff16}}
\and B.~Garilli\orcid{0000-0001-7455-8750}\inst{\ref{aff47}}
\and W.~Gillard\orcid{0000-0003-4744-9748}\inst{\ref{aff62}}
\and B.~Gillis\orcid{0000-0002-4478-1270}\inst{\ref{aff4}}
\and C.~Giocoli\orcid{0000-0002-9590-7961}\inst{\ref{aff2},\ref{aff63}}
\and P.~G\'omez-Alvarez\orcid{0000-0002-8594-5358}\inst{\ref{aff64},\ref{aff30}}
\and B.~R.~Granett\orcid{0000-0003-2694-9284}\inst{\ref{aff31}}
\and A.~Grazian\orcid{0000-0002-5688-0663}\inst{\ref{aff35}}
\and F.~Grupp\inst{\ref{aff14},\ref{aff37}}
\and L.~Guzzo\orcid{0000-0001-8264-5192}\inst{\ref{aff65},\ref{aff31}}
\and S.~V.~H.~Haugan\orcid{0000-0001-9648-7260}\inst{\ref{aff66}}
\and J.~Hoar\inst{\ref{aff30}}
\and H.~Hoekstra\orcid{0000-0002-0641-3231}\inst{\ref{aff27}}
\and M.~S.~Holliman\inst{\ref{aff67}}
\and W.~Holmes\inst{\ref{aff68}}
\and I.~Hook\orcid{0000-0002-2960-978X}\inst{\ref{aff69}}
\and F.~Hormuth\inst{\ref{aff70}}
\and A.~Hornstrup\orcid{0000-0002-3363-0936}\inst{\ref{aff71},\ref{aff72}}
\and P.~Hudelot\inst{\ref{aff73}}
\and K.~Jahnke\orcid{0000-0003-3804-2137}\inst{\ref{aff8}}
\and E.~Keih\"anen\orcid{0000-0003-1804-7715}\inst{\ref{aff74}}
\and S.~Kermiche\orcid{0000-0002-0302-5735}\inst{\ref{aff62}}
\and A.~Kiessling\orcid{0000-0002-2590-1273}\inst{\ref{aff68}}
\and M.~Kilbinger\orcid{0000-0001-9513-7138}\inst{\ref{aff75}}
\and T.~Kitching\orcid{0000-0002-4061-4598}\inst{\ref{aff32}}
\and R.~Kohley\inst{\ref{aff30}}
\and B.~Kubik\orcid{0009-0006-5823-4880}\inst{\ref{aff53}}
\and K.~Kuijken\orcid{0000-0002-3827-0175}\inst{\ref{aff27}}
\and M.~K\"ummel\orcid{0000-0003-2791-2117}\inst{\ref{aff37}}
\and M.~Kunz\orcid{0000-0002-3052-7394}\inst{\ref{aff76}}
\and H.~Kurki-Suonio\orcid{0000-0002-4618-3063}\inst{\ref{aff77},\ref{aff78}}
\and O.~Lahav\orcid{0000-0002-1134-9035}\inst{\ref{aff79}}
\and D.~Le~Mignant\orcid{0000-0002-5339-5515}\inst{\ref{aff80}}
\and P.~B.~Lilje\orcid{0000-0003-4324-7794}\inst{\ref{aff66}}
\and V.~Lindholm\orcid{0000-0003-2317-5471}\inst{\ref{aff77},\ref{aff78}}
\and I.~Lloro\inst{\ref{aff81}}
\and E.~Maiorano\orcid{0000-0003-2593-4355}\inst{\ref{aff2}}
\and O.~Mansutti\orcid{0000-0001-5758-4658}\inst{\ref{aff16}}
\and O.~Marggraf\orcid{0000-0001-7242-3852}\inst{\ref{aff82}}
\and K.~Markovic\orcid{0000-0001-6764-073X}\inst{\ref{aff68}}
\and N.~Martinet\orcid{0000-0003-2786-7790}\inst{\ref{aff80}}
\and F.~Marulli\orcid{0000-0002-8850-0303}\inst{\ref{aff83},\ref{aff2},\ref{aff34}}
\and R.~Massey\orcid{0000-0002-6085-3780}\inst{\ref{aff84}}
\and S.~Maurogordato\inst{\ref{aff85}}
\and H.~J.~McCracken\orcid{0000-0002-9489-7765}\inst{\ref{aff73}}
\and E.~Medinaceli\orcid{0000-0002-4040-7783}\inst{\ref{aff2}}
\and S.~Mei\orcid{0000-0002-2849-559X}\inst{\ref{aff86}}
\and Y.~Mellier\inst{\ref{aff87},\ref{aff73}}
\and M.~Meneghetti\orcid{0000-0003-1225-7084}\inst{\ref{aff2},\ref{aff34}}
\and E.~Merlin\orcid{0000-0001-6870-8900}\inst{\ref{aff18}}
\and G.~Meylan\inst{\ref{aff5}}
\and M.~Moresco\orcid{0000-0002-7616-7136}\inst{\ref{aff83},\ref{aff2}}
\and L.~Moscardini\orcid{0000-0002-3473-6716}\inst{\ref{aff83},\ref{aff2},\ref{aff34}}
\and E.~Munari\orcid{0000-0002-1751-5946}\inst{\ref{aff16},\ref{aff88}}
\and R.~Nakajima\inst{\ref{aff82}}
\and R.~C.~Nichol\inst{\ref{aff17}}
\and S.-M.~Niemi\inst{\ref{aff15}}
\and J.~W.~Nightingale\orcid{0000-0002-8987-7401}\inst{\ref{aff89},\ref{aff90}}
\and C.~Padilla\orcid{0000-0001-7951-0166}\inst{\ref{aff91}}
\and S.~Paltani\orcid{0000-0002-8108-9179}\inst{\ref{aff57}}
\and F.~Pasian\orcid{0000-0002-4869-3227}\inst{\ref{aff16}}
\and K.~Pedersen\inst{\ref{aff92}}
\and W.~J.~Percival\orcid{0000-0002-0644-5727}\inst{\ref{aff93},\ref{aff94},\ref{aff95}}
\and V.~Pettorino\inst{\ref{aff15}}
\and S.~Pires\orcid{0000-0002-0249-2104}\inst{\ref{aff3}}
\and G.~Polenta\orcid{0000-0003-4067-9196}\inst{\ref{aff96}}
\and M.~Poncet\inst{\ref{aff36}}
\and L.~A.~Popa\inst{\ref{aff97}}
\and L.~Pozzetti\orcid{0000-0001-7085-0412}\inst{\ref{aff2}}
\and G.~D.~Racca\inst{\ref{aff15}}
\and F.~Raison\orcid{0000-0002-7819-6918}\inst{\ref{aff14}}
\and R.~Rebolo\inst{\ref{aff25},\ref{aff26}}
\and A.~Refregier\inst{\ref{aff98}}
\and A.~Renzi\orcid{0000-0001-9856-1970}\inst{\ref{aff99},\ref{aff59}}
\and J.~Rhodes\inst{\ref{aff68}}
\and G.~Riccio\inst{\ref{aff10}}
\and E.~Romelli\orcid{0000-0003-3069-9222}\inst{\ref{aff16}}
\and M.~Roncarelli\orcid{0000-0001-9587-7822}\inst{\ref{aff2}}
\and E.~Rossetti\orcid{0000-0003-0238-4047}\inst{\ref{aff33}}
\and R.~Saglia\orcid{0000-0003-0378-7032}\inst{\ref{aff37},\ref{aff14}}
\and D.~Sapone\orcid{0000-0001-7089-4503}\inst{\ref{aff100}}
\and B.~Sartoris\orcid{0000-0003-1337-5269}\inst{\ref{aff37},\ref{aff16}}
\and P.~Schneider\orcid{0000-0001-8561-2679}\inst{\ref{aff82}}
\and T.~Schrabback\orcid{0000-0002-6987-7834}\inst{\ref{aff7}}
\and M.~Scodeggio\inst{\ref{aff47}}
\and A.~Secroun\orcid{0000-0003-0505-3710}\inst{\ref{aff62}}
\and G.~Seidel\orcid{0000-0003-2907-353X}\inst{\ref{aff8}}
\and S.~Serrano\orcid{0000-0002-0211-2861}\inst{\ref{aff60},\ref{aff101},\ref{aff102}}
\and C.~Sirignano\orcid{0000-0002-0995-7146}\inst{\ref{aff99},\ref{aff59}}
\and G.~Sirri\orcid{0000-0003-2626-2853}\inst{\ref{aff34}}
\and J.~Skottfelt\orcid{0000-0003-1310-8283}\inst{\ref{aff103}}
\and L.~Stanco\orcid{0000-0002-9706-5104}\inst{\ref{aff59}}
\and P.~Tallada-Cresp\'{i}\orcid{0000-0002-1336-8328}\inst{\ref{aff48},\ref{aff49}}
\and D.~Tavagnacco\orcid{0000-0001-7475-9894}\inst{\ref{aff16}}
\and A.~N.~Taylor\inst{\ref{aff4}}
\and H.~I.~Teplitz\orcid{0000-0002-7064-5424}\inst{\ref{aff104}}
\and I.~Tereno\inst{\ref{aff55},\ref{aff105}}
\and R.~Toledo-Moreo\orcid{0000-0002-2997-4859}\inst{\ref{aff106}}
\and F.~Torradeflot\orcid{0000-0003-1160-1517}\inst{\ref{aff49},\ref{aff48}}
\and I.~Tutusaus\orcid{0000-0002-3199-0399}\inst{\ref{aff107}}
\and E.~A.~Valentijn\inst{\ref{aff12}}
\and L.~Valenziano\orcid{0000-0002-1170-0104}\inst{\ref{aff2},\ref{aff108}}
\and T.~Vassallo\orcid{0000-0001-6512-6358}\inst{\ref{aff37},\ref{aff16}}
\and G.~Verdoes~Kleijn\orcid{0000-0001-5803-2580}\inst{\ref{aff12}}
\and A.~Veropalumbo\orcid{0000-0003-2387-1194}\inst{\ref{aff31},\ref{aff40},\ref{aff109}}
\and Y.~Wang\orcid{0000-0002-4749-2984}\inst{\ref{aff104}}
\and J.~Weller\orcid{0000-0002-8282-2010}\inst{\ref{aff37},\ref{aff14}}
\and O.~R.~Williams\orcid{0000-0003-0274-1526}\inst{\ref{aff110}}
\and G.~Zamorani\orcid{0000-0002-2318-301X}\inst{\ref{aff2}}
\and E.~Zucca\orcid{0000-0002-5845-8132}\inst{\ref{aff2}}
\and C.~Burigana\orcid{0000-0002-3005-5796}\inst{\ref{aff111},\ref{aff108}}
\and G.~De~Lucia\orcid{0000-0002-6220-9104}\inst{\ref{aff16}}
\and K.~George\orcid{0000-0002-1734-8455}\inst{\ref{aff37}}
\and V.~Scottez\inst{\ref{aff87},\ref{aff112}}
\and M.~Miluzio\inst{\ref{aff30}}
\and P.~Simon\inst{\ref{aff82}}
\and A.~Mora\orcid{0000-0002-1922-8529}\inst{\ref{aff113}}
\and J.~Mart\'{i}n-Fleitas\orcid{0000-0002-8594-569X}\inst{\ref{aff113}}
\and D.~Scott\orcid{0000-0002-6878-9840}\inst{\ref{aff114}}}
										   
\institute{INAF-Osservatorio Astrofisico di Arcetri, Largo E. Fermi 5, 50125, Firenze, Italy\label{aff1}
\and
INAF-Osservatorio di Astrofisica e Scienza dello Spazio di Bologna, Via Piero Gobetti 93/3, 40129 Bologna, Italy\label{aff2}
\and
Universit\'e Paris-Saclay, Universit\'e Paris Cit\'e, CEA, CNRS, AIM, 91191, Gif-sur-Yvette, France\label{aff3}
\and
Institute for Astronomy, University of Edinburgh, Royal Observatory, Blackford Hill, Edinburgh EH9 3HJ, UK\label{aff4}
\and
Institute of Physics, Laboratory of Astrophysics, Ecole Polytechnique F\'ed\'erale de Lausanne (EPFL), Observatoire de Sauverny, 1290 Versoix, Switzerland\label{aff5}
\and
Department of Astrophysics/IMAPP, Radboud University, PO Box 9010, 6500 GL Nijmegen, The Netherlands\label{aff6}
\and
Universit\"at Innsbruck, Institut f\"ur Astro- und Teilchenphysik, Technikerstr. 25/8, 6020 Innsbruck, Austria\label{aff7}
\and
Max-Planck-Institut f\"ur Astronomie, K\"onigstuhl 17, 69117 Heidelberg, Germany\label{aff8}
\and
Department of Physics, Universit\'{e} de Montr\'{e}al, 2900 Edouard Montpetit Blvd, Montr\'{e}al, Qu\'{e}bec H3T 1J4, Canada\label{aff9}
\and
INAF-Osservatorio Astronomico di Capodimonte, Via Moiariello 16, 80131 Napoli, Italy\label{aff10}
\and
Observatoire Astronomique de Strasbourg (ObAS), Universit\'e de Strasbourg - CNRS, UMR 7550, Strasbourg, France\label{aff11}
\and
Kapteyn Astronomical Institute, University of Groningen, PO Box 800, 9700 AV Groningen, The Netherlands\label{aff12}
\and
NRC Herzberg, 5071 West Saanich Rd, Victoria, BC V9E 2E7, Canada\label{aff13}
\and
Max Planck Institute for Extraterrestrial Physics, Giessenbachstr. 1, 85748 Garching, Germany\label{aff14}
\and
European Space Agency/ESTEC, Keplerlaan 1, 2201 AZ Noordwijk, The Netherlands\label{aff15}
\and
INAF-Osservatorio Astronomico di Trieste, Via G. B. Tiepolo 11, 34143 Trieste, Italy\label{aff16}
\and
School of Mathematics and Physics, University of Surrey, Guildford, Surrey, GU2 7XH, UK\label{aff17}
\and
INAF-Osservatorio Astronomico di Roma, Via Frascati 33, 00078 Monteporzio Catone, Italy\label{aff18}
\and
Observatorio Nacional, Rua General Jose Cristino, 77-Bairro Imperial de Sao Cristovao, Rio de Janeiro, 20921-400, Brazil\label{aff19}
\and
Universit\'e de Strasbourg, CNRS, Observatoire astronomique de Strasbourg, UMR 7550, 67000 Strasbourg, France\label{aff20}
\and
School of Physics \& Astronomy, University of Southampton, Highfield Campus, Southampton SO17 1BJ, UK\label{aff21}
\and
INFN-Sezione di Roma, Piazzale Aldo Moro, 2 - c/o Dipartimento di Fisica, Edificio G. Marconi, 00185 Roma, Italy\label{aff22}
\and
Institute of Astronomy, University of Cambridge, Madingley Road, Cambridge CB3 0HA, UK\label{aff23}
\and
Jodrell Bank Centre for Astrophysics, Department of Physics and Astronomy, University of Manchester, Oxford Road, Manchester M13 9PL, UK\label{aff24}
\and
Instituto de Astrof\'isica de Canarias, Calle V\'ia L\'actea s/n, 38204, San Crist\'obal de La Laguna, Tenerife, Spain\label{aff25}
\and
Departamento de Astrof\'isica, Universidad de La Laguna, 38206, La Laguna, Tenerife, Spain\label{aff26}
\and
Leiden Observatory, Leiden University, Einsteinweg 55, 2333 CC Leiden, The Netherlands\label{aff27}
\and
Departamento de F{\'\i}sica de la Tierra y Astrof{\'\i}sica, Universidad Complutense de Madrid, Plaza de las Ciencias 2, E-28040 Madrid, Spain\label{aff28}
\and
Universit\'e Paris-Saclay, CNRS, Institut d'astrophysique spatiale, 91405, Orsay, France\label{aff29}
\and
ESAC/ESA, Camino Bajo del Castillo, s/n., Urb. Villafranca del Castillo, 28692 Villanueva de la Ca\~nada, Madrid, Spain\label{aff30}
\and
INAF-Osservatorio Astronomico di Brera, Via Brera 28, 20122 Milano, Italy\label{aff31}
\and
Mullard Space Science Laboratory, University College London, Holmbury St Mary, Dorking, Surrey RH5 6NT, UK\label{aff32}
\and
Dipartimento di Fisica e Astronomia, Universit\`a di Bologna, Via Gobetti 93/2, 40129 Bologna, Italy\label{aff33}
\and
INFN-Sezione di Bologna, Viale Berti Pichat 6/2, 40127 Bologna, Italy\label{aff34}
\and
INAF-Osservatorio Astronomico di Padova, Via dell'Osservatorio 5, 35122 Padova, Italy\label{aff35}
\and
Centre National d'Etudes Spatiales -- Centre spatial de Toulouse, 18 avenue Edouard Belin, 31401 Toulouse Cedex 9, France\label{aff36}
\and
Universit\"ats-Sternwarte M\"unchen, Fakult\"at f\"ur Physik, Ludwig-Maximilians-Universit\"at M\"unchen, Scheinerstrasse 1, 81679 M\"unchen, Germany\label{aff37}
\and
INAF-Osservatorio Astrofisico di Torino, Via Osservatorio 20, 10025 Pino Torinese (TO), Italy\label{aff38}
\and
Dipartimento di Fisica, Universit\`a di Genova, Via Dodecaneso 33, 16146, Genova, Italy\label{aff39}
\and
INFN-Sezione di Genova, Via Dodecaneso 33, 16146, Genova, Italy\label{aff40}
\and
Department of Physics "E. Pancini", University Federico II, Via Cinthia 6, 80126, Napoli, Italy\label{aff41}
\and
INFN section of Naples, Via Cinthia 6, 80126, Napoli, Italy\label{aff42}
\and
Instituto de Astrof\'isica e Ci\^encias do Espa\c{c}o, Universidade do Porto, CAUP, Rua das Estrelas, PT4150-762 Porto, Portugal\label{aff43}
\and
Faculdade de Ci\^encias da Universidade do Porto, Rua do Campo de Alegre, 4150-007 Porto, Portugal\label{aff44}
\and
Dipartimento di Fisica, Universit\`a degli Studi di Torino, Via P. Giuria 1, 10125 Torino, Italy\label{aff45}
\and
INFN-Sezione di Torino, Via P. Giuria 1, 10125 Torino, Italy\label{aff46}
\and
INAF-IASF Milano, Via Alfonso Corti 12, 20133 Milano, Italy\label{aff47}
\and
Centro de Investigaciones Energ\'eticas, Medioambientales y Tecnol\'ogicas (CIEMAT), Avenida Complutense 40, 28040 Madrid, Spain\label{aff48}
\and
Port d'Informaci\'{o} Cient\'{i}fica, Campus UAB, C. Albareda s/n, 08193 Bellaterra (Barcelona), Spain\label{aff49}
\and
Institute for Theoretical Particle Physics and Cosmology (TTK), RWTH Aachen University, 52056 Aachen, Germany\label{aff50}
\and
Dipartimento di Fisica e Astronomia "Augusto Righi" - Alma Mater Studiorum Universit\`a di Bologna, Viale Berti Pichat 6/2, 40127 Bologna, Italy\label{aff51}
\and
European Space Agency/ESRIN, Largo Galileo Galilei 1, 00044 Frascati, Roma, Italy\label{aff52}
\and
Universit\'e Claude Bernard Lyon 1, CNRS/IN2P3, IP2I Lyon, UMR 5822, Villeurbanne, F-69100, France\label{aff53}
\and
UCB Lyon 1, CNRS/IN2P3, IUF, IP2I Lyon, 4 rue Enrico Fermi, 69622 Villeurbanne, France\label{aff54}
\and
Departamento de F\'isica, Faculdade de Ci\^encias, Universidade de Lisboa, Edif\'icio C8, Campo Grande, PT1749-016 Lisboa, Portugal\label{aff55}
\and
Instituto de Astrof\'isica e Ci\^encias do Espa\c{c}o, Faculdade de Ci\^encias, Universidade de Lisboa, Campo Grande, 1749-016 Lisboa, Portugal\label{aff56}
\and
Department of Astronomy, University of Geneva, ch. d'Ecogia 16, 1290 Versoix, Switzerland\label{aff57}
\and
INAF-Istituto di Astrofisica e Planetologia Spaziali, via del Fosso del Cavaliere, 100, 00100 Roma, Italy\label{aff58}
\and
INFN-Padova, Via Marzolo 8, 35131 Padova, Italy\label{aff59}
\and
Institut d'Estudis Espacials de Catalunya (IEEC),  Edifici RDIT, Campus UPC, 08860 Castelldefels, Barcelona, Spain\label{aff60}
\and
Institut de Ciencies de l'Espai (IEEC-CSIC), Campus UAB, Carrer de Can Magrans, s/n Cerdanyola del Vall\'es, 08193 Barcelona, Spain\label{aff61}
\and
Aix-Marseille Universit\'e, CNRS/IN2P3, CPPM, Marseille, France\label{aff62}
\and
Istituto Nazionale di Fisica Nucleare, Sezione di Bologna, Via Irnerio 46, 40126 Bologna, Italy\label{aff63}
\and
FRACTAL S.L.N.E., calle Tulip\'an 2, Portal 13 1A, 28231, Las Rozas de Madrid, Spain\label{aff64}
\and
Dipartimento di Fisica "Aldo Pontremoli", Universit\`a degli Studi di Milano, Via Celoria 16, 20133 Milano, Italy\label{aff65}
\and
Institute of Theoretical Astrophysics, University of Oslo, P.O. Box 1029 Blindern, 0315 Oslo, Norway\label{aff66}
\and
Higgs Centre for Theoretical Physics, School of Physics and Astronomy, The University of Edinburgh, Edinburgh EH9 3FD, UK\label{aff67}
\and
Jet Propulsion Laboratory, California Institute of Technology, 4800 Oak Grove Drive, Pasadena, CA, 91109, USA\label{aff68}
\and
Department of Physics, Lancaster University, Lancaster, LA1 4YB, UK\label{aff69}
\and
Felix Hormuth Engineering, Goethestr. 17, 69181 Leimen, Germany\label{aff70}
\and
Technical University of Denmark, Elektrovej 327, 2800 Kgs. Lyngby, Denmark\label{aff71}
\and
Cosmic Dawn Center (DAWN), Denmark\label{aff72}
\and
Institut d'Astrophysique de Paris, UMR 7095, CNRS, and Sorbonne Universit\'e, 98 bis boulevard Arago, 75014 Paris, France\label{aff73}
\and
Department of Physics and Helsinki Institute of Physics, Gustaf H\"allstr\"omin katu 2, 00014 University of Helsinki, Finland\label{aff74}
\and
AIM, CEA, CNRS, Universit\'{e} Paris-Saclay, Universit\'{e} de Paris, 91191 Gif-sur-Yvette, France\label{aff75}
\and
Universit\'e de Gen\`eve, D\'epartement de Physique Th\'eorique and Centre for Astroparticle Physics, 24 quai Ernest-Ansermet, CH-1211 Gen\`eve 4, Switzerland\label{aff76}
\and
Department of Physics, P.O. Box 64, 00014 University of Helsinki, Finland\label{aff77}
\and
Helsinki Institute of Physics, Gustaf H{\"a}llstr{\"o}min katu 2, University of Helsinki, Helsinki, Finland\label{aff78}
\and
Department of Physics and Astronomy, University College London, Gower Street, London WC1E 6BT, UK\label{aff79}
\and
Aix-Marseille Universit\'e, CNRS, CNES, LAM, Marseille, France\label{aff80}
\and
NOVA optical infrared instrumentation group at ASTRON, Oude Hoogeveensedijk 4, 7991PD, Dwingeloo, The Netherlands\label{aff81}
\and
Universit\"at Bonn, Argelander-Institut f\"ur Astronomie, Auf dem H\"ugel 71, 53121 Bonn, Germany\label{aff82}
\and
Dipartimento di Fisica e Astronomia "Augusto Righi" - Alma Mater Studiorum Universit\`a di Bologna, via Piero Gobetti 93/2, 40129 Bologna, Italy\label{aff83}
\and
Department of Physics, Centre for Extragalactic Astronomy, Durham University, South Road, DH1 3LE, UK\label{aff84}
\and
Universit\'e C\^{o}te d'Azur, Observatoire de la C\^{o}te d'Azur, CNRS, Laboratoire Lagrange, Bd de l'Observatoire, CS 34229, 06304 Nice cedex 4, France\label{aff85}
\and
Universit\'e Paris Cit\'e, CNRS, Astroparticule et Cosmologie, 75013 Paris, France\label{aff86}
\and
Institut d'Astrophysique de Paris, 98bis Boulevard Arago, 75014, Paris, France\label{aff87}
\and
IFPU, Institute for Fundamental Physics of the Universe, via Beirut 2, 34151 Trieste, Italy\label{aff88}
\and
School of Mathematics, Statistics and Physics, Newcastle University, Herschel Building, Newcastle-upon-Tyne, NE1 7RU, UK\label{aff89}
\and
Department of Physics, Institute for Computational Cosmology, Durham University, South Road, DH1 3LE, UK\label{aff90}
\and
Institut de F\'{i}sica d'Altes Energies (IFAE), The Barcelona Institute of Science and Technology, Campus UAB, 08193 Bellaterra (Barcelona), Spain\label{aff91}
\and
Department of Physics and Astronomy, University of Aarhus, Ny Munkegade 120, DK-8000 Aarhus C, Denmark\label{aff92}
\and
Waterloo Centre for Astrophysics, University of Waterloo, Waterloo, Ontario N2L 3G1, Canada\label{aff93}
\and
Department of Physics and Astronomy, University of Waterloo, Waterloo, Ontario N2L 3G1, Canada\label{aff94}
\and
Perimeter Institute for Theoretical Physics, Waterloo, Ontario N2L 2Y5, Canada\label{aff95}
\and
Space Science Data Center, Italian Space Agency, via del Politecnico snc, 00133 Roma, Italy\label{aff96}
\and
Institute of Space Science, Str. Atomistilor, nr. 409 M\u{a}gurele, Ilfov, 077125, Romania\label{aff97}
\and
Institute for Particle Physics and Astrophysics, Dept. of Physics, ETH Zurich, Wolfgang-Pauli-Strasse 27, 8093 Zurich, Switzerland\label{aff98}
\and
Dipartimento di Fisica e Astronomia "G. Galilei", Universit\`a di Padova, Via Marzolo 8, 35131 Padova, Italy\label{aff99}
\and
Departamento de F\'isica, FCFM, Universidad de Chile, Blanco Encalada 2008, Santiago, Chile\label{aff100}
\and
Satlantis, University Science Park, Sede Bld 48940, Leioa-Bilbao, Spain\label{aff101}
\and
Institute of Space Sciences (ICE, CSIC), Campus UAB, Carrer de Can Magrans, s/n, 08193 Barcelona, Spain\label{aff102}
\and
Centre for Electronic Imaging, Open University, Walton Hall, Milton Keynes, MK7~6AA, UK\label{aff103}
\and
Infrared Processing and Analysis Center, California Institute of Technology, Pasadena, CA 91125, USA\label{aff104}
\and
Instituto de Astrof\'isica e Ci\^encias do Espa\c{c}o, Faculdade de Ci\^encias, Universidade de Lisboa, Tapada da Ajuda, 1349-018 Lisboa, Portugal\label{aff105}
\and
Universidad Polit\'ecnica de Cartagena, Departamento de Electr\'onica y Tecnolog\'ia de Computadoras,  Plaza del Hospital 1, 30202 Cartagena, Spain\label{aff106}
\and
Institut de Recherche en Astrophysique et Plan\'etologie (IRAP), Universit\'e de Toulouse, CNRS, UPS, CNES, 14 Av. Edouard Belin, 31400 Toulouse, France\label{aff107}
\and
INFN-Bologna, Via Irnerio 46, 40126 Bologna, Italy\label{aff108}
\and
Dipartimento di Fisica, Universit\`a degli studi di Genova, and INFN-Sezione di Genova, via Dodecaneso 33, 16146, Genova, Italy\label{aff109}
\and
Centre for Information Technology, University of Groningen, P.O. Box 11044, 9700 CA Groningen, The Netherlands\label{aff110}
\and
INAF, Istituto di Radioastronomia, Via Piero Gobetti 101, 40129 Bologna, Italy\label{aff111}
\and
Junia, EPA department, 41 Bd Vauban, 59800 Lille, France\label{aff112}
\and
Aurora Technology for European Space Agency (ESA), Camino bajo del Castillo, s/n, Urbanizacion Villafranca del Castillo, Villanueva de la Ca\~nada, 28692 Madrid, Spain\label{aff113}
\and
Department of Physics and Astronomy, University of British Columbia, Vancouver, BC V6T 1Z1, Canada\label{aff114}}    


%
%
\abstract{\Euclid is poised to make significant advances in the study of nearby galaxies in the Local Universe.
Here we present a first look at six galaxies observed for the Nearby Galaxy Showcase as part of the \Euclid Early Release Observations
acquired between August and November, 2023.
These targets, three dwarf galaxies (Holmberg\,II, IC\,10, and NGC\,6822) and three spirals (IC\,342, NGC\,2403, and NGC\,6744), 
range in distance from 
about 0.5\,Mpc to 8.8\,Mpc.
We first assess the surface brightness depths in the stacked \Euclid images, and confirm previous estimates 
in 100\,arcsec$^2$ regions for VIS of $1\sigma$ limits of 
$30.5$\,\magarc,
but find deeper than previous estimates for NISP with $1\sigma\,=\,29.2$--$29.4$\,\magarc. 
By combining \Euclid \HE, \YE, and \IE\ into RGB images, we illustrate the large field-of-view (FoV) covered by a single
Reference Observing Sequence, together with exquisite detail on scales of $<1$--$4$ parsecs in these nearby galaxies.
Analysis of radial surface brightness and color profiles demonstrates that the photometric calibration
of \Euclid is consistent with what is expected for galaxy colors according to stellar synthesis models.
We perform standard source selection techniques for stellar photometry, and find 
approximately 1.3 million stars
across the six galaxy fields.
After subtracting foreground stars and background galaxies, and applying a color and magnitude selection, 
we extract stellar populations of different ages for the six galaxies.
The resolved stellar photometry obtained with \Euclid allows us to constrain the star-formation histories of these galaxies,
by disentangling the distributions of young stars, as well as asymptotic giant branch and red giant branch stellar populations. 
We finally examine two galaxies individually for surrounding systems of dwarf galaxy satellites and globular cluster populations.
Our analysis of the ensemble of dwarf satellites around NGC\,6744 recovers all the previously known dwarf satellites
within the \Euclid FoV, and also reveals a new system, EDwC1, a nucleated dwarf spheroidal at the end of a spiral arm.
Our new census of the globular clusters around NGC\,2403 yields nine new star-cluster candidates, eight of which with colors
indicative of evolved stellar populations.
In summary, our 
first investigation of the six Showcase galaxies
demonstrates that \Euclid is a powerful probe of stellar structure and stellar populations in nearby galaxies, 
and will provide vastly improved statistics on dwarf satellite systems
and extragalactic globular clusters in the local Universe, among many other exciting results.
}
%
%
%
%
   \titlerunning{\Euclid\/: ERO -- Deep anatomy of nearby galaxies}
   \authorrunning{L.~K. Hunt et al.}
   
   \maketitle
%
%
%
%
   
\section{Introduction}
\label{sec:intro}

Under the currently favored cosmological-constant-dominated 
cold dark matter (\lcdm) paradigm of structure formation, galaxies form 
hierarchically, through the accretion of lower mass systems.
Mergers of equal mass galaxies are catastrophic events that are expected to destroy altogether the pre-existing
stellar disks. 
However, such events are relatively rare, with massive galaxies, on average, participating in only one such
event over the last 
$10$\,Gyr \citep[e.g.,][]{mundy17,conselice22}.
On the other hand, minor mergers of a massive galaxy and a low-mass satellite,
or even of two low-mass dwarf galaxies,
are 
more common and occur even in the current epoch \citep[e.g.,][]{mihos94,hammer05,martinez10,lelli14,conselice22}. 
In such mergers, the disk structure of the parent galaxy may be conserved, but the lower mass accreted galaxy is completely
disrupted, leaving behind many faint structures, such as shells, streams, and plumes, in the parent stellar halo
\citep[e.g.,][]{bullock05}.

Observational verifications of such a scenario have been found in the Local Group of galaxies, 
with minor merger events occurring in the outskirts of the Milky Way 
\citep[e.g.,][]{ibata01,majewski03,belokurov06,juric08,carollo16,helmi18,martin22b},
around Andromeda \citep[M31, e.g.,][]{ferguson02,ibata07,carlberg11,komiyama18},
and the Triangulum galaxy 
\citep[M33, e.g.,][]{ibata07,mcconnachie09,mcconnachie10}.
In addition to the dark matter halo and stars, 
the globular cluster (GC) populations of the accreted galaxy also tend to merge with the GC
populations of the more massive parent \citep[e.g.,][]{forbes10,mackey19}.

The problem with observational confirmation of this `smoking gun' 
of hierarchical \lcdm\ galaxy formation
is that the tidal remnants are extremely faint with very low surface brightness 
\citep[LSB, $\mu_{R}\ga 27-28$\,\magarc,][]{johnston01,martinez08,martinez09,martin22a}.
In galaxies well beyond the Local Group ($D\ga 5\,{\rm Mpc}$), individual stars cannot be easily resolved so that
contrast enhancement techniques are used, and the intrinsic spatial resolution is degraded to obtain fainter
limits \citep[e.g.,][]{martinez10,trujillo16,merritt16,mihos19,martinez23,roman23b}.

The study of LSB emission in integrated light and resolved stars in nearby galaxies requires both high spatial resolution
(not achievable from the ground), and a wide field of view (FoV).
The first criterion is met by \HST (\hst) and by the James Webb Space Telecope (\jwst); \hst\ has revolutionized our 
understanding of star-formation histories (SFHs) through color-magnitude diagrams \citep[CMDs, e.g.,][]{mcquinn10,weisz11,cignoni19,annibali22}.
However, the FoV of \hst\ is limited to a few arcminutes, making it time consuming to perform large-scale photometric surveys
over entire nearby galaxy disks.

This limitation is now overcome by \Euclid, recently launched, commissioned, and currently taking data.
\Euclid will provide a new window on the stellar populations and LSB emission in nearby galaxies through its 
wide FoV of 
$0.67$\,deg$^2$ \citep[][]{Scaramella-EP1,EuclidSkyOverview}, using the
VIS camera with a broad visible filter \IE\ \citep{EuclidSkyVIS},
and NISP, \Euclid's near-infared (NIR) camera/spectrometer, endowed with three 
photometric filters \YE, \JE, and \HE\ \citep{Schirmer-EP18,EuclidSkyNISP}.
Detecting LSB emission also requires highly stable optics that minimize stray light, together with a 
well-defined point-spread function (PSF).
\Euclid's superb optics are designed to be thermally stable within a specific satellite orientation
\citep{laureijs11,EuclidSkyOverview}.
The unprecedented sensitivity of \Euclid to LSB emission is illustrated by \citet{Borlaff-EP16} and
\citet{Scaramella-EP1} who
predicted that \Euclid will enable detection of LSB emission down to \IE\,=\,$29.1$--$29.5$\,\magarc\ ($3\sigma$, $100$\,arcsec$^{2}$)
in the Wide Survey, and 2 magnitudes deeper in the Deep Survey.
More recently, similar limits have been demonstrated with ERO data in \citet{EROData}.

Another avenue of improvement offered by \Euclid comprises statistics of LSB and ultra-diffuse dwarf galaxies
\citep[UDGs,][]{vandokkum15}, as well as their compact dwarf counterparts.
Dwarf galaxies are the most abundant galaxy population at any redshift, but tend to be missed by large-scale
surveys that are not sensitive to LSB emission.
\Euclid's wide FoV and multi-band coverage will enable a new census of dwarf galaxies, both as satellites around
more massive hosts and as isolated galaxies in the field
\citep[e.g.,][]{mihos15,munoz15,marleau21,roman21,venhola22}.

In addition to LSB studies, \Euclid will also revolutionize investigations of nearby galaxies along many other avenues.
One of these will be a vast improvement of the demographics of extragalactic globular clusters (EGCs).
GCs are relics dating back to the earliest epochs of star formation in galaxies \citep[e.g.,][]{kruijssen15}.
Colors and other properties of EGCs provide strong constraints on hierarchical galaxy formation
\citep[e.g.,][]{brodie06,forbes10,harris13,roman23a}, and have been extensively studied both from the ground
\citep[e.g.,][]{harris79,forbes96,blakeslee97,pota13,cantiello18a}
and from space \citep[e.g.,][]{larsen01,harris09,peng09,pancino17}.
Euclid Collaboration:
Voggel et al. 2024 (in prep.) have shown that known GCs in galaxies within 20\,Mpc can be spatially resolved with \Euclid VIS,
and the NISP filters will constrain the stellar populations within the GCs
\citep[see also][]{EROFornaxGCs}.
\Euclid's wide FoV combined with its superb spatial resolution enables drastically improved statistics of EGCs in and 
around nearby galaxies. 

\begin{table*}[t!]
   \caption[]{{Showcase galaxy properties}} 
\label{tab:sample}
\resizebox{\linewidth}{!}{
\begin{centering}
\begin{tabular}{lcccccrcrr}
\hline
\hline
\multicolumn{1}{c}{\rule{0pt}{3ex} Galaxy} &
\multicolumn{1}{c}{Rank} &
\multicolumn{1}{c}{Morphological} &
\multicolumn{1}{c}{Major} &
\multicolumn{1}{c}{Distance} & 
\multicolumn{1}{c}{Foreground} & 
\multicolumn{1}{c}{Galactic} & 
\multicolumn{1}{c}{12$+$} &
\multicolumn{1}{c}{$\logten$} &
\multicolumn{1}{c}{$\logten$}
\\
\multicolumn{1}{c}{name} &
\multicolumn{1}{c}{order} &
\multicolumn{1}{c}{type$^\mathrm{a}$} & 
\multicolumn{1}{c}{diameter} &
\multicolumn{1}{c}{(Mpc)$^\mathrm{b}$} & 
\multicolumn{1}{c}{extinction} & 
\multicolumn{1}{c}{latitude} & 
\multicolumn{1}{c}{$\logten$(O/H)$^\mathrm{d}$} & 
\multicolumn{1}{c}{(\mstar/\msun)$^\mathrm{e}$} &
\multicolumn{1}{c}{(sSFR/yr$^{-1}$)$^\mathrm{e}$}  
\\ 
& \multicolumn{1}{c}{WXSC} & 
& 
\multicolumn{1}{c}{(arcmin)$^\mathrm{a}$} & 
& 
\multicolumn{1}{c}{\av\ (mag)$^\mathrm{c}$ } & 
\multicolumn{1}{c}{(deg)$^\mathrm{a}$} & & & \\ 
\noalign{\vskip 1ex}
\hline
\\ 
Holmberg\,II (UGC\,04305) & -- & Im          &  7.9 (--)    & 3.32 & 0.087 & $32.69$~~~~~  & 7.89  & 8.29  & $-9.43$ \\
IC\,10                    & 51 & IBm         &  6.3 (7.33)  & 0.72 & 4.299 & $-3.33$~~~~~  & 8.14 & 8.64  & $-9.25$ \\
IC\,342                   & 11 & SAB(rs)cd   & 21.4 (12.98) & 3.45 & 1.530 & $10.58$~~~~~  & 8.83 & 10.31 & $-9.70$ \\
NGC\,2403                 & 18 & SAB(s)cd    & 21.9 (10.94) & 3.20 & 0.110 & $29.19$~~~~~  & 8.48  & 9.47  & $-9.61$ \\
NGC\,6744                 & 29 & SAB(r)bc    & 20.0 (9.14)  & 8.80 & 0.118 & $-26.15$~~~~~ & 8.88  & 10.66 & $-10.34$ \\
NGC\,6822                 & 25 & IB(s)m      & 15.5 (9.53)  & 0.51 & 0.646 & $-18.40$~~~~~ & 8.11  & 8.16  & $-9.97$ \\
\\
\hline
\end{tabular}
\end{centering}
}
\vspace{0.5\baselineskip}
{\small
\begin{hangingpar}{0.8em}
$^\mathrm{a}$~Morphological types, galaxy major axis diameters (the blue isophotal values), and Galactic latitudes
are taken from the NASA/IPAC Extragalactic Database (NED);\footnote{The NASA/IPAC Extragalactic Database (NED) is funded by the National Aeronautics and Space Administration 
and operated by the California Institute of Technology.} the major diameters in parentheses are from \citet{jarrett19}.
\end{hangingpar}
\par
\begin{hangingpar}{0.8em}
$^\mathrm{b}$~Distance determinations: 
Holmberg\,II TRGB \citep{sabbi18};
IC\,10 TRGB \citep{gerbrandt15}; 
IC\,342 TRGB \citep{wu14}; 
NGC\,2403 TRGB \citep{radburn11};
NGC\,6744 TRGB \citep{sabbi18};
and NGC\,6822 CMD \citep{Fusco2012}. 
\end{hangingpar}
\par
\begin{hangingpar}{0.8em}
$^\mathrm{c}$~Foreground \av\ extinction values are 
calculated as described in Sect. \ref{sec:extinction}, based on the \citet{schlafly11} determinations.
\end{hangingpar}
\par
\begin{hangingpar}{0.8em}
$^\mathrm{d}$~Metallicity determinations: 
Holmberg\,II, IC\,10, IC\,342, NGC\,2403, and NGC\,6744 \citep{pilyugin14};
and NGC\,6822 \citep{lee06}.
\end{hangingpar}
\par
\begin{hangingpar}{0.8em}
$^\mathrm{e}$~Taken from \citet{nersesian19}, except for 
NGC\,6744 from \citet{leroy21}. 
All estimates use a similar technique, namely SED fitting with \texttt{CIGALE} \citep{boquien19}.
Distance-dependent quantities (\mstar) have been reported using the distances given here. 
\end{hangingpar}
}
\par
\end{table*}

In this paper, we explore the potential of \Euclid for studies of nearby galaxies 
provided by the Early Release Observations \citep[ERO,][]{EROcite}\footnote{\url{https://doi.org/10.57780/esa-qmocze3}}
taken in the context
of the ``\Euclid ERO Nearby Galaxy Showcase'' (hereafter Showcase).
These observations were acquired during the performance-verification (PV) phase of \Euclid operations
over a period of three months from August to November, 2023.

The preliminary results we present here focus on individual nearby galaxies and illustrate what will be possible with \Euclid
over the span of the Euclid Wide (EWS) and Deep Surveys (EDS).
The current analysis is confined to select key science themes including VIS and NISP integrated light and depth measured from
the images, resolved star photometry, and case studies of dwarf galaxy satellites and EGC demographics around
individual galaxies in the Showcase.
Future papers will discuss other science avenues for nearby galaxies with \Euclid, including
semi-resolved pixel-based fitting of spectral energy distributions \citep[SEDs, e.g.,][]{abdurrouf22},
and estimating distances with surface brightness fluctuations \citep[SBFs, e.g.,][]{tonry01,mei05,mei07,blakeslee09,cantiello18b}.
Section \ref{sec:sample} presents the Showcase targets and their selection criteria,
while Sect. \ref{sec:datareduction} briefly describes the \Euclid 
data 
processing and photometric calibration adopted for the ERO effort,
together with an estimate of surface brightness depth.
We report results for integrated light properties in Sect.~\ref{sec:integrated} and
for resolved stellar photometry and star counts in Sect.~\ref{sec:resolvedstars}. 
Case studies for 
dwarf satellites around NGC\,6744 are presented in Sect.~\ref{sec:dwarfs} 
and for EGCs of NGC\,2403 in Sect.~\ref{sec:egcs}. 
We summarize results and give conclusions in Sect.~\ref{sec:conclusions}.

\section{The ERO Showcase galaxies\label{sec:sample}}

The galaxies for the Showcase were selected from the WISE Extended Source Catalog (WXSC) that contains the
100 largest galaxies in the WISE survey in terms of angular size \citep{jarrett19}.
We required that the extent of the galaxy be smaller than the \Euclid FoV, so that the galaxy could be
properly imaged with one Reference Observation Sequence (ROS) typical of the EWS.
The other main selection criterion was visibility during the PV phase, that ultimately turned out to be
extremely stringent, 
given the spacecraft's strong pointing constraints, 
driven by thermal stability considerations and straylight suppression.
An additional consideration was the available ancillary data for the targets, including image cubes
of atomic and molecular gas.

With these criteria, the distances of the selected Showcase galaxies range from 
0.5\,Mpc within the Local
Group (NGC\,6822, IC\,10), to 
8.8\,Mpc (NGC\,6744).
The closest distances enable the comparison of limiting surface brightness
derived from resolved stellar photometry \citep[e.g.,][]{dejong07,barker12}
with that derived from integrated light.
This is a powerful approach, able to probe deeper surface brightness levels than integrated light alone,
and one that has been hampered so far by the small FoVs of space-borne facilities.

The final observed Showcase sample is given in Table \ref{tab:sample}, where
column (2) reports the WXSC rank order with the largest galaxy having rank 1 and the smallest galaxy 
in the WXSC ranked 100.
The Showcase galaxies are those with the largest apparent size observable during the PV phase of \Euclid observations.
The exception to this selection is Holmberg\,II, which  
had been selected as a possible target for another ERO proposal that could not be executed.
There are three dwarf irregulars and three late-type spiral galaxies in the Showcase, 
with individual descriptions given below.

\begin{itemize}
\renewcommand\labelitemi{--}
\setlength\itemsep{0.05\baselineskip}
\item
\textbf{Holmberg\,II}, a Magellanic dwarf irregular galaxy, was discovered by \citet{holmberg50} in the outskirts of the
M\,81 group of galaxies \citep{karach02}.
Star-formation activity in Holmberg\,II has been relatively constant over the past 100--200\,Myr,
with a recent peak at 10--20\,Myr \citep[e.g.,][]{mcquinn10,cignoni18}.
\citet{hodge94} identified 82 \hii\ regions in this galaxy, 
which capture the effects of triggered star formation on local and large scales \citep{stewart00,egorov17}.
On larger scales, a star count analysis has shown that, unusually,
the young stellar populations in Holmberg\,II have a more extended distribution than its older stars \citep{bernard12}.
Holmberg\,II is a `poster child' of \hi\ holes, shells, and bubbles,
possibly driven by stellar feedback from supernovae \citep[SNe;][]{puche92},
or from feedback over longer timescales \citep[e.g.,][]{rhode99,weisz09}. 
Holmberg\,II also hosts an ultra-luminous X-ray source, Ho\,II\,ULX-1, positioned along a chain of \hii\ regions bordering
one of the \hi\ cavities \citep{zezas99}, and 
probably associated with a stellar mass black hole \citep{goad06,barra23}.
\item 
\textbf{IC\,10} is a dwarf irregular member of the Local Group, considered to be the closest example of a starburst
galaxy, and a likely member of the Andromeda subgroup \citep{vandenbergh99}.
Its location at low Galactic latitude behind $\ga 4$ magnitudes of visual extinction (see Table \ref{tab:sample})
makes it challenging to study at UV and optical wavelengths.
IC\,10 hosts numerous star clusters \citep[e.g.,][]{hunter01,lim15} and possibly the most massive known
stellar-mass black hole, associated with a highly variable ultra-luminous X-ray binary, IC\,10\,X--1 \citep{silverman08}.
IC\,10\,X--1 may be powering a large non-thermal superbubble, 
possibly also associated with an \hi\ cavity \citep{heesen15,heesen18}.
The galaxy is embedded within a huge \hi\ envelope 
that shows signs of interaction and possibly late merger with another dwarf galaxy
\citep[e.g.,][]{wilcots98,nidever13,ashley14,namumba19}.
\item 
\textbf{IC\,342} is the dominant member of the IC\,342/Maffei group, one of the galaxy groups closest to the Milky Way
\citep{buta99}. 
It is a large spiral galaxy, close to face on, one of the apparently largest galaxies in the northern sky.
Historically, IC\,342 has been known as the `hidden galaxy' because of its low Galactic latitude, and like
IC\,10, suffers from a significant amount of foreground extinction (see Table \ref{tab:sample}).
The stellar populations in IC\,342 have not been extensively studied, because of its large size on the sky and 
its position behind the Milky Way disk.
Nevertheless, it is known to harbor a luminous nuclear star cluster \citep{boker99}, 
and \hst\ NIR imaging of the stellar populations in the galaxy's outskirts 
has allowed a distance determination \citep{wu14}.
Like Holmberg\,II and IC\,10, IC\,342 also hosts an ultra-luminous high-mass X-ray binary, IC\,342\,X--1,
considered to be a 
roughly $100$\,\msun\ black hole
\citep[e.g.,][]{cseh12,das21}, possibly coincident with a supernova remnant
\citep{roberts03}.
\item 
\textbf{NGC\,2403} is a late-type spiral, 
without a measurable bulge,
morphologically very similar to M33 and NGC\,300 \citep{williams13}.
Like Holmberg\,II,
it lies on the outskirts of the M81 group of galaxies \citep{karach02}.
The exponential disk of NGC\,2403 is extremely extended, out to 
18\,kpc,
with an additional stellar structural component reaching even larger distances \citep[$\la$\,40\,kpc,][]{barker12}. 
Recent deep Hyper Suprime-Cam imaging with Subaru reveals stellar streams in the direction of NGC\,2403 
emanating from a candidate dwarf satellite DDO\,40 \citep{carlin19}.
There is evidence of extraplanar \hi\ gas in NGC\,2403 \citep{fraternali02,walter08,deblok14}
that has been attributed to gas accretion caused by galactic fountains from stellar feedback
\citep{fraternali08,li23}, 
or an interaction with the nearby dwarf galaxy DDO\,40 \citep{veronese23}.
Despite its relatively low stellar mass, NGC\,2403 harbors a significant number of EGCs
with a wide range of ages \citep{forbes22}.
\item 
\textbf{NGC\,6744} is one of the physically largest spirals beyond the Local Group,
and the largest angular-extent barred ringed spiral in the southern sky \citep{devaucouleurs63}. 
An extensive multi-frequency study by \citet{yew18} 
found several point sources detected in both X-rays and radio, likely supernovae remnants,
and a luminous nuclear X-ray source thought to be associated with a super-massive black hole.
This central source is optically characterized as a very low-luminosity
active galactic nucleus \citep{dasilva18}. 
In 2005, a Type Ic SN exploded in the disk of NGC\,6744
\citep{kankare14}, adding evidence for a past star formation episode.
\hi\ observations show that the bulk of the atomic gas has a ring-like morphology,
associated with the spiral arms and the dwarf companion NGC\,6744\,A \citep{ryder99}.
NGC\,6744 may also possibly host a dwarf spheroidal satellite \citep{bedin19},
and several other LSB (candidate) dwarf satellites \citep{karach20}. 
\item 
\textbf{NGC\,6822} was first identified in 1925 by Hubble as a `very faint cluster of stars and nebulae'
well beyond the Milky Way \citep{hubble25}.
At 510\,kpc distance, NGC\,6822 is the closest galaxy in the Showcase sample, and its 
stellar populations have been heavily studied \citep[e.g.,][]{tantalo22}.
There are at least two distinct kinematic components seen in the \hi\
and stars of NGC\,6822 \citep[e.g.,][]{demers06}, although it may resemble
dynamically a late-type galaxy rather than a `polar ring' \citep{thompson16}.
NGC\,6822 shows a large \hi\ cavity, `supergiant shell' \citep{deblok00}, though with
fewer \hi\ features than Holmberg\,II.
Stellar age gradients around the \hii\ cavity point to a stellar feedback origin,
not necessarily related to star clusters \citep{deblok06}.
A recent panoramic view of NGC\,6822 in $g+i$ filters
shows no stellar overdensities in its outskirts, ruling out any recent interaction with a companion galaxy
\citep[][]{zhang21,mcconnachie21},
although it may have passed through the virial radius of the Milky Way about 3--4\,Gyr ago
\citep{teyssier12,zhang21}.
There are currently eight known GCs in NGC\,6822 \citep{huxor13,larsen18},
spread out over an extended region up to a projected radius of 
11\,kpc \citep{veljanoski15}. 
\end{itemize}

\section{\Euclid data reduction, photometric calibration, and surface brightness depth}
\label{sec:datareduction}


The ERO observations of the Showcase galaxies were obtained during \Euclid's PV phase, with the last
object Holmberg\,II obtained at the end of November, 2023.
\Euclid broadband coverage includes the VIS band \IE, and the 
three bands of NISP, \YE, \JE, and \HE.
With the exception of IC\,10, the Showcase galaxies were observed with one standard ROS, similar to the EWS 
\citep{Scaramella-EP1}, with four dithered images per band for a total
exposure time of roughly 1 hour, 
consisting of four repetitions of 560\,s for VIS and 87\,s for each NISP band.
For IC\,10, two ROS were acquired for a total of eight, rather than four,
exposures per band. 
The ROS exposures are dithered to mitigate cosmic rays and detector defects.
The NISP detector gaps are somewhat larger than those of VIS, 
and the photometric depth varies because of the interchip gaps.
More details of the payload and the instrumentation are given in \citet{EuclidSkyOverview}. 

The ERO data were not reduced with the standard Science Ground Segment pipeline,
but rather using a set of procedures optimized for LSB emission, developed ad hoc for the ERO program 
as described in \citet{EROData}.
The reduction starts with the calibrated Level 1 raw frames provided by the VIS and NISP 
processors \citep[e.g.,][]{EuclidSkyVIS,EuclidSkyNISP}.
Subsequent image processing considers:
(1) elimination of cosmic rays;
(2) astrometric distortion across the wide FoV;
(3) variation of the PSF full-width half maximum (FWHM) as a function of field position;
(4) 
modeling and subtraction of persistence effects that result from the preceding spectroscopic exposure imprinting remnant signal on the 
subsequent photometric exposures;
(5) developing a `super flat field' 
including the illumination pattern and low-level flux non-linearity.
Details of how these effects are treated are given in \citet{EROData}.

The pixel sizes for the VIS and NIR images are 0\farcs1 and 0\farcs3, respectively,
implying that for both instruments the PSF is slightly undersampled.
The final ERO stacked frames have a median PSF FWHM of 0\farcs16, 0\farcs47, 
0\farcs47, and  0\farcs49 (1.57, 1.57, 1.58, 1.65\,pixels) in \IE, \YE, \JE, and \HE, respectively 
\citep{EROData}.
Because of the rudimentary set of calibration data used by the ERO pipeline,
it was not possible to stringently constrain uncertainties, so that
the photometric calibration uncertainties were simply required to be $\la$\,10\%.
The ERO data were arbitrarily rescaled to have a nominal zero point 
of ZP\,=30\,AB mag;
this satisfies the uncertainty requirement for \YE, \JE, and \HE,
but subsequent checks against \textit{Gaia} showed that ZP\,=\,30.13\,AB mag
is a better estimate for \IE.
More details are provided by \citet{EROData}.

\subsection{Sky level and noise estimation}
\label{sec:sblim}

As described in Sect. \ref{sec:intro}, one of \Euclid's most important advantages is its sensitivity to LSB emission.
Following the metric used in previous studies
\citep[e.g.,][]{merritt16,trujillo16,borlaff19,roman20,Borlaff-EP16,Scaramella-EP1},
we quantify this sensitivity (image depth) $\sigma$ by considering sky surface brightness variations over areas of 100\,arcsec$^{2}$ 
in empty regions of the images with only sky emission. 
We have adopted the common scaling \citep[see, e.g.,][]{akhlaghi19a,roman20} for converting $\sigma$ 
(in units of counts or ADU per pixel) to a limiting surface brightness \mulim\ (AB \magarc) within a region of area $b^2$:
\begin{equation}
\mu_\mathrm{lim}\,=\,\mathrm{ZP} - 2.5\,\log_{10}(n\,\sigma) + 2.5\,\log_{10}(b p)\, \quad \mathrm{mag\,arcsec}^{-2}\ ,
\label{eqn:sblim}
\end{equation}
where $n$ is the signal-to-noise of the detection,
$b$ is the square root of the area of the region in arcsec, and $p$ is the pixel scale of the image (arcsec\,pixel$^{-1}$).
This scaling can be understood in several ways, in particular by considering that 
uncorrelated noise measured by $\sigma$ adds in quadrature within a region of area $b^2$,
and that within a 100\,arcsec$^2$ region there are $(b/p)^2$ pixels (see Appendix \ref{app:sky}).\footnote{See also 
\url{https://www.gnu.org/software/gnuastro/manual/html_node/Surface-brightness-limit-of-image.html}.}

We adopted three approaches to estimate $\sigma$:
(1) \texttt{gnuastro/noisechisel} \citep{akhlaghi15,akhlaghi19a,akhlaghi19b};
(2) Gaussian fitting on the sky-only masked image following the 
scheme of \citet{roman20}, with the mask provided
by \texttt{noisechisel} in the previous step; and
(3) \texttt{AutoProf} \citep{stone21}.
Details of these calculations are given in Appendix \ref{app:sky}. 
 
A caveat of 
our calculations is that the scaling to convert $\sigma$ to a limiting surface brightness \mulim\
assumes that the noise is uncorrelated, and that the noise per pixel ($\sigma$) can be accurately
scaled to a limiting $\mu$ for an arbitrary region size.
In any stacked mosaic, the noise is correlated because of rebinning, so our estimates assuming
Eq. \eqref{eqn:sblim} are lower (fainter) than the true SB limits.
We have assessed this effect in some detail, as described in Appendix \ref{app:sky},
and estimate that it would make our SB limits over 100\,arcsec$^2$ regions brighter at most by $\la$\,0.15\,mag in VIS,
and $\la$\,0.3\,mag in NISP.

The results given in Appendix \ref{app:sky} show that \Euclid's sensitivity to LSB emission on 100\,arcsec$^2$ scales is superb,
with $1\sigma$ limits $\ga\,30.5$\,AB \magarc\ in \IE, and slightly brighter, $29.2$--$29.4$\,AB \magarc, in \YE, \JE, and \HE.
Our measured LSB performance of \Euclid for VIS 
is roughly consistent with the predictions of \citet{Borlaff-EP16} and \citet{Scaramella-EP1},
but nominally $\sim0.5$ mag better (fainter) than the NISP estimates given in \citet{Scaramella-EP1}.
This comparison takes into account (see Table \ref{tab:sky}) the 
asinh scaling used by \citet[][equivalent to $-0.5$\,mag]{Scaramella-EP1};
however, it is possible that their background models of zodiacal light for the NIR emission were overly pessimistic.
Our SB limits are also 
consistent with those given in \citet{EROData}, once
the additional factors applied there to the noise measurements are taken into account:
the asinh factor ($-0.52$\,mag); and the scaling factors that consider the \texttt{SWarp} stacking, 1.32 for VIS ($-0.30$\,mag),
and 1.69 for NISP ($-0.57$\,mag).
These scaling factors for stacking are somewhat larger than what we inferred for the rebinning correction as discussed above
(see Appendix \ref{app:sky}).
Converting these $1\sigma$ limits to $3\sigma$ would reduce them by 1.19 mag. 
Such limits are particularly striking, given the relatively short exposure time of 
less than 1\,h for a single ROS, and the wide FoV covered in a single pointing.


\section{Integrated light properties}
\label{sec:integrated}

\begin{figure*}[t!]
\centering
\includegraphics[width=0.40\textwidth]{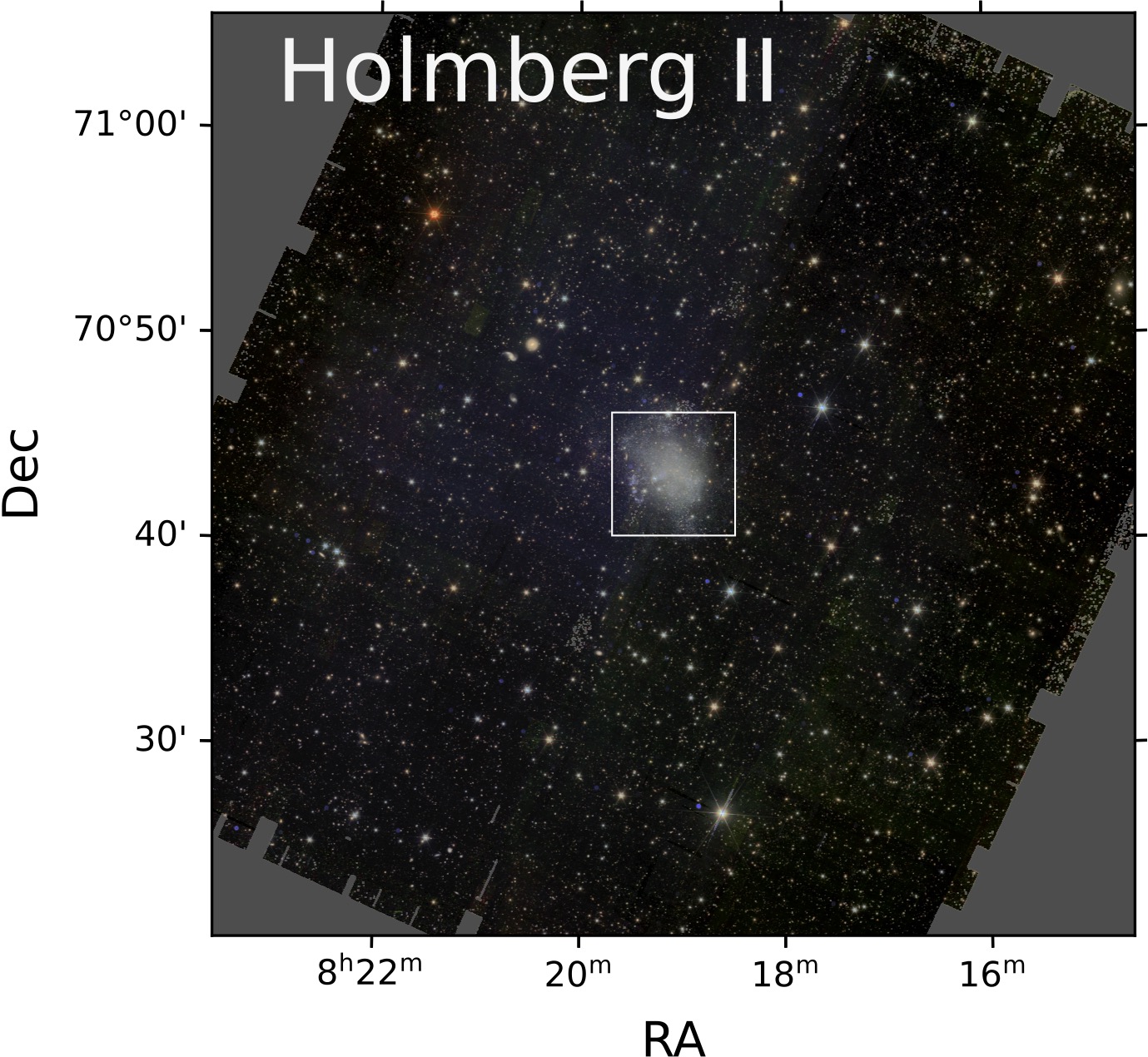}
\vspace{\baselineskip}
\includegraphics[width=0.85\textwidth]{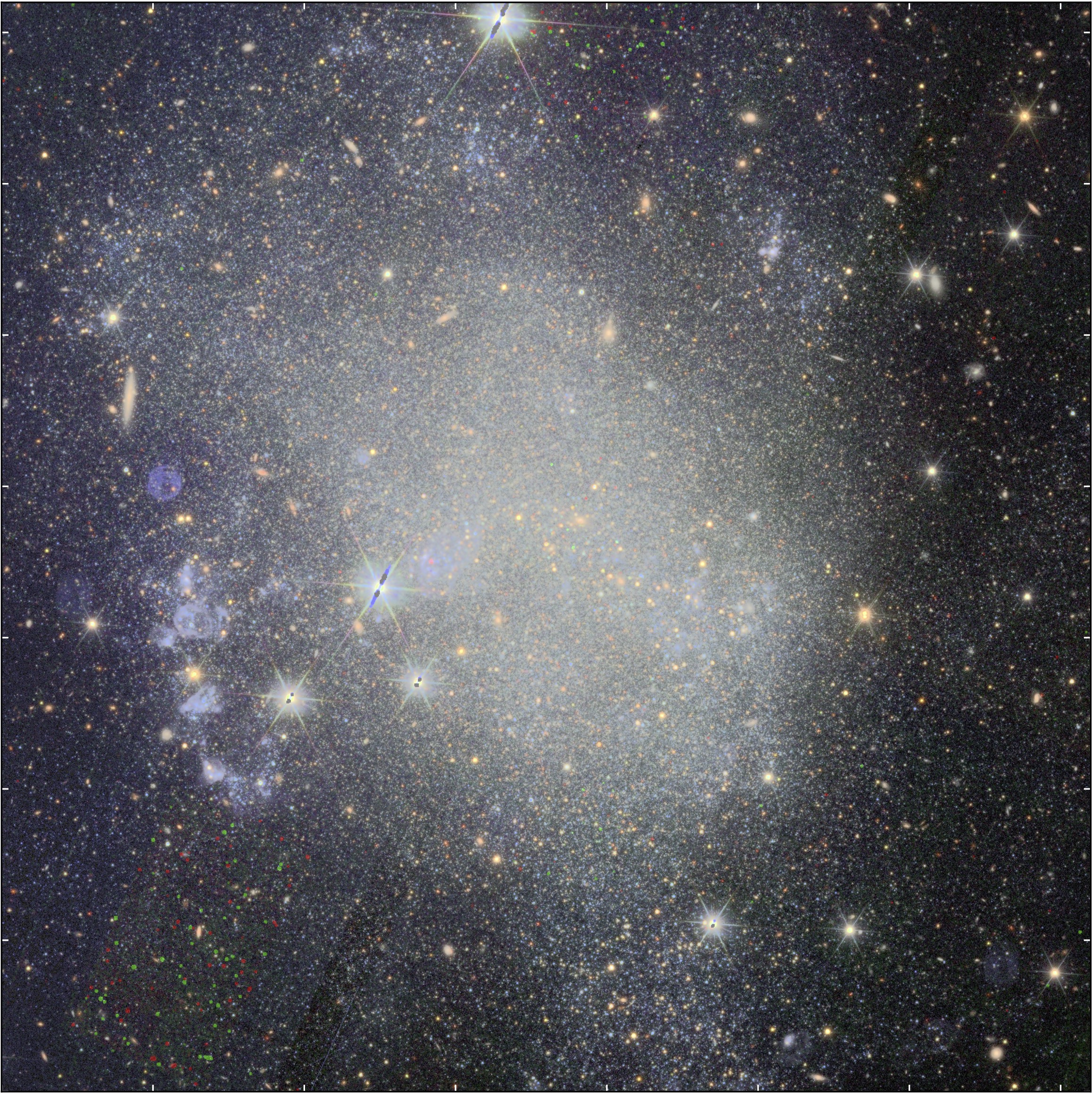}
\vspace{-\baselineskip}
\caption{RGB image of Holmberg\,II with \HE\ red, \YE\ green, and \IE\ blue.
Foreground extinction has been corrected and sky subtracted as described in the text (Sect. \ref{sec:extinction}).
In the top panel, the full FoV of $0\fdg7\,\times\,0\fdg7$ 
is shown,
while the bottom one displays the inner 6\arcmin\,$\times$\,6\,\arcmin\ 
region corresponding to the white box in the upper panel. 
In the lower panel, to the east, there is an extensive north-south chain of \hii\ regions \citep[e.g.,][]{hodge94}, that harbors
the ultraluminous X-ray source Ho\,II\,ULX-1 \citep[e.g.,][]{zezas99,kaaret04}, visible as a triangular-shaped blue
\hii\ region at $\alpha\,=\,$08:19:28.98, $\delta\,=\,+$70:42:19.3 (J2000).
Also visible as a blue circular structure to the north of the \hii-region chain is an artefact dichroic ghost
(see also Sect. \ref{sec:dwarfs}, Fig. \ref{fig:dwarf1-2}).
}
\label{fig:pretty_holmbergii}
\vspace{-\baselineskip}
\end{figure*}

\begin{figure*}[t!]
\centering
\hbox{
\includegraphics[width=0.40\textwidth]{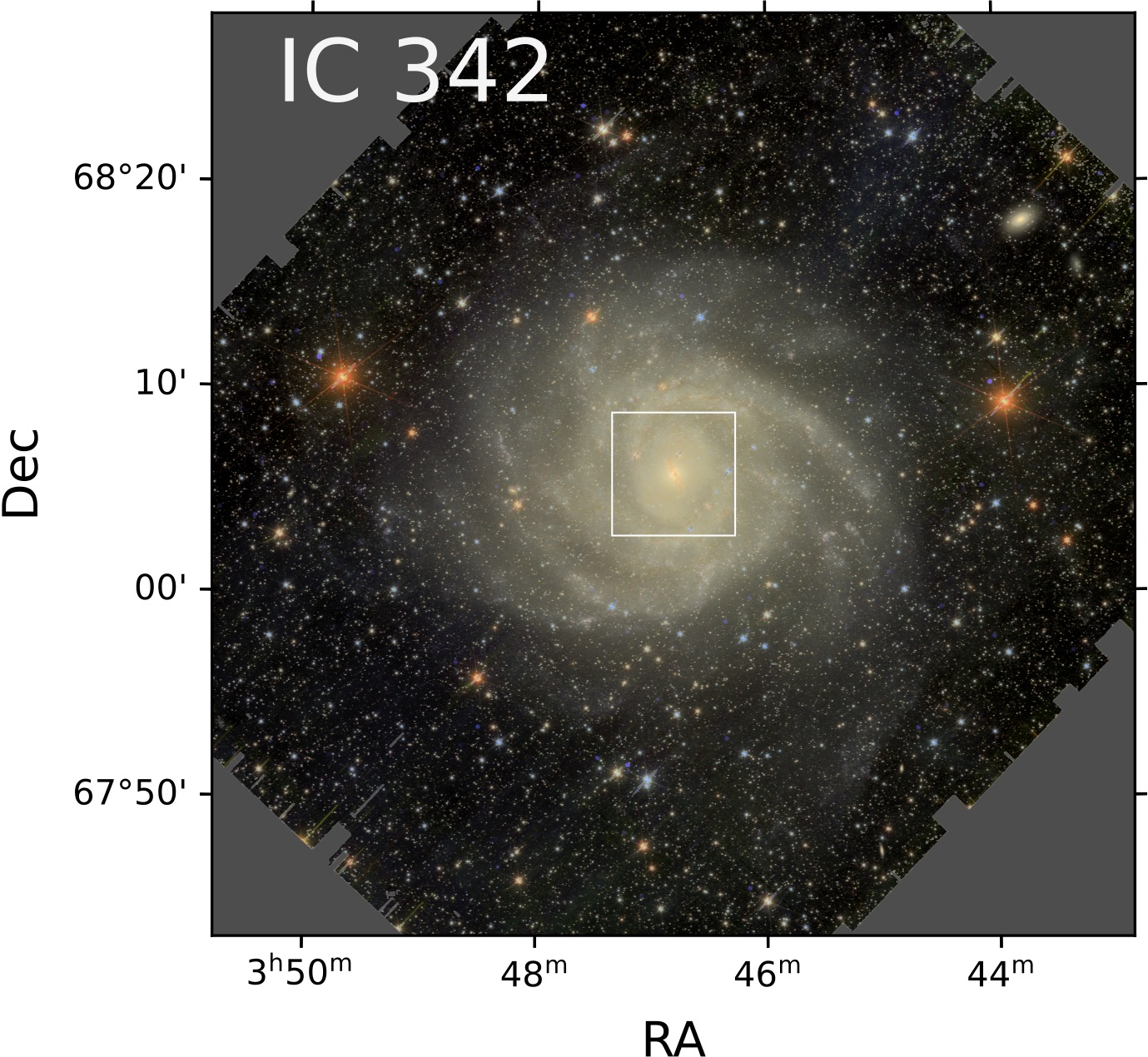}
\hspace{0.11\textwidth}
\includegraphics[width=0.40\textwidth]{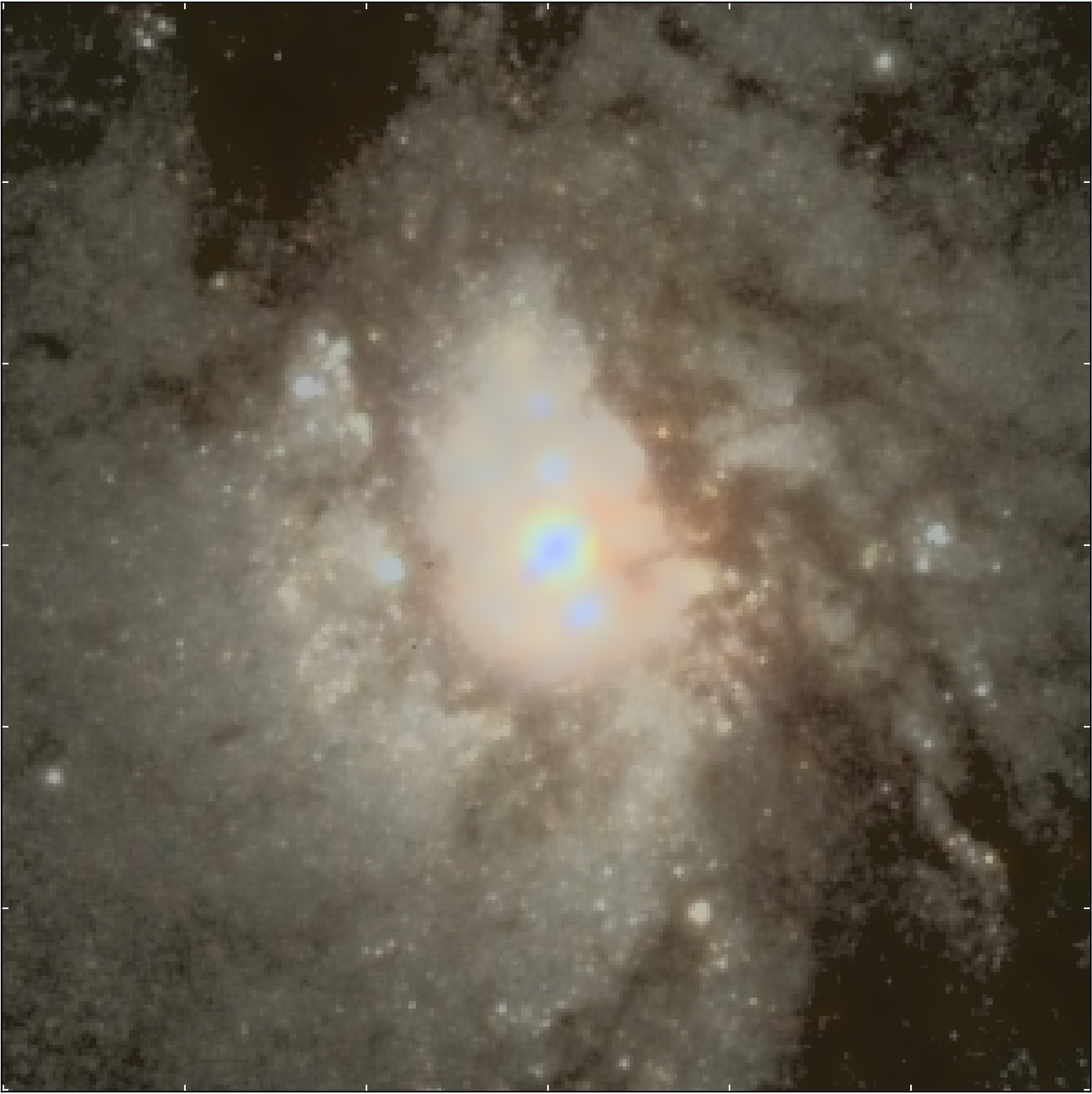}
}
\vspace{0.5\baselineskip}
\includegraphics[width=0.85\textwidth]{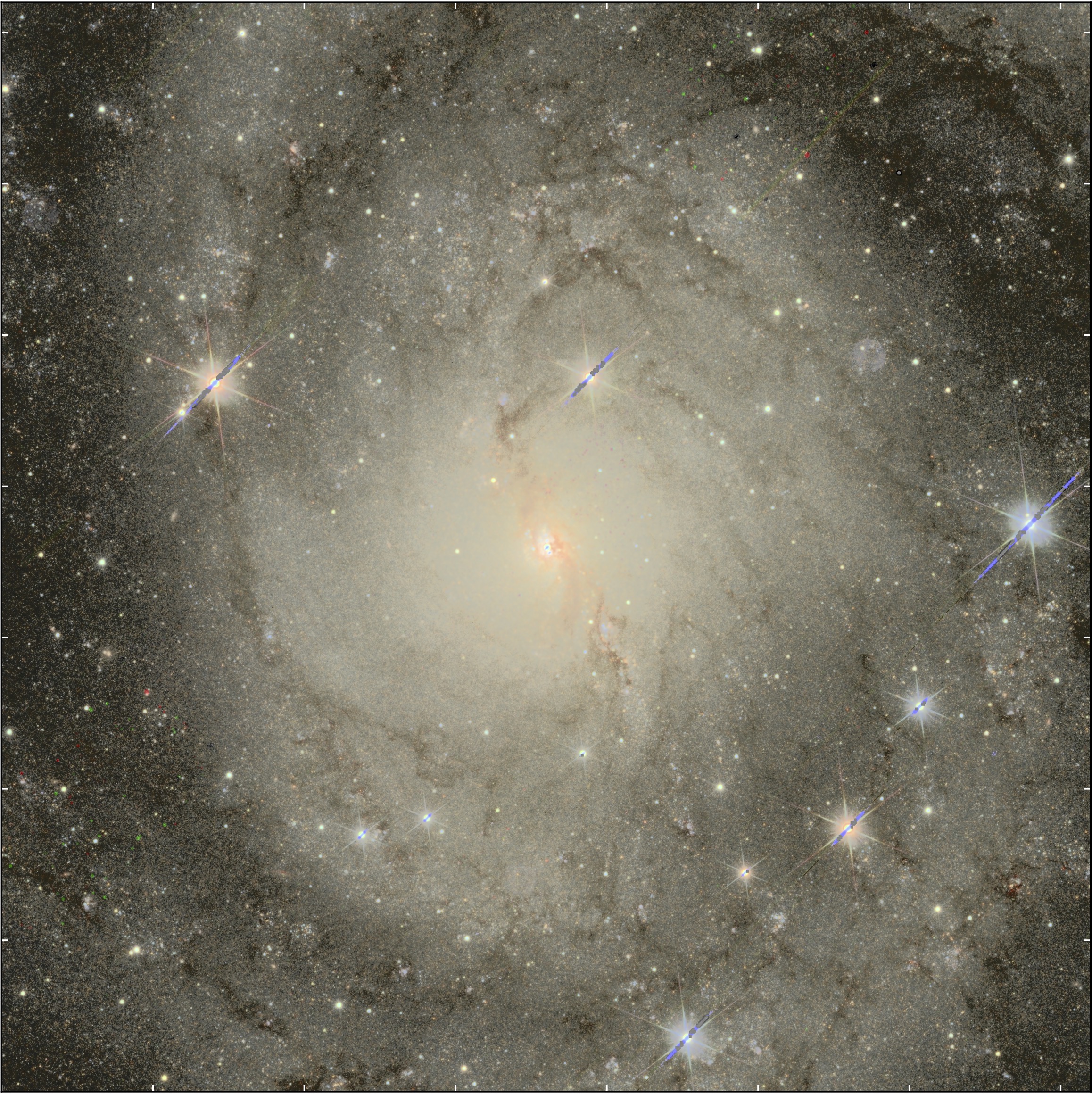}
\caption{Same as for Fig. \ref{fig:pretty_holmbergii}, but for IC\,342, and with the top left panel
showing the full FoV of $0\fdg7\,\times\,0\fdg7$\, and the bottom panel 
the inner 6\arcmin\,$\times\,$\,6\arcmin\ region corresponding to the white box in the upper left. 
The top right panel shows the zoomed-in 30\arcsec\,$\times$\,30\arcsec\ RGB image of the blue nucleus,
 also revealed in the radial color profiles (see Sect. \ref{sec:profiles}, Fig. \ref{fig:profiles_1}).
}
\label{fig:pretty_ic342}
\vspace{-2\baselineskip}
\end{figure*}

\begin{figure*}[t!]
\centering
\includegraphics[width=0.40\textwidth]{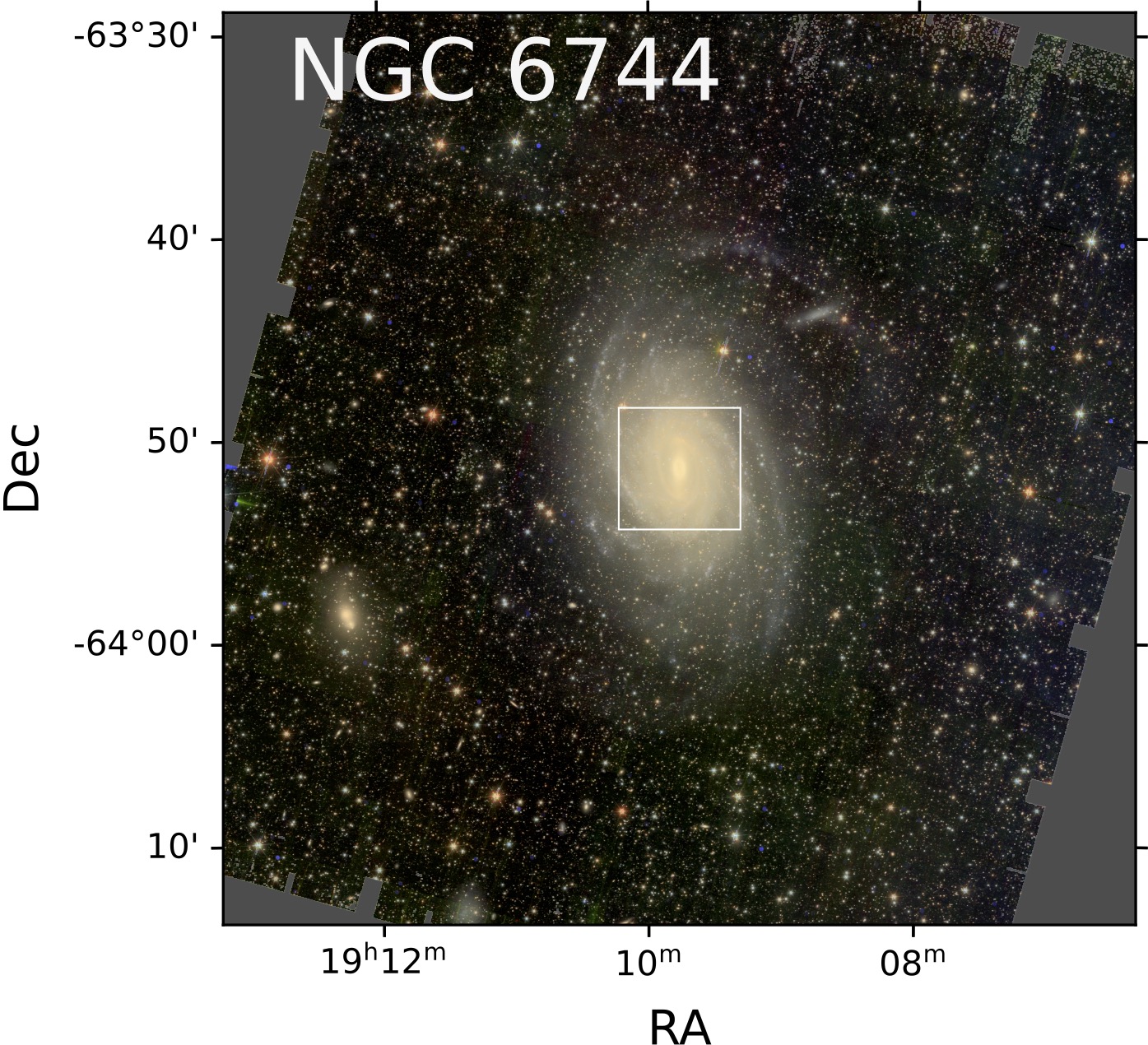}
\vspace{\baselineskip}
\includegraphics[width=0.85\textwidth]{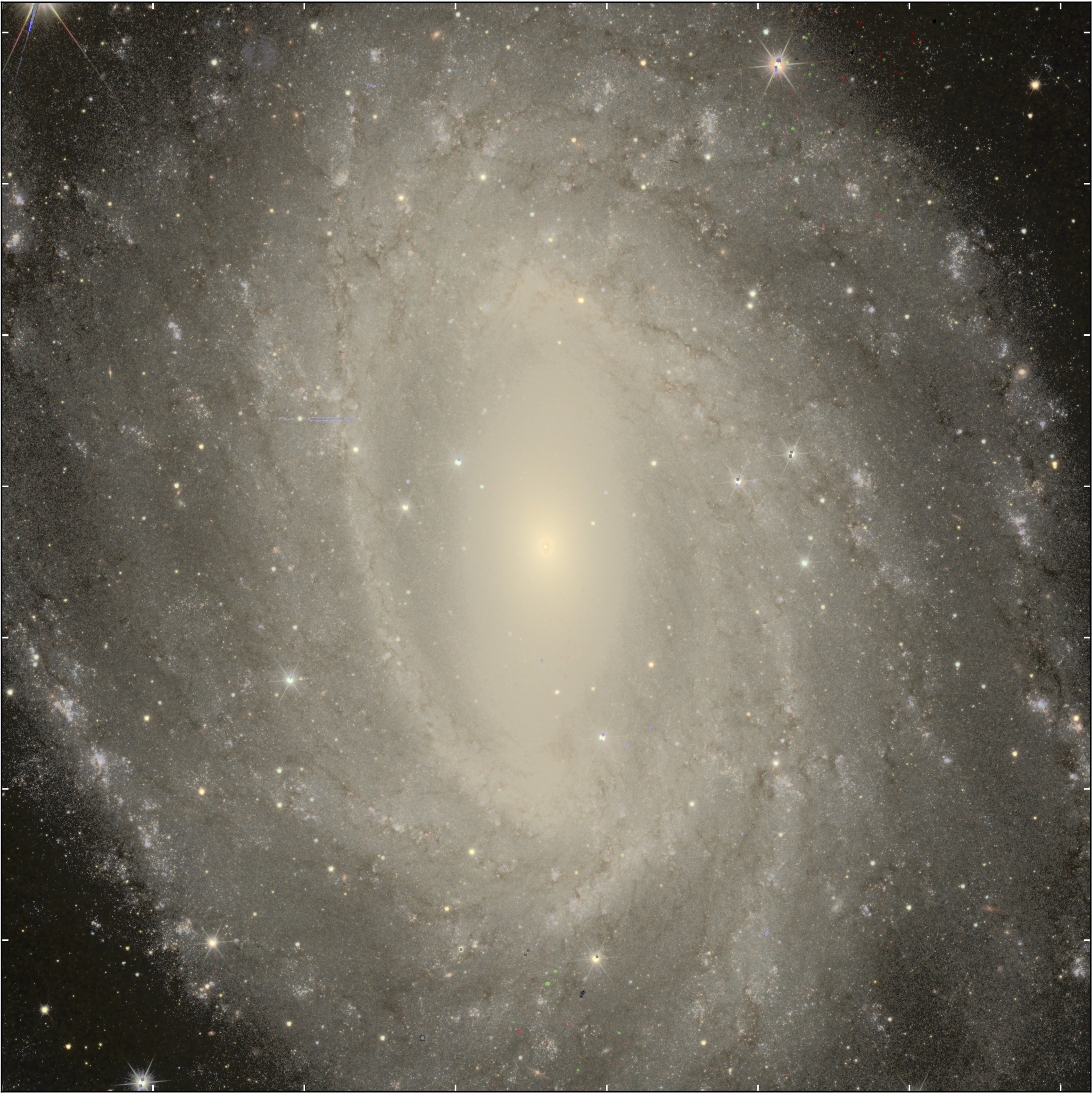}
\caption{As in Fig. \ref{fig:pretty_holmbergii}, but for NGC\,6744, the most distant galaxy of the Showcase.
In the top panel, the full FoV of $0\fdg7\,\times\,0\fdg7$ 
is shown,
while the lower panel gives the inner 6\arcmin\,$\times$\,6\,\arcmin\ 
region corresponding to the white box in the upper panel. 
The filamentary dust lanes within the spiral arms are delineated with exquisite detail.
}
\label{fig:pretty_ngc6744}
\vspace{-2\baselineskip}
\end{figure*}

To combine and compare the multi-band images for each galaxy, the images were aligned astrometrically
and rebinned to a common 0\farcs3 pixel size (the same as for NISP) using \texttt{gnuastro} routines.
Sky background emission was subtracted globally, adopting the sky level determined from Gaussian fitting
using Approach (2) (see Sect. \ref{sec:sblim} and Table \ref{tab:sky}). 

\subsection{Correction for foreground extinction}
\label{sec:extinction}

It is also necessary to correct for foreground extinction by the Galaxy.
Foreground extinction for each target has been estimated from the \citet{schlegel98} dust maps recalibrated
to the scale of \citet{schlafly11}, as implemented in the publicly available Python package 
\texttt{dustmaps}.\footnote{\texttt{dustmaps} is found at 
\url{https://dustmaps.readthedocs.io/en/latest/maps.html} and the dust maps themselves can be
accessed and downloaded in the context of this package.}
For a given location on the sky, the module returns the corresponding \ebv\ value derived by linearly interpolating the dust maps. 
We have used \rv\,=\,3.1 to convert \ebv\ to \av.
Values of \av\ for each galaxy are given in Table \ref{tab:sample}, and agree with the \av\ values from
\citet{schlafly11} tabulated by NED.
For the integrated light, we have adopted a single value of \av\ for each galaxy; 
instead for the resolved stellar photometry, 
we implemented a spatially variable foreground extinction, as described in Sect. \ref{sec:resolvedstars}. 

We corrected the images for foreground reddening according to the extinction curve from 
\citet[][G23]{gordon23}, implemented through \texttt{dust-extinction},\footnote{Available at 
\url{https://dust-extinction.readthedocs.io/en/stable/}.}
an affiliated package of \texttt{astropy}. 
Because of the difficulties in knowing the source spectrum 
a priori, and
its variation across the FoV, for the integrated light, we assume a flat source spectrum in wavelength. 
Thus, to compute the effective wavelength across the \Euclid filters, we took the bandwidths from \citet{laureijs11}
and computed the mean across the bandwidth.
This also assumes that the filters have a flat transmission curve, which is not far from the true transmission,
as shown in 
\citet{laureijs11} and \citet{Schirmer-EP18}.
The effective wavelengths obtained in this way are
0.725\,\micron, 1.033\,\micron, 1.259\,\micron, and 1.686\,\micron, respectively, 
for \IE, \YE, \JE, and \HE.
The G23 extinction curve gives
relative ratios of $A_\lambda$/\av\,=\,0.678, 0.366, 0.261, and 0.160, for \IE, \YE, \JE, and \HE, respectively.
Corrections for each \Euclid band have then been applied to the images using the G23 models 
within the \texttt{dust-extinction} package.
As mentioned above,
we have assumed a single value of \ebv\ for each galaxy (see Table \ref{tab:sample}), so
that the extinction correction is constant across the image.
Future papers will delve more deeply into the question of the effects of spatially variable foreground extinction
for the integrated light, as well as investigate the color-dependence of the extinction coefficients.

Our central wavelengths for the \Euclid bands are not exactly coincident with those given in \citet{Scaramella-EP1}:
$\lambda$\,=\,0.72\,\micron\ (\IE); 1.10\,\micron\ (\YE); 1.40\,\micron\ (\JE); and 1.80\,\micron\ (\HE).
However, their final $A_\lambda$ extinction corrections are quite close to our estimates, despite
their different $\lambda$ and adopted extinction curve \citep[][]{gordon03}:
$A_\lambda$/\av\,=\,0.68, 0.34, 0.23, 0.16, for \IE, \YE, \JE, and \HE, respectively.

Figures \ref{fig:pretty_holmbergii}, \ref{fig:pretty_ic342}, and \ref{fig:pretty_ngc6744} show the
aligned, sky-subtracted, extinction-corrected images 
of representative Showcase galaxies combined into RGB format,
with \IE\ as blue, \YE\ green, and \HE\ red;
Holmberg\,II, IC\,342, and NGC\,6744 are shown here, while 
the remaining galaxies are shown in Appendix \ref{app:imaging}. 
Figures \ref{fig:pretty_holmbergii}--\ref{fig:pretty_ngc6744} (and Appendix \ref{app:imaging})
illustrate the capability of \Euclid
to image extremely wide regions over the sky, but also to probe the fine, highly spatially resolved, details of stellar
content and background objects.
The close proximity of IC\,10 
and NGC\,6822 enables careful assessment of star counts
and resolved stellar populations (see Sect. \ref{sec:resolvedstars}).
Stellar populations are still resolved in slightly more distant galaxies 
(out to about 3\,Mpc)
such as Holmberg\,II, IC\,342, NGC\,2403, and even NGC\,6744 at 
9\,Mpc.
\Euclid's superb resolution probes the central regions of IC\,10, IC\,342, and NGC\,2403 at 
1--4\,pc scales,
revealing young star clusters and dusty filaments across their nuclei. 
The more distant spiral, NGC\,6744 at 
9\,Mpc, can be examined on slightly coarser 
4--13\,pc scales,
ideal for comparing stellar populations with the distribution of molecular gas
\citep[e.g.,][]{leroy21} and other tracers of the interstellar medium (ISM).

\begin{figure}[h!]
\centerline{
\includegraphics[width=0.48\textwidth]{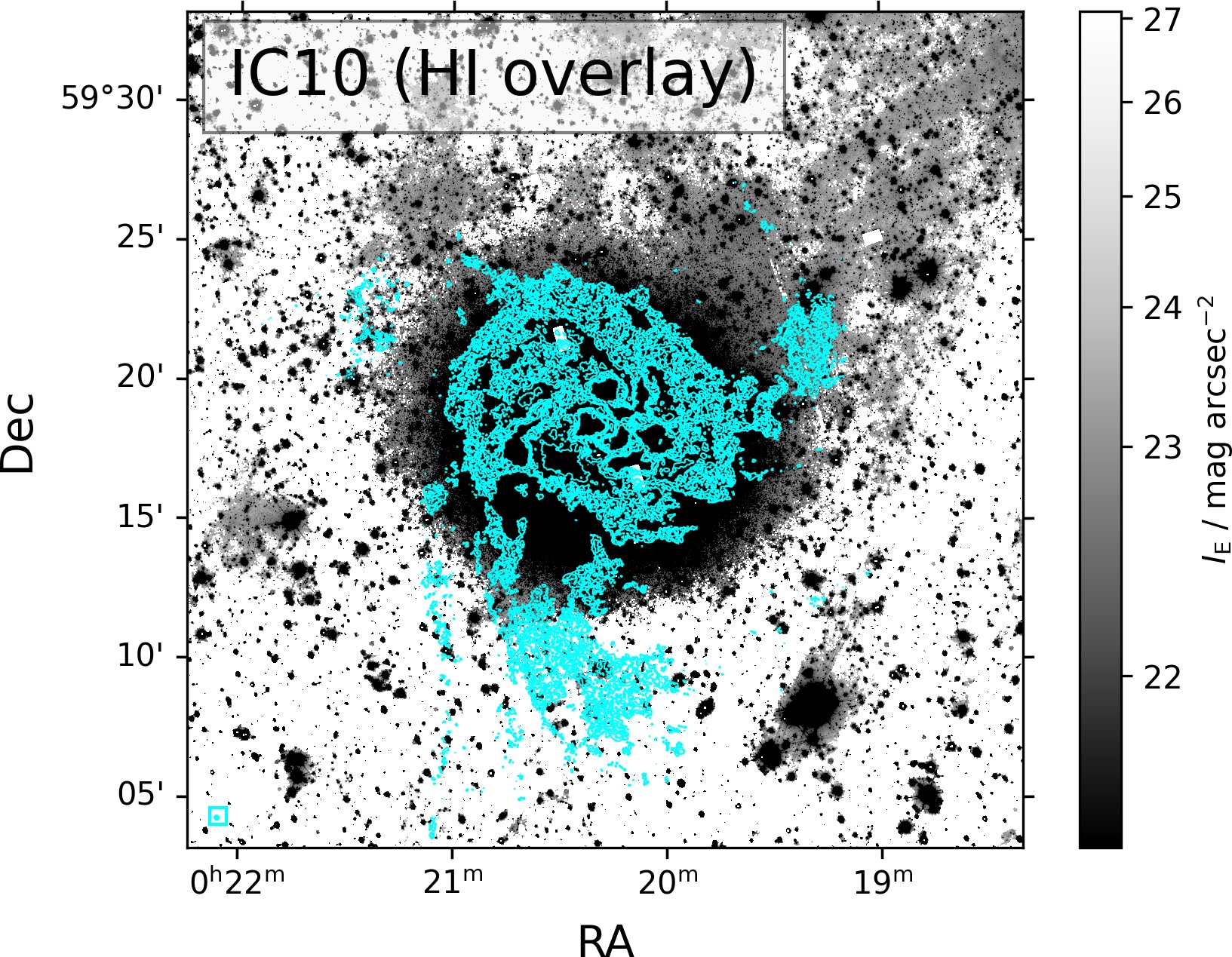} 
}
\vspace{0.5\baselineskip}
\centerline{
\includegraphics[width=0.48\textwidth]{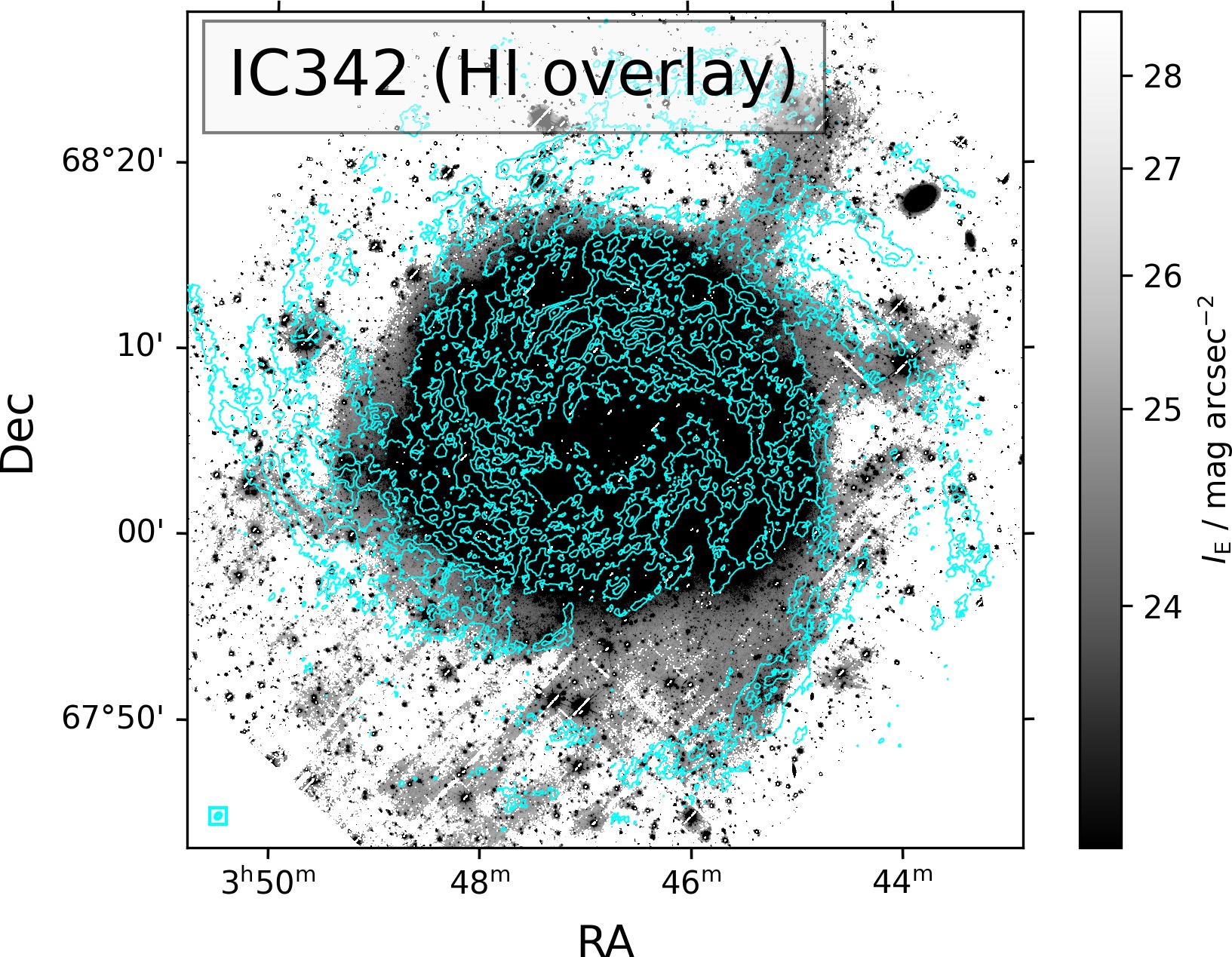}
}
\vspace{0.5\baselineskip}
\centerline{
\includegraphics[width=0.48\textwidth]{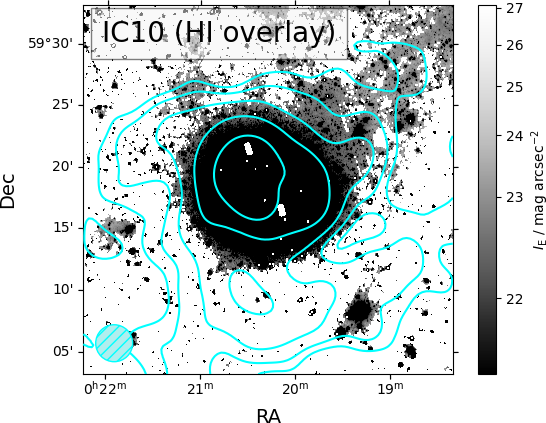} 
}
\caption{\hi\ overlays on high-contrast \Euclid \IE\ images: 
IC\,10 (top panel); and IC\,342 (middle). 
The \hi\ beam size is shown in the lower left corner,
and contours are at 2$\sigma$, 4$\sigma$, 7$\sigma$, 10$\sigma$, and 20$\sigma$.
The bottom panel gives the \hi\ overlay for IC\,10 as in the top panel, but
using the 3-arcmin beam-smoothed \hi\ image from \citet{namumba19};
contours are at 2$\sigma$, 3$\sigma$, 5$\sigma$, 10$\sigma$, 10$\sigma$, and 70$\sigma$.
}
\label{fig:hioverlay}
\vspace{-\baselineskip}
\end{figure}

\subsection{Comparison of stellar and H\,I morphologies}
\label{sec:hi}

Stellar content and \hi\ gas properties are intimately related.
In luminous galaxies not dominated by dark matter (DM), the stars dominate the gravitational potential
\citep[e.g.,][]{vanderkruit81,mancera22}, and for galaxies of all types,
the combination of stars and \hi\ is fundamental for determining the characteristics of the DM.
It is commonly thought that \hi\ gas tends to be more extended than the stellar disk
\citep[e.g.,][]{bosma17}, possibly because of dwarf galaxy satellites being disrupted in the process
of a minor merger \citep[e.g.,][]{kamphuis91,meyer06,boselli14,zemaitis23}, or through
cold accretion episodes \citep[e.g.,][]{blandhawthorn17}, or both.
However, deep optical imaging suggests that stellar substructures can extend as far as the \hi\
disk \citep[e.g.,][]{lewis13,okamoto15}.
Recent work on \hi\ demographics finds that the extent of the \hi\ disk depends on star-formation
activity, and that more massive galaxies tend to have less extended \hi\ disks
\citep{reynolds23}.
The extent of \hi\ also depends on environment, since there is a decrease in the \hi-to-optical diameter 
in cluster environments \citep[e.g.,][]{reynolds22}.

\begin{figure*}[h!]
\centerline{
\includegraphics[width=0.48\textwidth]{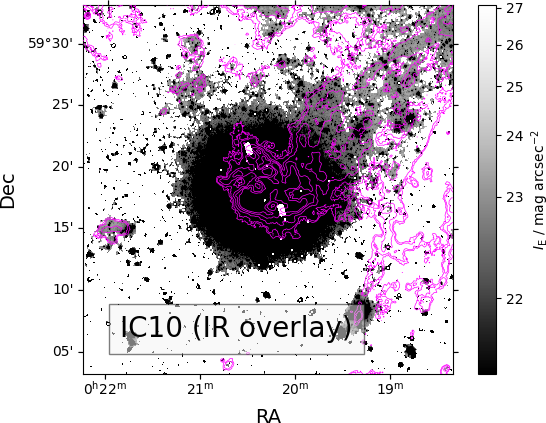}
\hspace{0.01\textwidth}
\includegraphics[width=0.48\textwidth]{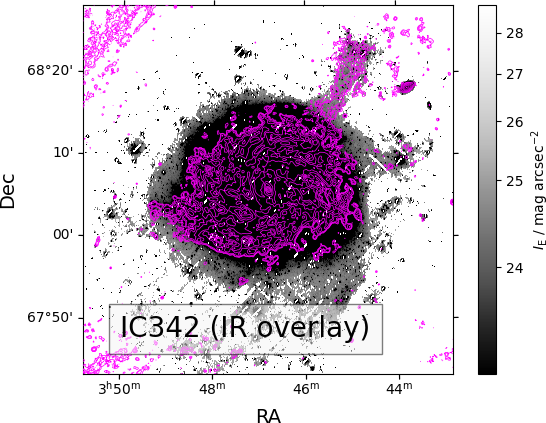} 
}
\caption{\hers\ SPIRE\,250-\micron\ overlays on high-contrast \Euclid \IE\ images 
for IC\,10 (left panel) and IC\,342 (right).
}
\label{fig:iroverlay}
\end{figure*}

Here, for illustration, in Fig. \ref{fig:hioverlay} we compare the \hi\ morphology in IC\,10 and IC\,342 
to the stellar content as traced by \Euclid\ imaging.\footnote{The \magarc\ units have not been  
rescaled by Eq.\,\eqref{eqn:sblim} within the 100\,arcsec$^2$ regions discussed in Sect. \ref{sec:sblim}.}
The \hi\ data for IC\,10 are taken from
\citet{wilcots98},\footnote{These archival data have been reprocessed by F. Walter et al. (priv. comm.).} 
and for IC\,342 from \citet{chiang21}. 
The beam sizes are shown in the lower-left corner of the overlay.
It can be seen from Fig. \ref{fig:hioverlay} that the stars in IC\,10 are more spatially extended than this \hi\ map,
and even parts of the \hi\ streams toward the south and the west are within the stellar disk
as traced by \Euclid.
The outer \hi\ spiral arms in IC\,342 do not fall within the stellar disk, but the bulk of the \hi\ distribution
is closely mirrored by the stars.
The \Euclid \IE\ `spur' toward the northwest is not reflected in the \hi\ morphology tracing spiral arms in the gas.

Measurement sensitivity in terms of \hi\ beam size and limiting column density, and 
the SB limits that can be achieved for the stellar component, are arguably the most important 
discriminators for determining the relative sizes of the \hi\ and stellar distributions 
\citep[e.g.,][]{xu22}.
The bottom panel of Fig. \ref{fig:hioverlay} shows an \hi\ moment map taken from \citet{namumba19}
with a larger beam than that shown in Fig. \ref{fig:hioverlay} (top panel).
With this larger beam, sensitive to fainter \hi\ column densities, the \hi\ extends beyond the stellar
disk, extending to the northwest where there is a putative stellar stream culminating beyond the \Euclid\ FoV
\citep[e.g.,][]{nidever13,namumba19}.
We examine whether this extension seen with \Euclid\ can be associated with stars or foreground cirrus in the next
section.
In any case, it seems clear that the interplay of stars and \hi\ morphology in galaxies
can be reassessed on a statistical basis with the sensitivity of \Euclid.

\subsection{Comparison of stellar, ISM, and cirrus emission}
\label{sec:ir}

Atomic gas and dust tend to be spatially correlated within a typical ISM.
However, in nearby galaxies it is not always straightforward to separate the foreground dust
emission in the Milky Way (MW) from dust emission originating within the nearby galaxy itself.
Conversely, \hi\ enables such a separation because of the spectral resolution and corresponding velocity measurements.
Figure \ref{fig:iroverlay} overlays far-infrared (FIR) dust emission from \hers\ SPIRE/250\,\micron\
over high-contrast \Euclid \IE\ images of IC\,10 and IC\,342.
The FIR images are taken from the Dwarf Galaxy Survey \citep{madden13}
and the Key Insights on Nearby Galaxies: A FIR Survey with \hers\ \citep{kennicutt11}.

The dust emission in IC\,10 roughly follows the \IE\ filament to the northwest, but it is not
altogether possible to distinguish the dust morphology from that of the \hi\ gas shown in 
Fig.~\ref{fig:hioverlay} (bottom panel).
Although \hi\ and dust tend to be cospatial, 
identifying the origin of \hi\ and FIR is problematic in IC\,10 because of its proximity,
and thus low recession velocity, relative to the MW. 
IC\,10 has a recession velocity of $-348$\,\kms, so is somewhat more blue-shifted than the highest \hi\
velocities ($-150$\,\kms) considered as belonging to the MW in \citet{planck11}.
Those authors also found that the emissivity of Galactic dust in these high-velocity clouds is low,
so that relatively little dust emission would be expected from such clouds
\citep[see also][who analyzed the Virgo cluster]{bianchi17}.
The \hi\ gas around IC\,10 toward the northwest extension is found at 
about $-400$\,\kms\
\citep[e.g.,][]{nidever13}, a higher velocity than expected for gas belonging to the MW,
and consistent with being intrinsic to IC\,10.
It is thus likely that the filamentary dust traced by the FIR in IC\,10 belongs indeed to IC\,10.
Galactic cirrus tends to have blue optical colors \citep{roman20}, but the severe foreground
extinction of \av$\ga4$ toward IC\,10, and its possible spatial variation, makes an accurate 
color determination difficult, as well as the separation of the \IE\ stellar emission from potential cirrus,
either belonging to the MW or IC\,10.

In contrast, the case of IC\,342 is unambiguous.
Dust emission toward 
IC\,342 follows the \IE\ emission in the spur feature,
but the \hi\ at the recession velocity of IC\,342 does not.
The conclusion is that in IC\,342, the \IE\ emission in the spur is due to foreground cirrus,
rather than a stellar stream.
Its morphology is mirrored exactly by the FIR 250\,\micron-emission, but not in \hi.
Consequently, in addition to tracing stars,
sensitive \Euclid \IE\ images, compared with other wavelengths, will be a powerful diagnostic
for the Galactic ISM, in particular for assessing the importance of the diffuse cirrus component. 

\subsection{Surface brightness profiles}
\label{sec:profiles}

\begin{figure*}[h!]
\includegraphics[width=0.48\textwidth]{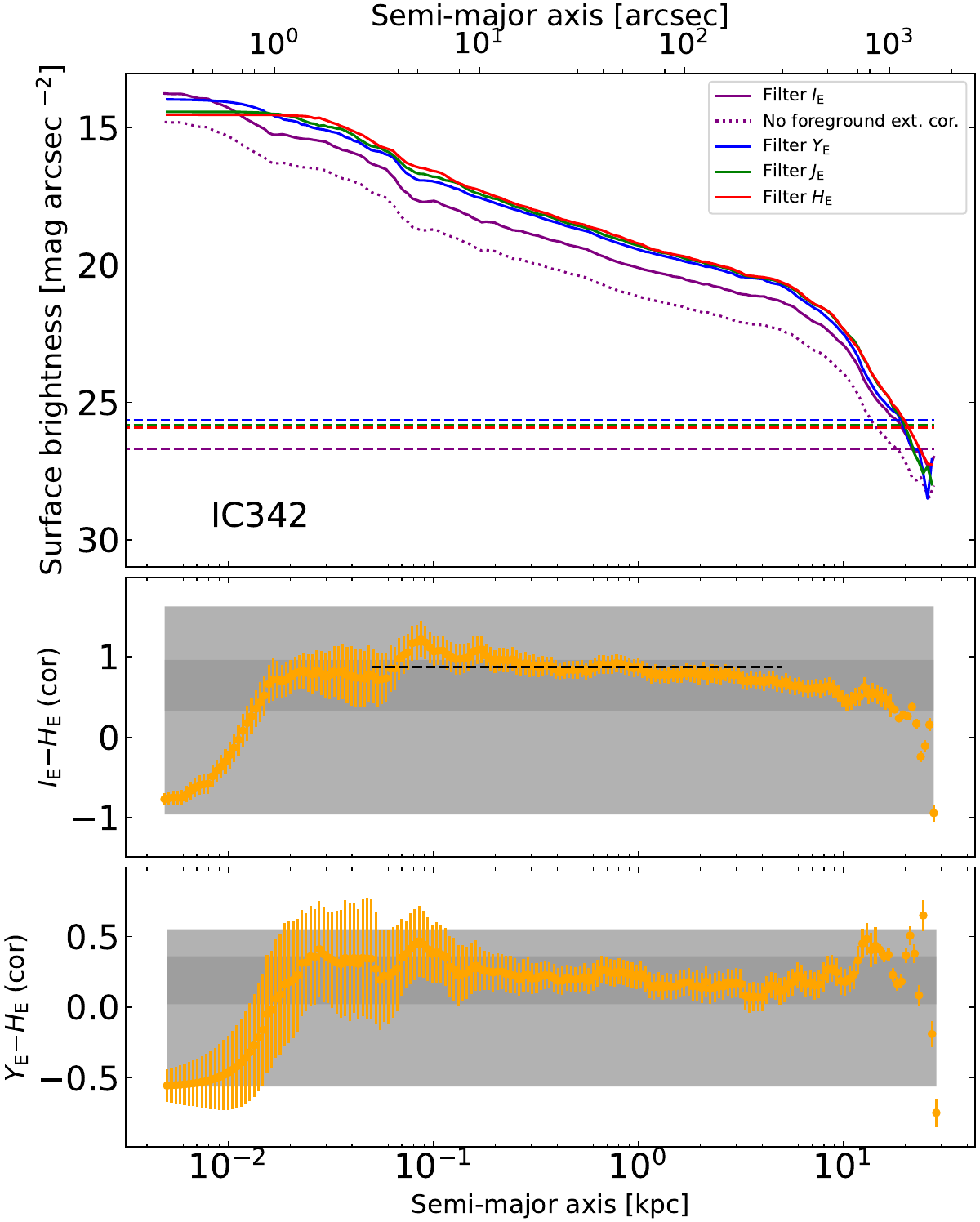}
\hspace{0.04\textwidth}
\includegraphics[width=0.48\textwidth]{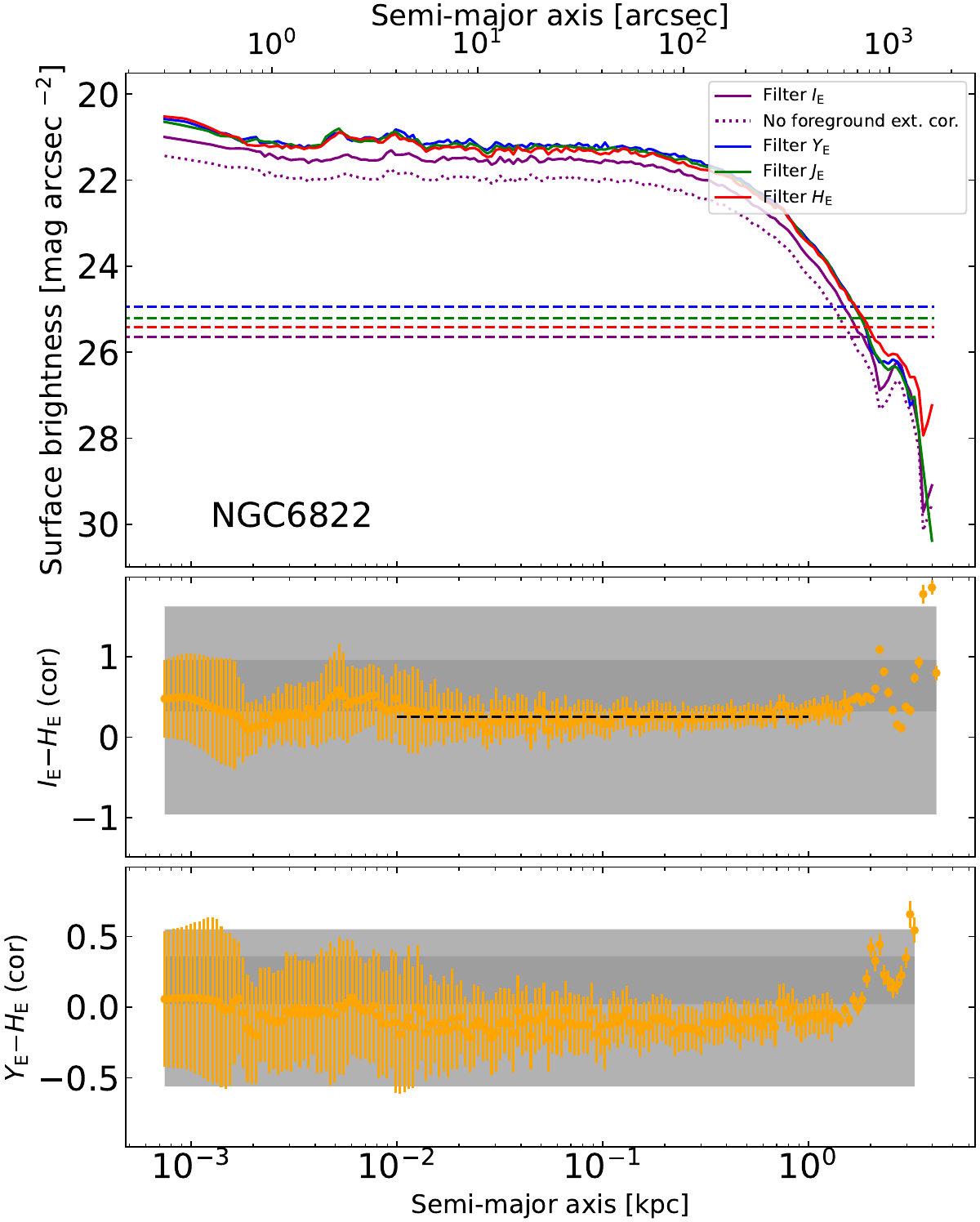}
\caption{\textit{Top}: Surface brightness profiles extracted by \texttt{AutoProf} as described
in the text for IC\,342 (left panel) and NGC\,6822 (right). 
The four bands are given by purple, blue, green, and red curves for \IE, \YE, \JE, and \HE, respectively.
The $1\sigma$ SB limits from \texttt{AutoProf} (not rescaled to 100\,arcsec$^2$ regions)
in units of \magarc\ 
are shown as dashed horizontal lines, with colors corresponding to the \Euclid bands.
The fluxes have 
been corrected for foreground extinction (Sect. \ref{sec:extinction}; 
the uncorrected \IE\ profile is shown as a dotted (purple) curve in the top panel.
\textit{Middle and bottom}:  \IE$-$\HE\ and  \YE$-$\HE\ radial color profiles.
The top axis corresponds to galactocentric radii in units of arcsec, and the bottom in units of kpc.
The mean \IE$-$\HE\ color over typically a factor of $100$ in radius 
is shown as a horizontal dashed line in the middle panel; 
the light gray rectangular regions illustrate the full spread in model colors (see Fig. \ref{fig:colors})
and the dark gray one the standard deviation about the mean of the models.
The mean galaxy \IE$-$\HE\ with its standard deviation is also shown as a gray rectangular region in Fig. \ref{fig:colors}.
}
\label{fig:profiles_1}
\end{figure*}

We have generated surface brightness (SB) profiles for the Showcase galaxies using \texttt{AutoProf}.
\texttt{AutoProf} is a Python-based pipeline for non-parametric profile extraction, 
that includes masking, sky determination, centroiding, and isophotal fitting \citep{stone21}.
For this paper, we use \texttt{AutoProf} in the default mode but with $5\sigma$ clipping.
Because centers are difficult to determine, particularly in nearby dwarf galaxies with resolved
stars, we have fixed the profile centers to the NED coordinates for the galaxy.
Results are shown in Fig. \ref{fig:profiles_1} for IC\,342 and NGC\,6822; the profiles 
of the remaining galaxies appear in Appendix \ref{app:profiles} (Fig. \ref{fig:profiles_2}).
The sky values determined from Approach (2),
as given in Table \ref{tab:sky} and shown in Fig. \ref{fig:sblims},
are consistent to within a few percent with the sky levels from \texttt{AutoProf}.

In Fig. \ref{fig:profiles_1} (and Fig. \ref{fig:profiles_2}),
the surface brightness profiles corrected for foreground extinction are shown as solid lines, 
while  for \IE\ the uncorrected profile is shown as a dotted line. 
Because of their low Galactic latitude (see Table \ref{tab:sample}),
for IC\,10, IC\,342, and NGC\,6822, these corrections can be significant, up
to 
$3$\,\IE\,mag in the case of IC\,10.
The top horizontal axis reports the angular galactocentric distance, while the bottom
gives the physical radii in kpc.
Figures \ref{fig:profiles_1} and \ref{fig:profiles_2} show that \Euclid is able to trace
galaxy emission out to 20$-$30\,kpc in radius in a single ROS.

\begin{figure*}[h!]
\includegraphics[width=\textwidth]{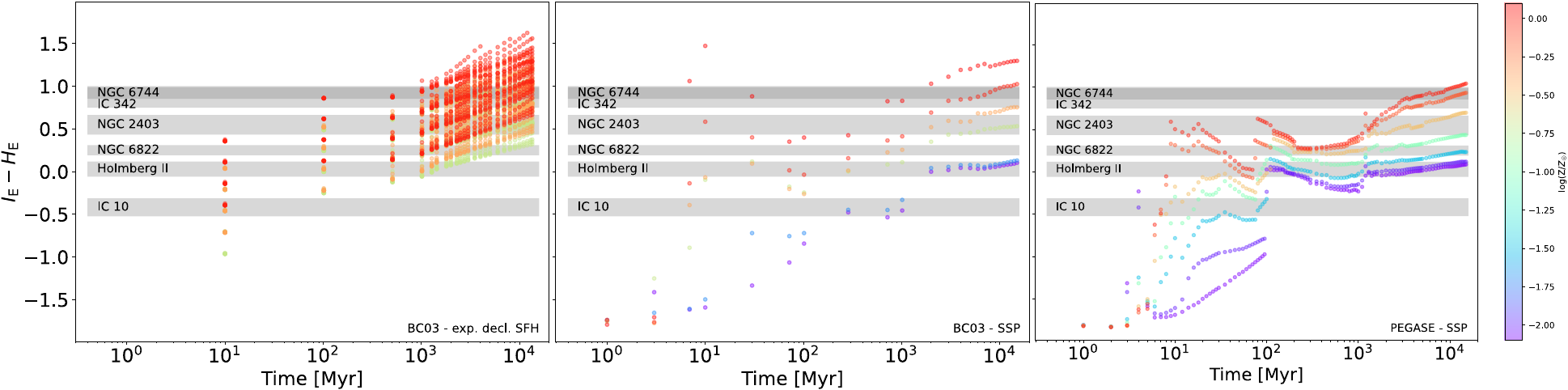}
\caption{Synthetic \IE$-$\HE\ from the \citet{BC03} models with exponentially declining SFH (left panel); 
the same \citet{BC03} models but SSPs (middle);
and PEGASE SSP models from \citet[][right]{fioc97} and \citet{leborgne04}.
Also shown as gray regions are the mean \IE$-$\HE\ colors and their standard deviations,
as reported in the middle panel of Figs. \ref{fig:profiles_1} and \ref{fig:profiles_2},
evaluated over a factor of 100 in galactocentric radius.
As discussed in Sect. \ref{sec:extinction}, these colors have been corrected for foreground extinction
from our Galaxy.
The different behavior of the models is due to the smoothed-out SFH in the left panel,
and the different treatment of red supergiants (RSGs) that begin to dominate around 
$10$\,Myr, and the most massive AGB stars around 
$100$\,Myr.
}
\label{fig:colors}
\end{figure*}

Figures \ref{fig:profiles_1} and \ref{fig:profiles_2} illustrate the difficulty of determining the sky value when
the galaxy fills the image.
IC\,342's profile extends smoothly out to the limits of the \Euclid FoV (1600\arcsec\ on the diagonal),
but does not quite achieve the SB limits expected from Sect. \ref{sec:sblim}. 
The implication is that the galaxy emission could have been measured at even larger radii,
were it not limited by the (already large) FoV.
On the other hand, the profile of NGC\,6822, an apparently smaller galaxy,
approximates the SB limits in the very outer regions.
The problem of large galaxies will be mitigated in the EWS, 
because of more continuous coverage. 

\subsection{The diagnostic power of \Euclid colors}
\label{sec:colors}

The radial trends of selected \Euclid colors, \IE$-$\HE\ and \YE$-$\HE\ are
also shown in Figs. \ref{fig:profiles_1} and \ref{fig:profiles_2}. 
As an initial evaluation of the photometric calibration (see also Sect. \ref{sec:resolvedstars}), 
we compare the colors of the Showcase galaxies with those obtained using synthetic templates. 
In particular, we 
use \cite{BC03} synthetic models, and 
calculate the magnitudes using the SED-fitting code \texttt{lephare} 
\citep{arnouts99,ilbert06}. 
We assume an exponentially declining SFH, with an SFR duration $\tau$ in the range $0.1$--$30$\,Gyr, 
ages up to 14\,Gyr, metallicity from sub-Solar to Solar 
($0.2\,$\zsun, $0.4\,$\zsun, and \zsun),
and vary the internal extinction with \ebv\,=\,$0, 0.1, 0.2$, and $0.3$ using the attenuation curve of \citet{calzetti94}.
From the combination of these parameters, we 
generate a library of 2300 synthetic magnitude sets at $z\,=\,0$. 
Given the wide range of parameters explored, most of the galaxies in the local Universe would be expected to possess colors 
within the model predictions.

From the distribution of the \Euclid colors in the library, we determine the median and the 16$-$84th 
quantile of the distributions, and find the following color ranges: 
\IE$-$\YE\,=\,$0.45\,\pm\,0.24$; 
\YE$-$\JE\,=\,$0.10\,\pm\,0.08$; and 
\JE$-$\HE\,=\,$0.09\,\pm\,0.09$. 
In Figs. \ref{fig:profiles_1} and \ref{fig:profiles_2}, these are shown as dark gray rectangular regions.
The full color ranges spanned by the models are illustrated as light gray regions. 

Virtually all of the colors shown in Figs. \ref{fig:profiles_1} and \ref{fig:profiles_2} fall within
the ranges predicted by these models.
In the spirals, IC\,342, NGC\,2403, and NGC\,6744, there is a trend for the 
outer regions to be bluer than the bulk of the inner disk, possibly implying an inside-out
disk formation scenario \citep[e.g.,][]{williams09b,gogarten10,wang11},
consistent with radial metallicity gradients in nearby spirals \citep[e.g.,][]{sanchez14}
Conversely, the central regions of IC\,342 and IC\,10 are extremely blue,
challenging the spread of allowable colors predicted by the models.
However, as shown in Figs. \ref{fig:pretty_ic342} and \ref{fig:pretty_ic10}, the centers of both galaxies are unusual.
IC\,342 has an extremely young star luminous cluster complex in its nucleus \citep{boker99,carson15,balser17},
the brightest of those examined by \citet{carson15}.
Figure \ref{fig:pretty_ic342} and \hst\ colors show that it is extremely blue, associated with a massive \hii\ region
and an X-ray source \citep{mak08}.
In IC\,10, the NED center position corresponds to a complex of \hii\ regions \citep[e.g.,][]{hodge90,polles19},
and there are several more located near the nucleus \citep[e.g.,][]{vacca07}.
Thus, extremely blue nuclear colors are expected for both IC\,342 and IC\,10.

We explore this further in Fig. \ref{fig:colors}, where \IE$-$\HE\ is plotted as a function
of age, and color coded by metallicity $Z$.
We compare the predictions of the \citet{BC03} models described above and shown in the left panel,
with those of \citet{BC03} for single stellar population (SSP) models and 
with \texttt{PEGASE} SSPs by \citet{fioc97} and \citet{leborgne04}.
Also shown as gray regions are the mean \IE$-$\HE\ color ranges of the Showcase galaxies
that are reported as horizontal dashed lines in Figs. \ref{fig:profiles_1} and \ref{fig:profiles_2}.
The models in the left panel of Fig. \ref{fig:colors} were generated with a limited range of metallicities
and ages, as can be seen from the comparison with the SSPs in the right two panels.
In addition to the different parameter ranges, the BC03 models behave differently due to the smoothed-out SFH in the left panel,
compared to the SSP in the middle one.
The BC03 and \texttt{PEGASE} SSP models also 
differ in their treatment of red supergiants (RSGs) that begin to dominate at $\sim 10$\,Myr,
and the most massive asymptotic giant branch (AGB) stars at $\sim100$\,Myr.

The bluest colors are found at young ages, $\la 10$\,Myr, consistent with the
properties of IC\,10 and IC\,342 in their central regions.
Moreover, sub-Solar metallicity makes these colors even bluer, so appropriate for IC\,10
at 
about 0.3\,\zsun\ ($\log_{10}(Z/Z_\odot)\,=\,-0.55$ for the color coding).
At a slightly super-Solar metallicity,
the age of the IC\,342 nuclear star cluster is estimated to be $\sim\,$5\,Myr \citep[e.g.,][]{carson15},
so the limits of the SSP \texttt{PEGASE} models constrain well the observed colors at this young age.

In summary, \Euclid colors are diagnostic of the age and metallicity of the stellar populations
in galaxies, and will provide an important tool for the exploration of broader galaxy populations.
At \Euclid's resolution, in the centers of these nearby galaxies, we are essentially just probing the brightness of 
small star clusters or even bright stars. 
At the same time, \Euclid's sensitive SB limits allow the examination 
of galaxy disks to depths that can reveal disk breaks and faint external features of galaxies
that could be signatures of interaction
\citep[e.g.,][]{peters17,sanchez23}.
Details of the SB profiles and color gradients, and the disk properties
of the Showcase galaxies, will be discussed in a future paper.


\section{Resolved stellar photometry and star counts }
\label{sec:resolvedstars}


Going beyond the integrated light described in Sect. \ref{sec:integrated},
photometry of resolved stars in nearby galaxies is a powerful tool, not only for understanding
stellar content and galaxy formation scenarios, but also for probing the outer regions of galaxy disks
and disk formation
\citep[e.g.,][]{barker12,crnojevic16,hillis16,jang20a}.
The surface brightness of resolved stellar populations, once corrected for completeness and projection effects,
can reach fainter SB limits than integrated light alone \citep[e.g.,][]{barker12}.
Thus, through stellar photometry in nearby galaxies,
\Euclid opens a new perspective also on resolved stellar populations and their diagnostic capabilities.
Here we present a first look at resolved stellar photometry in the Showcase galaxies.

\subsection{Stellar photometry}
\label{sec:starphot}

For all galaxies, point-source photometry was performed with \texttt{SourceExtractor} \citep{bertin96}. 
Detections were considered inependently in the four \Euclid bands, 
adopting a \texttt{mexhat} (wavelet) filter before detection, and then  
considering as valid detections all sources having even a single pixel 1.5$\sigma$ above the background. 
The filtering step performed by \texttt{SourceExtractor} prior to source identification has the effect
of ``smoothing'' the images, thus minimizing 
spurious detections despite the low 1.5$\sigma$ detection threshold. 
The photometric analysis of the identified sources is performed on the original images.
For the photometry, we adopted a 5-pixel diameter aperture, corresponding to 0\farcs5 and 1\farcs5 in the VIS and NISP images, 
respectively. 
This aperture, which totals 
approximately 3 times the PSF FWHM in all \Euclid bands, is sufficiently small to guarantee accurate photometry 
in moderately crowded regions of the galaxies. 
Aperture corrections from 5-pixel to large apertures  of 6\arcsec\ for VIS and 18\arcsec\ for the NISP images, totaling about 40 times 
the FWHM of the PSFs, were computed from the most isolated, bright, unsaturated stars.
The corrections amount to 
$-0.27$, $-0.11$, $-0.12$, and $-0.15$\,mag in \IE, \YE, \JE, and \HE, respectively. 
Finally, magnitudes were calibrated applying the zero points of ZP$_{\IE}$\,=\,30.13, and ZP$_\mathrm{NIR}$\,=\,30.0,
as discussed in Sect. \ref{sec:datareduction}.  
PSF-fitting photometry aimed at characterizing the resolved stellar content of the innermost star-forming regions will be presented 
in subsequent papers.

The photometric catalogs in the VIS and NISP bands were 
cross-matched 
by assigning a 1\arcsec\ maximum tolerance in separation between sources.
For every galaxy except IC\,342 and NGC\,6744, we were able to
produce a final master catalog containing only sources with photometric detections in all four bands. 
For IC\,342 and NGC\,6744, in order to achieve sufficient statistics, 
we adopted a less conservative approach, and cross-matched the \IE\ VIS band 
with only the \JE\ and \HE\ NISP bands. 
In any case, the cross-match removes the majority of spurious detections, such as cosmic rays, emission peaks on bright star spikes, 
or residual artefacts from the image reduction pipeline described in Sect. \ref{sec:datareduction}. 

\begin{figure}[!t]
\centering
\includegraphics[width=\columnwidth]{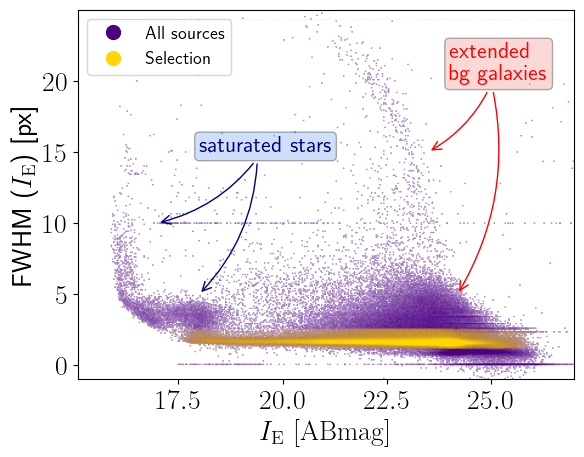} \\
\vspace{\baselineskip}
\includegraphics[width=\columnwidth]{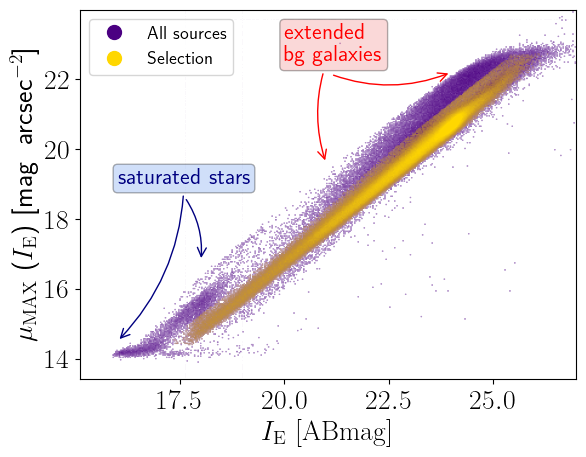}
\caption{\texttt{SourceExtractor} output parameters for NGC\,6822, intended to illustrate the typical selection cuts applied to our photometric catalogs. 
Purple points are the sources matched in all four \Euclid bands, while yellow points correspond to our selections. In the top panel, 
we retain all sources with a measured FWHM in VIS between 1.2 and 2.5 pixels, while in the bottom panel,
we show our adopted selection in the plane defined by central surface brightness versus aperture magnitude. 
Sources with a FWHM smaller than 1.2 pixels (namely smaller than the PSF) are likely artefacts, while 
sources with large FWHM values and/or large values of $\mu_\mathrm{max}$ compared to aperture photometry are either 
saturated stars or extended objects (background galaxies or resolved star clusters).}
\label{fig:selection}
\end{figure}

Additional selection cuts based on some of the \texttt{SourceExtractor} output parameters are then applied to remove saturated stars 
and extended background galaxies. 
More specifically, we retain sources that: (i)~have a measured FWHM in VIS between 1.2 and 2.5 pixels; and 
(ii)~lie within the locus populated by compact sources in the plane defined by central surface brightness ($\mu_\mathrm{max}$) 
versus aperture magnitude, as illustrated in Fig. \ref{fig:selection}. 
Objects with a FWHM smaller than 1.2 pixels (namely smaller than the PSF) are likely artefacts,
while values of the FWHM larger than 2.5 pixels have a high probability of being associated with extended objects.
Indeed, extended systems 
(such as background galaxies or resolved star clusters), as well as saturated stars, 
tend to have fainter central surface brightnesses compared to point sources with the same aperture flux. 
Nevertheless, such selection criteria are not always effective in removing very compact background galaxies from the final catalog. 
With these cuts, we are left with: 332\,900, 323\,260, 116\,551 and 30\,755 sources in the \IE--\YE--\JE--\HE\ matched catalogs of 
NGC\,6822, IC\,10, NGC\,2403, and Holmberg\,II, respectively; 
and 318\,366 and 162\,286 sources, respectively, in the \IE--\JE--\HE\ matched catalogs of IC\,342 and NGC\,6744.

\subsection{Reddening correction}
\label{sec:stellar_reddening}

Individual source magnitudes were corrected for spatially variable foreground reddening,
as described in Sect. \ref{sec:extinction} but for each source position, rather than assuming a single
value for the entire galaxy.
Also, rather than using a flat spectrum as for the integrated light, here we assume
a 5700\,K blackbody to approximate a G2V stellar spectrum in order to better emulate the emission from individual stars. 
This assumption provides relative ratios of A$_{\lambda}$/A$_V$=0.726, 0.375, 0.266, and 0.173 for \IE, \YE, \JE, and \HE, respectively. The correction is modest in the case of Holmberg\,II, NGC\,2403, and NGC\,6744, while it has a major impact on the CMDs of 
IC\,10, IC\,342, and NGC\,6822 which suffer the strongest extinctions. 
Indeed, the reddening-corrected CMDs of these galaxies exhibit, besides a global shift toward brighter magnitudes and bluer colors,  
narrower and cleaner stellar evolutionary sequences compared to the non-corrected CMDs. 

\begin{figure}[!t]
\centering
\includegraphics[width=\columnwidth]{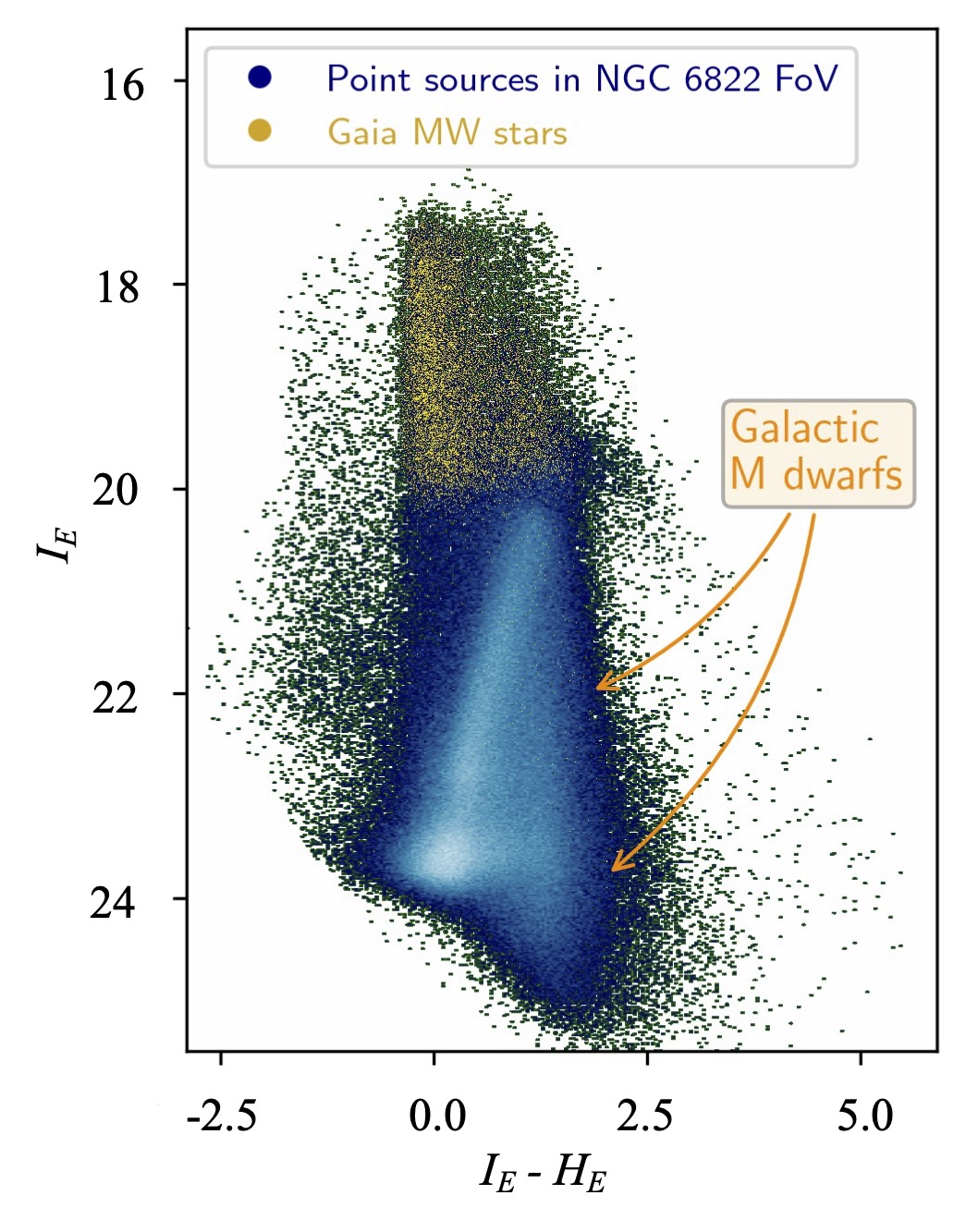}
\caption{\IE \ versus \IE-\HE\ color-magnitude diagram of all sources 
within the FoV of NGC\,6822, after applying the selection cuts described 
in Sect. \ref{sec:starphot} and illustrated in Fig. \ref{fig:selection}, 
and after correction for foreground extinction (Sect. \ref{sec:stellar_reddening}).
Yellow points indicate bright MW and background galaxy contaminants, 
namely
sources cross-matched with the \textit{Gaia} DR3 catalog that 
have a measured proper motion PM larger than $3\sigma_\mathrm{PM}$. 
The vertical feature at $\IE-\HE\simeq1.3$, $\IE\gtrsim21$ is due to the MW M dwarf population.}
\label{fig:cmd_n6822}
\end{figure}

\begin{figure}[!h]
\centering
\includegraphics[width=\columnwidth]{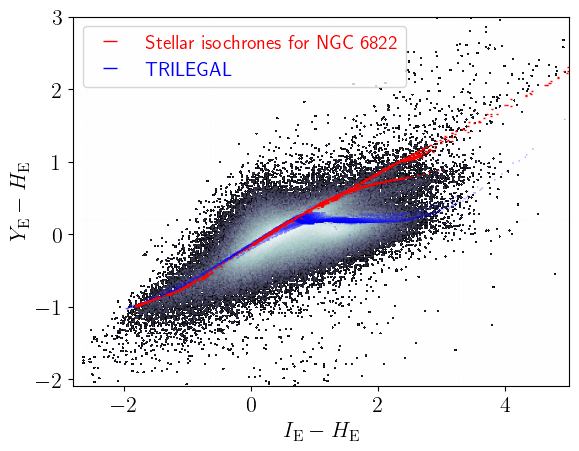} \\
\vspace{\baselineskip}
\includegraphics[width=\columnwidth]{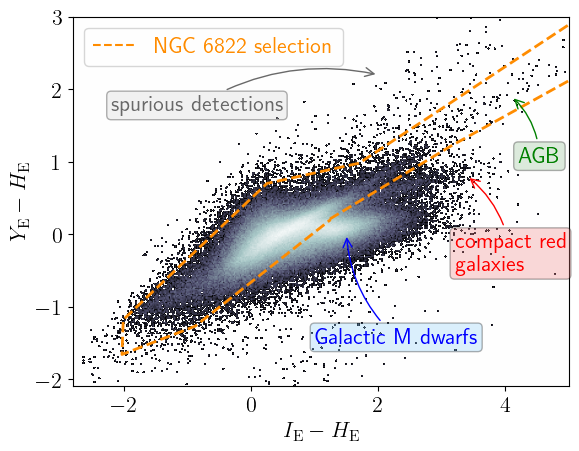}
\caption{Distribution in the \YE$-$\HE\ versus \IE$-$\HE\ plane of sources in the NGC\,6822 photometric catalog, after removal of bright \IE$\lesssim\,20$ MW disk stars in \textit{Gaia} DR3. In the top panel, the PARSEC stellar isochrones \citep{Bressan2012,Marigo2017} in the \Euclid bands are superimposed in red for different ages (from 10\,Myr to 10\,Gyr) and metallicities of $Z\lesssim\,0.006$ (about one third Solar); in blue is the TRILEGAL Galaxy model. Giant and dwarf stars in NGC\,6822 and in the MW overlap at \IE$-$\HE$\lesssim\,1$, while the two populations diverge at redder colors. 
In the bottom panel, we denote the location of Galactic M dwarf stars, AGB stars in NGC\,6822, background compact red galaxies, and residual spurious detections. 
The dashed orange polygon outlines our final selection,
which provides a reasonable compromise between the need of retaining the largest possible number of stars belonging to NGC\,6822,
while removing Galactic M dwarfs, compact red galaxies, and residual spurious detections (see text for details).
}
\label{fig:color_color}
\end{figure}

\subsection{Foreground star removal}
\label{sec:foreground_removal}

\begin{figure*}[!h]
\centering
\includegraphics[height=0.41\textwidth]{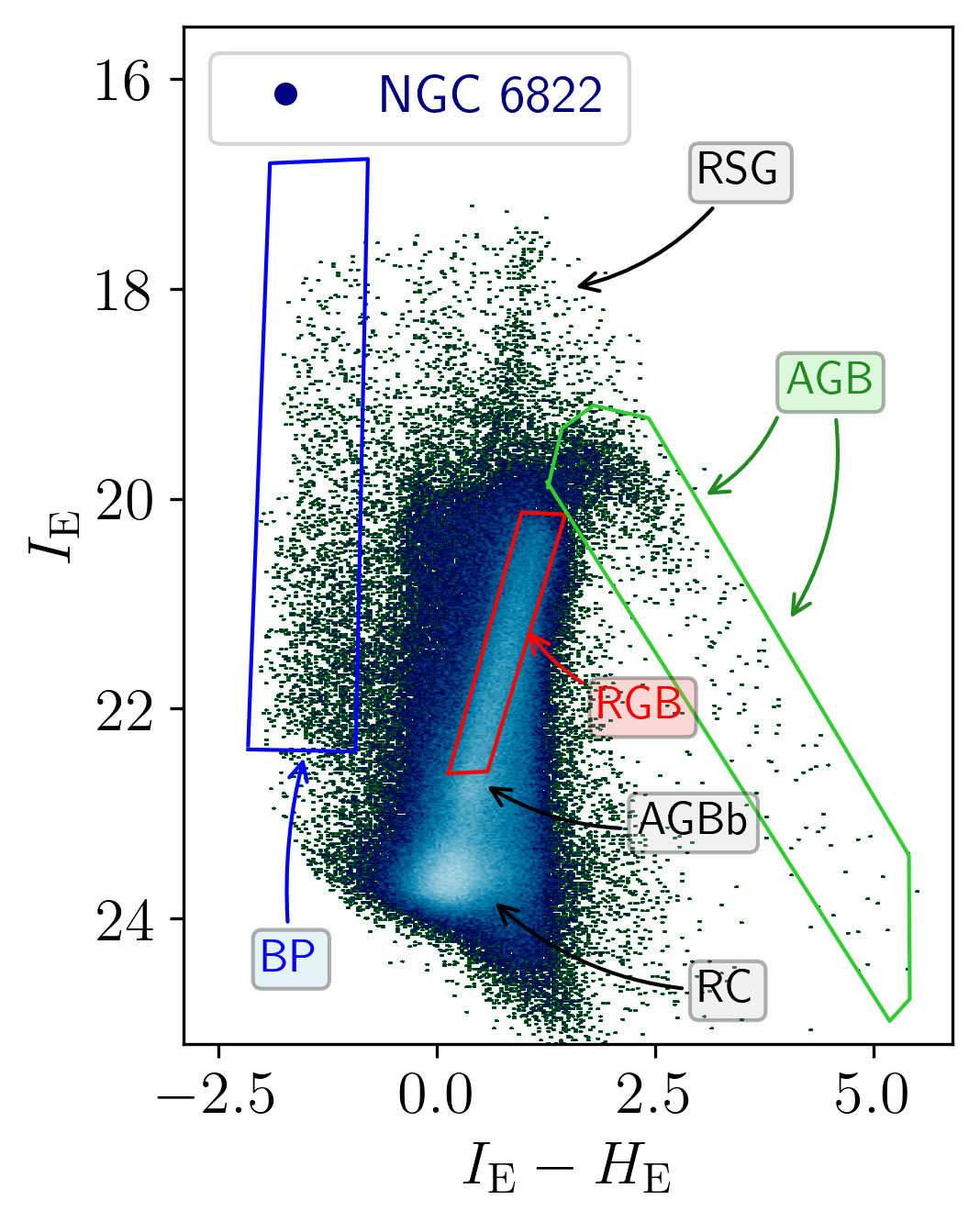}
\includegraphics[height=0.41\textwidth]{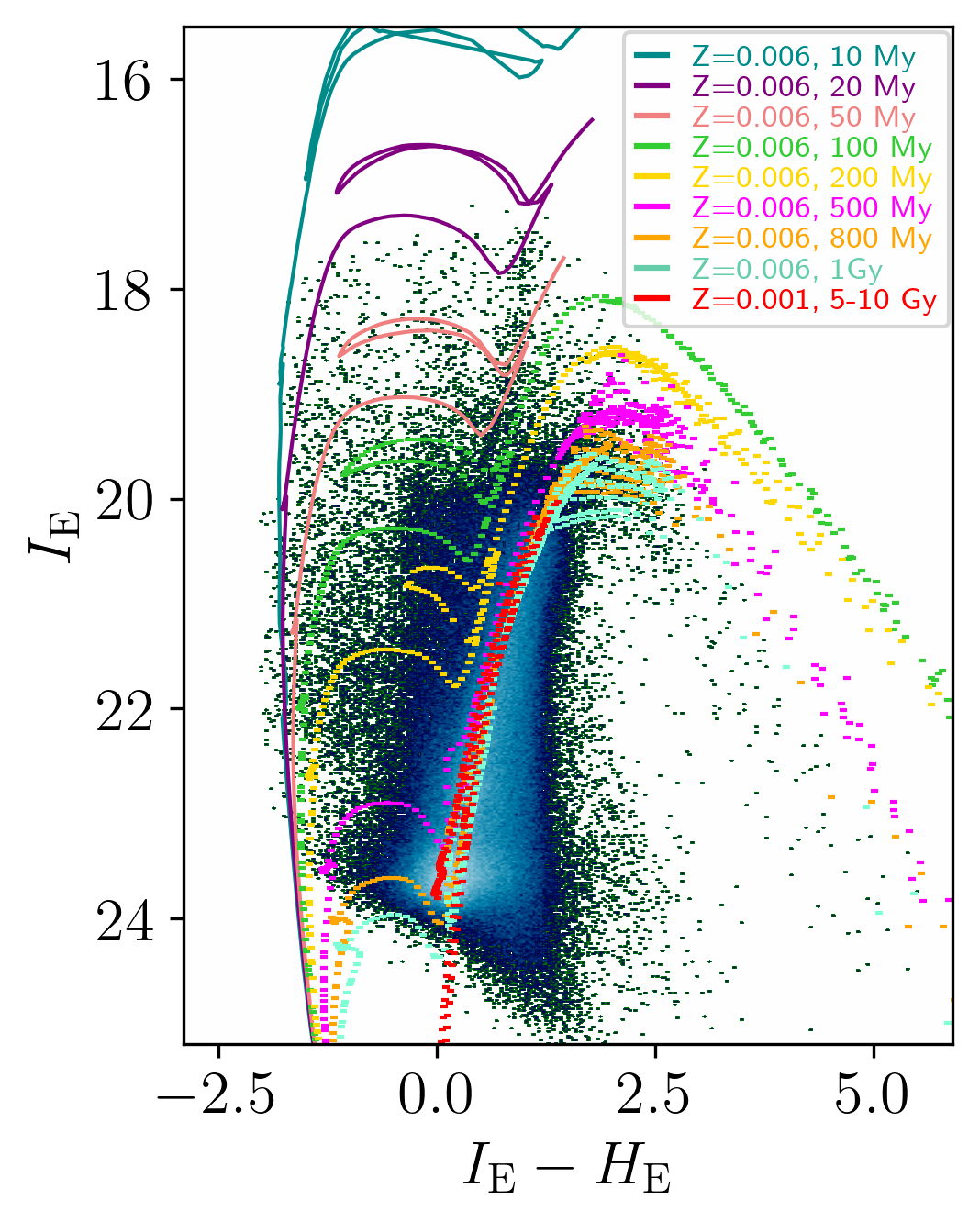}
\includegraphics[height=0.41\textwidth]{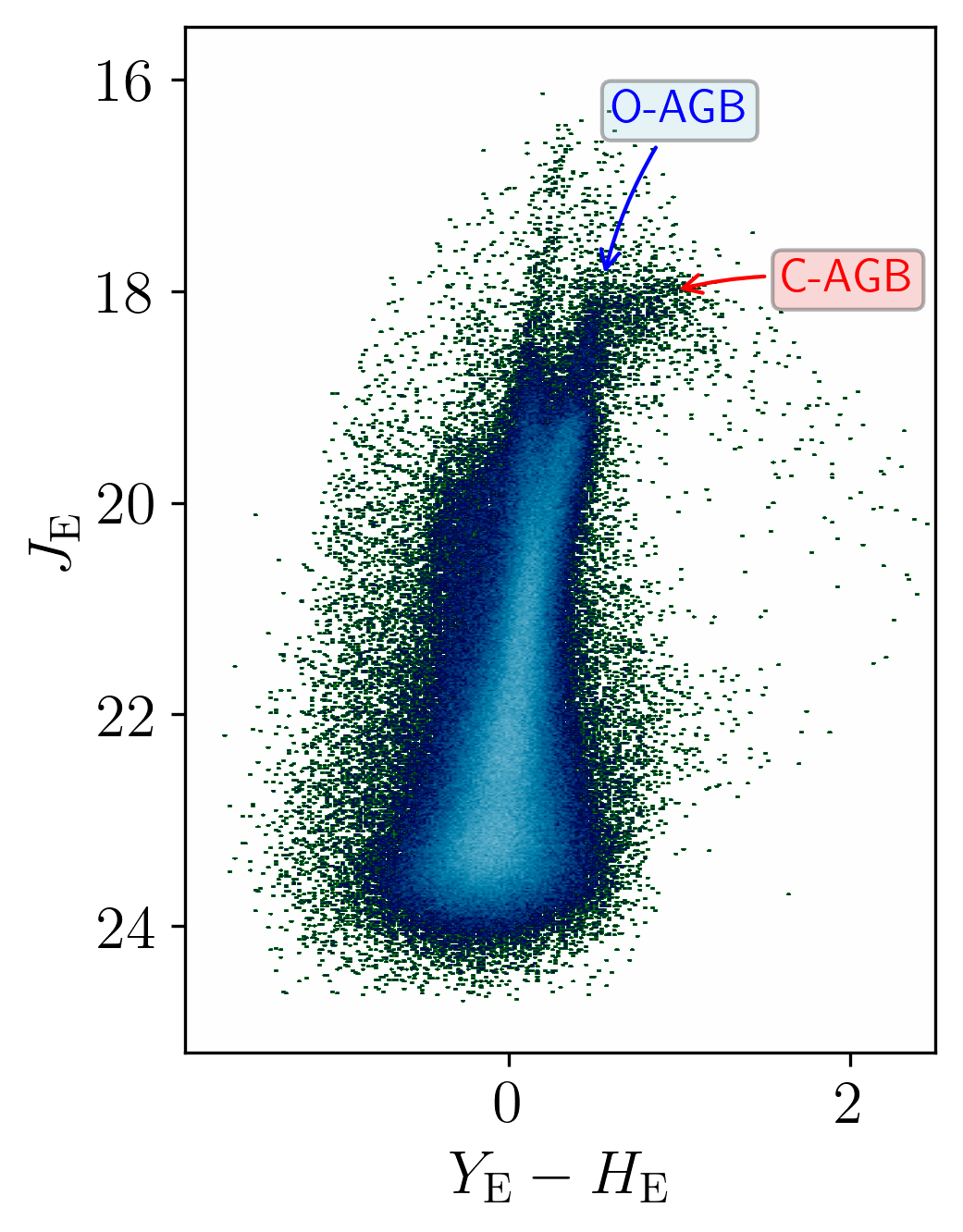}
\caption{\textit{Left}: Final calibrated, reddening-corrected, \IE\ versus \IE$-$\HE\ 
CMD of NGC~6822 
after removal of extended background galaxies, bright MW contaminants and faint Galactic M dwarf stars.
Indicated are the main stellar evolutionary sequences: 
the blue plume (BP), populated by massive MS and post-MS stars in the hot core helium-burning phase (ages $\lesssim\,100$\,Myr); 
red supergiants (RSG), with ages from 
about 20\,Myr to 50\,Myr; bright and red AGB stars with ages from $0.1$ to $2$\,Gyr; red giant branch (RGB) stars, 
with ages older than 1$-$2\,Gyr; the red clump (RC) of low-mass stars in the core-helium burning phase; and the AGB bump (AGBb). 
The blue, green, and red polygons, driven by the comparison with stellar evolutionary models, 
indicate the selection regions used to create the star count maps in Sect. \ref{sec:star_count_maps}.
as indicated in the legend. 
\textit{Middle}: \IE\ versus $\IE-\HE$ CMD (same as left panel) with 
superimposed PARSEC stellar isochrones for different ages (10\,Myr to 10\,Gyr) 
and for two metallciity values, $Z=0.006$ and $0.001$ (40\% and 6\% solar, respectively.).
\textit{Right}: \JE\ versus $\YE-\HE$ CMD. 
In this diagram, the oxygen-rich and the carbon-rich AGB stars (O-AGB and C-AGB) appear well separated and define vertical 
and horizontal sequences, respectively.}
\label{fig:3cmd_n6822}
\end{figure*}

Although the selection cuts described in Sect. \ref{sec:starphot} are effective in removing a substantial fraction 
of extended background galaxies, our photometric catalogs still suffer from major contamination due to foreground Galactic stars. 
This is evident in Fig. \ref{fig:cmd_n6822} where we show, as an illustrative example, the final calibrated, reddening-corrected \IE\ versus 
$\IE-\HE$ CMD of NGC\,6822. In the diagram, the vertical band of sources delineating a sharp edge at $\IE-\HE\simeq-0.4$ and 
extending toward the red up to $\IE-\HE\simeq1.3$ are main sequence stars from the Galactic disk, with the vertical feature at 
$\IE-\HE\simeq1.3$, $\IE\ga21$ due to the M dwarf population. 
In order to remove these contaminants, we adopt two complementary steps using 
(i)~the constraints provided by \textit{Gaia} proper motions (PMs), and 
(ii)~the implementation of additional selections based on color-color diagrams in the \Euclid bands.

In step (i), we cross-correlate our photometric catalogs with the \textit{Gaia} DR3 release \citep{gaiadr3},
adopting a 1\arcsec\ maximum tolerance in RA, Dec coordinates. 
Since the ERO Showcase galaxies have PMs compatible with zero within the errors \citep[e.g.,][]{mcconnachie21,Bennet2023}, 
likely MW members are identified, and then removed from our catalogs, 
as those having a measured proper motion PM larger than $3\sigma_\mathrm{PM}$, where $\sigma_\mathrm{PM}$ is the PM uncertainty. 
With this approach, 
we effectively remove bright foreground stars with $\IE\la20$ from our CMD. 
The removed sources are indicated as yellow points in the CMD of 
Fig. \ref{fig:cmd_n6822}. Nonetheless, it is evident that the vertical sequence of MW contaminants at 
$\IE-\HE\sim-0.4$ and $\IE\gtrsim20$ is still present in the CMD.

Next, we implement step (ii) to remove some foreground contaminants fainter than $\IE\sim20$, which do not have a counterpart in \textit{Gaia}. 
More specifically, we apply a selection in the $\YE-\HE$ versus $\IE-\HE$ plane, as shown in Fig. \ref{fig:color_color} for NGC\,6822. As illustrated in the top  panel of the figure, stars in the MW and in NGC\,6822 populate the same locus of the diagram at $\IE-\HE\lesssim1$,
because the colors of giant and dwarf stars are degenerate for early spectral types.  
Indeed, stellar isochrones \citep{Bressan2012,Marigo2017} displayed for a wide range of ages (10\,Myr to 10\,Gyr) and 
metallicities of $\lesssim40\%$\,Solar,\footnote{The PARSEC isochrones in the \Euclid bands were downloaded 
from \url{http://stev.oapd.inaf.it/cgi-bin/cmd}.} 
compatible with NGC\,6822's chemical abundance estimates \citep{Venn2001,lee06,Patrick2015}, 
completely overlap at blue colors with the TRILEGAL Galaxy model \citep{Trilegal_A, Trilegal_B}, 
so that a separation between the two components is not possible in this regime.
On the other hand, the colors of dwarf and giant stars start to diverge at $\IE-\HE\gtrsim1$, and 
Galactic M dwarfs depart from giants in 
NGC\,6822, forming a relatively bluer sequence with $-0.15\lesssim\YE-\HE\lesssim0.3$ \citep[see e.g.,][for similar classifications]{majewski03,Bentley2019}. 
The selection outlined in the bottom panel of Fig.\ref{fig:color_color} 
therefore provides a sensible strategy for the removal of a large number of MW M-dwarf contaminants.
M dwarfs belonging to NGC\,6822 are not present in our catalog because, at the galaxy distance of 
0.5\,Mpc, they are too faint to be detected.

The selection illustrated in Fig. \ref{fig:color_color} also enables the removal of a few residual spurious detections (typically located at the edge of detectors) and the contribution from compact red galaxies that survived the initial cuts based on the \texttt{SourceExtractor} parameters in Sect. \ref{sec:starphot}; 
these sources have IR colors typically redder than Galactic M dwarfs \citep[see, e.g., Fig.~2 of ][]{Bell2019} and form a separate sequence with $\YE-\HE$ colors intermediate between those defined by Galactic M dwarfs and AGB stars in NGC\,6822. 
A visual inspection of these sources in the VIS image confirms that they are compact background galaxies.  

After removal of MW contaminants, we are left with 233\,900,
199\,260, 65\,296, and 16\,928 sources in the \IE--\YE--\JE--\HE\ matched catalogs of NGC\,6822, IC\,10, 
NGC\,2403, and Holmberg\,II, respectively; 
and with 120\,747 and 112\,872 sources, respectively, in the \IE--\JE--\HE\ matched catalogs of IC\,342 and NGC\,6744.  
The surviving stars after removal of these contaminants are typically 56$-$70\% of the original sample. 
The exception is IC\,342, where the fraction drops to 
38\% due to the high foreground star contamination for this low-Galactic latitude galaxy, 
coupled with its relatively large distance (see Table \ref{tab:sample}),
which hampers the detection of its resolved stellar population.

\subsection{Identifying individual stellar populations}
\label{sec:stellar_pop}

As an illustrative example, we show in Fig. \ref{fig:3cmd_n6822} the final \IE\ versus $\IE-\HE$ and \JE versus 
$\YE-\HE$ CMDs of NGC\,6822. These diagrams present a dramatic improvement when compared to the CMD of Fig. \ref{fig:cmd_n6822}, 
since the removal of foreground and background contaminants 
unveils the presence of well-defined stellar evolutionary sequences within NGC\,6822, 
which is indicated in the left panel of Fig. \ref{fig:3cmd_n6822}: 
a blue plume (BP) at $-2\la\IE-\HE\la-1$, populated by massive main sequence (MS) stars 
and post-MS stars in the hot core helium-burning phase, with ages $\la$100 Myr; 
a vertical sequence of red supergiants (RSG) at $\IE-\HE\simeq1$, $17.5\la\IE\la19.5$, with ages from 
about 20\,Myr to 50\,Myr; bright and red ($1.2\la\IE-\HE\la5$) AGB stars with ages from about 0.1 to 2\,Gyr; 
and red giant branch (RGB) stars 
with $0\la\IE-\HE\la1.5$, $\IE\gtrsim20$ and ages older than 1--2 Gyr (and potentially as old as $\sim13$\,Gyr). Also visible at 
$-0.4\la\IE-\HE\la0.6$, $\IE\ga23.5$, towards our detection limit, is the red clump (RC) of low-mass stars in the core-helium 
burning phase, with ages $>$1--2\,Gyr. 
At $\IE-\HE\simeq0.3$, $\IE\simeq22.7$ we detect the AGB bump (AGBb). 

\begin{figure*}[h!]
\includegraphics[width=0.49\textwidth]{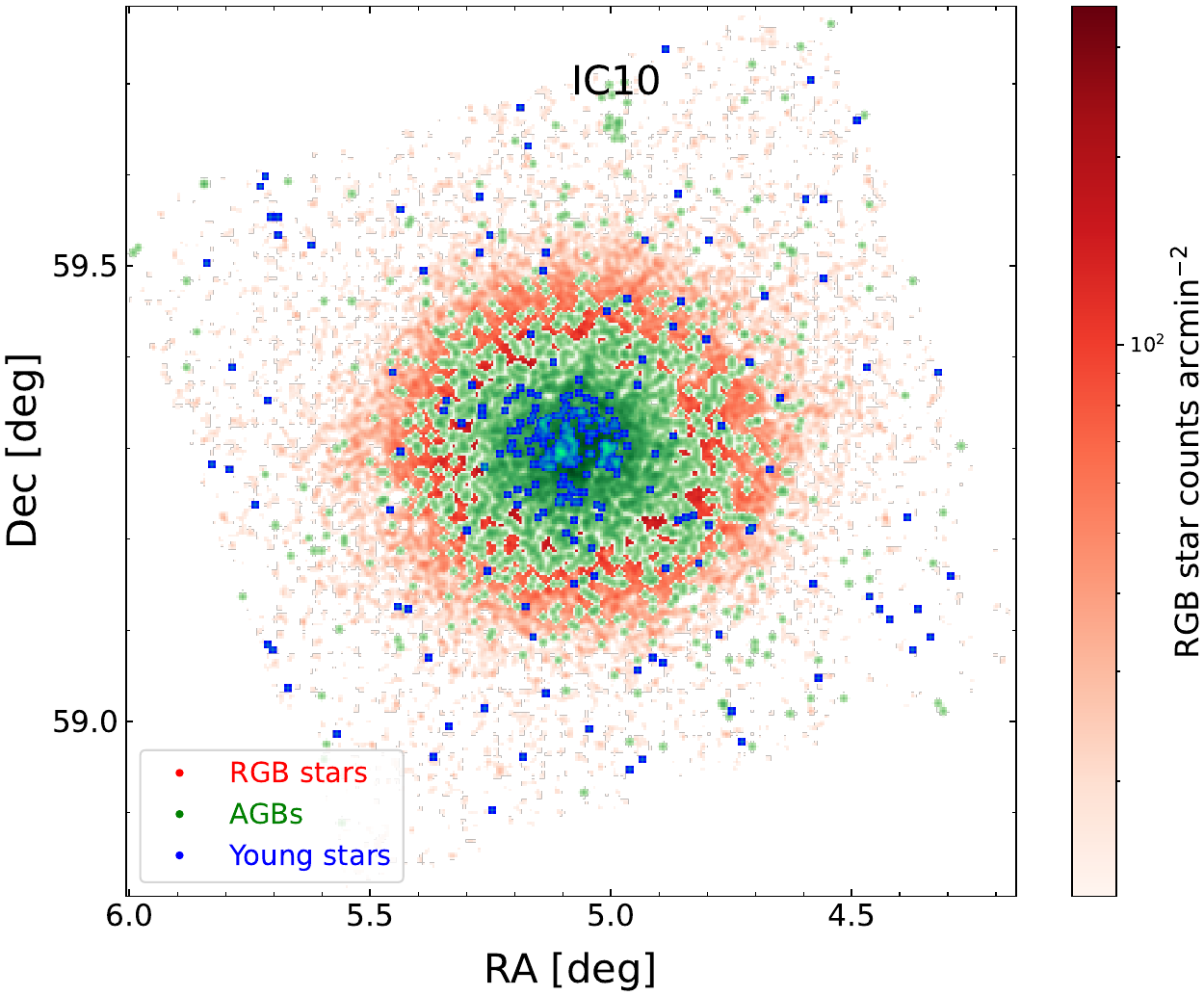}
\hspace{0.02\textwidth}
\includegraphics[width=0.49\textwidth]{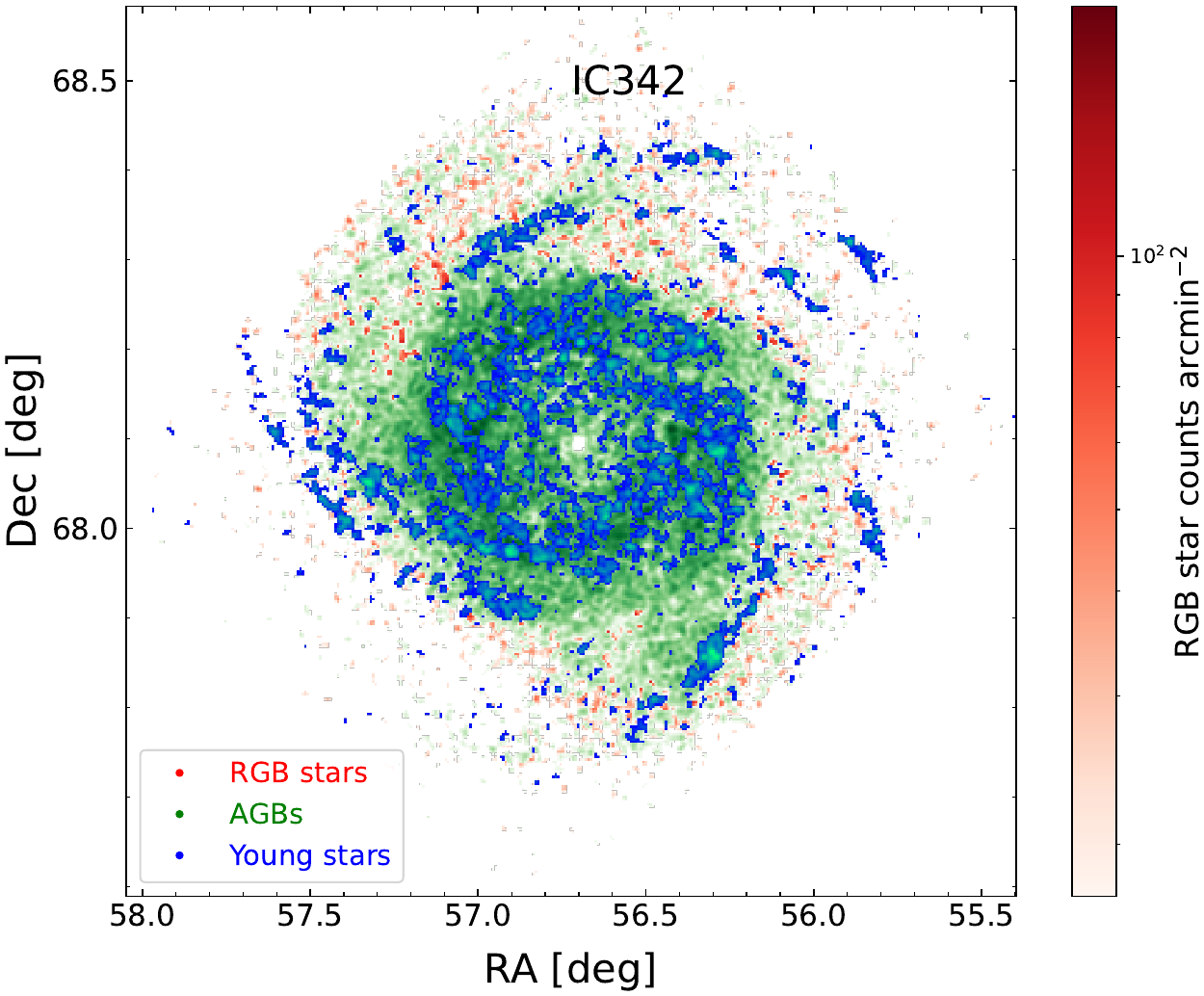}
\vspace{-1.5\baselineskip}
\caption{Maps of star counts of two Showcase galaxies divided into individual stellar populations as described in the text:
IC\,10 (left panel); and IC\,342 (right). 
Young stars are shown as blue points, AGB stars as green, and evolved RGB stars as red.
More details are given in the text.
The star-counts maps for Holmberg\,II, NGC\,2403, and NGC\,6744 are given in Appendix \ref{app:starcounts},
while those for NGC\,6822 are shown in Fig. \ref{fig:starcounts_2}.
In IC\,10 (left panel), young stars are clearly more concentrated than the older ones, and in IC\,342
(right), the young stars clearly delineate the spiral arms out to large galactocentric radii.
}
\label{fig:starcounts_1}
\end{figure*}

A direct comparison between the observed CMD and the predictions of stellar models is presented in the middle panel of 
Fig.~\ref{fig:3cmd_n6822}, 
where we overplot the PARSEC stellar isochrones \citep{Bressan2012,Marigo2017} in the \Euclid bands for stellar ages in the range 10\,Myr$-$10\,Gyr; 
isochrones younger than $\sim$1\,Gyr are displayed for a $Z=0.006$ metallicity  
(about 30\% Solar), compatible with estimates from \hii\ regions or young supergiants in 
NGC\,6822 \citep{Venn2001,lee06,Patrick2015},
while a lower metallicity of $Z=0.001$ is adopted for older populations. 
The models were shifted by applying a distance modulus of $(m-M)_0=23.54$ 
from \cite{Fusco2012}, corresponding to a distance of 510\,kpc. 
This distance is compatible with the observed RGB tip (TRGB) at $\IE\simeq20.2$; 
indeed, a recent calibration of the TRGB in the \Euclid bands based on \textit{Gaia}-DR3 synthetic photometry 
predicts an absolute value of $M_{\IE,\mathrm{TRGB}}=-3.3$ (Bellazzini \& Pascale, A\&A submitted) 
that translates into a distance modulus of $23.5$ in \IE, consistent with the 
distance from \cite{Fusco2012}, but somewhat larger than the 470-kpc distance found by \citet{weisz14}.

The right panel in Fig. \ref{fig:3cmd_n6822} presents the \JE\ versus $\YE-\HE$ CMD of NGC\,6822. In this CMD, the same stellar evolutionary sequences 
described for the \IE\ versus $\IE-\HE$ CMD can be easily identified, with the exception of the less evident RC and AGBb features. 
On the other hand, 
there is a clearer separation between O-rich versus C-rich AGB stars, with the former delineating an almost vertical sequence with 
colors $0.35\la\YE-\HE\la0.55$, and the latter forming a horizontal feature 
at  $17.9\la\JE\la18.5$ \citep[see also][for a similar classification based on JWST data]{Nally2024}. 

\begin{figure*}[h!]
\centerline{
\includegraphics[width=0.49\textwidth]{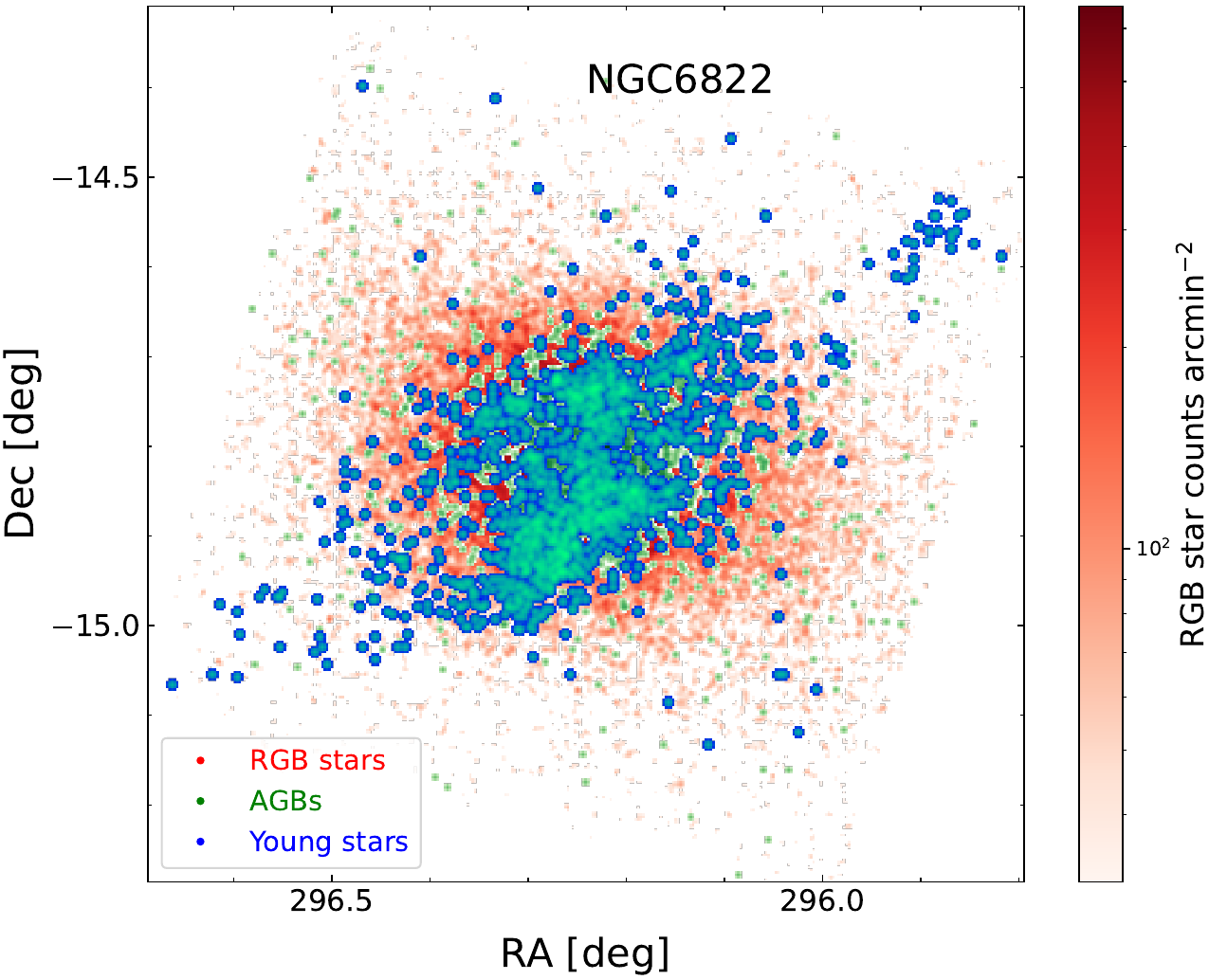}
\hspace{0.04\textwidth}
\includegraphics[width=0.49\textwidth]{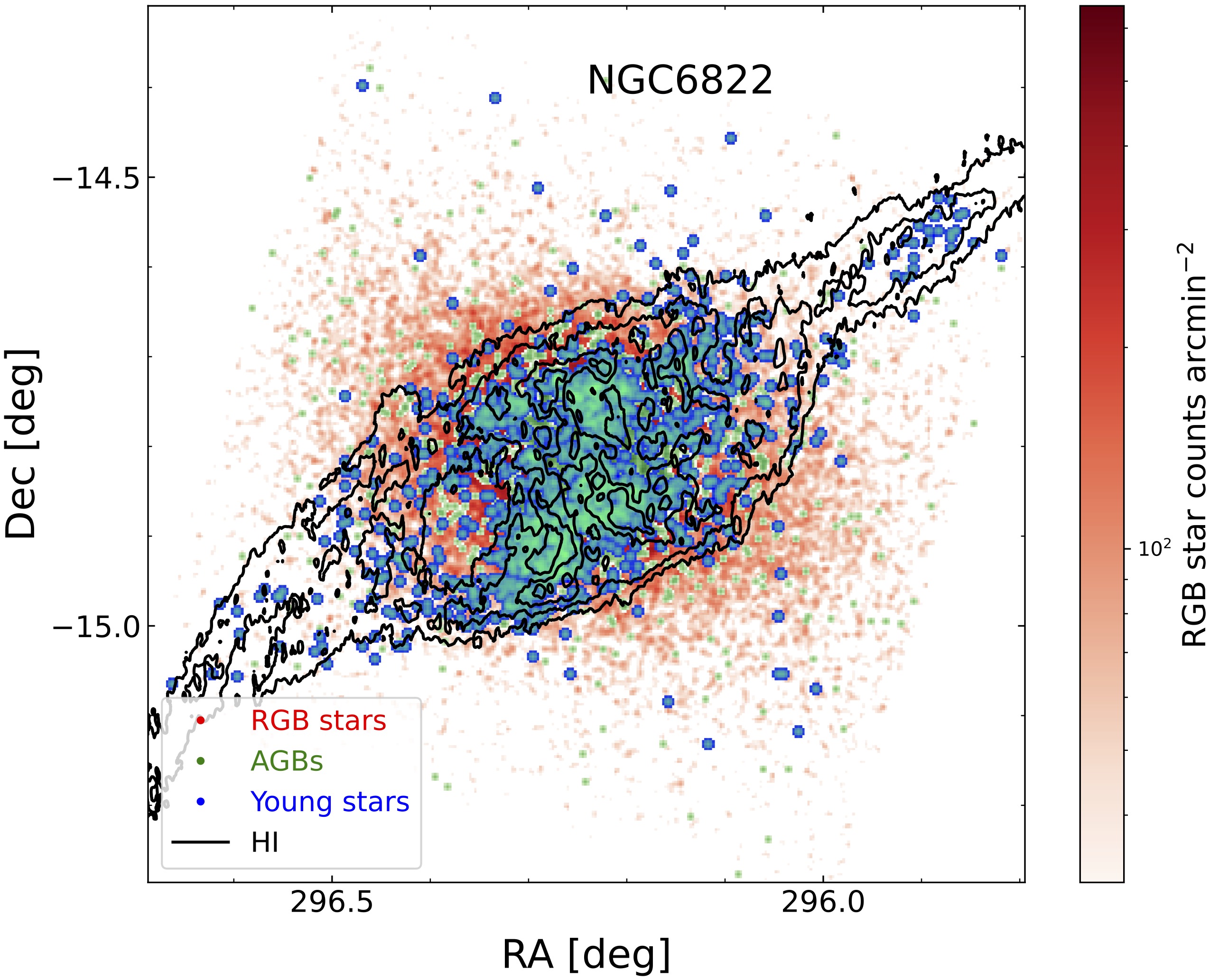} \\
}
\caption{Star-count maps obtained as for Fig. \ref{fig:starcounts_1}, but for the closest Showcase
galaxy, NGC\,6822 (left panel), and with \hi\ contours overlaid on the maps (right). 
\hi\ data are taken from \citet{deblok00}; 
contours are at 2$\sigma$, 4$\sigma$, 7$\sigma$, 10$\sigma$, and 20$\sigma$.
The young stars in NGC\,6822 are clearly aligned with the atomic gas, as discussed in the text.
}
\label{fig:starcounts_2}
\end{figure*}

\subsection{Star-count maps}
\label{sec:star_count_maps}

The features outlined by the polygons in the left panel of Fig.~\ref{fig:3cmd_n6822} were used to select stellar populations 
for different age intervals in the six Showcase galaxies. 
These selections were slightly adapted to account for the different depths sampled, 
depending on the galaxies' distances. 
The 
age selection thus enabled the construction of
star count maps in different age intervals.    

The final maps were smoothed by convolving with a Gaussian kernel the two-dimensional histograms.
Subtraction, in a statistical sense, of foreground stars or background objects that do not belong to the 
galaxies was performed by 
removing the density of counts computed in  regions at large galactocentric distance, where it is reasonable to assume 
a negligible presence of the intrinsic stellar populations from the galaxy. 
However, this may lead to an over subtraction of the background in the case of NGC\,6822, for which  \citet{zhang21} 
demonstrated that the stellar component extends at faint levels to beyond the Euclid FoV. 
The star counts within these regions were calculated, together with their standard deviation, $\sigma_\mathrm{bck}$.
The final maps were constructed by only considering the star counts that exceeded $\sigma_\mathrm{bck}$ 
by a given signal-to-noise: S/N\,=\,5 for young stars; and S/N\,=\,3 for RGB and AGB stellar populations, 
in order to better highlight low surface brightness structures in  the old stellar component. 

The stellar populations in the Showcase galaxies will be examined in detail in
future papers, with a careful analysis of completeness limits, local extinction corrections,
and galaxy membership. 
Here, we present the results of the preliminary analysis described above.
Figure~\ref{fig:starcounts_1} shows smoothed maps of the star counts for IC\,10 and IC\,342, while
NGC\,6822 is presented in Fig.~\ref{fig:starcounts_2}.
The remaining maps can be found in Appendix \ref{app:starcounts}; 
we have obtained star counts with \Euclid even for NGC\,6744, 
which has the largest distance in the Showcase sample (9\,Mpc),
although there we mostly detect young stars, AGB stars, and blends of bright RGB stars.

The maps of the giant spirals, IC\,342 and NGC\,6744, show that young stars 
follow closely the spiral structure well into the outer disk, exemplifying the notion that spiral arms tend to be the sites
of recent star formation \citep[e.g.,][]{gerola78,roberts84,wada11}.
Similar behavior is also seen in M\,33, a flocculent spiral \citep{lazzarini22}, 
M\,81 \citep{williams09a,okamoto15}, and  NGC\,6946 \citep{tran23}. 
The AGBs in IC\,342 also follow the spiral arms, but tend to be more broadly distributed,
possibly implying the lack of a systematic time delay in the SFH across the arm,
similar to the case of M\,81 \citep[e.g.,][]{choi15}.

Conversely, rather than tracing the flocculent spiral structure,
the young stars in NGC\,2403 (see Appendix \ref{app:starcounts}) are more
uniformly distributed across the disk \citep[see also][]{barker12}.
This would imply that the young stellar disk in NGC\,2403 has been relatively undisturbed
out to a galactocentric radius $\ga 10$\,kpc, 
similar to its morphological twin, NGC\,300
\citep[e.g.,][]{hillis16,jang20b}. 

Figures \ref{fig:starcounts_1} and \ref{fig:starcounts_2} show that in the dwarf galaxies, IC\,10 and NGC\,6822, 
the AGB stars are more centrally concentrated than the RGBs,
a common \citep[e.g.,][]{gerbrandt15}, 
but not inevitable \citep[e.g.,][]{bernard12},
feature in dwarf irregular galaxies.
It is more difficult to characterize the stellar populations in 
IC\,10 than the other galaxies in the Showcase,
because of large foreground extinction and contamination by foreground stars \citep[e.g.,][]{massey07}.
We find that the stellar distribution in IC\,10 is quite extended
in roughly a circular morphology (see also Fig. \ref{fig:hioverlay}),
consistent with \citet{gerbrandt15}.
The young-star counts are even more centrally concentrated than the AGB stars,
with the AGB stars possibly showing 
possibly showing more of a flattened distribution in the central regions.
It is interesting to speculate that this feature could be a signature of
a past star-formation event.
According to \citet{weisz14}, in IC\,10, like other dIrr galaxies,
the majority have stars have formed over the last 2--3\,Gyr;
some of these are most likely the progenitors of the AGB population in IC\,10
\citep[see also][]{dellagli18}.
Maps of the atomic gas (see Fig. \ref{fig:hioverlay}) 
suggest that IC\,10 has undergone an interaction in the
past, or is currently accreting gas \citep[e.g.,][]{shostak89,nidever13,ashley14,namumba19}. 

\begin{figure*}[t!]
\centering
\includegraphics[width=0.245\textwidth]{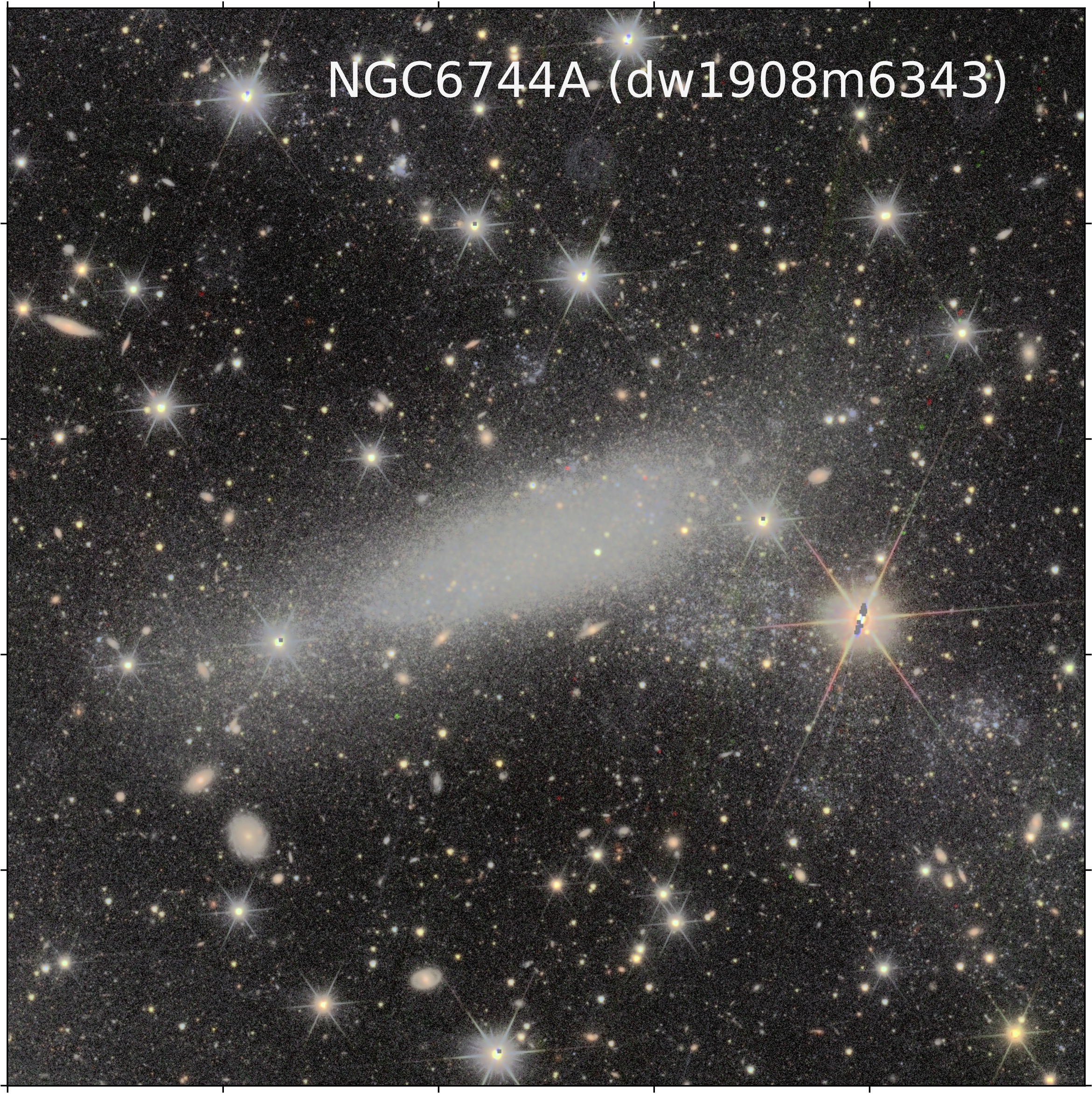}
\includegraphics[width=0.245\textwidth]{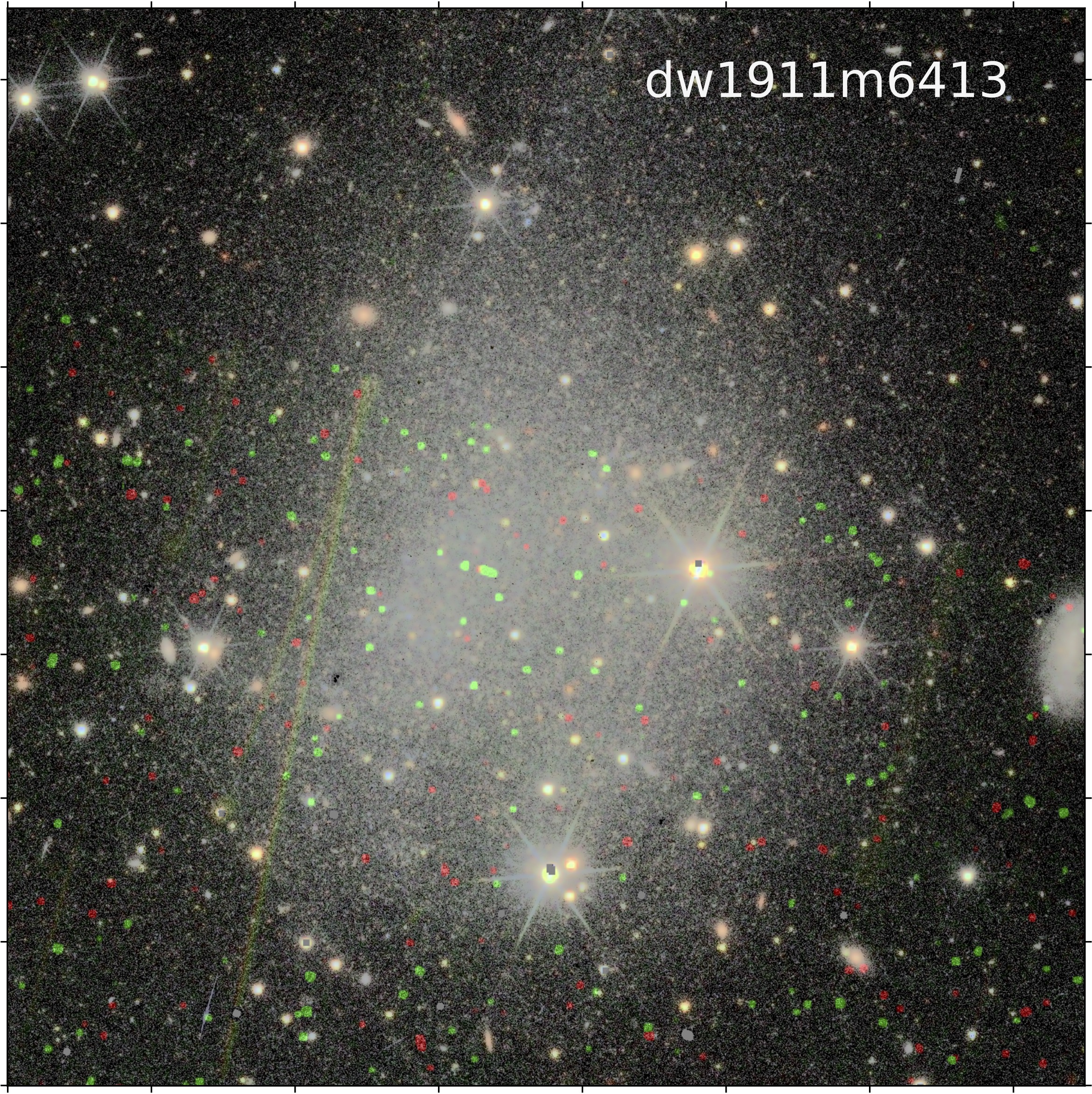}
\includegraphics[width=0.245\textwidth]{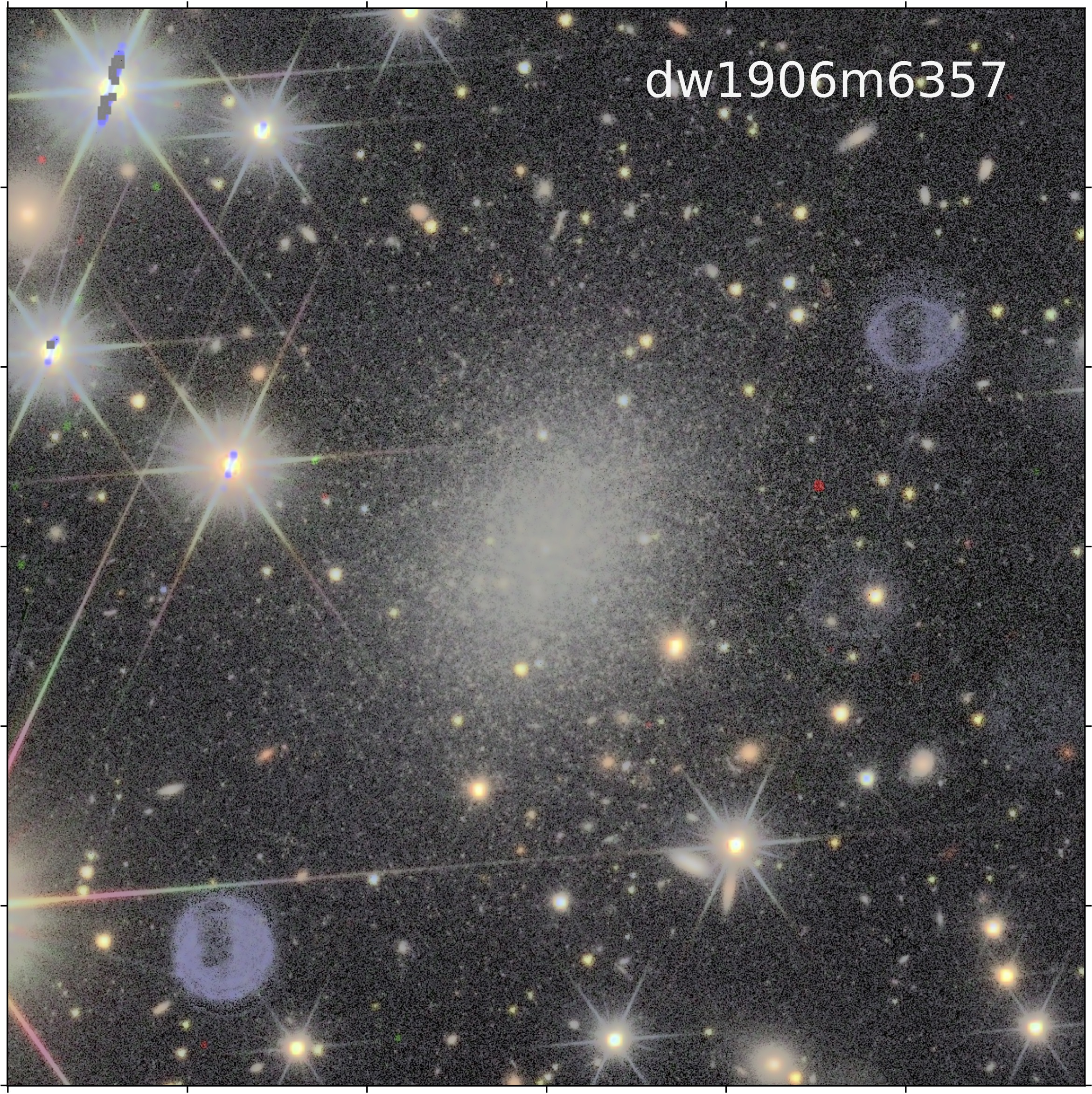}
\includegraphics[width=0.245\textwidth]{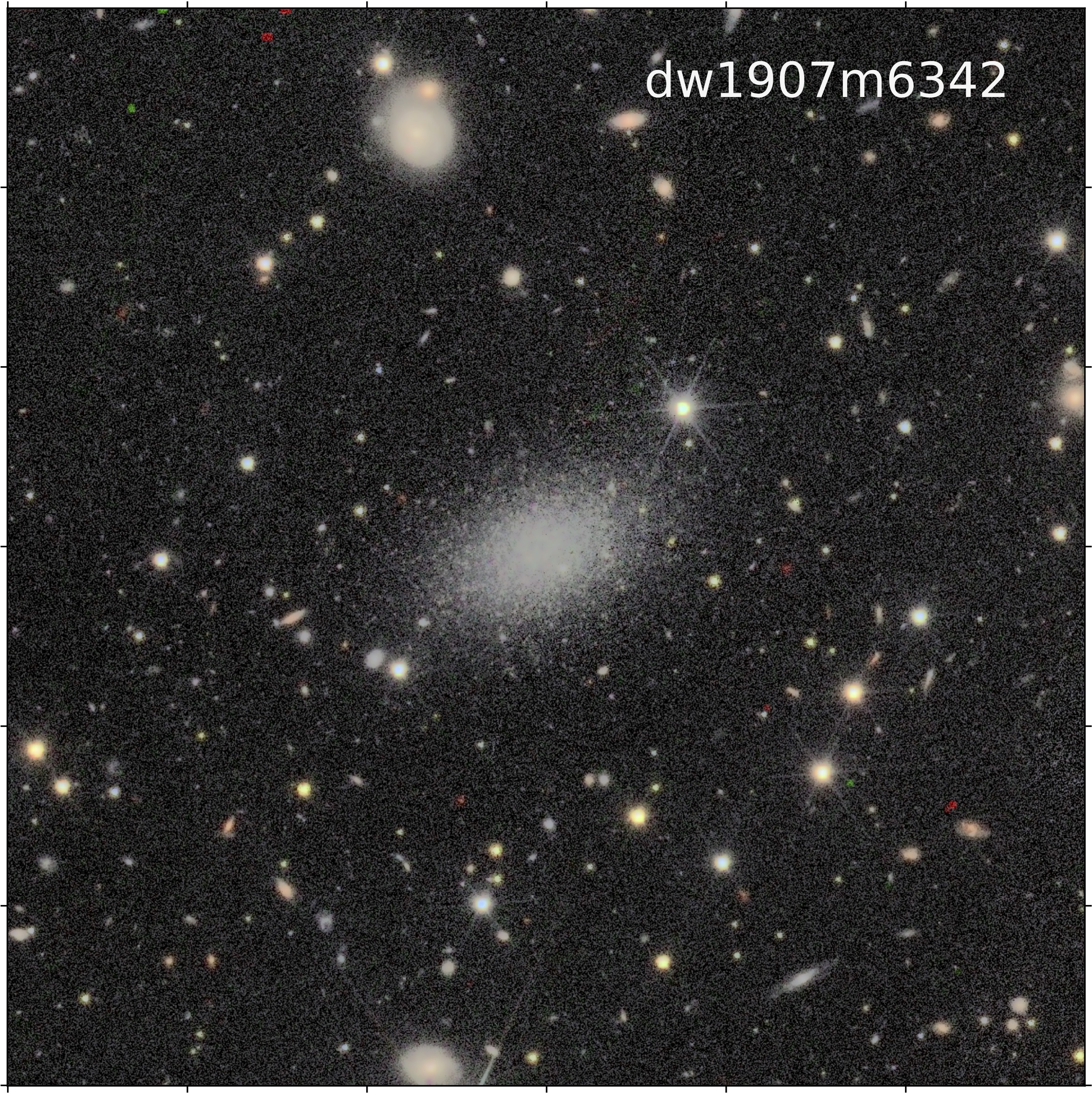}
\caption{\Euclid RGB images of the four previously known dwarf satellites from \citet{carlsten22} that fall in the field of view of the 
\Euclid image of NGC\,6744. Red is given by \HE, green by \YE, and blue by \IE.
The sizes of the images are 250\arcsec, 150\arcsec, 120\arcsec, and 120\arcsec\ on a side, from left to right; north is up, east to the left.
With the exception of NGC\,6744A, a well-known dwarf satellite lying along spiral arms to the northwest of the nucleus,
name designations are from \citet{carlsten22}.
In the cut-out of dw1906m6357 (third panel from the left), there are also two
blue circular structures that are dichroic ghost artefacts (see also Fig. \ref{fig:pretty_holmbergii}).
}
\label{fig:dwarf1-2}
\end{figure*}

The configuration and orientation of the RGB and AGB populations in
NGC\,6822 shown in Fig. \ref{fig:starcounts_2} agree with previous maps
\citep[e.g.,][]{demers06,sibbons12,tantalo22}.
In NGC\,6822, the young stars are oriented roughly along the \hi\ emission, as illustrated
by the overlay of \hi\ in the right panel of Fig. \ref{fig:starcounts_2}. 
Young stars are found where the \hi\ has higher column density, but are also 
present in the \hi\ cavities, such as the `hole' toward the southeast of the nucleus.
This orientation of the young population was also found 
previously \citep[e.g.,][]{komiyama03,deblok03,zhang21},
implying that high \hi\ column density may foster star formation.
However, NGC\,6822 is kinematically complex, 
with a potentially counter-rotating component both in the gas and the stars
\citep[e.g.,][]{deblok06,belland20},
so the connection of the \hi\ with star formation may also be influenced by kinematics.
Figure \ref{fig:starcounts_1} shows that the combined stellar morphology produces
an `X'-like configuration of the overall stellar content, with the young stars
oriented along a NW-SE direction, like the \hi, and the older RGBs elongated along a NE-SW direction;
such a configuration is consistent with that found in previous studies \citep[e.g.,][]{komiyama03,deblok03,zhang21}.
There is still debate about whether the unusual properties of NGC\,6822 have been caused
by a prior merger or stellar feedback \citep[e.g.,][]{deblok00,demers06,cannon12,belland20,zhang21},
but the case of NGC\,6822 illustrates the potential of \Euclid to contribute to this debate.
In summary,
\Euclid\ will be a powerful tool for further constraining the past history of IC\,10, NGC\,6822,
and other dwarf galaxies that will be observed during its lifetime, as well as 
assessing the SFH of more massive disk galaxies and the origin and longevity of spiral arms. 


\section{Dwarf satellites around NGC\,6744 }
\label{sec:dwarfs}


The unprecedented combination of low surface brightness sensitivity, high spatial resolution with a pristine PSF, 
and wide-area coverage of \Euclid enables the detection and characterization of the low surface brightness dwarf 
satellites around their host galaxies, as well as the simultaneous study of their nuclear star clusters and globular cluster systems. 
To demonstrate the capability of \Euclid to investigate the satellite systems of nearby galaxies, we visually identified the dwarf galaxies in the \Euclid Showcase fields. Here, we present some highlights for NGC\,6744. 

\subsection{Known dwarf satellites of NGC\,6744}

It is common to find ensembles of dwarf satellites around nearby galaxies \citep[e.g.,][]{karach14}.
The dwarf satellite system of the host galaxy NGC\,6744 was explored
in the context of the ELVES (Exploration of Local VolumE Satellites)
survey by \citet{carlsten22}, which confirmed 338 satellites with
absolute magnitude $M_V < -9$~mag and central surface brightness
$\mu_{0,V} < 26.5$\,\magarc\ in the vicinity (the majority within
$300$\,kpc) of 
30 host galaxies in the local volume ($D < 12$\,Mpc). 
In particular, the galaxy NGC\,6744 was found to have 15 dwarf satellite candidates. 
Of these, 
five were confirmed via SBF measurements \citep{carlsten19} or 
other methods, 
four were rejected via SBF, and six remained unconfirmed.

\subsection{Visual identification of new satellites}

In the ERO Showcase field of NGC\,6744, we first identified the dwarf candidates using a combination of the high-resolution 
VIS image and the lower resolution (by a factor of 3) VIS+NISP color image. 
\texttt{Jafar}, an on-line visualization and annotation tool that makes use of the CDS Aladin 
lite facility,\footnote{\url{https://aladin.cds.unistra.fr/AladinLite/}} was used for 
identifying and labelling the dwarfs \cite[see][]{sola22}.
The use of the color image was crucial as artefacts of the optical system, 
the so-called `optical ghosts', appear as faint small round regions in the VIS image and look very similar to dwarf galaxies. 
However, since they are more prominent in the VIS image than in NISP, they have a very distinctive fuzzy blue color in the image, 
unlike the real dwarfs (see Fig. \ref{fig:dwarf1-2}). 

Of the five confirmed dwarfs of \citet{carlsten22}, 
only four fall in the ERO field of view and thus are also found in our \Euclid NGC\,6744 dwarf catalogue, as shown 
as RGB images in Fig. \ref{fig:dwarf1-2}. 
The second galaxy from the left is close to the edge of the image, resulting in slightly more artefacts in NISP.

\begin{figure}[h!]
\centering
\includegraphics[width=\columnwidth]{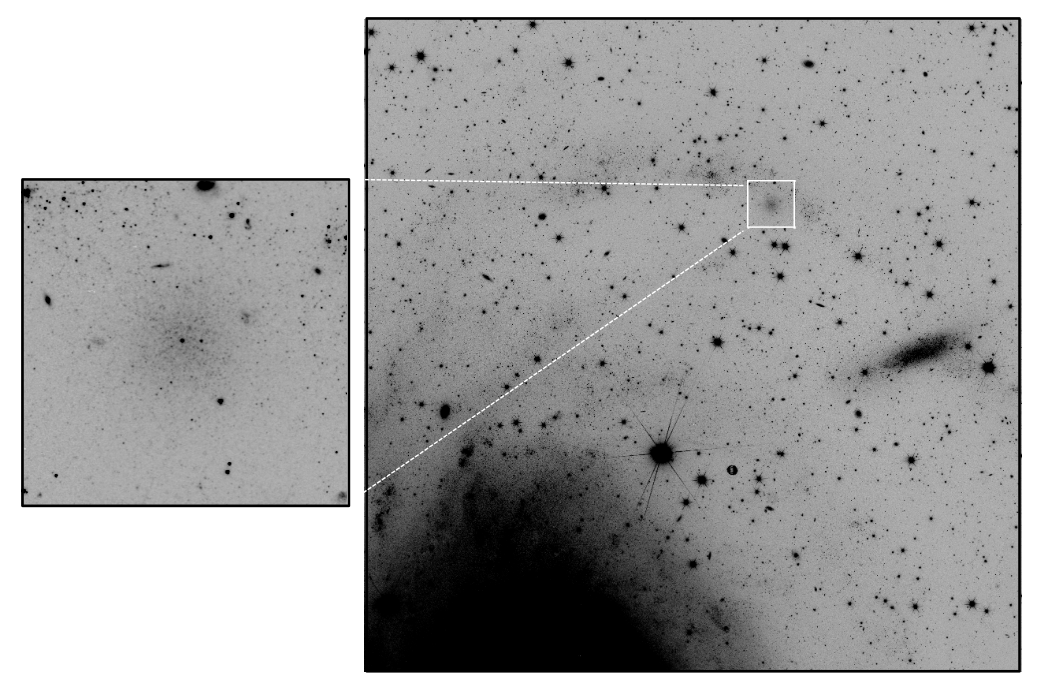}
\caption{Newly discovered dwarf satellite of NGC\,6744; the image on the left is 50\arcsec$\times$50\arcsec. 
The larger 12\arcmin\,$\times$\,12\arcmin\ image on the right
shows its location near a dense stellar region that belongs to one of the spiral arms of the galaxy. 
The images are oriented with north up, and east to the left.
}
\label{fig:dwarf3}
\end{figure}

The capability of \Euclid to detect and characterize new populations of dwarf (satellite) galaxies is highlighted by 
the detection of a new dwarf satellite candidate in this ERO field, shown in Fig. \ref{fig:dwarf3}. 
The identification of a previously undiscovered companion
was unexpected, since NGC\,6744 is a nearby spiral whose dwarf satellite system has been well studied.
The new candidate galaxy is designated as EDwC1 (\Euclid Dwarf galaxy Candidate, number 1),
and located at 
$\alpha\,=\,$\ra{19;09;08.33}, 
$\delta\,=\,$\ang{-63;41;07.9} 
(J2000).  
EDwC1 is a nucleated dwarf elliptical with 
an absolute \IE-band magnitude of $-12.2$,
assuming a distance of 8.8\,Mpc, 
with an effective radius of 0.4\,kpc and a surface brightness within one effective radius of 24.1\,\magarc, 
placing it clearly within the scaling relations of the classical dwarf regime. 
It is located at the end of a spiral arm and therefore has avoided past detection, most likely because of the 
high stellar density in this region. 
However, the combination of low surface brightness sensitivity and high spatial resolution of \Euclid 
allows us to easily detect such galaxies, even in crowded regions.

\subsection{Resolved stellar populations of dwarf satellites}

\begin{figure}[h!]
\centering
\includegraphics[width=\columnwidth]{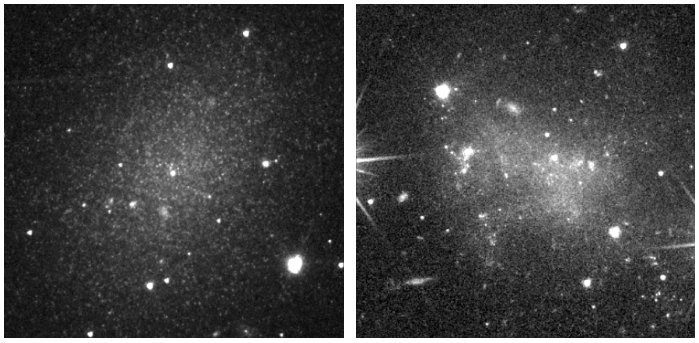}
\caption{Comparison of the zoomed-in images of one of the known satellites (\textit{left panel}, dw1906m6357) 
shown in Fig. \ref{fig:dwarf1-2} 
and a galaxy (dw1912m6351, \textit{right panel}) identified as a background object based on 
surface brightness fluctuation measurements by \citet{carlsten22}.
The 
sharp PSF of \Euclid allows us to clearly see that the stellar population is resolved in the image 
on the left (satellite), but not in the image on the right (background galaxy). The images are 40\arcsec\ on a side,
with north up, and east to the left.}
\label{fig:dwarf4}
\end{figure}

We compare the visual appearance of the images of a known satellite, dw1906m6357 \citep[see Fig. \ref{fig:dwarf1-2},][]{carlsten22}, 
and one of the galaxies identified by \citet{carlsten22} to be a background contaminant, dw1912m6351.  
An example of a zoomed-in image of dw1906m6357 
is shown in the left panel of Fig. \ref{fig:dwarf4},
compared with 
dw1912m6351 (right panel),
identified as a background object based on surface brightness fluctuation measurements.
The background object is shown at the same zoomed-in scale, but is characterized by unresolved stellar light.
This illustrates the capability of \Euclid to identify clearly satellites of massive galaxies at the distance of NGC\,6744. 
As can be seen in Fig. \ref{fig:dwarf3}, the stellar population of the new dwarf candidate is also resolved, 
suggesting that the new dwarf is indeed a satellite of NGC\,6744.  


\section{Extragalactic globular cluster candidates in NGC\,2403 }
\label{sec:egcs}


The investigation of EGCs with \Euclid  
will revolutionize our understanding of their properties, and the constraints they impose on hierarchical galaxy formation.
Here we present preliminary results for the EGCs around NGC\,2403,
while the GC and star-cluster populations of other Showcase galaxies will be discussed in future papers
(Howell et al. 2024, in prep; Larsen et al. 2024 in prep).

\subsection{Known clusters and cluster candidates}

The first discussion of star clusters in NGC\,2403 dates back to \citet{Tammann1968}, 
who presented a list of four candidates that they had identified on photographic plates from the Hale 200-inch telescope.
They noted that one of their candidates, C4, `could well be a globular cluster such as $\omega$ Cen or 47\,Tuc', 
while C1--C3 had blue colors resembling those of young star clusters in the Milky Way. A fifth object, C5, was associated with an 
\hii\ region.
The current list of spectroscopically confirmed old ($>$ several Gyr) GCs in NGC\,2403 consists of seven objects: 
C4, D6, F1, F16, F46, JD1, and JC15 \citep{forbes22,Larsen2022}.  
Objects with IDs starting with a `C' are from the original list by \citet{Tammann1968}, 
with a `D' referring to candidates identified by \citet{Davidge2007}, and 
`F' to \citet{Battistini1984}, while clusters JD1 and JC15 were identified by \citet{forbes22}.

\begin{figure}
\noindent Known GCs: \\
\includegraphics[width=\columnwidth]{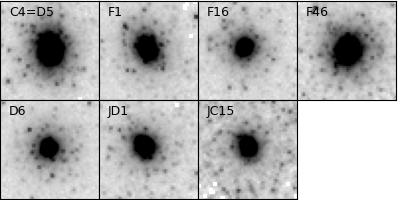} 
\noindent Known young clusters / HII regions: \\
\includegraphics[width=\columnwidth]{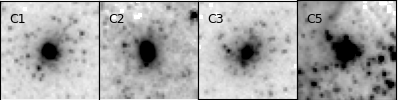} 
\noindent Objects identified as non-clusters: \\
\includegraphics[width=\columnwidth]{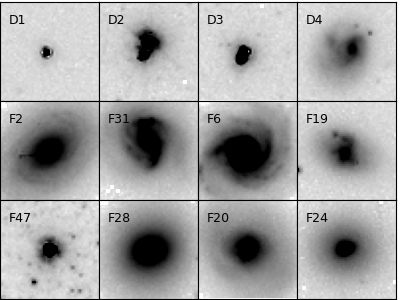}
\caption{\Euclid VIS images of known GCs in NGC\,2403 (top), young clusters (middle), and non-clusters (bottom). Each 
image shows a $5\arcsec\times5\arcsec$ cut-out around each object. North is up, and east to the left in all images.}
\label{fig:littgcs}
\end{figure}

The top panel in Fig.~\ref{fig:littgcs} shows $5\arcsec\times5\arcsec$ cut-outs from the \Euclid VIS image around the seven 
confirmed NGC\,2403 GCs. The cut-outs show that all of these clusters are resolved into individual stars in their outer parts, 
demonstrating \Euclid's potential for revealing additional candidates. 

A number of other objects have been discussed as potential GCs in the literature. 
The middle panel in Fig.~\ref{fig:littgcs} shows the four bluer objects from \citet{Tammann1968}.
The spectra of C1 and C3 exhibit strong Balmer absorption lines, confirming these objects as young clusters, 
with ages of a few hundred Myr \citep{Battistini1984,forbes22}, while \citet{Larsen1998} find an age of about 250\,Myr 
for C2 based on its $UBV$ colors. 
By comparing the images of these young clusters with those of the old GCs, 
it is evident that morphology alone is not sufficient to identify old GCs as such. 
The bottom group of cut-outs in Fig.~\ref{fig:littgcs} shows various candidates that clearly are not stellar clusters. 
The spectra obtained by \citet{forbes22} already revealed D2 to be a foreground star. 
The \Euclid VIS image shows that this source (as well as D3) is actually composed of two stars separated by less than $1\arcsec$, 
which explains the non-stellar appearance in 
the ground-based images from \citet{Davidge2007}.
The VIS image also shows that D1 is a single star, while D4 is a background galaxy.  
None of the additional non-confirmed GC candidates from \citet{Battistini1984} are stellar clusters. 
F47 has a non-zero parallax of $0\farcs88\pm0\farcs12$ according to \textit{Gaia} DR3, 
which demonstrates that it is a Milky Way foreground star, while the remaining objects are background galaxies. 
All coordinates listed in this paper are based on the \Euclid VIS astrometry.\footnote{There are notable discrepancies 
for many of their candidates between the coordinates listed by \citet{Battistini1984} 
and the locations given on their finding chart.}

\subsection{Searching for new clusters}

Having established that the VIS images allow star clusters in NGC\,2403 to be identified based on their resolution into individual stars, 
we carried out a search for additional cluster candidates. 
While we expected clusters to be identifiable from a careful, visual inspection of the images, 
the sheer size of the \Euclid VIS image
($36\,000\,\mathrm{pixels}\times36\,000$\,pixels)\footnote{The original VIS detector has  $6\times 6$ chips with $4000^2$ pixels each,
but the ROS dithers and the resulting resampled stacks increase the size of these images from the original $24\,000\times 24\,000$ pixels.}
made it necessary to first apply a preselection 
to reduce the number of candidate sources that were then subjected to visual inspection. 
To this end, we first generated a source catalogue by running \texttt{SourceExtractor} on the VIS images with relatively 
conservative extraction settings: \texttt{DETECT\_MINAREA}=6 and \texttt{DETECT\_THRESH}=8, meaning that a source 
detection requires six connected pixels, each with a signal of at least 8$\sigma$ above the background noise.
This produced an initial catalogue of 
392\,099 sources.
We also ran \texttt{SourceExtractor} on the NISP images, using the \YE-band data 
for source detection 
while carrying out photometry on all three NISP images. 
The magnitudes were measured within circular apertures with diameters of 20 pixels on the VIS frame and 
6.7 
pixels on the NISP frames 
(in both cases corresponding to an aperture radius of $1\arcsec$), as well as in Kron-like (AUTO) apertures. 

According to \citet{EROData}, the encircled energy fractions in the \IE\ and \HE\ filters are $0.936$ and $0.883$ for a point source measured in a 1\arcsec\ aperture, corresponding to a aperture color correction of $\Delta(\IE-\HE)\,=\,0.06$. 
While the corrections may differ for more extended sources, we assume that a similar correction is applicable to the colors of the cluster (candidates) 
in NGC\,2403. We verified for two clusters, D6 and F1 (which are relatively isolated), 
that similar color corrections are indeed obtained. 
Specifically, we found a mean correction of 0.05 from 1\arcsec\ to 8\arcsec\ for these two clusters, 
essentially the same value reported by \citet{EROData} for point sources.

Next, we used the \texttt{ISHAPE} software \citep{Larsen1999} to measure PSF-corrected sizes for all sources brighter than 
$\IE(\texttt{AUTO})\,=\,22$\,mag. 
\texttt{ISHAPE} requires a PSF subsampled by a factor of 10 with respect to the pixel size of the VIS images, 
which was produced with the \texttt{PSF} task in the \texttt{IRAF} version of \texttt{DAOPHOT} \citep{Stetson1987,Stetson1994} 
from about 100 stars. 
\texttt{ISHAPE} then measures the sizes by convolving the PSF with \citet{King1962} models with a concentration parameter 
$c=r_{\rm t}/r_{\rm c}=30$ (for tidal- and core radii $r_{\rm t}$ and
$r_{\rm c}$) until the best fit was obtained for each source. 
A list of cluster candidates to be inspected visually was then produced by applying the following selection criteria.
\begin{itemize}
\item Magnitude: $17 < \IE < 21.5$.
\item Size: $\mathrm{FWHM} > [0.7 - (\IE-17)/10]$~pixels (for $\IE < 19$) 
and $\mathrm{FWHM}>0.5$~pixels for $19<\IE< 21.5$. 
Here the FWHM is the intrinsic size of the object (namely, corrected for the PSF, so that a point source should have FWHM$\approx$0).
\item 
Color: $\IE-\HE<0.9$ (1\arcsec), or equivalently, $\IE-\HE<0.96$ (infinite aperture).
\end{itemize}
At the distance of NGC\,2403, the size cut of $\mathrm{FWHM}\,=\,0.5$ VIS pixels corresponds to a linear 
$\mathrm{FWHM}\,=\,0.78$\,pc 
or a half-light radius of $r_{\rm h}\,=\,1.1$\,pc for the adopted King models. 
We found this cut to effectively eliminate the vast majority of individual stars, 
while still comfortably allowing the inclusion of GCs that have typical half-light radii $\ga3$~pc \citep{Harris1996}. 
The magnitude-dependent size cut for $\IE<19$ accounts for the fact that individual stars start saturating in the VIS images 
above this limit, and thus no longer appear point-like to \texttt{ISHAPE}. 
The magnitude cuts take into account that the known GCs are all
fainter than \IE\,=\,17, as can be seen in Fig.~\ref{fig:IH_I_clusters}, and that the ratio of
contaminants to GC candidates becomes unmanageable at \IE\,$>$\,21.5.
The color cut includes all of the known GCs in NGC\,2403, but excludes many background sources with redder colors.

\begin{figure}
\centering
\includegraphics[width=\columnwidth]{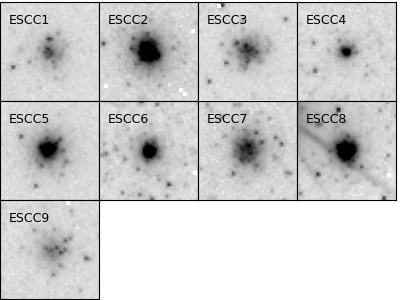}
\caption{New GC candidates identified in the \Euclid VIS images. Each stamp image shows a $5\arcsec\times5\arcsec$ cut-out around each object.
North is up, and east to the left in all images.}
\label{fig:newGC}
\end{figure}

The preselection described above left 
1227 objects to be visually inspected. 
The inspection was done on the \Euclid VIS image using the \texttt{SAOImage DS9} tool with the preselected sources marked. 
In this paper we restrict the discussion to candidates located more than $7\arcmin$ from the center (taken from NED) of NGC\,2403 
(about 6.5\,kpc in projection), since the higher proportion of younger objects and increased crowding in the inner 
regions of the galaxy require a more comprehensive analysis that is deferred to a follow-up paper. 
Outside the $7\arcmin$ radius a total of 
866
objects fulfill the selection criteria outlined above. 
Based on visual inspection, most of these 
(781)
were found to be background galaxies, 
while a smaller number of objects were \hii\ regions, individual stars or 
young clusters in crowded regions of the NGC\,2403 outer disk. 
Nine objects were identified as probable star-cluster candidates based on their morphological appearance
and are labeled here as ESCC$n$ (\Euclid Star Cluster Candidate $n$).
One of these (ESCC9) is fainter than the $\IE\,=\,21.5$ magnitude cut, but was noticed during the inspection of the images 
and added back to the list manually. 
It is quite possible that other candidates fainter than $\IE\,=\,21.5$ are present in the image, but have been missed. 

Cut-outs of the nine new cluster candidates are shown in Fig.~\ref{fig:newGC}, and
Table \ref{tab:clusters} lists the coordinates and photometry for all of the sources discussed in this paper. 
For $\IE$ we give both the AUTO magnitudes and the magnitudes within an $r\,=\,1\arcsec$ circular aperture, 
while only the $r=1\arcsec$ magnitudes are given for the NISP bands. 
In most cases, the AUTO magnitudes capture a larger fraction of the total light, 
while we use the fixed-aperture measurements to define colors.

\begin{figure}
\centering
\includegraphics[width=\columnwidth]{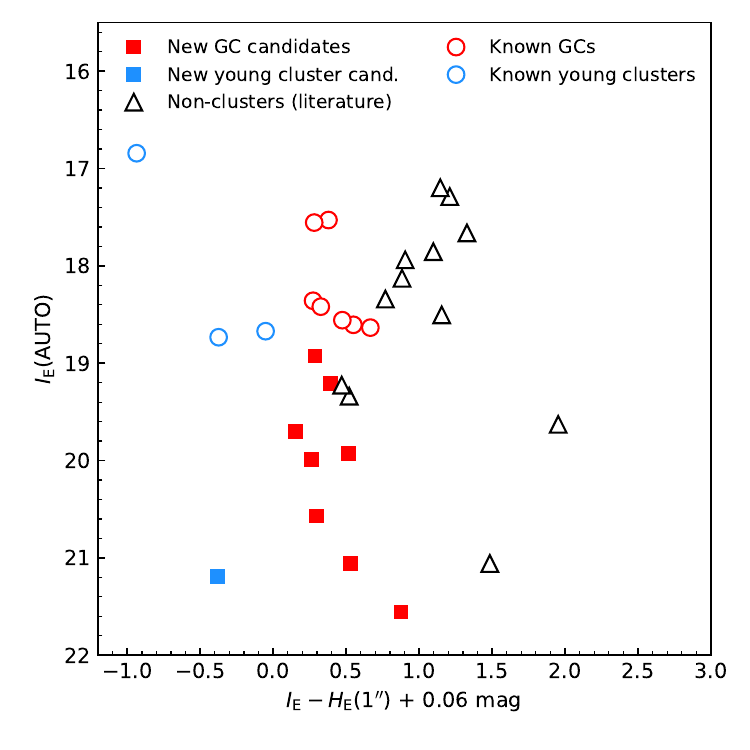}
\caption{Color-magnitude diagram for known young and old clusters in
  NGC\,2403 and our newly identified cluster candidates. 
}
\label{fig:IH_I_clusters}
\end{figure}

A CMD of the previously known and new cluster candidates is shown in Fig.~\ref{fig:IH_I_clusters}. 
For reference we also include the literature candidates identified as non-clusters. 
In this figure, we have corrected the $\IE-\HE$ colors by 0.06 to account for the aperture corrections from 1\arcsec\ to infinity,
as described above.
We first note that most of the new cluster candidates have colors similar to the already known GCs, 
spanning the range $0 < \IE-\HE < 0.9$, 
and we tentatively identify them as GC candidates.
The younger clusters (C1, C3, and C5) are generally bluer than the old GCs (C2 was not detected by \texttt{SourceExtractor} in the NISP images), 
while the background galaxies tend to be redder although there is some overlap in color with the GCs.
One of our new candidates, ESCC4, has blue colors similar to those of the known young clusters, suggesting that this cluster too may be relatively young. 

\begin{figure*}[!t]
\centering
\includegraphics[width=\textwidth]{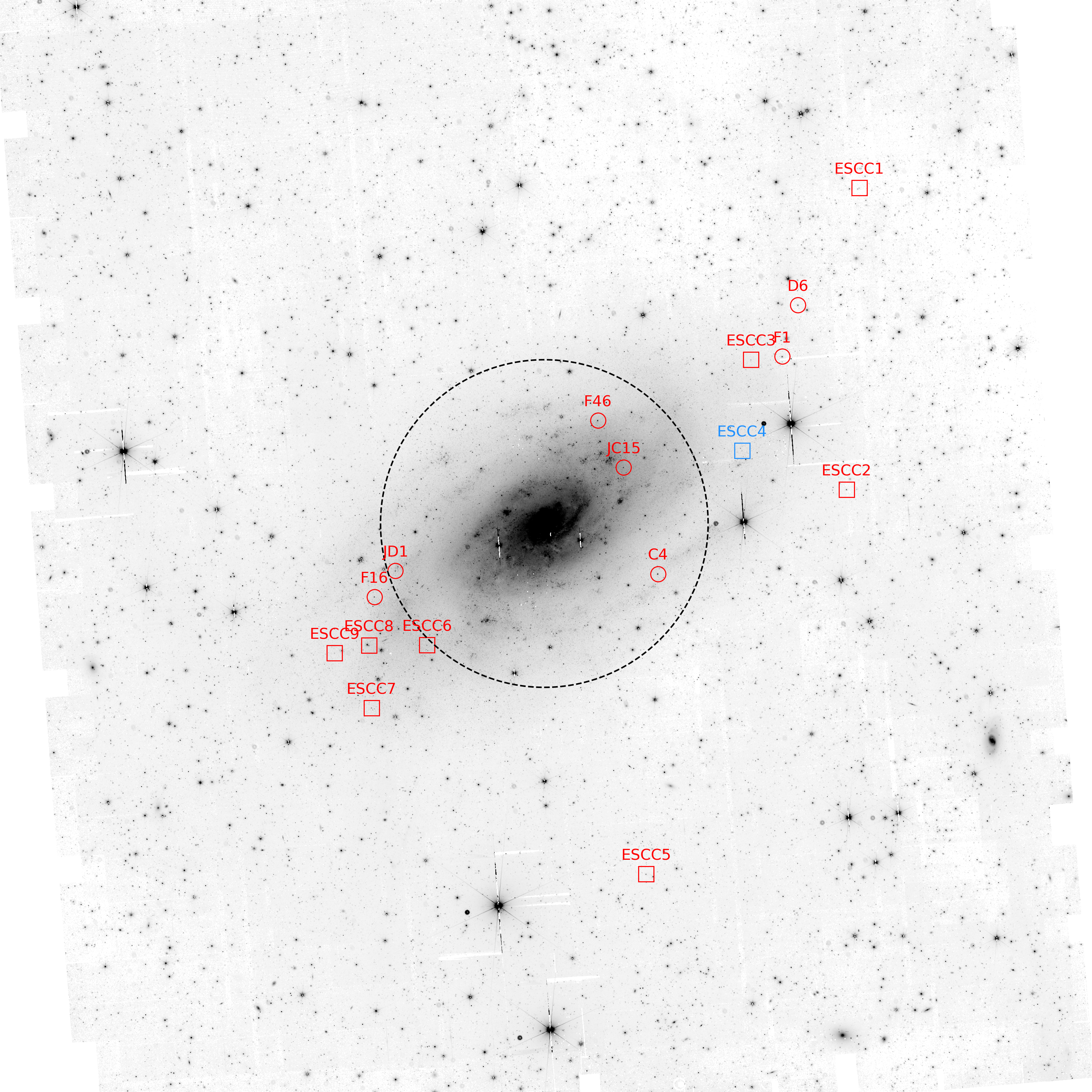}
\caption{\Euclid VIS image of NGC\,2403 with previously known GCs (magenta, see also Fig. \ref{fig:littgcs}), 
new GC candidates (red), and the new young cluster candidate ESCC4 (blue) marked. 
The dashed circle indicates the inner $7\arcmin$ limit of our search, beyond which we focus our discussion in this paper.
North is up, and east to the left.}
\label{fig:n2403_with_gcs}
\end{figure*}

The turn-over of the GC luminosity function (GCLF) is expected at $M_{V} \approx -7.5$ 
\citep{Rejkuba2012}.
From simple stellar population models based on PARSEC isochrones \citep{Marigo2017}, 
the $V-\IE$ color of an old, moderately metal-poor ($\mathrm{[Fe/H]}\approx-1.5$) GC is about $V-\IE \approx 0.5$;
hence we expect the GCLF turn-over at $M_{\IE} \approx -8.0$ or $\IE\approx19.6$ for the distance and $A_V$ value of NGC\,2403. 
It is evident, then, that the confirmed GC sample from the literature only probes the brighter part of the GCLF, 
to about 1\,mag brighter than the GCLF turn-over, and we should expect a number of fainter GCs to be present in NGC\,2403. 
Our new candidates are all fainter than those in the existing list, so that it may be assumed that the census of GCs in 
NGC\,2403 was already fairly complete down to the $\IE\approx18.5$ limit of the previous studies, at least for galactocentric radii $>7\arcmin$. With \Euclid we can now probe GCs that are significantly fainter than the GCLF turn-over. 
A detailed discussion of the NGC\,2403 GCLF is complicated by 
differences in the spatial distributions of the samples from the literature compared with our new candidates, 
with four of the known GCs (C4, F46, JD1, JC15) lying within the $7\arcmin$ limit imposed here.
Nevertheless, defining a combined sample of confirmed GCs located outside $7\arcmin$ (D6, F1, F16) and our new candidates (excluding ESCC4), 
the mean apparent magnitude is $\IE\,=\,19.7\,\pm\,0.3$  
with a dispersion $\sigma_{\IE}\,=\,1.06\pm0.17$, 
which is very close to the expected GCLF turn-over magnitude and dispersion.

The positions of the known and new GC candidates are indicated on the VIS image of NGC\,2403 in Fig.~\ref{fig:n2403_with_gcs}. 
The two outermost new candidates, 
ESCC1 and ESCC5, 
are located at projected galactocentric distances of $r_\mathrm{GC}=15\farcm6$ 
(14.5\,kpc) and $r_\mathrm{GC}=19\farcm7$ (18.3\,kpc), respectively, beyond the most distant cluster previously known (D6 at $r_\mathrm{GC}=14\farcm3$). This illustrates the power of the combination of a wide field and excellent image quality provided by \Euclid.

\begin{table*}[t!]
   \caption[]{{Star cluster candidates in NGC\,2403}} 
\label{tab:clusters}
\resizebox{\linewidth}{!}{
\begin{centering}
\begin{tabular}{lcccccccc}
\hline
\hline
\rule{0pt}{3ex} ID & Note & RA & Dec & $\IE$(AUTO) & $\IE$($r=1''$) & $\YE$($r=1''$) & $\JE$($r=1''$) & 
\multicolumn{1}{c}{$\HE$($r=1''$)} \\
\noalign{\vskip 1ex}
\hline
\\
C4=D5 & Known GC & \ra{07;36;04.43} & \ang{65;33;59.76}     & $17.53$ & $17.66$ & $17.28$ & $17.30$ & $17.34$ \\
D6 & Known GC & \ra{07;35;05.77} & \ang{65;45;27.22}        & $18.56$ & $18.67$ & $18.24$ & $18.23$ & $18.26$ \\
F1 & Known GC & \ra{07;35;12.46} & \ang{65;43;15.55}        & $18.42$ & $18.63$ & $18.29$ & $18.30$ & $18.36$ \\
F16 & Known GC & \ra{07;38;01.55} & \ang{65;33;00.26}       & $18.61$ & $18.70$ & $18.25$ & $18.22$ & $18.21$ \\
F46 & Known GC & \ra{07;36;29.17} & \ang{65;40;33.47}       & $17.56$ & $17.68$ & $17.34$ & $17.39$ & $17.46$ \\
JD1 & Known GC & \ra{07;37;52.92} & \ang{65;34;07.29}       & $18.36$ & $18.43$ & $18.11$ & $18.14$ & $18.21$ \\
JC15 & Known GC & \ra{07;36;18.56} & \ang{65;38;32.98}      & $18.63$ & $18.71$ & $18.14$ & $18.09$ & $18.11$ \\
C1 & Young & \ra{07;36;10.40} & \ang{65;39;34.58}           & $18.67$ & $18.93$ & $18.82$ & $18.94$ & $19.05$ \\
C2 & Young & \ra{07;37;03.18} & \ang{65;36;44.93}           & $18.12$ & $18.36$ & \ldots & \ldots & \ldots \\
C3 & Young & \ra{07;35;42.94} & \ang{65;35;31.91}           & $18.73$ & $19.09$ & $19.12$ & $19.32$ & $19.52$ \\
C5 & Young & \ra{07;36;19.80} & \ang{65;37;09.32}           & $16.84$ & $17.14$ & $17.31$ & $17.76$ & $18.13$ \\
D1 & Star & \ra{07;38;16.83} & \ang{65;42;41.28}            & $21.06$ & $20.97$ & $19.68$ & $19.58$ & $19.55$ \\
D2 & Two stars & \ra{07;37;27.01} & \ang{65;28;20.87}       & $18.34$ & $18.32$ & $17.33$ & $17.42$ & $17.61$ \\
D3 & Two stars & \ra{07;37;21.33} & \ang{65;28;16.19}       & $19.63$ & $19.57$ & $17.74$ & $17.66$ & $17.68$ \\
D4 & Galaxy & \ra{07;36;17.25} & \ang{65;29;00.59}          & $19.34$ & $19.87$ & $19.49$ & $19.49$ & $19.41$ \\
F2 & Galaxy & \ra{07;35;42.21} & \ang{65;51;45.54}          & $17.66$ & $18.65$ & $17.79$ & $17.59$ & $17.38$ \\
F6 & Galaxy & \ra{07;36;11.34} & \ang{65;19;42.25}          & $17.29$ & $18.17$ & $17.42$ & $17.24$ & $17.02$ \\
F19 & Galaxy & \ra{07;38;12.17} & \ang{65;41;08.50}         & $19.23$ & $19.63$ & $19.30$ & $19.31$ & $19.22$ \\
F20 & Galaxy & \ra{07;38;20.37} & \ang{65;14;24.86}         & $17.85$ & $18.62$ & $17.93$ & $17.76$ & $17.58$ \\
F24 & Galaxy & \ra{07;39;14.86} & \ang{65;33;51.70}         & $18.50$ & $18.94$ & $18.17$ & $18.02$ & $17.84$ \\
F28 & Galaxy & \ra{07;39;29.86} & \ang{65;23;55.38}         & $17.20$ & $18.00$ & $17.24$ & $17.10$ & $16.92$ \\
F31 & Galaxy & \ra{07;40;23.93} & \ang{65;50;00.19}         & $17.94$ & $18.64$ & $18.10$ & $17.97$ & $17.80$ \\
F47 & Star & \ra{07;36;40.47} & \ang{65;40;35.11}           & $18.13$ & $18.10$ & $17.47$ & $17.36$ & $17.28$ \\
ESCC1 & New GC cand. & \ra{07;34;39.84} & \ang{65;50;26.32} & $21.06$ & $21.33$ & $20.83$ & $20.83$ & $20.86$ \\
ESCC2 & New GC cand. & \ra{07;34;46.17} & \ang{65;37;33.29} & $19.21$ & $19.39$ & $19.02$ & $19.02$ & $19.06$ \\
ESCC3 & New GC cand. & \ra{07;35;25.46} & \ang{65;43;07.53} & $20.57$ & $20.76$ & $20.40$ & $20.43$ & $20.52$ \\
ESCC4 & Young cluster cand. & \ra{07;35;29.34} & \ang{65;39;14.91}    & $21.19$ & $21.18$ & $21.19$ & $21.43$ & $21.62$ \\
ESCC5 & New GC cand. & \ra{07;36;09.64} & \ang{65;21;09.95} & $19.70$ & $19.83$ & $19.57$ & $19.65$ & $19.74$ \\
ESCC6 & New GC cand. & \ra{07;37;39.77} & \ang{65;30;56.92} & $19.93$ & $19.93$ & $19.53$ & $19.50$ & $19.48$ \\
ESCC7 & New GC cand. & \ra{07;38;02.51} & \ang{65;28;15.52} & $19.99$ & $20.29$ & $19.97$ & $20.00$ & $20.08$ \\
ESCC8 & New GC cand. & \ra{07;38;03.68} & \ang{65;30;55.43} & $18.93$ & $19.38$ & $19.07$ & $19.09$ & $19.16$ \\
ESCC9 & New GC cand. & \ra{07;38;17.86} & \ang{65;30;35.65} & $21.56$ & $21.83$ & $21.05$ & $20.99$ & $21.01$ \\
\\
\hline
\end{tabular}
\end{centering}
}
\tablefoot{The photometry here has not been corrected for Galactic foreground extinction.
The formal photometric errors are all $\leq0.01$\,mag, but do not include the uncertainties due to background subtraction, 
choice of aperture, and other factors that affect the true uncertainties in the quantities listed.
We estimate total photometric errors of $0.2$--$0.3$\,mag once these unknown contributions to the uncertainties are taken into account.
}
\end{table*}

\section{Summary and conclusions }
\label{sec:conclusions}


We have presented the first-look analysis of \Euclid Early Release Observations with VIS and NISP imaging of the 
Nearby Galaxy Showcase.
Galaxies in the Showcase range in distance from $0.5$\,Mpc (NGC\,6822) to 8.8\,Mpc (NGC\,6744), and
include three dwarf galaxies (Holmberg\,II, IC\,10, and NGC\,6822),
and three spirals (IC\,342, NGC\,2403, and NGC\,6744).
The galaxies were selected to be among the apparently largest galaxies on the sky,
in order to guarantee their photogenic nature, but also to enable an in-depth scientific analysis.
The sample is described in Sect.~\ref{sec:sample}.

The surface brightness limits of the VIS and NISP stacked images are calculated in Appendix \ref{app:sky},
with a summary given in Sect. \ref{sec:sblim}.
Confirming previous estimates based on simulations for VIS, and exceeding previous expectations for NISP,
we find that in 1 ROS, \Euclid can probe $1\sigma$ surface brightness depths in 100\,arcsec$^2$ regions of 
$30.5$\,AB\,\magarc\ in VIS,
and $29.2$--$29.4$\,AB\,\magarc\ in NISP.

In Sect. \ref{sec:integrated}, we assessed the properties of the integrated light in the Showcase galaxies,
and presented RGB images in Figs. \ref{fig:pretty_holmbergii}, \ref{fig:pretty_ngc6744} in the main text,
and Figs. \ref{fig:pretty_ic10}, \ref{fig:pretty_ic342}, \ref{fig:pretty_ngc2403}, and
\ref{fig:pretty_ngc6822} in Appendix \ref{app:imaging}.
These composite images illustrate \Euclid's unique capacity to probe a large FoV, but also
to provide exquisite detail on parsec scales in nearby galaxies.

Sections \ref{sec:hi} and \ref{sec:ir} compared high-contrast \Euclid \IE\ images with representative
galaxies having \hi\ and FIR data.
The characteristic blue colors of diffuse cirrus emission combined with multiwavelength FIR data
can disentangle foreground cirrus emission originating in the MW from potential stellar streams. 
This will be important not only for the studies of nearby galaxies, but also for probing the ISM
of the Galaxy.
Radial surface brightness and color profiles were discussed in
Sects.~\ref{sec:profiles} and~\ref{sec:colors}, where we also compared the observed \Euclid
colors $\IE-\HE$ with stellar population synthesis models as seen in Fig. \ref{fig:colors}.
This comparison is a validation of measurements and models and shows that
\Euclid colors provide a powerful diagnostic of the age and metallicity of galaxies in the local Universe.
Moreover, \Euclid colors have identified extremely blue young nuclear star clusters in
IC\,342 and IC\,10.

Section~\ref{sec:resolvedstars} gave a detailed analysis of resolved stellar photometry,
finding altogether 
$1.3\times10^6$ stars in the Showcase galaxy images.
After carefully removing as well as possible foreground stars and background compact galaxies
(see Figs. \ref{fig:cmd_n6822}, \ref{fig:color_color}), CMDs for NGC\,6822
demonstrate the resulting well-sampled stellar statistics provided by a single \Euclid ROS.
By selecting young stars, AGBs, and RGBs from regions defined within the CMDs as in Fig. \ref{fig:3cmd_n6822}, 
we were able to construct
star-count maps that probe the spiral structure, and the age differentiation across dwarf galaxies.
In particular, Fig. \ref{fig:starcounts_1} demonstrates that young stars clearly trace spiral structure in IC\,342,
and  Fig. \ref{fig:starcounts_2} shows that in NGC\,6822 the distribution of young stars is 
perfectly matched by the morphology of the \hi\ emission.

Ensembles of dwarf galaxy satellites around a parent galaxy are common, and 
we have investigated the dwarf galaxy system of NGC\,6744 in Sect. \ref{sec:dwarfs}.
Although the satellites of NGC\,6744 have been previously studied, not only do we recover
the four known dwarf satellites within the \Euclid FoV, we also identify a previously unknown
dwarf satellite candidate, EDwC1, as shown in Fig. \ref{fig:dwarf3}.
This new discovery sets the stage for the new dwarf satellite demographics that can be obtained
with \Euclid.

In Sect. \ref{sec:egcs}, our analysis of the \Euclid imaging of NGC\,2403 revealed nine new
star cluster candidates as shown in Fig. \ref{fig:newGC}, eight of them almost certainly evolved clusters, thus true GCs.
This new census more than doubles the number of known GCs around NGC\,2403, and extends the galactocentric
radius at which they have been found out to 18.3\,kpc, 
$40$\% further away than previously identified.
\Euclid is poised to 
transform the study of extragalactic GCs in nearby galaxies.

In conclusion, the ERO Nearby Galaxy Showcase demonstrates the power of \Euclid to probe
large areas of sky, but at the same time, with exquisite spatial scales and sensitivity.
This unique combination will revolutionize studies of: 
resolved stellar populations and star-formation histories;
the extent and origin of galaxy disks, together with their spiral structure;
the demographics of dwarf satellites;
star clusters within and around galaxies;
and the measurement of the ISM via dust extinction and cirrus emission on scales of a few parsecs.
Using data from the EWS and EDS, 
it will be possible to pursue not only the core
cosmological science that was the main aim of the mission, but also to examine galaxy populations
in the nearby Universe with unprecedented detail.


\begin{acknowledgements}
We wish to acknowledge our colleague Mario Nonino, who recently passed away, for his important contribution to the \Euclid mission. 
Mario was the coordinator of the OU-MER pipeline, one of the central organizational units of the Science Ground Segment, and an important
member of the \Euclid Local Universe Science Working Group.
We are grateful to Fabian Walter, I-Da Chiang, and Karin Sandstrom for passing on \hi\ moment maps,
and are deeply indebted to Michele Bellazzini for enlightening discussions. 
LKH, PD, RS, and CT acknowledge funding from the Italian INAF Large Grant 12-2022.
AMNF is grateful for support from the UK STFC via grant ST/Y001281/1.
JHK and JR acknowledge grant PID2022-136505NB-I00 funded by MCIN/AEI/10.13039/501100011033 and EU, ERDF; 
their work is co-funded by the European Union, however views and opinions expressed are those of the author(s) only and do not necessarily reflect those of the European Union. Neither the European Union nor the granting authority can be held responsible for them.
This work was partly carried out using GNU Astronomy Utilities (Gnuastro, ascl.net/1801.009) version 0.21.43-3101. Work on Gnuastro has been funded by the Japanese Ministry of Education, Culture, Sports, Science, and Technology (MEXT) scholarship and its Grant-in-Aid for Scientific Research (21244012, 24253003), the European Research Council (ERC) advanced grant 339659-MUSICOS, the Spanish Ministry of Economy and Competitiveness (MINECO, grant number AYA2016-76219-P), and the NextGenerationEU grant through the Recovery and Resilience Facility project ICTS-MRR-2021-03-CEFCA.
This research has made use of the NASA/IPAC Extragalactic Database (NED), which is funded by the National Aeronautics and Space Administration and operated by the California Institute of Technology.
This research has also relied on \texttt{AutoProf}, a package for galaxy image photometry \citep{stone21}, and 
on Astropy (\url{http://www.astropy.org}), a community-developed core Python package 
and an ecosystem of tools and resources for astronomy \citep{astropy:2013, astropy:2018, astropy:2022}. 
\AckERO
  \AckEC
\end{acknowledgements}

%
%

\bibliography{Euclid,EROplus,Showcase,Resolved_plus}

\begin{thebibliography}{236}
\expandafter\ifx\csname natexlab\endcsname\relax\def\natexlab#1{#1}\fi

\bibitem[{{Abdurro'uf} {et~al.}(2022){Abdurro'uf}, {Lin}, {Hirashita},
  {Morishita}, {Tacchella}, {Akiyama}, {Takeuchi}, \& {Wu}}]{abdurrouf22}
{Abdurro'uf}, {Lin}, Y.-T., {Hirashita}, H., {et~al.} 2022, \apj, 926, 81

\bibitem[{{Akhlaghi}(2019a)}]{akhlaghi19a}
{Akhlaghi}, M. 2019a, arXiv e-prints, arXiv:1909.11230

\bibitem[{{Akhlaghi}(2019b)}]{akhlaghi19b}
{Akhlaghi}, M. 2019b, in Astronomical Society of the Pacific Conference Series,
  Vol. 521, Astronomical Data Analysis Software and Systems XXVI, ed.
  M.~{Molinaro}, K.~{Shortridge}, \& F.~{Pasian}, 299

\bibitem[{{Akhlaghi} \& {Ichikawa}(2015)}]{akhlaghi15}
{Akhlaghi}, M. \& {Ichikawa}, T. 2015, \apjs, 220, 1

\bibitem[{{Annibali} \& {Tosi}(2022)}]{annibali22}
{Annibali}, F. \& {Tosi}, M. 2022, Nature Astronomy, 6, 48

\bibitem[{{Arnouts} {et~al.}(1999){Arnouts}, {Cristiani}, {Moscardini},
  {Matarrese}, {Lucchin}, {Fontana}, \& {Giallongo}}]{arnouts99}
{Arnouts}, S., {Cristiani}, S., {Moscardini}, L., {et~al.} 1999, \mnras, 310,
  540

\bibitem[{{Ashley} {et~al.}(2014){Ashley}, {Elmegreen}, {Johnson}, {Nidever},
  {Simpson}, \& {Pokhrel}}]{ashley14}
{Ashley}, T., {Elmegreen}, B.~G., {Johnson}, M., {et~al.} 2014, \aj, 148, 130

\bibitem[{{Astropy Collaboration} {et~al.}(2022){Astropy Collaboration},
  {Price-Whelan}, {Lim}, {Earl}, {Starkman}, {Bradley}, {Shupe}, {Patil},
  {Corrales}, {Brasseur}, {N{\"o}the}, {Donath}, {Tollerud}, {Morris},
  {Ginsburg}, {Vaher}, {Weaver}, {Tocknell}, {Jamieson}, {van Kerkwijk},
  {Robitaille}, {Merry}, {Bachetti}, {G{\"u}nther}, {Aldcroft},
  {Alvarado-Montes}, {Archibald}, {B{\'o}di}, {Bapat}, {Barentsen},
  {Baz{\'a}n}, {Biswas}, {Boquien}, {Burke}, {Cara}, {Cara}, {Conroy},
  {Conseil}, {Craig}, {Cross}, {Cruz}, {D'Eugenio}, {Dencheva}, {Devillepoix},
  {Dietrich}, {Eigenbrot}, {Erben}, {Ferreira}, {Foreman-Mackey}, {Fox},
  {Freij}, {Garg}, {Geda}, {Glattly}, {Gondhalekar}, {Gordon}, {Grant},
  {Greenfield}, {Groener}, {Guest}, {Gurovich}, {Handberg}, {Hart},
  {Hatfield-Dodds}, {Homeier}, {Hosseinzadeh}, {Jenness}, {Jones}, {Joseph},
  {Kalmbach}, {Karamehmetoglu}, {Ka{\l}uszy{\'n}ski}, {Kelley}, {Kern},
  {Kerzendorf}, {Koch}, {Kulumani}, {Lee}, {Ly}, {Ma}, {MacBride}, {Maljaars},
  {Muna}, {Murphy}, {Norman}, {O'Steen}, {Oman}, {Pacifici}, {Pascual},
  {Pascual-Granado}, {Patil}, {Perren}, {Pickering}, {Rastogi}, {Roulston},
  {Ryan}, {Rykoff}, {Sabater}, {Sakurikar}, {Salgado}, {Sanghi}, {Saunders},
  {Savchenko}, {Schwardt}, {Seifert-Eckert}, {Shih}, {Jain}, {Shukla}, {Sick},
  {Simpson}, {Singanamalla}, {Singer}, {Singhal}, {Sinha}, {Sip{\H{o}}cz},
  {Spitler}, {Stansby}, {Streicher}, {{\v{S}}umak}, {Swinbank}, {Taranu},
  {Tewary}, {Tremblay}, {de Val-Borro}, {Van Kooten}, {Vasovi{\'c}}, {Verma},
  {de Miranda Cardoso}, {Williams}, {Wilson}, {Winkel}, {Wood-Vasey}, {Xue},
  {Yoachim}, {Zhang}, {Zonca}, \& {Astropy Project
  Contributors}}]{astropy:2022}
{Astropy Collaboration}, {Price-Whelan}, A.~M., {Lim}, P.~L., {et~al.} 2022,
  \apj, 935, 167

\bibitem[{{Astropy Collaboration} {et~al.}(2018){Astropy Collaboration},
  {Price-Whelan}, {Sip{\H{o}}cz}, {G{\"u}nther}, {Lim}, {Crawford}, {Conseil},
  {Shupe}, {Craig}, {Dencheva}, {Ginsburg}, {VanderPlas}, {Bradley},
  {P{\'e}rez-Su{\'a}rez}, {de Val-Borro}, {Aldcroft}, {Cruz}, {Robitaille},
  {Tollerud}, {Ardelean}, {Babej}, {Bach}, {Bachetti}, {Bakanov}, {Bamford},
  {Barentsen}, {Barmby}, {Baumbach}, {Berry}, {Biscani}, {Boquien}, {Bostroem},
  {Bouma}, {Brammer}, {Bray}, {Breytenbach}, {Buddelmeijer}, {Burke},
  {Calderone}, {Cano Rodr{\'\i}guez}, {Cara}, {Cardoso}, {Cheedella}, {Copin},
  {Corrales}, {Crichton}, {D'Avella}, {Deil}, {Depagne}, {Dietrich}, {Donath},
  {Droettboom}, {Earl}, {Erben}, {Fabbro}, {Ferreira}, {Finethy}, {Fox},
  {Garrison}, {Gibbons}, {Goldstein}, {Gommers}, {Greco}, {Greenfield},
  {Groener}, {Grollier}, {Hagen}, {Hirst}, {Homeier}, {Horton}, {Hosseinzadeh},
  {Hu}, {Hunkeler}, {Ivezi{\'c}}, {Jain}, {Jenness}, {Kanarek}, {Kendrew},
  {Kern}, {Kerzendorf}, {Khvalko}, {King}, {Kirkby}, {Kulkarni}, {Kumar},
  {Lee}, {Lenz}, {Littlefair}, {Ma}, {Macleod}, {Mastropietro}, {McCully},
  {Montagnac}, {Morris}, {Mueller}, {Mumford}, {Muna}, {Murphy}, {Nelson},
  {Nguyen}, {Ninan}, {N{\"o}the}, {Ogaz}, {Oh}, {Parejko}, {Parley}, {Pascual},
  {Patil}, {Patil}, {Plunkett}, {Prochaska}, {Rastogi}, {Reddy Janga},
  {Sabater}, {Sakurikar}, {Seifert}, {Sherbert}, {Sherwood-Taylor}, {Shih},
  {Sick}, {Silbiger}, {Singanamalla}, {Singer}, {Sladen}, {Sooley},
  {Sornarajah}, {Streicher}, {Teuben}, {Thomas}, {Tremblay}, {Turner},
  {Terr{\'o}n}, {van Kerkwijk}, {de la Vega}, {Watkins}, {Weaver}, {Whitmore},
  {Woillez}, {Zabalza}, \& {Astropy Contributors}}]{astropy:2018}
{Astropy Collaboration}, {Price-Whelan}, A.~M., {Sip{\H{o}}cz}, B.~M., {et~al.}
  2018, \aj, 156, 123

\bibitem[{{Astropy Collaboration} {et~al.}(2013){Astropy Collaboration},
  {Robitaille}, {Tollerud}, {Greenfield}, {Droettboom}, {Bray}, {Aldcroft},
  {Davis}, {Ginsburg}, {Price-Whelan}, {Kerzendorf}, {Conley}, {Crighton},
  {Barbary}, {Muna}, {Ferguson}, {Grollier}, {Parikh}, {Nair}, {Unther},
  {Deil}, {Woillez}, {Conseil}, {Kramer}, {Turner}, {Singer}, {Fox}, {Weaver},
  {Zabalza}, {Edwards}, {Azalee Bostroem}, {Burke}, {Casey}, {Crawford},
  {Dencheva}, {Ely}, {Jenness}, {Labrie}, {Lim}, {Pierfederici}, {Pontzen},
  {Ptak}, {Refsdal}, {Servillat}, \& {Streicher}}]{astropy:2013}
{Astropy Collaboration}, {Robitaille}, T.~P., {Tollerud}, E.~J., {et~al.} 2013,
  \aap, 558, A33

\bibitem[{{Balser} {et~al.}(2017){Balser}, {Wenger}, {Goss}, {Johnson}, \&
  {Kepley}}]{balser17}
{Balser}, D.~S., {Wenger}, T.~V., {Goss}, W.~M., {Johnson}, K.~E., \& {Kepley},
  A.~A. 2017, \apj, 844, 73

\bibitem[{{Barker} {et~al.}(2012){Barker}, {Ferguson}, {Irwin}, {Arimoto}, \&
  {Jablonka}}]{barker12}
{Barker}, M.~K., {Ferguson}, A. M.~N., {Irwin}, M.~J., {Arimoto}, N., \&
  {Jablonka}, P. 2012, \mnras, 419, 1489

\bibitem[{{Barra} {et~al.}(2023){Barra}, {Pinto}, {Middleton}, {Di Salvo},
  {Walton}, {G{\'u}rpide}, \& {Roberts}}]{barra23}
{Barra}, F., {Pinto}, C., {Middleton}, M., {et~al.} 2023, arXiv e-prints,
  arXiv:2311.16243

\bibitem[{Battistini {et~al.}(1984)Battistini, Bonoli, Federici, Fusi~Pecci, \&
  Kron}]{Battistini1984}
Battistini, P., Bonoli, F., Federici, L., Fusi~Pecci, F., \& Kron, R.~G. 1984,
  A\&A, 130, 162

\bibitem[{{Bedin} {et~al.}(2019){Bedin}, {Salaris}, {Rich}, {Richer},
  {Anderson}, {Bettoni}, {Nardiello}, {Milone}, {Marino}, {Libralato},
  {Bellini}, {Dieball}, {Bergeron}, {Burgasser}, \& {Apai}}]{bedin19}
{Bedin}, L.~R., {Salaris}, M., {Rich}, R.~M., {et~al.} 2019, \mnras, 484, L54

\bibitem[{{Bell} {et~al.}(2019){Bell}, {Cioni}, {Wright}, {Rubele}, {Nidever},
  {Tatton}, {van Loon}, {Ivanov}, {Subramanian}, {Oliveira}, {de Grijs},
  {Pennock}, {Choi}, {Zaritsky}, {Olsen}, {Niederhofer}, {Choudhury},
  {Mart{\'\i}nez-Delgado}, \& {Mu{\~n}oz}}]{Bell2019}
{Bell}, C. P.~M., {Cioni}, M.-R.~L., {Wright}, A.~H., {et~al.} 2019, \mnras,
  489, 3200

\bibitem[{{Belland} {et~al.}(2020){Belland}, {Kirby}, {Boylan-Kolchin}, \&
  {Wheeler}}]{belland20}
{Belland}, B., {Kirby}, E., {Boylan-Kolchin}, M., \& {Wheeler}, C. 2020, \apj,
  903, 10

\bibitem[{{Belokurov} {et~al.}(2006){Belokurov}, {Zucker}, {Evans}, {Gilmore},
  {Vidrih}, {Bramich}, {Newberg}, {Wyse}, {Irwin}, {Fellhauer}, {Hewett},
  {Walton}, {Wilkinson}, {Cole}, {Yanny}, {Rockosi}, {Beers}, {Bell},
  {Brinkmann}, {Ivezi{\'c}}, \& {Lupton}}]{belokurov06}
{Belokurov}, V., {Zucker}, D.~B., {Evans}, N.~W., {et~al.} 2006, \apjl, 642,
  L137

\bibitem[{{Bennet} {et~al.}(2023){Bennet}, {Patel}, {Sohn}, {del Pino}, {van
  der Marel}, {Libralato}, {Watkins}, {Aparicio}, {Besla}, {Gallart}, {Fardal},
  {Monelli}, {Sacchi}, {Tollerud}, \& {Weisz}}]{Bennet2023}
{Bennet}, P., {Patel}, E., {Sohn}, S.~T., {et~al.} 2023, arXiv e-prints,
  arXiv:2312.09276

\bibitem[{{Bentley} {et~al.}(2019){Bentley}, {Tinney}, {Sharma}, \&
  {Wright}}]{Bentley2019}
{Bentley}, J., {Tinney}, C.~G., {Sharma}, S., \& {Wright}, D. 2019, \mnras,
  490, 4107

\bibitem[{{Bernard} {et~al.}(2012){Bernard}, {Ferguson}, {Barker}, {Irwin},
  {Jablonka}, \& {Arimoto}}]{bernard12}
{Bernard}, E.~J., {Ferguson}, A. M.~N., {Barker}, M.~K., {et~al.} 2012, \mnras,
  426, 3490

\bibitem[{{Bertin} \& {Arnouts}(1996)}]{bertin96}
{Bertin}, E. \& {Arnouts}, S. 1996, \aaps, 117, 393

\bibitem[{{Bianchi} {et~al.}(2017){Bianchi}, {Giovanardi}, {Smith}, {Fritz},
  {Davies}, {Haynes}, {Giovanelli}, {Baes}, {Bocchio}, {Boissier}, {Boquien},
  {Boselli}, {Casasola}, {Clark}, {De Looze}, {di Serego Alighieri}, {Grossi},
  {Jones}, {Hughes}, {Hunt}, {Madden}, {Magrini}, {Pappalardo}, {Ysard}, \&
  {Zibetti}}]{bianchi17}
{Bianchi}, S., {Giovanardi}, C., {Smith}, M.~W.~L., {et~al.} 2017, \aap, 597,
  A130

\bibitem[{{Blakeslee} {et~al.}(2009){Blakeslee}, {Jord{\'a}n}, {Mei},
  {C{\^o}t{\'e}}, {Ferrarese}, {Infante}, {Peng}, {Tonry}, \&
  {West}}]{blakeslee09}
{Blakeslee}, J.~P., {Jord{\'a}n}, A., {Mei}, S., {et~al.} 2009, \apj, 694, 556

\bibitem[{{Blakeslee} {et~al.}(1997){Blakeslee}, {Tonry}, \&
  {Metzger}}]{blakeslee97}
{Blakeslee}, J.~P., {Tonry}, J.~L., \& {Metzger}, M.~R. 1997, \aj, 114, 482

\bibitem[{{Bland-Hawthorn} {et~al.}(2017){Bland-Hawthorn}, {Maloney},
  {Stephens}, {Zovaro}, \& {Popping}}]{blandhawthorn17}
{Bland-Hawthorn}, J., {Maloney}, P.~R., {Stephens}, A., {Zovaro}, A., \&
  {Popping}, A. 2017, \apj, 849, 51

\bibitem[{{B{\"o}ker} {et~al.}(1999){B{\"o}ker}, {van der Marel}, \&
  {Vacca}}]{boker99}
{B{\"o}ker}, T., {van der Marel}, R.~P., \& {Vacca}, W.~D. 1999, \aj, 118, 831

\bibitem[{{Boquien} {et~al.}(2019){Boquien}, {Burgarella}, {Roehlly}, {Buat},
  {Ciesla}, {Corre}, {Inoue}, \& {Salas}}]{boquien19}
{Boquien}, M., {Burgarella}, D., {Roehlly}, Y., {et~al.} 2019, \aap, 622, A103

\bibitem[{{Borlaff} {et~al.}(2019){Borlaff}, {Trujillo}, {Rom{\'a}n},
  {Beckman}, {Eliche-Moral}, {Infante-S{\'a}inz}, {Lumbreras-Calle}, {de
  Almagro}, {G{\'o}mez-Guijarro}, {Cebri{\'a}n}, {Dorta}, {Cardiel},
  {Akhlaghi}, \& {Mart{\'\i}nez-Lombilla}}]{borlaff19}
{Borlaff}, A., {Trujillo}, I., {Rom{\'a}n}, J., {et~al.} 2019, \aap, 621, A133

\bibitem[{{Boselli} {et~al.}(2014){Boselli}, {Voyer}, {Boissier}, {Cucciati},
  {Consolandi}, {Cortese}, {Fumagalli}, {Gavazzi}, {Heinis}, {Roehlly}, \&
  {Toloba}}]{boselli14}
{Boselli}, A., {Voyer}, E., {Boissier}, S., {et~al.} 2014, \aap, 570, A69

\bibitem[{{Bosma}(2017)}]{bosma17}
{Bosma}, A. 2017, in Astrophysics and Space Science Library, Vol. 434,
  Outskirts of Galaxies, ed. J.~H. {Knapen}, J.~C. {Lee}, \& A.~{Gil de Paz},
  209

\bibitem[{{Bressan} {et~al.}(2012){Bressan}, {Marigo}, {Girardi}, {Salasnich},
  {Dal Cero}, {Rubele}, \& {Nanni}}]{Bressan2012}
{Bressan}, A., {Marigo}, P., {Girardi}, L., {et~al.} 2012, \mnras, 427, 127

\bibitem[{{Brodie} \& {Strader}(2006)}]{brodie06}
{Brodie}, J.~P. \& {Strader}, J. 2006, \araa, 44, 193

\bibitem[{{Bruzual} \& {Charlot}(2003)}]{BC03}
{Bruzual}, G. \& {Charlot}, S. 2003, \mnras, 344, 1000

\bibitem[{{Bullock} \& {Johnston}(2005)}]{bullock05}
{Bullock}, J.~S. \& {Johnston}, K.~V. 2005, \apj, 635, 931

\bibitem[{{Buta} \& {McCall}(1999)}]{buta99}
{Buta}, R.~J. \& {McCall}, M.~L. 1999, \apjs, 124, 33

\bibitem[{{Calzetti} {et~al.}(1994){Calzetti}, {Kinney}, \&
  {Storchi-Bergmann}}]{calzetti94}
{Calzetti}, D., {Kinney}, A.~L., \& {Storchi-Bergmann}, T. 1994, \apj, 429, 582

\bibitem[{{Cannon} {et~al.}(2012){Cannon}, {O'Leary}, {Weisz}, {Skillman},
  {Dolphin}, {Bigiel}, {Cole}, {de Blok}, \& {Walter}}]{cannon12}
{Cannon}, J.~M., {O'Leary}, E.~M., {Weisz}, D.~R., {et~al.} 2012, \apj, 747,
  122

\bibitem[{{Cantiello} {et~al.}(2018b){Cantiello}, {Blakeslee}, {Ferrarese},
  {C{\^o}t{\'e}}, {Roediger}, {Raimondo}, {Peng}, {Gwyn}, {Durrell}, \&
  {Cuillandre}}]{cantiello18b}
{Cantiello}, M., {Blakeslee}, J.~P., {Ferrarese}, L., {et~al.} 2018b, \apj,
  856, 126

\bibitem[{{Cantiello} {et~al.}(2018a){Cantiello}, {Grado}, {Rejkuba},
  {Arnaboldi}, {Capaccioli}, {Greggio}, {Iodice}, \& {Limatola}}]{cantiello18a}
{Cantiello}, M., {Grado}, A., {Rejkuba}, M., {et~al.} 2018a, \aap, 611, A21

\bibitem[{{Carlberg} {et~al.}(2011){Carlberg}, {Richer}, {McConnachie},
  {Irwin}, {Ibata}, {Dotter}, {Chapman}, {Fardal}, {Ferguson}, {Lewis},
  {Navarro}, {Puzia}, \& {Valls-Gabaud}}]{carlberg11}
{Carlberg}, R.~G., {Richer}, H.~B., {McConnachie}, A.~W., {et~al.} 2011, \apj,
  731, 124

\bibitem[{{Carlin} {et~al.}(2019){Carlin}, {Garling}, {Peter}, {Crnojevi{\'c}},
  {Forbes}, {Hargis}, {Mutlu-Pakdil}, {Pucha}, {Romanowsky}, {Sand},
  {Spekkens}, {Strader}, \& {Willman}}]{carlin19}
{Carlin}, J.~L., {Garling}, C.~T., {Peter}, A. H.~G., {et~al.} 2019, \apj, 886,
  109

\bibitem[{{Carlsten} {et~al.}(2019){Carlsten}, {Beaton}, {Greco}, \&
  {Greene}}]{carlsten19}
{Carlsten}, S.~G., {Beaton}, R.~L., {Greco}, J.~P., \& {Greene}, J.~E. 2019,
  \apj, 879, 13

\bibitem[{{Carlsten} {et~al.}(2022){Carlsten}, {Greene}, {Beaton}, \&
  {Greco}}]{carlsten22}
{Carlsten}, S.~G., {Greene}, J.~E., {Beaton}, R.~L., \& {Greco}, J.~P. 2022,
  \apj, 927, 44

\bibitem[{{Carollo} {et~al.}(2016){Carollo}, {Beers}, {Placco}, {Santucci},
  {Denissenkov}, {Tissera}, {Lentner}, {Rossi}, {Lee}, \&
  {Tumlinson}}]{carollo16}
{Carollo}, D., {Beers}, T.~C., {Placco}, V.~M., {et~al.} 2016, Nature Physics,
  12, 1170

\bibitem[{{Carson} {et~al.}(2015){Carson}, {Barth}, {Seth}, {den Brok},
  {Cappellari}, {Greene}, {Ho}, \& {Neumayer}}]{carson15}
{Carson}, D.~J., {Barth}, A.~J., {Seth}, A.~C., {et~al.} 2015, \aj, 149, 170

\bibitem[{{Chiang} {et~al.}(2021){Chiang}, {Sandstrom}, {Chastenet}, {Herrera},
  {Koch}, {Kreckel}, {Leroy}, {Pety}, {Schruba}, {Utomo}, \&
  {Williams}}]{chiang21}
{Chiang}, I.-D., {Sandstrom}, K.~M., {Chastenet}, J., {et~al.} 2021, \apj, 907,
  29

\bibitem[{{Choi} {et~al.}(2015){Choi}, {Dalcanton}, {Williams}, {Weisz},
  {Skillman}, {Fouesneau}, \& {Dolphin}}]{choi15}
{Choi}, Y., {Dalcanton}, J.~J., {Williams}, B.~F., {et~al.} 2015, \apj, 810, 9

\bibitem[{{Cignoni} {et~al.}(2018){Cignoni}, {Sacchi}, {Aloisi}, {Tosi},
  {Calzetti}, {Lee}, {Sabbi}, {Adamo}, {Cook}, {Dale}, {Elmegreen},
  {Gallagher}, {Gouliermis}, {Grasha}, {Grebel}, {Hunter}, {Johnson}, {Messa},
  {Smith}, {Thilker}, {Ubeda}, \& {Whitmore}}]{cignoni18}
{Cignoni}, M., {Sacchi}, E., {Aloisi}, A., {et~al.} 2018, \apj, 856, 62

\bibitem[{{Cignoni} {et~al.}(2019){Cignoni}, {Sacchi}, {Tosi}, {Aloisi},
  {Cook}, {Calzetti}, {Lee}, {Sabbi}, {Thilker}, {Adamo}, {Dale}, {Elmegreen},
  {Gallagher}, {Grebel}, {Johnson}, {Messa}, {Smith}, \& {Ubeda}}]{cignoni19}
{Cignoni}, M., {Sacchi}, E., {Tosi}, M., {et~al.} 2019, \apj, 887, 112

\bibitem[{{Conselice} {et~al.}(2022){Conselice}, {Mundy}, {Ferreira}, \&
  {Duncan}}]{conselice22}
{Conselice}, C.~J., {Mundy}, C.~J., {Ferreira}, L., \& {Duncan}, K. 2022, \apj,
  940, 168

\bibitem[{{Crnojevi{\'c}} {et~al.}(2016){Crnojevi{\'c}}, {Sand}, {Spekkens},
  {Caldwell}, {Guhathakurta}, {McLeod}, {Seth}, {Simon}, {Strader}, \&
  {Toloba}}]{crnojevic16}
{Crnojevi{\'c}}, D., {Sand}, D.~J., {Spekkens}, K., {et~al.} 2016, \apj, 823,
  19

\bibitem[{{Cseh} {et~al.}(2012){Cseh}, {Corbel}, {Kaaret}, {Lang}, {Gris{\'e}},
  {Paragi}, {Tzioumis}, {Tudose}, \& {Feng}}]{cseh12}
{Cseh}, D., {Corbel}, S., {Kaaret}, P., {et~al.} 2012, \apj, 749, 17

\bibitem[{{Cuillandre} {et~al.}(2024){Cuillandre}, {Bertin}, {Bolzonella},
  {et~al.}}]{EROData}
{Cuillandre}, J.-C., {Bertin}, E., {Bolzonella}, M., {et~al.} 2024, \aap, this
  issue

\bibitem[{{da Silva} {et~al.}(2018){da Silva}, {Steiner}, \&
  {Menezes}}]{dasilva18}
{da Silva}, P., {Steiner}, J.~E., \& {Menezes}, R.~B. 2018, \apj, 861, 83

\bibitem[{{Das} {et~al.}(2021){Das}, {Nandi}, {Agrawal}, {Dihingia}, \&
  {Majumder}}]{das21}
{Das}, S., {Nandi}, A., {Agrawal}, V.~K., {Dihingia}, I.~K., \& {Majumder}, S.
  2021, \mnras, 507, 2777

\bibitem[{Davidge(2007)}]{Davidge2007}
Davidge, T.~J. 2007, ApJ, 664, 820

\bibitem[{{de Blok} {et~al.}(2014){de Blok}, {Keating}, {Pisano}, {Fraternali},
  {Walter}, {Oosterloo}, {Brinks}, {Bigiel}, \& {Leroy}}]{deblok14}
{de Blok}, W.~J.~G., {Keating}, K.~M., {Pisano}, D.~J., {et~al.} 2014, \aap,
  569, A68

\bibitem[{{de Blok} \& {Walter}(2000)}]{deblok00}
{de Blok}, W.~J.~G. \& {Walter}, F. 2000, \apjl, 537, L95

\bibitem[{{de Blok} \& {Walter}(2003)}]{deblok03}
{de Blok}, W.~J.~G. \& {Walter}, F. 2003, \mnras, 341, L39

\bibitem[{{de Blok} \& {Walter}(2006)}]{deblok06}
{de Blok}, W.~J.~G. \& {Walter}, F. 2006, \aj, 131, 343

\bibitem[{{de Jong} {et~al.}(2007){de Jong}, {Seth}, {Bell}, {Brown},
  {Bullock}, {Courteau}, {Dalcanton}, {Ferguson}, {Goudfrooij}, {Holfeltz},
  {Purcell}, {Radburn-Smith}, \& {Zucker}}]{dejong07}
{de Jong}, R.~S., {Seth}, A.~C., {Bell}, E.~F., {et~al.} 2007, in Stellar
  Populations as Building Blocks of Galaxies, ed. A.~{Vazdekis} \&
  R.~{Peletier}, Vol. 241, 503--504

\bibitem[{{de Vaucouleurs}(1963)}]{devaucouleurs63}
{de Vaucouleurs}, G. 1963, \apj, 138, 934

\bibitem[{{Dell'Agli} {et~al.}(2018){Dell'Agli}, {Di Criscienzo}, {Ventura},
  {Limongi}, {Garc{\'\i}a-Hern{\'a}ndez}, {Marini}, \& {Rossi}}]{dellagli18}
{Dell'Agli}, F., {Di Criscienzo}, M., {Ventura}, P., {et~al.} 2018, \mnras,
  479, 5035

\bibitem[{{Demers} {et~al.}(2006){Demers}, {Battinelli}, \&
  {Kunkel}}]{demers06}
{Demers}, S., {Battinelli}, P., \& {Kunkel}, W.~E. 2006, \apjl, 636, L85

\bibitem[{{Egorov} {et~al.}(2017){Egorov}, {Lozinskaya}, {Moiseev}, \&
  {Shchekinov}}]{egorov17}
{Egorov}, O.~V., {Lozinskaya}, T.~A., {Moiseev}, A.~V., \& {Shchekinov}, Y.~A.
  2017, \mnras, 464, 1833

\bibitem[{{Euclid Collaboration: Borlaff} {et~al.}(2022){Euclid Collaboration:
  Borlaff}, {G{\'o}mez-Alvarez}, {Altieri}, {Marcum}, {Vavrek}, {Laureijs},
  {Kohley}, {Buitrago}, {Cuillandre}, {Duc}, {Gaspar Venancio}, {Amara},
  {Andreon}, {Auricchio}, {Azzollini}, {Baccigalupi},
  {Balaguera-Antol{\'\i}nez}, {Baldi}, {Bardelli}, {Bender}, {Biviano},
  {Bodendorf}, {Bonino}, {Bozzo}, {Branchini}, {Brescia}, {Brinchmann},
  {Burigana}, {Cabanac}, {Camera}, {Candini}, {Capobianco}, {Cappi}, {Carbone},
  {Carretero}, {Carvalho}, {Casas}, {Castander}, {Castellano}, {Castignani},
  {Cavuoti}, {Cimatti}, {Cledassou}, {Colodro-Conde}, {Congedo}, {Conselice},
  {Conversi}, {Copin}, {Corcione}, {Coupon}, {Courtois}, {Cropper}, {Da Silva},
  {Degaudenzi}, {Di Ferdinando}, {Douspis}, {Dubath}, {Duncan}, {Dupac},
  {Dusini}, {Ealet}, {Fabricius}, {Farina}, {Farrens}, {Ferreira}, {Ferriol},
  {Finelli}, {Flose-Reimberg}, {Fosalba}, {Frailis}, {Franceschi}, {Fumana},
  {Galeotta}, {Ganga}, {Garilli}, {Gillis}, {Giocoli}, {Gozaliasl},
  {Graci{\'a}-Carpio}, {Grazian}, {Grupp}, {Haugan}, {Holmes}, {Hormuth},
  {Jahnke}, {Keihanen}, {Kermiche}, {Kiessling}, {Kilbinger}, {Kirkpatrick},
  {Kitching}, {Knapen}, {Kubik}, {K{\"u}mmel}, {Kunz}, {Kurki-Suonio},
  {Liebing}, {Ligori}, {Lilje}, {Lindholm}, {Lloro}, {Mainetti}, {Maino},
  {Mansutti}, {Marggraf}, {Markovic}, {Martinelli}, {Martinet},
  {Mart{\'\i}nez-Delgado}, {Marulli}, {Massey}, {Maturi}, {Maurogordato},
  {Medinaceli}, {Mei}, {Meneghetti}, {Merlin}, {Metcalf}, {Meylan}, {Moresco},
  {Morgante}, {Moscardini}, {Munari}, {Nakajima}, {Neissner}, {Niemi},
  {Nightingale}, {Nucita}, {Padilla}, {Paltani}, {Pasian}, {Patrizii},
  {Pedersen}, {Percival}, {Pettorino}, {Pires}, {Poncet}, {Popa}, {Potter},
  {Pozzetti}, {Raison}, {Rebolo}, {Renzi}, {Rhodes}, {Riccio}, {Romelli},
  {Roncarelli}, {Rosset}, {Rossetti}, {Saglia}, {S{\'a}nchez}, {Sapone},
  {Sauvage}, {Schneider}, {Scottez}, {Secroun}, {Seidel}, {Serrano},
  {Sirignano}, {Sirri}, {Skottfelt}, {Stanco}, {Starck}, {Sureau},
  {Tallada-Cresp{\'\i}}, {Taylor}, {Tenti}, {Tereno}, {Teyssier},
  {Toledo-Moreo}, {Torradeflot}, {Tutusaus}, {Valentijn}, {Valenziano},
  {Valiviita}, {Vassallo}, {Viel}, {Wang}, {Weller}, {Whittaker}, {Zacchei},
  {Zamorani}, \& {Zucca}}]{Borlaff-EP16}
{Euclid Collaboration: Borlaff}, A.~S., {G{\'o}mez-Alvarez}, P., {Altieri}, B.,
  {et~al.} 2022, \aap, 657, A92

\bibitem[{{Euclid Collaboration: Cropper} {et~al.}(2024){Euclid Collaboration:
  Cropper}, {Al Bahlawan}, {Amiaux}, {et~al.}}]{EuclidSkyVIS}
{Euclid Collaboration: Cropper}, M., {Al Bahlawan}, A., {Amiaux}, J., {et~al.}
  2024, \aap, this issue

\bibitem[{{Euclid Collaboration: Jahnke} {et~al.}(2024){Euclid Collaboration:
  Jahnke}, {Gillard}, {Schirmer}, {et~al.}}]{EuclidSkyNISP}
{Euclid Collaboration: Jahnke}, K., {Gillard}, W., {Schirmer}, M., {et~al.}
  2024, \aap, this issue

\bibitem[{{Euclid Collaboration: Mellier} {et~al.}(2024){Euclid Collaboration:
  Mellier}, {Abdurro'uf}, {Acevedo~Barroso}, {Ach\'ucarro},
  {et~al.}}]{EuclidSkyOverview}
{Euclid Collaboration: Mellier}, Y., {Abdurro'uf}, {Acevedo~Barroso}, J.,
  {Ach\'ucarro}, A., {et~al.} 2024, \aap, this issue

\bibitem[{{Euclid Collaboration: Scaramella} {et~al.}(2022){Euclid
  Collaboration: Scaramella}, {Amiaux}, {Mellier}, {Burigana}, {Carvalho},
  {Cuillandre}, {Da Silva}, {Derosa}, {Dinis}, {Maiorano}, {Maris}, {Tereno},
  {Laureijs}, {Boenke}, {Buenadicha}, {Dupac}, {Gaspar Venancio},
  {G{\'o}mez-{\'A}lvarez}, {Hoar}, {Lorenzo Alvarez}, {Racca},
  {Saavedra-Criado}, {Schwartz}, {Vavrek}, {Schirmer}, {Aussel}, {Azzollini},
  {Cardone}, {Cropper}, {Ealet}, {Garilli}, {Gillard}, {Granett}, {Guzzo},
  {Hoekstra}, {Jahnke}, {Kitching}, {Maciaszek}, {Meneghetti}, {Miller},
  {Nakajima}, {Niemi}, {Pasian}, {Percival}, {Pottinger}, {Sauvage},
  {Scodeggio}, {Wachter}, {Zacchei}, {Aghanim}, {Amara}, {Auphan}, {Auricchio},
  {Awan}, {Balestra}, {Bender}, {Bodendorf}, {Bonino}, {Branchini},
  {Brau-Nogue}, {Brescia}, {Candini}, {Capobianco}, {Carbone}, {Carlberg},
  {Carretero}, {Casas}, {Castander}, {Castellano}, {Cavuoti}, {Cimatti},
  {Cledassou}, {Congedo}, {Conselice}, {Conversi}, {Copin}, {Corcione},
  {Costille}, {Courbin}, {Degaudenzi}, {Douspis}, {Dubath}, {Duncan}, {Dusini},
  {Farrens}, {Ferriol}, {Fosalba}, {Fourmanoit}, {Frailis}, {Franceschi},
  {Franzetti}, {Fumana}, {Gillis}, {Giocoli}, {Grazian}, {Grupp}, {Haugan},
  {Holmes}, {Hormuth}, {Hudelot}, {Kermiche}, {Kiessling}, {Kilbinger},
  {Kohley}, {Kubik}, {K{\"u}mmel}, {Kunz}, {Kurki-Suonio}, {Lahav}, {Ligori},
  {Lilje}, {Lloro}, {Mansutti}, {Marggraf}, {Markovic}, {Marulli}, {Massey},
  {Maurogordato}, {Melchior}, {Merlin}, {Meylan}, {Mohr}, {Moresco}, {Morin},
  {Moscardini}, {Munari}, {Nichol}, {Padilla}, {Paltani}, {Peacock},
  {Pedersen}, {Pettorino}, {Pires}, {Poncet}, {Popa}, {Pozzetti}, {Raison},
  {Rebolo}, {Rhodes}, {Rix}, {Roncarelli}, {Rossetti}, {Saglia}, {Schneider},
  {Schrabback}, {Secroun}, {Seidel}, {Serrano}, {Sirignano}, {Sirri},
  {Skottfelt}, {Stanco}, {Starck}, {Tallada-Cresp{\'\i}}, {Tavagnacco},
  {Taylor}, {Teplitz}, {Toledo-Moreo}, {Torradeflot}, {Trifoglio}, {Valentijn},
  {Valenziano}, {Verdoes Kleijn}, {Wang}, {Welikala}, {Weller}, {Wetzstein},
  {Zamorani}, {Zoubian}, {Andreon}, {Baldi}, {Bardelli}, {Boucaud}, {Camera},
  {Di Ferdinando}, {Fabbian}, {Farinelli}, {Galeotta}, {Graci{\'a}-Carpio},
  {Maino}, {Medinaceli}, {Mei}, {Neissner}, {Polenta}, {Renzi}, {Romelli},
  {Rosset}, {Sureau}, {Tenti}, {Vassallo}, {Zucca}, {Baccigalupi},
  {Balaguera-Antol{\'\i}nez}, {Battaglia}, {Biviano}, {Borgani}, {Bozzo},
  {Cabanac}, {Cappi}, {Casas}, {Castignani}, {Colodro-Conde}, {Coupon},
  {Courtois}, {Cuby}, {de la Torre}, {Desai}, {Dole}, {Fabricius}, {Farina},
  {Ferreira}, {Finelli}, {Flose-Reimberg}, {Fotopoulou}, {Ganga}, {Gozaliasl},
  {Hook}, {Keihanen}, {Kirkpatrick}, {Liebing}, {Lindholm}, {Mainetti},
  {Martinelli}, {Martinet}, {Maturi}, {McCracken}, {Metcalf}, {Morgante},
  {Nightingale}, {Nucita}, {Patrizii}, {Potter}, {Riccio}, {S{\'a}nchez},
  {Sapone}, {Schewtschenko}, {Schultheis}, {Scottez}, {Teyssier}, {Tutusaus},
  {Valiviita}, {Viel}, {Vriend}, \& {Whittaker}}]{Scaramella-EP1}
{Euclid Collaboration: Scaramella}, R., {Amiaux}, J., {Mellier}, Y., {et~al.}
  2022, \aap, 662, A112

\bibitem[{{Euclid Collaboration: Schirmer} {et~al.}(2022){Euclid Collaboration:
  Schirmer}, {Jahnke}, {Seidel}, {Aussel}, {Bodendorf}, {Grupp}, {Hormuth},
  {Wachter}, {Appleton}, {Barbier}, {Brinchmann}, {Carrasco}, {Castander},
  {Coupon}, {De Paolis}, {Franco}, {Ganga}, {Hudelot}, {Jullo}, {Lan{\c{c}}on},
  {Nucita}, {Paltani}, {Smadja}, {Strafella}, {Venancio}, {Weiler}, {Amara},
  {Auphan}, {Auricchio}, {Balestra}, {Bender}, {Bonino}, {Branchini},
  {Brescia}, {Capobianco}, {Carbone}, {Carretero}, {Casas}, {Castellano},
  {Cavuoti}, {Cimatti}, {Cledassou}, {Congedo}, {Conselice}, {Conversi},
  {Copin}, {Corcione}, {Costille}, {Courbin}, {Da Silva}, {Degaudenzi},
  {Douspis}, {Dubath}, {Dupac}, {Dusini}, {Ealet}, {Farrens}, {Ferriol},
  {Fosalba}, {Frailis}, {Franceschi}, {Franzetti}, {Fumana}, {Garilli},
  {Gillard}, {Gillis}, {Giocoli}, {Grazian}, {Guzzo}, {Haugan}, {Hoekstra},
  {Holmes}, {Hornstrup}, {K{\"u}mmel}, {Kermiche}, {Kiessling}, {Kilbinger},
  {Kitching}, {Kohley}, {Kunz}, {Kurki-Suonio}, {Laureijs}, {Ligori}, {Lilje},
  {Lloro}, {Maciaszek}, {Maiorano}, {Mansutti}, {Marggraf}, {Markovic},
  {Marulli}, {Massey}, {Maurogordato}, {Mellier}, {Meneghetti}, {Merlin},
  {Meylan}, {Moresco}, {Moscardini}, {Munari}, {Nakajima}, {Nichol}, {Niemi},
  {Padilla}, {Pasian}, {Pedersen}, {Percival}, {Pettorino}, {Pires}, {Poncet},
  {Popa}, {Pozzetti}, {Prieto}, {Raison}, {Rhodes}, {Rix}, {Roncarelli},
  {Rossetti}, {Saglia}, {Sartoris}, {Scaramella}, {Schneider}, {Secroun},
  {Serrano}, {Sirignano}, {Sirri}, {Stanco}, {Tallada-Cresp{\'\i}}, {Taylor},
  {Teplitz}, {Tereno}, {Toledo-Moreo}, {Torradeflot}, {Trifoglio}, {Valentijn},
  {Valenziano}, {Wang}, {Weller}, {Zamorani}, {Zoubian}, {Andreon}, {Bardelli},
  {Boucaud}, {Camera}, {Farinelli}, {Graci{\'a}-Carpio}, {Maino}, {Medinaceli},
  {Mei}, {Morisset}, {Polenta}, {Renzi}, {Romelli}, {Tenti}, {Vassallo},
  {Zacchei}, {Zucca}, {Baccigalupi}, {Balaguera-Antol{\'\i}nez}, {Biviano},
  {Blanchard}, {Borgani}, {Bozzo}, {Burigana}, {Cabanac}, {Cappi}, {Carvalho},
  {Casas}, {Castignani}, {Colodro-Conde}, {Cooray}, {Courtois}, {Crocce},
  {Cuby}, {Davini}, {de la Torre}, {Di Ferdinando}, {Escartin}, {Farina},
  {Ferreira}, {Finelli}, {Fotopoulou}, {Galeotta}, {Garcia-Bellido},
  {Gaztanaga}, {George}, {Gozaliasl}, {Hook}, {Ili{\'c}}, {Kansal},
  {Kashlinsky}, {Keihanen}, {Kirkpatrick}, {Lindholm}, {Mainetti}, {Maoli},
  {Martinelli}, {Martinet}, {Maturi}, {Mauri}, {McCracken}, {Metcalf},
  {Monaco}, {Morgante}, {Nightingale}, {Patrizii}, {Peel}, {Popa}, {Porciani},
  {Potter}, {Reimberg}, {Riccio}, {S{\'a}nchez}, {Sapone}, {Scottez},
  {Sefusatti}, {Teyssier}, {Tutusaus}, {Valieri}, {Valiviita}, {Viel}, \&
  {Hildebrandt}}]{Schirmer-EP18}
{Euclid Collaboration: Schirmer}, M., {Jahnke}, K., {Seidel}, G., {et~al.}
  2022, \aap, 662, A92

\bibitem[{{Euclid Early Release Observations}(2024)}]{EROcite}
{Euclid Early Release Observations}. 2024,
  \url{https://doi.org/10.57780/esa-qmocze3}

\bibitem[{{Ferguson} {et~al.}(2002){Ferguson}, {Irwin}, {Ibata}, {Lewis}, \&
  {Tanvir}}]{ferguson02}
{Ferguson}, A. M.~N., {Irwin}, M.~J., {Ibata}, R.~A., {Lewis}, G.~F., \&
  {Tanvir}, N.~R. 2002, \aj, 124, 1452

\bibitem[{{Fioc} \& {Rocca-Volmerange}(1997)}]{fioc97}
{Fioc}, M. \& {Rocca-Volmerange}, B. 1997, \aap, 326, 950

\bibitem[{{Forbes} \& {Bridges}(2010)}]{forbes10}
{Forbes}, D.~A. \& {Bridges}, T. 2010, \mnras, 404, 1203

\bibitem[{{Forbes} {et~al.}(2022){Forbes}, {Ferr{\'e}-Mateu}, {Gannon},
  {Romanowsky}, {Carlin}, {Brodie}, \& {Day}}]{forbes22}
{Forbes}, D.~A., {Ferr{\'e}-Mateu}, A., {Gannon}, J.~S., {et~al.} 2022, \mnras,
  512, 802

\bibitem[{{Forbes} {et~al.}(1996){Forbes}, {Franx}, {Illingworth}, \&
  {Carollo}}]{forbes96}
{Forbes}, D.~A., {Franx}, M., {Illingworth}, G.~D., \& {Carollo}, C.~M. 1996,
  \apj, 467, 126

\bibitem[{{Fraternali} \& {Binney}(2008)}]{fraternali08}
{Fraternali}, F. \& {Binney}, J.~J. 2008, \mnras, 386, 935

\bibitem[{{Fraternali} {et~al.}(2002){Fraternali}, {van Moorsel}, {Sancisi}, \&
  {Oosterloo}}]{fraternali02}
{Fraternali}, F., {van Moorsel}, G., {Sancisi}, R., \& {Oosterloo}, T. 2002,
  \aj, 123, 3124

\bibitem[{{Fusco} {et~al.}(2012){Fusco}, {Buonanno}, {Bono}, {Cassisi},
  {Monelli}, \& {Pietrinferni}}]{Fusco2012}
{Fusco}, F., {Buonanno}, R., {Bono}, G., {et~al.} 2012, \aap, 548, A129

\bibitem[{{Gaia Collaboration} {et~al.}(2021){Gaia Collaboration}, {Brown},
  {Vallenari}, {Prusti}, {de Bruijne}, {Babusiaux}, {Biermann}, {Creevey},
  {Evans}, {Eyer}, {Hutton}, {Jansen}, {Jordi}, {Klioner}, {Lammers},
  {Lindegren}, {Luri}, {Mignard}, {Panem}, {Pourbaix}, {Randich}, {Sartoretti},
  {Soubiran}, {Walton}, {Arenou}, {Bailer-Jones}, {Bastian}, {Cropper},
  {Drimmel}, {Katz}, {Lattanzi}, {van Leeuwen}, {Bakker}, {Cacciari},
  {Casta{\~n}eda}, {De Angeli}, {Ducourant}, {Fabricius}, {Fouesneau},
  {Fr{\'e}mat}, {Guerra}, {Guerrier}, {Guiraud}, {Jean-Antoine Piccolo},
  {Masana}, {Messineo}, {Mowlavi}, {Nicolas}, {Nienartowicz}, {Pailler},
  {Panuzzo}, {Riclet}, {Roux}, {Seabroke}, {Sordo}, {Tanga}, {Th{\'e}venin},
  {Gracia-Abril}, {Portell}, {Teyssier}, {Altmann}, {Andrae}, {Bellas-Velidis},
  {Benson}, {Berthier}, {Blomme}, {Brugaletta}, {Burgess}, {Busso}, {Carry},
  {Cellino}, {Cheek}, {Clementini}, {Damerdji}, {Davidson}, {Delchambre},
  {Dell'Oro}, {Fern{\'a}ndez-Hern{\'a}ndez}, {Galluccio}, {Garc{\'\i}a-Lario},
  {Garcia-Reinaldos}, {Gonz{\'a}lez-N{\'u}{\~n}ez}, {Gosset}, {Haigron},
  {Halbwachs}, {Hambly}, {Harrison}, {Hatzidimitriou}, {Heiter},
  {Hern{\'a}ndez}, {Hestroffer}, {Hodgkin}, {Holl}, {Jan{\ss}en}, {Jevardat de
  Fombelle}, {Jordan}, {Krone-Martins}, {Lanzafame}, {L{\"o}ffler}, {Lorca},
  {Manteiga}, {Marchal}, {Marrese}, {Moitinho}, {Mora}, {Muinonen}, {Osborne},
  {Pancino}, {Pauwels}, {Petit}, {Recio-Blanco}, {Richards}, {Riello},
  {Rimoldini}, {Robin}, {Roegiers}, {Rybizki}, {Sarro}, {Siopis}, {Smith},
  {Sozzetti}, {Ulla}, {Utrilla}, {van Leeuwen}, {van Reeven}, {Abbas}, {Abreu
  Aramburu}, {Accart}, {Aerts}, {Aguado}, {Ajaj}, {Altavilla}, {{\'A}lvarez},
  {{\'A}lvarez Cid-Fuentes}, {Alves}, {Anderson}, {Anglada Varela}, {Antoja},
  {Audard}, {Baines}, {Baker}, {Balaguer-N{\'u}{\~n}ez}, {Balbinot}, {Balog},
  {Barache}, {Barbato}, {Barros}, {Barstow}, {Bartolom{\'e}}, {Bassilana},
  {Bauchet}, {Baudesson-Stella}, {Becciani}, {Bellazzini}, {Bernet}, {Bertone},
  {Bianchi}, {Blanco-Cuaresma}, {Boch}, {Bombrun}, {Bossini}, {Bouquillon},
  {Bragaglia}, {Bramante}, {Breedt}, {Bressan}, {Brouillet}, {Bucciarelli},
  {Burlacu}, {Busonero}, {Butkevich}, {Buzzi}, {Caffau}, {Cancelliere},
  {C{\'a}novas}, {Cantat-Gaudin}, {Carballo}, {Carlucci}, {Carnerero},
  {Carrasco}, {Casamiquela}, {Castellani}, {Castro-Ginard}, {Castro Sampol},
  {Chaoul}, {Charlot}, {Chemin}, {Chiavassa}, {Cioni}, {Comoretto}, {Cooper},
  {Cornez}, {Cowell}, {Crifo}, {Crosta}, {Crowley}, {Dafonte}, {Dapergolas},
  {David}, {David}, {de Laverny}, {De Luise}, {De March}, {De Ridder}, {de
  Souza}, {de Teodoro}, {de Torres}, {del Peloso}, {del Pozo}, {Delbo},
  {Delgado}, {Delgado}, {Delisle}, {Di Matteo}, {Diakite}, {Diener},
  {Distefano}, {Dolding}, {Eappachen}, {Edvardsson}, {Enke}, {Esquej}, {Fabre},
  {Fabrizio}, {Faigler}, {Fedorets}, {Fernique}, {Fienga}, {Figueras},
  {Fouron}, {Fragkoudi}, {Fraile}, {Franke}, {Gai}, {Garabato},
  {Garcia-Gutierrez}, {Garc{\'\i}a-Torres}, {Garofalo}, {Gavras}, {Gerlach},
  {Geyer}, {Giacobbe}, {Gilmore}, {Girona}, {Giuffrida}, {Gomel}, {Gomez},
  {Gonzalez-Santamaria}, {Gonz{\'a}lez-Vidal}, {Granvik},
  {Guti{\'e}rrez-S{\'a}nchez}, {Guy}, {Hauser}, {Haywood}, {Helmi}, {Hidalgo},
  {Hilger}, {H{\l}adczuk}, {Hobbs}, {Holland}, {Huckle}, {Jasniewicz},
  {Jonker}, {Juaristi Campillo}, {Julbe}, {Karbevska}, {Kervella}, {Khanna},
  {Kochoska}, {Kontizas}, {Kordopatis}, {Korn}, {Kostrzewa-Rutkowska},
  {Kruszy{\'n}ska}, {Lambert}, {Lanza}, {Lasne}, {Le Campion}, {Le Fustec},
  {Lebreton}, {Lebzelter}, {Leccia}, {Leclerc}, {Lecoeur-Taibi}, {Liao},
  {Licata}, {Lindstr{\o}m}, {Lister}, {Livanou}, {Lobel}, {Madrero Pardo},
  {Managau}, {Mann}, {Marchant}, {Marconi}, {Marcos Santos}, {Marinoni},
  {Marocco}, {Marshall}, {Martin Polo}, {Mart{\'\i}n-Fleitas}, {Masip},
  {Massari}, {Mastrobuono-Battisti}, {Mazeh}, {McMillan}, {Messina},
  {Michalik}, {Millar}, {Mints}, {Molina}, {Molinaro}, {Moln{\'a}r},
  {Montegriffo}, {Mor}, {Morbidelli}, {Morel}, {Morris}, {Mulone}, {Munoz},
  {Muraveva}, {Murphy}, {Musella}, {Noval}, {Ord{\'e}novic}, {Orr{\`u}},
  {Osinde}, {Pagani}, {Pagano}, {Palaversa}, {Palicio}, {Panahi}, {Pawlak},
  {Pe{\~n}alosa Esteller}, {Penttil{\"a}}, {Piersimoni}, {Pineau}, {Plachy},
  {Plum}, {Poggio}, {Poretti}, {Poujoulet}, {Pr{\v{s}}a}, {Pulone}, {Racero},
  {Ragaini}, {Rainer}, {Raiteri}, {Rambaux}, {Ramos}, {Ramos-Lerate}, {Re
  Fiorentin}, {Regibo}, {Reyl{\'e}}, {Ripepi}, {Riva}, {Rixon}, {Robichon},
  {Robin}, {Roelens}, {Rohrbasser}, {Romero-G{\'o}mez}, {Rowell}, {Royer},
  {Rybicki}, {Sadowski}, {Sagrist{\`a} Sell{\'e}s}, {Sahlmann}, {Salgado},
  {Salguero}, {Samaras}, {Sanchez Gimenez}, {Sanna}, {Santove{\~n}a},
  {Sarasso}, {Schultheis}, {Sciacca}, {Segol}, {Segovia}, {S{\'e}gransan},
  {Semeux}, {Shahaf}, {Siddiqui}, {Siebert}, {Siltala}, {Slezak}, {Smart},
  {Solano}, {Solitro}, {Souami}, {Souchay}, {Spagna}, {Spoto}, {Steele},
  {Steidelm{\"u}ller}, {Stephenson}, {S{\"u}veges}, {Szabados}, {Szegedi-Elek},
  {Taris}, {Tauran}, {Taylor}, {Teixeira}, {Thuillot}, {Tonello}, {Torra},
  {Torra}, {Turon}, {Unger}, {Vaillant}, {van Dillen}, {Vanel}, {Vecchiato},
  {Viala}, {Vicente}, {Voutsinas}, {Weiler}, {Wevers}, {Wyrzykowski}, {Yoldas},
  {Yvard}, {Zhao}, {Zorec}, {Zucker}, {Zurbach}, \& {Zwitter}}]{gaiadr3}
{Gaia Collaboration}, {Brown}, A.~G.~A., {Vallenari}, A., {et~al.} 2021, \aap,
  649, A1

\bibitem[{{Gerbrandt} {et~al.}(2015){Gerbrandt}, {McConnachie}, \&
  {Irwin}}]{gerbrandt15}
{Gerbrandt}, S. A.~N., {McConnachie}, A.~W., \& {Irwin}, M. 2015, \mnras, 454,
  1000

\bibitem[{{Gerola} \& {Seiden}(1978)}]{gerola78}
{Gerola}, H. \& {Seiden}, P.~E. 1978, \apj, 223, 129

\bibitem[{{Girardi} {et~al.}(2012){Girardi}, {Barbieri}, {Groenewegen},
  {Marigo}, {Bressan}, {Rocha-Pinto}, {Santiago}, {Camargo}, \& {da
  Costa}}]{Trilegal_B}
{Girardi}, L., {Barbieri}, M., {Groenewegen}, M. A.~T., {et~al.} 2012, in
  Astrophysics and Space Science Proceedings, Vol.~26, Red Giants as Probes of
  the Structure and Evolution of the Milky Way, 165

\bibitem[{{Girardi} {et~al.}(2005){Girardi}, {Groenewegen}, {Hatziminaoglou},
  \& {da Costa}}]{Trilegal_A}
{Girardi}, L., {Groenewegen}, M.~A.~T., {Hatziminaoglou}, E., \& {da Costa}, L.
  2005, \aap, 436, 895

\bibitem[{{Goad} {et~al.}(2006){Goad}, {Roberts}, {Reeves}, \&
  {Uttley}}]{goad06}
{Goad}, M.~R., {Roberts}, T.~P., {Reeves}, J.~N., \& {Uttley}, P. 2006, \mnras,
  365, 191

\bibitem[{{Gogarten} {et~al.}(2010){Gogarten}, {Dalcanton}, {Williams},
  {Ro{\v{s}}kar}, {Holtzman}, {Seth}, {Dolphin}, {Weisz}, {Cole}, {Debattista},
  {Gilbert}, {Olsen}, {Skillman}, {de Jong}, {Karachentsev}, \&
  {Quinn}}]{gogarten10}
{Gogarten}, S.~M., {Dalcanton}, J.~J., {Williams}, B.~F., {et~al.} 2010, \apj,
  712, 858

\bibitem[{{Gordon} {et~al.}(2023){Gordon}, {Clayton}, {Decleir}, {Fitzpatrick},
  {Massa}, {Misselt}, \& {Tollerud}}]{gordon23}
{Gordon}, K.~D., {Clayton}, G.~C., {Decleir}, M., {et~al.} 2023, \apj, 950, 86

\bibitem[{{Gordon} {et~al.}(2003){Gordon}, {Clayton}, {Misselt}, {Landolt}, \&
  {Wolff}}]{gordon03}
{Gordon}, K.~D., {Clayton}, G.~C., {Misselt}, K.~A., {Landolt}, A.~U., \&
  {Wolff}, M.~J. 2003, \apj, 594, 279

\bibitem[{{Hammer} {et~al.}(2005){Hammer}, {Flores}, {Elbaz}, {Zheng}, {Liang},
  \& {Cesarsky}}]{hammer05}
{Hammer}, F., {Flores}, H., {Elbaz}, D., {et~al.} 2005, \aap, 430, 115

\bibitem[{{Harris}(1996)}]{Harris1996}
{Harris}, W.~E. 1996, \aj, 112, 1487

\bibitem[{{Harris}(2009)}]{harris09}
{Harris}, W.~E. 2009, \apj, 699, 254

\bibitem[{{Harris} {et~al.}(2013){Harris}, {Harris}, \& {Alessi}}]{harris13}
{Harris}, W.~E., {Harris}, G. L.~H., \& {Alessi}, M. 2013, \apj, 772, 82

\bibitem[{{Harris} \& {Racine}(1979)}]{harris79}
{Harris}, W.~E. \& {Racine}, R. 1979, \araa, 17, 241

\bibitem[{{Heesen} {et~al.}(2015){Heesen}, {Brinks}, {Krause}, {Harwood},
  {Rau}, {Rupen}, {Hunter}, {Chyzy}, \& {Kitchener}}]{heesen15}
{Heesen}, V., {Brinks}, E., {Krause}, M.~G.~H., {et~al.} 2015, \mnras, 447, L1

\bibitem[{{Heesen} {et~al.}(2018){Heesen}, {Rafferty}, {Horneffer}, {Beck},
  {Basu}, {Westcott}, {Hindson}, {Brinks}, {Chy{\.Z}y}, {Scaife},
  {Br{\"u}ggen}, {Heald}, {Fletcher}, {Horellou}, {Tabatabaei}, {Paladino},
  {Nikiel-Wroczy{\'n}ski}, {Hoeft}, \& {Dettmar}}]{heesen18}
{Heesen}, V., {Rafferty}, D.~A., {Horneffer}, A., {et~al.} 2018, \mnras, 476,
  1756

\bibitem[{{Helmi} {et~al.}(2018){Helmi}, {Babusiaux}, {Koppelman}, {Massari},
  {Veljanoski}, \& {Brown}}]{helmi18}
{Helmi}, A., {Babusiaux}, C., {Koppelman}, H.~H., {et~al.} 2018, \nat, 563, 85

\bibitem[{{Hillis} {et~al.}(2016){Hillis}, {Williams}, {Dolphin}, {Dalcanton},
  \& {Skillman}}]{hillis16}
{Hillis}, T.~J., {Williams}, B.~F., {Dolphin}, A.~E., {Dalcanton}, J.~J., \&
  {Skillman}, E.~D. 2016, \apj, 831, 191

\bibitem[{{Hodge} \& {Lee}(1990)}]{hodge90}
{Hodge}, P. \& {Lee}, M.~G. 1990, \pasp, 102, 26

\bibitem[{{Hodge} {et~al.}(1994){Hodge}, {Strobel}, \& {Kennicutt}}]{hodge94}
{Hodge}, P., {Strobel}, N.~V., \& {Kennicutt}, R.~C. 1994, \pasp, 106, 309

\bibitem[{{Holmberg}(1950)}]{holmberg50}
{Holmberg}, E. 1950, Meddelanden fran Lunds Astronomiska Observatorium Serie
  II, 128, 5

\bibitem[{{Hubble}(1925)}]{hubble25}
{Hubble}, E.~P. 1925, \apj, 62, 409

\bibitem[{{Hunter}(2001)}]{hunter01}
{Hunter}, D.~A. 2001, \apj, 559, 225

\bibitem[{{Huxor} {et~al.}(2013){Huxor}, {Ferguson}, {Veljanoski}, {Mackey}, \&
  {Tanvir}}]{huxor13}
{Huxor}, A.~P., {Ferguson}, A.~M.~N., {Veljanoski}, J., {Mackey}, A.~D., \&
  {Tanvir}, N.~R. 2013, \mnras, 429, 1039

\bibitem[{{Ibata} {et~al.}(2001){Ibata}, {Lewis}, {Irwin}, {Totten}, \&
  {Quinn}}]{ibata01}
{Ibata}, R., {Lewis}, G.~F., {Irwin}, M., {Totten}, E., \& {Quinn}, T. 2001,
  \apj, 551, 294

\bibitem[{{Ibata} {et~al.}(2007){Ibata}, {Martin}, {Irwin}, {Chapman},
  {Ferguson}, {Lewis}, \& {McConnachie}}]{ibata07}
{Ibata}, R., {Martin}, N.~F., {Irwin}, M., {et~al.} 2007, \apj, 671, 1591

\bibitem[{{Ilbert} {et~al.}(2006){Ilbert}, {Arnouts}, {McCracken},
  {Bolzonella}, {Bertin}, {Le F{\`e}vre}, {Mellier}, {Zamorani}, {Pell{\`o}},
  {Iovino}, {Tresse}, {Le Brun}, {Bottini}, {Garilli}, {Maccagni}, {Picat},
  {Scaramella}, {Scodeggio}, {Vettolani}, {Zanichelli}, {Adami}, {Bardelli},
  {Cappi}, {Charlot}, {Ciliegi}, {Contini}, {Cucciati}, {Foucaud}, {Franzetti},
  {Gavignaud}, {Guzzo}, {Marano}, {Marinoni}, {Mazure}, {Meneux}, {Merighi},
  {Paltani}, {Pollo}, {Pozzetti}, {Radovich}, {Zucca}, {Bondi}, {Bongiorno},
  {Busarello}, {de La Torre}, {Gregorini}, {Lamareille}, {Mathez}, {Merluzzi},
  {Ripepi}, {Rizzo}, \& {Vergani}}]{ilbert06}
{Ilbert}, O., {Arnouts}, S., {McCracken}, H.~J., {et~al.} 2006, \aap, 457, 841

\bibitem[{{Jang} {et~al.}(2020a){Jang}, {de Jong}, {Holwerda}, {Monachesi},
  {Bell}, \& {Bailin}}]{jang20a}
{Jang}, I.~S., {de Jong}, R.~S., {Holwerda}, B.~W., {et~al.} 2020a, \aap, 637,
  A8

\bibitem[{{Jang} {et~al.}(2020b){Jang}, {de Jong}, {Minchev}, {Bell},
  {Monachesi}, {Holwerda}, {Bailin}, {Smercina}, \& {D'Souza}}]{jang20b}
{Jang}, I.~S., {de Jong}, R.~S., {Minchev}, I., {et~al.} 2020b, \aap, 640, L19

\bibitem[{{Jarrett} {et~al.}(2019){Jarrett}, {Cluver}, {Brown}, {Dale}, {Tsai},
  \& {Masci}}]{jarrett19}
{Jarrett}, T.~H., {Cluver}, M.~E., {Brown}, M.~J.~I., {et~al.} 2019, \apjs,
  245, 25

\bibitem[{{Johnston} {et~al.}(2001){Johnston}, {Sackett}, \&
  {Bullock}}]{johnston01}
{Johnston}, K.~V., {Sackett}, P.~D., \& {Bullock}, J.~S. 2001, \apj, 557, 137

\bibitem[{{Juri{\'c}} {et~al.}(2008){Juri{\'c}}, {Ivezi{\'c}}, {Brooks},
  {Lupton}, {Schlegel}, {Finkbeiner}, {Padmanabhan}, {Bond}, {Sesar},
  {Rockosi}, {Knapp}, {Gunn}, {Sumi}, {Schneider}, {Barentine}, {Brewington},
  {Brinkmann}, {Fukugita}, {Harvanek}, {Kleinman}, {Krzesinski}, {Long},
  {Neilsen}, {Nitta}, {Snedden}, \& {York}}]{juric08}
{Juri{\'c}}, M., {Ivezi{\'c}}, {\v{Z}}., {Brooks}, A., {et~al.} 2008, \apj,
  673, 864

\bibitem[{{Kaaret} {et~al.}(2004){Kaaret}, {Ward}, \& {Zezas}}]{kaaret04}
{Kaaret}, P., {Ward}, M.~J., \& {Zezas}, A. 2004, \mnras, 351, L83

\bibitem[{{Kamphuis} \& {Briggs}(1992)}]{kamphuis91}
{Kamphuis}, J. \& {Briggs}, F. 1992, \aap, 253, 335

\bibitem[{{Kankare} {et~al.}(2014){Kankare}, {Fraser}, {Ryder},
  {Romero-Ca{\~n}izales}, {Mattila}, {Kotak}, {Laursen}, {Monard}, {Salvo}, \&
  {V{\"a}is{\"a}nen}}]{kankare14}
{Kankare}, E., {Fraser}, M., {Ryder}, S., {et~al.} 2014, \aap, 572, A75

\bibitem[{{Karachentsev} {et~al.}(2002){Karachentsev}, {Dolphin}, {Geisler},
  {Grebel}, {Guhathakurta}, {Hodge}, {Karachentseva}, {Sarajedini}, {Seitzer},
  \& {Sharina}}]{karach02}
{Karachentsev}, I.~D., {Dolphin}, A.~E., {Geisler}, D., {et~al.} 2002, \aap,
  383, 125

\bibitem[{{Karachentsev} {et~al.}(2014){Karachentsev}, {Kaisina}, \&
  {Makarov}}]{karach14}
{Karachentsev}, I.~D., {Kaisina}, E.~I., \& {Makarov}, D.~I. 2014, \aj, 147, 13

\bibitem[{{Karachentsev} {et~al.}(2020){Karachentsev}, {Riepe}, \&
  {Zilch}}]{karach20}
{Karachentsev}, I.~D., {Riepe}, P., \& {Zilch}, T. 2020, Astrophysics, 63, 5

\bibitem[{{Kennicutt} {et~al.}(2011){Kennicutt}, {Calzetti}, {Aniano},
  {Appleton}, {Armus}, {Beir{\~a}o}, {Bolatto}, {Brandl}, {Crocker}, {Croxall},
  {Dale}, {Donovan Meyer}, {Draine}, {Engelbracht}, {Galametz}, {Gordon},
  {Groves}, {Hao}, {Helou}, {Hinz}, {Hunt}, {Johnson}, {Koda}, {Krause},
  {Leroy}, {Li}, {Meidt}, {Montiel}, {Murphy}, {Rahman}, {Rix}, {Roussel},
  {Sandstrom}, {Sauvage}, {Schinnerer}, {Skibba}, {Smith}, {Srinivasan},
  {Vigroux}, {Walter}, {Wilson}, {Wolfire}, \& {Zibetti}}]{kennicutt11}
{Kennicutt}, R.~C., {Calzetti}, D., {Aniano}, G., {et~al.} 2011, \pasp, 123,
  1347

\bibitem[{{King}(1962)}]{King1962}
{King}, I. 1962, \aj, 67, 471

\bibitem[{{Komiyama} {et~al.}(2018){Komiyama}, {Chiba}, {Tanaka}, {Tanaka},
  {Kirihara}, {Miki}, {Mori}, {Lupton}, {Guhathakurta}, {Kalirai}, {Gilbert},
  {Kirby}, {Lee}, {Jang}, {Sharma}, \& {Hayashi}}]{komiyama18}
{Komiyama}, Y., {Chiba}, M., {Tanaka}, M., {et~al.} 2018, \apj, 853, 29

\bibitem[{{Komiyama} {et~al.}(2003){Komiyama}, {Okamura}, {Yagi}, {Furusawa},
  {Doi}, {Hamabe}, {Imi}, {Kimura}, {Miyazaki}, {Nakata}, {Okada}, {Ouchi},
  {Sekiguchi}, {Shimasaku}, {Yasuda}, {Arimoto}, \& {Ikuta}}]{komiyama03}
{Komiyama}, Y., {Okamura}, S., {Yagi}, M., {et~al.} 2003, \apjl, 590, L17

\bibitem[{{Kruijssen}(2015)}]{kruijssen15}
{Kruijssen}, J.~M.~D. 2015, \mnras, 454, 1658

\bibitem[{{Larsen} \& {Richtler}(1998)}]{Larsen1998}
{Larsen}, S. \& {Richtler}, T. 1998, in Astronomical Society of the Pacific
  Conference Series, Vol. 136, Galactic Halos, ed. D.~{Zaritsky}, 67

\bibitem[{{Larsen}(1999)}]{Larsen1999}
{Larsen}, S.~S. 1999, \aaps, 139, 393

\bibitem[{{Larsen} {et~al.}(2001){Larsen}, {Brodie}, {Huchra}, {Forbes}, \&
  {Grillmair}}]{larsen01}
{Larsen}, S.~S., {Brodie}, J.~P., {Huchra}, J.~P., {Forbes}, D.~A., \&
  {Grillmair}, C.~J. 2001, \aj, 121, 2974

\bibitem[{{Larsen} {et~al.}(2018){Larsen}, {Brodie}, {Wasserman}, \&
  {Strader}}]{larsen18}
{Larsen}, S.~S., {Brodie}, J.~P., {Wasserman}, A., \& {Strader}, J. 2018, \aap,
  613, A56

\bibitem[{{Larsen} {et~al.}(2022){Larsen}, {Eitner}, {Magg}, {Bergemann},
  {Moltzer}, {Brodie}, {Romanowsky}, \& {Strader}}]{Larsen2022}
{Larsen}, S.~S., {Eitner}, P., {Magg}, E., {et~al.} 2022, \aap, 660, A88

\bibitem[{{Laureijs} {et~al.}(2011){Laureijs}, {Amiaux}, {Arduini},
  {Augu{\`e}res}, {Brinchmann}, {Cole}, {Cropper}, {Dabin}, {Duvet}, {Ealet},
  {Garilli}, {Gondoin}, {Guzzo}, {Hoar}, {Hoekstra}, {Holmes}, {Kitching},
  {Maciaszek}, {Mellier}, {Pasian}, {Percival}, {Rhodes}, {Saavedra Criado},
  {Sauvage}, {Scaramella}, {Valenziano}, {Warren}, {Bender}, {Castander},
  {Cimatti}, {Le F{\`e}vre}, {Kurki-Suonio}, {Levi}, {Lilje}, {Meylan},
  {Nichol}, {Pedersen}, {Popa}, {Rebolo Lopez}, {Rix}, {Rottgering},
  {Zeilinger}, {Grupp}, {Hudelot}, {Massey}, {Meneghetti}, {Miller}, {Paltani},
  {Paulin-Henriksson}, {Pires}, {Saxton}, {Schrabback}, {Seidel}, {Walsh},
  {Aghanim}, {Amendola}, {Bartlett}, {Baccigalupi}, {Beaulieu}, {Benabed},
  {Cuby}, {Elbaz}, {Fosalba}, {Gavazzi}, {Helmi}, {Hook}, {Irwin}, {Kneib},
  {Kunz}, {Mannucci}, {Moscardini}, {Tao}, {Teyssier}, {Weller}, {Zamorani},
  {Zapatero Osorio}, {Boulade}, {Foumond}, {Di Giorgio}, {Guttridge}, {James},
  {Kemp}, {Martignac}, {Spencer}, {Walton}, {Bl{\"u}mchen}, {Bonoli},
  {Bortoletto}, {Cerna}, {Corcione}, {Fabron}, {Jahnke}, {Ligori}, {Madrid},
  {Martin}, {Morgante}, {Pamplona}, {Prieto}, {Riva}, {Toledo}, {Trifoglio},
  {Zerbi}, {Abdalla}, {Douspis}, {Grenet}, {Borgani}, {Bouwens}, {Courbin},
  {Delouis}, {Dubath}, {Fontana}, {Frailis}, {Grazian}, {Koppenh{\"o}fer},
  {Mansutti}, {Melchior}, {Mignoli}, {Mohr}, {Neissner}, {Noddle}, {Poncet},
  {Scodeggio}, {Serrano}, {Shane}, {Starck}, {Surace}, {Taylor},
  {Verdoes-Kleijn}, {Vuerli}, {Williams}, {Zacchei}, {Altieri}, {Escudero
  Sanz}, {Kohley}, {Oosterbroek}, {Astier}, {Bacon}, {Bardelli}, {Baugh},
  {Bellagamba}, {Benoist}, {Bianchi}, {Biviano}, {Branchini}, {Carbone},
  {Cardone}, {Clements}, {Colombi}, {Conselice}, {Cresci}, {Deacon}, {Dunlop},
  {Fedeli}, {Fontanot}, {Franzetti}, {Giocoli}, {Garcia-Bellido}, {Gow},
  {Heavens}, {Hewett}, {Heymans}, {Holland}, {Huang}, {Ilbert}, {Joachimi},
  {Jennins}, {Kerins}, {Kiessling}, {Kirk}, {Kotak}, {Krause}, {Lahav}, {van
  Leeuwen}, {Lesgourgues}, {Lombardi}, {Magliocchetti}, {Maguire}, {Majerotto},
  {Maoli}, {Marulli}, {Maurogordato}, {McCracken}, {McLure}, {Melchiorri},
  {Merson}, {Moresco}, {Nonino}, {Norberg}, {Peacock}, {Pello}, {Penny},
  {Pettorino}, {Di Porto}, {Pozzetti}, {Quercellini}, {Radovich}, {Rassat},
  {Roche}, {Ronayette}, {Rossetti}, {Sartoris}, {Schneider}, {Semboloni},
  {Serjeant}, {Simpson}, {Skordis}, {Smadja}, {Smartt}, {Spano}, {Spiro},
  {Sullivan}, {Tilquin}, {Trotta}, {Verde}, {Wang}, {Williger}, {Zhao},
  {Zoubian}, \& {Zucca}}]{laureijs11}
{Laureijs}, R., {Amiaux}, J., {Arduini}, S., {et~al.} 2011, ESA/SRE(2011)12,
  arXiv:1110.3193

\bibitem[{{Lazzarini} {et~al.}(2022){Lazzarini}, {Williams}, {Durbin},
  {Dalcanton}, {Smercina}, {Bell}, {Choi}, {Dolphin}, {Gilbert},
  {Guhathakurta}, {Rosolowsky}, {Skillman}, {Telford}, \&
  {Weisz}}]{lazzarini22}
{Lazzarini}, M., {Williams}, B.~F., {Durbin}, M.~J., {et~al.} 2022, \apj, 934,
  76

\bibitem[{{Le Borgne} {et~al.}(2004){Le Borgne}, {Rocca-Volmerange},
  {Prugniel}, {Lan{\c{c}}on}, {Fioc}, \& {Soubiran}}]{leborgne04}
{Le Borgne}, D., {Rocca-Volmerange}, B., {Prugniel}, P., {et~al.} 2004, \aap,
  425, 881

\bibitem[{{Lee} {et~al.}(2006){Lee}, {Skillman}, \& {Venn}}]{lee06}
{Lee}, H., {Skillman}, E.~D., \& {Venn}, K.~A. 2006, \apj, 642, 813

\bibitem[{{Lelli} {et~al.}(2014){Lelli}, {Verheijen}, \&
  {Fraternali}}]{lelli14}
{Lelli}, F., {Verheijen}, M., \& {Fraternali}, F. 2014, \mnras, 445, 1694

\bibitem[{{Leroy} {et~al.}(2021){Leroy}, {Schinnerer}, {Hughes}, {Rosolowsky},
  {Pety}, {Schruba}, {Usero}, {Blanc}, {Chevance}, {Emsellem}, {Faesi},
  {Herrera}, {Liu}, {Meidt}, {Querejeta}, {Saito}, {Sandstrom}, {Sun},
  {Williams}, {Anand}, {Barnes}, {Behrens}, {Belfiore}, {Benincasa},
  {Be{\v{s}}li{\'c}}, {Bigiel}, {Bolatto}, {den Brok}, {Cao}, {Chandar},
  {Chastenet}, {Chiang}, {Congiu}, {Dale}, {Deger}, {Eibensteiner}, {Egorov},
  {Garc{\'\i}a-Rodr{\'\i}guez}, {Glover}, {Grasha}, {Henshaw}, {Ho}, {Kepley},
  {Kim}, {Klessen}, {Kreckel}, {Koch}, {Kruijssen}, {Larson}, {Lee}, {Lopez},
  {Machado}, {Mayker}, {McElroy}, {Murphy}, {Ostriker}, {Pan}, {Pessa},
  {Puschnig}, {Razza}, {S{\'a}nchez-Bl{\'a}zquez}, {Santoro}, {Sardone},
  {Scheuermann}, {Sliwa}, {Sormani}, {Stuber}, {Thilker}, {Turner}, {Utomo},
  {Watkins}, \& {Whitmore}}]{leroy21}
{Leroy}, A.~K., {Schinnerer}, E., {Hughes}, A., {et~al.} 2021, \apjs, 257, 43

\bibitem[{{Lewis} {et~al.}(2013){Lewis}, {Braun}, {McConnachie}, {Irwin},
  {Ibata}, {Chapman}, {Ferguson}, {Martin}, {Fardal}, {Dubinski}, {Widrow},
  {Mackey}, {Babul}, {Tanvir}, \& {Rich}}]{lewis13}
{Lewis}, G.~F., {Braun}, R., {McConnachie}, A.~W., {et~al.} 2013, \apj, 763, 4

\bibitem[{{Li} {et~al.}(2023){Li}, {Fraternali}, {Marasco}, {Trager},
  {Pezzulli}, {Mancera Pi{\~n}a}, \& {Verheijen}}]{li23}
{Li}, A., {Fraternali}, F., {Marasco}, A., {et~al.} 2023, \mnras, 520, 147

\bibitem[{{Lim} \& {Lee}(2015)}]{lim15}
{Lim}, S. \& {Lee}, M.~G. 2015, \apj, 804, 123

\bibitem[{{Mackey} {et~al.}(2019){Mackey}, {Ferguson}, {Huxor}, {Veljanoski},
  {Lewis}, {McConnachie}, {Martin}, {Ibata}, {Irwin}, {C{\^o}t{\'e}},
  {Collins}, {Tanvir}, \& {Bate}}]{mackey19}
{Mackey}, A.~D., {Ferguson}, A.~M.~N., {Huxor}, A.~P., {et~al.} 2019, \mnras,
  484, 1756

\bibitem[{{Madden} {et~al.}(2013){Madden}, {R{\'e}my-Ruyer}, {Galametz},
  {Cormier}, {Lebouteiller}, {Galliano}, {Hony}, {Bendo}, {Smith}, {Pohlen},
  {Roussel}, {Sauvage}, {Wu}, {Sturm}, {Poglitsch}, {Contursi}, {Doublier},
  {Baes}, {Barlow}, {Boselli}, {Boquien}, {Carlson}, {Ciesla}, {Cooray},
  {Cortese}, {de Looze}, {Irwin}, {Isaak}, {Kamenetzky}, {Karczewski}, {Lu},
  {MacHattie}, {O'Halloran}, {Parkin}, {Rangwala}, {Schirm}, {Schulz},
  {Spinoglio}, {Vaccari}, {Wilson}, \& {Wozniak}}]{madden13}
{Madden}, S.~C., {R{\'e}my-Ruyer}, A., {Galametz}, M., {et~al.} 2013, \pasp,
  125, 600

\bibitem[{{Majewski} {et~al.}(2003){Majewski}, {Skrutskie}, {Weinberg}, \&
  {Ostheimer}}]{majewski03}
{Majewski}, S.~R., {Skrutskie}, M.~F., {Weinberg}, M.~D., \& {Ostheimer}, J.~C.
  2003, \apj, 599, 1082

\bibitem[{{Mak} {et~al.}(2008){Mak}, {Pun}, \& {Kong}}]{mak08}
{Mak}, D. S.~Y., {Pun}, C. S.~J., \& {Kong}, A. K.~H. 2008, \apj, 686, 995

\bibitem[{{Mancera Pi{\~n}a} {et~al.}(2022){Mancera Pi{\~n}a}, {Fraternali},
  {Oosterloo}, {Adams}, {di Teodoro}, {Bacchini}, \& {Iorio}}]{mancera22}
{Mancera Pi{\~n}a}, P.~E., {Fraternali}, F., {Oosterloo}, T., {et~al.} 2022,
  \mnras, 514, 3329

\bibitem[{{Marigo} {et~al.}(2017){Marigo}, {Girardi}, {Bressan}, {Rosenfield},
  {Aringer}, {Chen}, {Dussin}, {Nanni}, {Pastorelli}, {Rodrigues}, {Trabucchi},
  {Bladh}, {Dalcanton}, {Groenewegen}, {Montalb{\'a}n}, \& {Wood}}]{Marigo2017}
{Marigo}, P., {Girardi}, L., {Bressan}, A., {et~al.} 2017, \apj, 835, 77

\bibitem[{{Marleau} {et~al.}(2021){Marleau}, {Habas}, {Poulain}, {Duc},
  {M{\"u}ller}, {Lim}, {Durrell}, {S{\'a}nchez-Janssen}, {Paudel}, {Ahad},
  {Chougule}, {B{\'\i}lek}, \& {Fensch}}]{marleau21}
{Marleau}, F.~R., {Habas}, R., {Poulain}, M., {et~al.} 2021, \aap, 654, A105

\bibitem[{{Martin} {et~al.}(2022a){Martin}, {Bazkiaei}, {Spavone}, {Iodice},
  {Mihos}, {Montes}, {Benavides}, {Brough}, {Carlin}, {Collins}, {Duc},
  {G{\'o}mez}, {Galaz}, {Hern{\'a}ndez-Toledo}, {Jackson}, {Kaviraj}, {Knapen},
  {Mart{\'\i}nez-Lombilla}, {McGee}, {O'Ryan}, {Prole}, {Rich}, {Rom{\'a}n},
  {Shah}, {Starkenburg}, {Watkins}, {Zaritsky}, {Pichon}, {Armus}, {Bianconi},
  {Buitrago}, {Bus{\'a}}, {Davis}, {Demarco}, {Desmons}, {Garc{\'\i}a},
  {Graham}, {Holwerda}, {Hon}, {Khalid}, {Klehammer}, {Klutse}, {Lazar},
  {Nair}, {Noakes-Kettel}, {Rutkowski}, {Saha}, {Sahu}, {Sola},
  {V{\'a}zquez-Mata}, {Vera-Casanova}, \& {Yoon}}]{martin22a}
{Martin}, G., {Bazkiaei}, A.~E., {Spavone}, M., {et~al.} 2022a, \mnras, 513,
  1459

\bibitem[{{Martin} {et~al.}(2022b){Martin}, {Ibata}, {Starkenburg}, {Yuan},
  {Malhan}, {Bellazzini}, {Viswanathan}, {Aguado}, {Arentsen}, {Bonifacio},
  {Carlberg}, {Gonz{\'a}lez Hern{\'a}ndez}, {Hill}, {Jablonka}, {Kordopatis},
  {Lardo}, {McConnachie}, {Navarro}, {S{\'a}nchez-Janssen}, {Sestito},
  {Thomas}, {Venn}, {Vitali}, \& {Voggel}}]{martin22b}
{Martin}, N.~F., {Ibata}, R.~A., {Starkenburg}, E., {et~al.} 2022b, \mnras,
  516, 5331

\bibitem[{{Mart{\'\i}nez-Delgado} {et~al.}(2023){Mart{\'\i}nez-Delgado},
  {Cooper}, {Rom{\'a}n}, {Pillepich}, {Erkal}, {Pearson}, {Moustakas},
  {Laporte}, {Laine}, {Akhlaghi}, {Lang}, {Makarov}, {Borlaff}, {Donatiello},
  {Pearson}, {Mir{\'o}-Carretero}, {Cuillandre}, {Dom{\'\i}nguez},
  {Roca-F{\`a}brega}, {Frenk}, {Schmidt}, {G{\'o}mez-Flechoso}, {Guzman},
  {Libeskind}, {Dey}, {Weaver}, {Schlegel}, {Myers}, \& {Valdes}}]{martinez23}
{Mart{\'\i}nez-Delgado}, D., {Cooper}, A.~P., {Rom{\'a}n}, J., {et~al.} 2023,
  \aap, 671, A141

\bibitem[{{Mart{\'\i}nez-Delgado} {et~al.}(2010){Mart{\'\i}nez-Delgado},
  {Gabany}, {Crawford}, {Zibetti}, {Majewski}, {Rix}, {Fliri},
  {Carballo-Bello}, {Bardalez-Gagliuffi}, {Pe{\~n}arrubia}, {Chonis}, {Madore},
  {Trujillo}, {Schirmer}, \& {McDavid}}]{martinez10}
{Mart{\'\i}nez-Delgado}, D., {Gabany}, R.~J., {Crawford}, K., {et~al.} 2010,
  \aj, 140, 962

\bibitem[{{Mart{\'\i}nez-Delgado} {et~al.}(2008){Mart{\'\i}nez-Delgado},
  {Pe{\~n}arrubia}, {Gabany}, {Trujillo}, {Majewski}, \& {Pohlen}}]{martinez08}
{Mart{\'\i}nez-Delgado}, D., {Pe{\~n}arrubia}, J., {Gabany}, R.~J., {et~al.}
  2008, \apj, 689, 184

\bibitem[{{Mart{\'\i}nez-Delgado} {et~al.}(2009){Mart{\'\i}nez-Delgado},
  {Pohlen}, {Gabany}, {Majewski}, {Pe{\~n}arrubia}, \& {Palma}}]{martinez09}
{Mart{\'\i}nez-Delgado}, D., {Pohlen}, M., {Gabany}, R.~J., {et~al.} 2009,
  \apj, 692, 955

\bibitem[{{Massey} {et~al.}(2007){Massey}, {Olsen}, {Hodge}, {Jacoby},
  {McNeill}, {Smith}, \& {Strong}}]{massey07}
{Massey}, P., {Olsen}, K.~A.~G., {Hodge}, P.~W., {et~al.} 2007, \aj, 133, 2393

\bibitem[{{Mayer} {et~al.}(2006){Mayer}, {Mastropietro}, {Wadsley}, {Stadel},
  \& {Moore}}]{meyer06}
{Mayer}, L., {Mastropietro}, C., {Wadsley}, J., {Stadel}, J., \& {Moore}, B.
  2006, \mnras, 369, 1021

\bibitem[{{McConnachie} {et~al.}(2010){McConnachie}, {Ferguson}, {Irwin},
  {Dubinski}, {Widrow}, {Dotter}, {Ibata}, \& {Lewis}}]{mcconnachie10}
{McConnachie}, A.~W., {Ferguson}, A. M.~N., {Irwin}, M.~J., {et~al.} 2010,
  \apj, 723, 1038

\bibitem[{{McConnachie} {et~al.}(2021){McConnachie}, {Higgs}, {Thomas}, {Venn},
  {C{\^o}t{\'e}}, {Battaglia}, \& {Lewis}}]{mcconnachie21}
{McConnachie}, A.~W., {Higgs}, C.~R., {Thomas}, G.~F., {et~al.} 2021, \mnras,
  501, 2363

\bibitem[{{McConnachie} {et~al.}(2009){McConnachie}, {Irwin}, {Ibata},
  {Dubinski}, {Widrow}, {Martin}, {C{\^o}t{\'e}}, {Dotter}, {Navarro},
  {Ferguson}, {Puzia}, {Lewis}, {Babul}, {Barmby}, {Bienaym{\'e}}, {Chapman},
  {Cockcroft}, {Collins}, {Fardal}, {Harris}, {Huxor}, {Mackey},
  {Pe{\~n}arrubia}, {Rich}, {Richer}, {Siebert}, {Tanvir}, {Valls-Gabaud}, \&
  {Venn}}]{mcconnachie09}
{McConnachie}, A.~W., {Irwin}, M.~J., {Ibata}, R.~A., {et~al.} 2009, \nat, 461,
  66

\bibitem[{{McQuinn} {et~al.}(2010){McQuinn}, {Skillman}, {Cannon}, {Dalcanton},
  {Dolphin}, {Hidalgo-Rodr{\'\i}guez}, {Holtzman}, {Stark}, {Weisz}, \&
  {Williams}}]{mcquinn10}
{McQuinn}, K. B.~W., {Skillman}, E.~D., {Cannon}, J.~M., {et~al.} 2010, \apj,
  721, 297

\bibitem[{{Mei} {et~al.}(2007){Mei}, {Blakeslee}, {C{\^o}t{\'e}}, {Tonry},
  {West}, {Ferrarese}, {Jord{\'a}n}, {Peng}, {Anthony}, \& {Merritt}}]{mei07}
{Mei}, S., {Blakeslee}, J.~P., {C{\^o}t{\'e}}, P., {et~al.} 2007, \apj, 655,
  144

\bibitem[{{Mei} {et~al.}(2005){Mei}, {Blakeslee}, {Tonry}, {Jord{\'a}n},
  {Peng}, {C{\^o}t{\'e}}, {Ferrarese}, {West}, {Merritt}, \&
  {Milosavljevi{\'c}}}]{mei05}
{Mei}, S., {Blakeslee}, J.~P., {Tonry}, J.~L., {et~al.} 2005, \apj, 625, 121

\bibitem[{{Merritt} {et~al.}(2016){Merritt}, {van Dokkum}, {Abraham}, \&
  {Zhang}}]{merritt16}
{Merritt}, A., {van Dokkum}, P., {Abraham}, R., \& {Zhang}, J. 2016, \apj, 830,
  62

\bibitem[{{Mihos}(2019)}]{mihos19}
{Mihos}, J.~C. 2019, arXiv e-prints, arXiv:1909.09456

\bibitem[{{Mihos} {et~al.}(2015){Mihos}, {Durrell}, {Ferrarese}, {Feldmeier},
  {C{\^o}t{\'e}}, {Peng}, {Harding}, {Liu}, {Gwyn}, \& {Cuillandre}}]{mihos15}
{Mihos}, J.~C., {Durrell}, P.~R., {Ferrarese}, L., {et~al.} 2015, \apjl, 809,
  L21

\bibitem[{{Mihos} \& {Hernquist}(1994)}]{mihos94}
{Mihos}, J.~C. \& {Hernquist}, L. 1994, \apjl, 425, L13

\bibitem[{{Mu{\~n}oz} {et~al.}(2015){Mu{\~n}oz}, {Eigenthaler}, {Puzia},
  {Taylor}, {Ordenes-Brice{\~n}o}, {Alamo-Mart{\'\i}nez}, {Ribbeck},
  {{\'A}ngel}, {Capaccioli}, {C{\^o}t{\'e}}, {Ferrarese}, {Galaz}, {Hempel},
  {Hilker}, {Jord{\'a}n}, {Lan{\c{c}}on}, {Mieske}, {Paolillo}, {Richtler},
  {S{\'a}nchez-Janssen}, \& {Zhang}}]{munoz15}
{Mu{\~n}oz}, R.~P., {Eigenthaler}, P., {Puzia}, T.~H., {et~al.} 2015, \apjl,
  813, L15

\bibitem[{{Mundy} {et~al.}(2017){Mundy}, {Conselice}, {Duncan}, {Almaini},
  {H{\"a}u{\ss}ler}, \& {Hartley}}]{mundy17}
{Mundy}, C.~J., {Conselice}, C.~J., {Duncan}, K.~J., {et~al.} 2017, \mnras,
  470, 3507

\bibitem[{{Nally} {et~al.}(2023){Nally}, {Jones}, {Lenki{\'c}}, {Habel},
  {Hirschauer}, {Meixner}, {Kavanagh}, {Boyer}, {Ferguson}, {Sargent}, {Nayak},
  \& {Temim}}]{Nally2024}
{Nally}, C., {Jones}, O.~C., {Lenki{\'c}}, L., {et~al.} 2023, arXiv e-prints,
  arXiv:2309.13521

\bibitem[{{Namumba} {et~al.}(2019){Namumba}, {Carignan}, {Foster}, \&
  {Deg}}]{namumba19}
{Namumba}, B., {Carignan}, C., {Foster}, T., \& {Deg}, N. 2019, \mnras, 490,
  3365

\bibitem[{{Nersesian} {et~al.}(2019){Nersesian}, {Xilouris}, {Bianchi},
  {Galliano}, {Jones}, {Baes}, {Casasola}, {Cassar{\`a}}, {Clark}, {Davies},
  {Decleir}, {Dobbels}, {De Looze}, {De Vis}, {Fritz}, {Galametz}, {Madden},
  {Mosenkov}, {Tr{\v{c}}ka}, {Verstocken}, {Viaene}, \& {Lianou}}]{nersesian19}
{Nersesian}, A., {Xilouris}, E.~M., {Bianchi}, S., {et~al.} 2019, \aap, 624,
  A80

\bibitem[{{Nidever} {et~al.}(2013){Nidever}, {Ashley}, {Slater}, {Ott},
  {Johnson}, {Bell}, {Stanimirovi{\'c}}, {Putman}, {Majewski}, {Simpson},
  {J{\"u}tte}, {Oosterloo}, \& {Butler Burton}}]{nidever13}
{Nidever}, D.~L., {Ashley}, T., {Slater}, C.~T., {et~al.} 2013, \apjl, 779, L15

\bibitem[{{Okamoto} {et~al.}(2015){Okamoto}, {Arimoto}, {Ferguson}, {Bernard},
  {Irwin}, {Yamada}, \& {Utsumi}}]{okamoto15}
{Okamoto}, S., {Arimoto}, N., {Ferguson}, A. M.~N., {et~al.} 2015, \apjl, 809,
  L1

\bibitem[{{Pancino} {et~al.}(2017){Pancino}, {Romano}, {Tang},
  {Tautvai{\v{s}}ien{\.{e}}}, {Casey}, {Gruyters}, {Geisler}, {San Roman},
  {Randich}, {Alfaro}, {Bragaglia}, {Flaccomio}, {Korn}, {Recio-Blanco},
  {Smiljanic}, {Carraro}, {Bayo}, {Costado}, {Damiani}, {Jofr{\'e}}, {Lardo},
  {de Laverny}, {Monaco}, {Morbidelli}, {Sbordone}, {Sousa}, \&
  {Villanova}}]{pancino17}
{Pancino}, E., {Romano}, D., {Tang}, B., {et~al.} 2017, \aap, 601, A112

\bibitem[{{Patrick} {et~al.}(2015){Patrick}, {Evans}, {Davies}, {Kudritzki},
  {Gazak}, {Bergemann}, {Plez}, \& {Ferguson}}]{Patrick2015}
{Patrick}, L.~R., {Evans}, C.~J., {Davies}, B., {et~al.} 2015, \apj, 803, 14

\bibitem[{{Peng} {et~al.}(2009){Peng}, {Jord{\'a}n}, {Blakeslee}, {Mieske},
  {C{\^o}t{\'e}}, {Ferrarese}, {Harris}, {Madrid}, \& {Meurer}}]{peng09}
{Peng}, E.~W., {Jord{\'a}n}, A., {Blakeslee}, J.~P., {et~al.} 2009, \apj, 703,
  42

\bibitem[{{Peters} {et~al.}(2017){Peters}, {van der Kruit}, {Knapen},
  {Trujillo}, {Fliri}, {Cisternas}, \& {Kelvin}}]{peters17}
{Peters}, S.~P.~C., {van der Kruit}, P.~C., {Knapen}, J.~H., {et~al.} 2017,
  \mnras, 470, 427

\bibitem[{{Pilyugin} {et~al.}(2014){Pilyugin}, {Grebel}, \&
  {Kniazev}}]{pilyugin14}
{Pilyugin}, L.~S., {Grebel}, E.~K., \& {Kniazev}, A.~Y. 2014, \aj, 147, 131

\bibitem[{{Planck Collaboration} {et~al.}(2011){Planck Collaboration},
  {Abergel}, {Ade}, {Aghanim}, {Arnaud}, {Ashdown}, {Aumont}, {Baccigalupi},
  {Balbi}, {Banday}, {Barreiro}, {Bartlett}, {Battaner}, {Benabed},
  {Beno{\^\i}t}, {Bernard}, {Bersanelli}, {Bhatia}, {Blagrave}, {Bock},
  {Bonaldi}, {Bond}, {Borrill}, {Bouchet}, {Boulanger}, {Bucher}, {Burigana},
  {Cabella}, {Cantalupo}, {Cardoso}, {Catalano}, {Cay{\'o}n}, {Challinor},
  {Chamballu}, {Chiang}, {Chiang}, {Christensen}, {Clements}, {Colombi},
  {Couchot}, {Coulais}, {Crill}, {Cuttaia}, {Danese}, {Davies}, {Davis}, {de
  Bernardis}, {de Gasperis}, {de Rosa}, {de Zotti}, {Delabrouille}, {Delouis},
  {D{\'e}sert}, {Dickinson}, {Donzelli}, {Dor{\'e}}, {D{\"o}rl}, {Douspis},
  {Dupac}, {Efstathiou}, {En{\ss}lin}, {Eriksen}, {Finelli}, {Forni},
  {Frailis}, {Franceschi}, {Galeotta}, {Ganga}, {Giard}, {Giardino},
  {Giraud-H{\'e}raud}, {Gonz{\'a}lez-Nuevo}, {G{\'o}rski}, {Gratton},
  {Gregorio}, {Gruppuso}, {Hansen}, {Harrison}, {Helou}, {Henrot-Versill{\'e}},
  {Herranz}, {Hildebrandt}, {Hivon}, {Hobson}, {Holmes}, {Hovest}, {Hoyland},
  {Huffenberger}, {Jaffe}, {Joncas}, {Jones}, {Jones}, {Juvela},
  {Keih{\"a}nen}, {Keskitalo}, {Kisner}, {Kneissl}, {Knox}, {Kurki-Suonio},
  {Lagache}, {Lamarre}, {Lasenby}, {Laureijs}, {Lawrence}, {Leach}, {Leonardi},
  {Leroy}, {Linden-V{\o}rnle}, {Lockman}, {L{\'o}pez-Caniego}, {Lubin},
  {Mac{\'\i}as-P{\'e}rez}, {MacTavish}, {Maffei}, {Maino}, {Mandolesi}, {Mann},
  {Maris}, {Marshall}, {Martin}, {Mart{\'\i}nez-Gonz{\'a}lez}, {Masi},
  {Matarrese}, {Matthai}, {Mazzotta}, {McGehee}, {Meinhold}, {Melchiorri},
  {Mendes}, {Mennella}, {Miville-Desch{\^e}nes}, {Moneti}, {Montier},
  {Morgante}, {Mortlock}, {Munshi}, {Murphy}, {Naselsky}, {Nati}, {Natoli},
  {Netterfield}, {N{\o}rgaard-Nielsen}, {Noviello}, {Novikov}, {Novikov},
  {O'Dwyer}, {Osborne}, {Pajot}, {Paladini}, {Pasian}, {Patanchon},
  {Perdereau}, {Perotto}, {Perrotta}, {Piacentini}, {Piat}, {Pinheiro
  Gon{\c{c}}alves}, {Plaszczynski}, {Pointecouteau}, {Polenta}, {Ponthieu},
  {Poutanen}, {Pr{\'e}zeau}, {Prunet}, {Puget}, {Rachen}, {Reach}, {Reinecke},
  {Renault}, {Ricciardi}, {Riller}, {Ristorcelli}, {Rocha}, {Rosset},
  {Rowan-Robinson}, {Rubi{\~n}o-Mart{\'\i}n}, {Rusholme}, {Sandri}, {Santos},
  {Savini}, {Scott}, {Seiffert}, {Shellard}, {Smoot}, {Starck}, {Stivoli},
  {Stolyarov}, {Stompor}, {Sudiwala}, {Sygnet}, {Tauber}, {Terenzi},
  {Toffolatti}, {Tomasi}, {Torre}, {Tristram}, {Tuovinen}, {Umana},
  {Valenziano}, {Vielva}, {Villa}, {Vittorio}, {Wade}, {Wandelt}, {Wilkinson},
  {Yvon}, {Zacchei}, \& {Zonca}}]{planck11}
{Planck Collaboration}, {Abergel}, A., {Ade}, P.~A.~R., {et~al.} 2011, \aap,
  536, A24

\bibitem[{{Polles} {et~al.}(2019){Polles}, {Madden}, {Lebouteiller}, {Cormier},
  {Abel}, {Galliano}, {Hony}, {Karczewski}, {Lee}, {Chevance}, {Galametz}, \&
  {Lianou}}]{polles19}
{Polles}, F.~L., {Madden}, S.~C., {Lebouteiller}, V., {et~al.} 2019, \aap, 622,
  A119

\bibitem[{{Pota} {et~al.}(2013){Pota}, {Forbes}, {Romanowsky}, {Brodie},
  {Spitler}, {Strader}, {Foster}, {Arnold}, {Benson}, {Blom}, {Hargis},
  {Rhode}, \& {Usher}}]{pota13}
{Pota}, V., {Forbes}, D.~A., {Romanowsky}, A.~J., {et~al.} 2013, \mnras, 428,
  389

\bibitem[{{Puche} {et~al.}(1992){Puche}, {Westpfahl}, {Brinks}, \&
  {Roy}}]{puche92}
{Puche}, D., {Westpfahl}, D., {Brinks}, E., \& {Roy}, J.-R. 1992, \aj, 103,
  1841

\bibitem[{{Radburn-Smith} {et~al.}(2011){Radburn-Smith}, {de Jong}, {Seth},
  {Bailin}, {Bell}, {Brown}, {Bullock}, {Courteau}, {Dalcanton}, {Ferguson},
  {Goudfrooij}, {Holfeltz}, {Holwerda}, {Purcell}, {Sick}, {Streich}, {Vlajic},
  \& {Zucker}}]{radburn11}
{Radburn-Smith}, D.~J., {de Jong}, R.~S., {Seth}, A.~C., {et~al.} 2011, \apjs,
  195, 18

\bibitem[{{Rejkuba}(2012)}]{Rejkuba2012}
{Rejkuba}, M. 2012, \apss, 341, 195

\bibitem[{{Reynolds} {et~al.}(2023){Reynolds}, {Catinella}, {Cortese}, {Deg},
  {D{\'e}nes}, {Elagali}, {For}, {Kamphuis}, {Kleiner}, {Koribalski},
  {Lee-Waddell}, {Murugeshan}, {Raja}, {Rhee}, {Spekkens}, {Staveley-Smith},
  {van der Hulst}, {Wang}, {Westmeier}, {Wong}, {Bigiel}, {Bosma}, {Holwerda},
  {Leahy}, \& {Meyer}}]{reynolds23}
{Reynolds}, T.~N., {Catinella}, B., {Cortese}, L., {et~al.} 2023, \pasa, 40,
  e032

\bibitem[{{Reynolds} {et~al.}(2022){Reynolds}, {Catinella}, {Cortese},
  {Westmeier}, {Meurer}, {Shao}, {Obreschkow}, {Rom{\'a}n},
  {Verdes-Montenegro}, {Deg}, {D{\'e}nes}, {For}, {Kleiner}, {Koribalski},
  {Lee-Waddell}, {Murugeshan}, {Oh}, {Rhee}, {Spekkens}, {Staveley-Smith},
  {Stevens}, {van der Hulst}, {Wang}, {Wong}, {Holwerda}, {Bosma}, {Madrid}, \&
  {Bekki}}]{reynolds22}
{Reynolds}, T.~N., {Catinella}, B., {Cortese}, L., {et~al.} 2022, \mnras, 510,
  1716

\bibitem[{{Rhode} {et~al.}(1999){Rhode}, {Salzer}, {Westpfahl}, \&
  {Radice}}]{rhode99}
{Rhode}, K.~L., {Salzer}, J.~J., {Westpfahl}, D.~J., \& {Radice}, L.~A. 1999,
  \aj, 118, 323

\bibitem[{{Roberts} \& {Hausman}(1984)}]{roberts84}
{Roberts}, W.~W., J. \& {Hausman}, M.~A. 1984, \apj, 277, 744

\bibitem[{{Roberts} {et~al.}(2003){Roberts}, {Goad}, {Ward}, \&
  {Warwick}}]{roberts03}
{Roberts}, T.~P., {Goad}, M.~R., {Ward}, M.~J., \& {Warwick}, R.~S. 2003,
  \mnras, 342, 709

\bibitem[{{Rom{\'a}n} {et~al.}(2021){Rom{\'a}n}, {Castilla}, \&
  {Pascual-Granado}}]{roman21}
{Rom{\'a}n}, J., {Castilla}, A., \& {Pascual-Granado}, J. 2021, \aap, 656, A44

\bibitem[{{Rom{\'a}n} {et~al.}(2023b){Rom{\'a}n}, {Rich}, {Ahvazi}, {Sales},
  {Li}, {Golini}, {Trujillo}, {Knapen}, {Peletier}, \&
  {S{\'a}nchez-Alarc{\'o}n}}]{roman23b}
{Rom{\'a}n}, J., {Rich}, R.~M., {Ahvazi}, N., {et~al.} 2023b, \aap, 679, A157

\bibitem[{{Rom{\'a}n} {et~al.}(2023a){Rom{\'a}n}, {S{\'a}nchez-Alarc{\'o}n},
  {Knapen}, \& {Peletier}}]{roman23a}
{Rom{\'a}n}, J., {S{\'a}nchez-Alarc{\'o}n}, P.~M., {Knapen}, J.~H., \&
  {Peletier}, R. 2023a, \aap, 671, L7

\bibitem[{{Rom{\'a}n} {et~al.}(2020){Rom{\'a}n}, {Trujillo}, \&
  {Montes}}]{roman20}
{Rom{\'a}n}, J., {Trujillo}, I., \& {Montes}, M. 2020, \aap, 644, A42

\bibitem[{{Ryder} {et~al.}(1999){Ryder}, {Walsh}, \& {Malin}}]{ryder99}
{Ryder}, S.~D., {Walsh}, W., \& {Malin}, D. 1999, \pasa, 16, 84

\bibitem[{{Sabbi} {et~al.}(2018){Sabbi}, {Calzetti}, {Ubeda}, {Adamo},
  {Cignoni}, {Thilker}, {Aloisi}, {Elmegreen}, {Elmegreen}, {Gouliermis},
  {Grebel}, {Messa}, {Smith}, {Tosi}, {Dolphin}, {Andrews}, {Ashworth},
  {Bright}, {Brown}, {Chandar}, {Christian}, {Clayton}, {Cook}, {Dale}, {de
  Mink}, {Dobbs}, {Evans}, {Fumagalli}, {Gallagher}, {Grasha}, {Herrero},
  {Hunter}, {Johnson}, {Kahre}, {Kennicutt}, {Kim}, {Krumholz}, {Lee},
  {Lennon}, {Martin}, {Nair}, {Nota}, {{\"O}stlin}, {Pellerin}, {Prieto},
  {Regan}, {Ryon}, {Sacchi}, {Schaerer}, {Schiminovich}, {Shabani}, {Van Dyk},
  {Walterbos}, {Whitmore}, \& {Wofford}}]{sabbi18}
{Sabbi}, E., {Calzetti}, D., {Ubeda}, L., {et~al.} 2018, \apjs, 235, 23

\bibitem[{{Saifollahi} {et~al.}(2024){Saifollahi}, {Voggel}, {Lan\c{c}on},
  {et~al.}}]{EROFornaxGCs}
{Saifollahi}, T., {Voggel}, K., {Lan\c{c}on}, A., {et~al.} 2024, \aap, this
  issue

\bibitem[{{S{\'a}nchez} {et~al.}(2014){S{\'a}nchez}, {Rosales-Ortega},
  {Iglesias-P{\'a}ramo}, {Moll{\'a}}, {Barrera-Ballesteros}, {Marino},
  {P{\'e}rez}, {S{\'a}nchez-Blazquez}, {Gonz{\'a}lez Delgado}, {Cid Fernandes},
  {de Lorenzo-C{\'a}ceres}, {Mendez-Abreu}, {Galbany}, {Falcon-Barroso},
  {Miralles-Caballero}, {Husemann}, {Garc{\'\i}a-Benito}, {Mast}, {Walcher},
  {Gil de Paz}, {Garc{\'\i}a-Lorenzo}, {Jungwiert}, {V{\'\i}lchez},
  {J{\'\i}lkov{\'a}}, {Lyubenova}, {Cortijo-Ferrero}, {D{\'\i}az}, {Wisotzki},
  {M{\'a}rquez}, {Bland-Hawthorn}, {Ellis}, {van de Ven}, {Jahnke},
  {Papaderos}, {Gomes}, {Mendoza}, \& {L{\'o}pez-S{\'a}nchez}}]{sanchez14}
{S{\'a}nchez}, S.~F., {Rosales-Ortega}, F.~F., {Iglesias-P{\'a}ramo}, J.,
  {et~al.} 2014, \aap, 563, A49

\bibitem[{{S{\'a}nchez-Alarc{\'o}n} {et~al.}(2023){S{\'a}nchez-Alarc{\'o}n},
  {Rom{\'a}n}, {Knapen}, {Verdes-Montenegro}, {Comer{\'o}n}, {Rich}, {Beckman},
  {Argudo-Fern{\'a}ndez}, {Ram{\'\i}rez-Moreta}, {Blasco}, {Unda-Sanzana},
  {Garrido}, \& {S{\'a}nchez-Exposito}}]{sanchez23}
{S{\'a}nchez-Alarc{\'o}n}, P.~M., {Rom{\'a}n}, J., {Knapen}, J.~H., {et~al.}
  2023, \aap, 677, A117

\bibitem[{{Schlafly} \& {Finkbeiner}(2011)}]{schlafly11}
{Schlafly}, E.~F. \& {Finkbeiner}, D.~P. 2011, \apj, 737, 103

\bibitem[{{Schlegel} {et~al.}(1998){Schlegel}, {Finkbeiner}, \&
  {Davis}}]{schlegel98}
{Schlegel}, D.~J., {Finkbeiner}, D.~P., \& {Davis}, M. 1998, \apj, 500, 525

\bibitem[{{Shostak} \& {Skillman}(1989)}]{shostak89}
{Shostak}, G.~S. \& {Skillman}, E.~D. 1989, \aap, 214, 33

\bibitem[{{Sibbons} {et~al.}(2012){Sibbons}, {Ryan}, {Cioni}, {Irwin}, \&
  {Napiwotzki}}]{sibbons12}
{Sibbons}, L.~F., {Ryan}, S.~G., {Cioni}, M. R.~L., {Irwin}, M., \&
  {Napiwotzki}, R. 2012, \aap, 540, A135

\bibitem[{{Silverman} \& {Filippenko}(2008)}]{silverman08}
{Silverman}, J.~M. \& {Filippenko}, A.~V. 2008, \apjl, 678, L17

\bibitem[{{Sola} {et~al.}(2022){Sola}, {Duc}, {Richards}, {Paiement}, {Urbano},
  {Klehammer}, {B{\'\i}lek}, {Cuillandre}, {Gwyn}, \& {McConnachie}}]{sola22}
{Sola}, E., {Duc}, P.-A., {Richards}, F., {et~al.} 2022, \aap, 662, A124

\bibitem[{{Stetson}(1987)}]{Stetson1987}
{Stetson}, P.~B. 1987, \pasp, 99, 191

\bibitem[{{Stetson}(1994)}]{Stetson1994}
{Stetson}, P.~B. 1994, \pasp, 106, 250

\bibitem[{{Stewart} {et~al.}(2000){Stewart}, {Fanelli}, {Byrd}, {Hill},
  {Westpfahl}, {Cheng}, {O'Connell}, {Roberts}, {Neff}, {Smith}, \&
  {Stecher}}]{stewart00}
{Stewart}, S.~G., {Fanelli}, M.~N., {Byrd}, G.~G., {et~al.} 2000, \apj, 529,
  201

\bibitem[{{Stone} {et~al.}(2021){Stone}, {Arora}, {Courteau}, \&
  {Cuillandre}}]{stone21}
{Stone}, C.~J., {Arora}, N., {Courteau}, S., \& {Cuillandre}, J.-C. 2021,
  \mnras, 508, 1870

\bibitem[{Tammann \& Sandage(1968)}]{Tammann1968}
Tammann, G.~A. \& Sandage, A. 1968, ApJ, 151, 825

\bibitem[{{Tantalo} {et~al.}(2022){Tantalo}, {Dall'Ora}, {Bono}, {Stetson},
  {Fabrizio}, {Ferraro}, {Nonino}, {Braga}, {da Silva}, {Fiorentino},
  {Iannicola}, {Marengo}, {Monelli}, {Mullen}, {Pietrinferni}, \&
  {Salaris}}]{tantalo22}
{Tantalo}, M., {Dall'Ora}, M., {Bono}, G., {et~al.} 2022, \apj, 933, 197

\bibitem[{{Teyssier} {et~al.}(2012){Teyssier}, {Johnston}, \&
  {Kuhlen}}]{teyssier12}
{Teyssier}, M., {Johnston}, K.~V., \& {Kuhlen}, M. 2012, \mnras, 426, 1808

\bibitem[{{Thompson} {et~al.}(2016){Thompson}, {Ryan}, \&
  {Sibbons}}]{thompson16}
{Thompson}, G.~P., {Ryan}, S.~G., \& {Sibbons}, L.~F. 2016, \mnras, 462, 3376

\bibitem[{{Tonry} {et~al.}(2001){Tonry}, {Dressler}, {Blakeslee}, {Ajhar},
  {Fletcher}, {Luppino}, {Metzger}, \& {Moore}}]{tonry01}
{Tonry}, J.~L., {Dressler}, A., {Blakeslee}, J.~P., {et~al.} 2001, \apj, 546,
  681

\bibitem[{{Tran} {et~al.}(2023){Tran}, {Williams}, {Levesque}, {Lazzarini},
  {Dalcanton}, {Dolphin}, {Koplitz}, {Smercina}, \& {Telford}}]{tran23}
{Tran}, D., {Williams}, B., {Levesque}, E., {et~al.} 2023, \apj, 954, 211

\bibitem[{{Trujillo} \& {Fliri}(2016)}]{trujillo16}
{Trujillo}, I. \& {Fliri}, J. 2016, \apj, 823, 123

\bibitem[{{Vacca} {et~al.}(2007){Vacca}, {Sheehy}, \& {Graham}}]{vacca07}
{Vacca}, W.~D., {Sheehy}, C.~D., \& {Graham}, J.~R. 2007, \apj, 662, 272

\bibitem[{{van den Bergh}(1999)}]{vandenbergh99}
{van den Bergh}, S. 1999, \aapr, 9, 273

\bibitem[{{van der Kruit}(1981)}]{vanderkruit81}
{van der Kruit}, P.~C. 1981, \aap, 99, 298

\bibitem[{{van Dokkum} {et~al.}(2015){van Dokkum}, {Abraham}, {Merritt},
  {Zhang}, {Geha}, \& {Conroy}}]{vandokkum15}
{van Dokkum}, P.~G., {Abraham}, R., {Merritt}, A., {et~al.} 2015, \apjl, 798,
  L45

\bibitem[{{Veljanoski} {et~al.}(2015){Veljanoski}, {Ferguson}, {Mackey},
  {Huxor}, {Hurley}, {Bernard}, {C{\^o}t{\'e}}, {Irwin}, {Martin}, {Burgett},
  {Chambers}, {Flewelling}, {Kudritzki}, \& {Waters}}]{veljanoski15}
{Veljanoski}, J., {Ferguson}, A.~M.~N., {Mackey}, A.~D., {et~al.} 2015, \mnras,
  452, 320

\bibitem[{{Venhola} {et~al.}(2022){Venhola}, {Peletier}, {Salo}, {Laurikainen},
  {Janz}, {Haigh}, {Wilkinson}, {Iodice}, {Hilker}, {Mieske}, {Cantiello}, \&
  {Spavone}}]{venhola22}
{Venhola}, A., {Peletier}, R.~F., {Salo}, H., {et~al.} 2022, \aap, 662, A43

\bibitem[{{Venn} {et~al.}(2001){Venn}, {Lennon}, {Kaufer}, {McCarthy},
  {Przybilla}, {Kudritzki}, {Lemke}, {Skillman}, \& {Smartt}}]{Venn2001}
{Venn}, K.~A., {Lennon}, D.~J., {Kaufer}, A., {et~al.} 2001, \apj, 547, 765

\bibitem[{{Veronese} {et~al.}(2023){Veronese}, {de Blok}, \&
  {Walter}}]{veronese23}
{Veronese}, S., {de Blok}, W.~J.~G., \& {Walter}, F. 2023, \aap, 672, A55

\bibitem[{{{\v{Z}}emaitis} {et~al.}(2023){{\v{Z}}emaitis}, {Ferguson},
  {Okamoto}, {Cuillandre}, {Stone}, {Arimoto}, \& {Irwin}}]{zemaitis23}
{{\v{Z}}emaitis}, R., {Ferguson}, A. M.~N., {Okamoto}, S., {et~al.} 2023,
  \mnras, 518, 2497

\bibitem[{{Wada} {et~al.}(2011){Wada}, {Baba}, \& {Saitoh}}]{wada11}
{Wada}, K., {Baba}, J., \& {Saitoh}, T.~R. 2011, \apj, 735, 1

\bibitem[{{Walter} {et~al.}(2008){Walter}, {Brinks}, {de Blok}, {Bigiel},
  {Kennicutt}, {Thornley}, \& {Leroy}}]{walter08}
{Walter}, F., {Brinks}, E., {de Blok}, W.~J.~G., {et~al.} 2008, \aj, 136, 2563

\bibitem[{{Wang} {et~al.}(2011){Wang}, {Kauffmann}, {Overzier}, {Catinella},
  {Schiminovich}, {Heckman}, {Moran}, {Haynes}, {Giovanelli}, \&
  {Kong}}]{wang11}
{Wang}, J., {Kauffmann}, G., {Overzier}, R., {et~al.} 2011, \mnras, 412, 1081

\bibitem[{{Weisz} {et~al.}(2011){Weisz}, {Dalcanton}, {Williams}, {Gilbert},
  {Skillman}, {Seth}, {Dolphin}, {McQuinn}, {Gogarten}, {Holtzman}, {Rosema},
  {Cole}, {Karachentsev}, \& {Zaritsky}}]{weisz11}
{Weisz}, D.~R., {Dalcanton}, J.~J., {Williams}, B.~F., {et~al.} 2011, \apj,
  739, 5

\bibitem[{{Weisz} {et~al.}(2014){Weisz}, {Dolphin}, {Skillman}, {Holtzman},
  {Gilbert}, {Dalcanton}, \& {Williams}}]{weisz14}
{Weisz}, D.~R., {Dolphin}, A.~E., {Skillman}, E.~D., {et~al.} 2014, \apj, 789,
  147

\bibitem[{{Weisz} {et~al.}(2009){Weisz}, {Skillman}, {Cannon}, {Dolphin},
  {Kennicutt}, {Lee}, \& {Walter}}]{weisz09}
{Weisz}, D.~R., {Skillman}, E.~D., {Cannon}, J.~M., {et~al.} 2009, \apj, 704,
  1538

\bibitem[{{Wilcots} \& {Miller}(1998)}]{wilcots98}
{Wilcots}, E.~M. \& {Miller}, B.~W. 1998, \aj, 116, 2363

\bibitem[{{Williams} {et~al.}(2009b){Williams}, {Dalcanton}, {Dolphin},
  {Holtzman}, \& {Sarajedini}}]{williams09b}
{Williams}, B.~F., {Dalcanton}, J.~J., {Dolphin}, A.~E., {Holtzman}, J., \&
  {Sarajedini}, A. 2009b, \apjl, 695, L15

\bibitem[{{Williams} {et~al.}(2009a){Williams}, {Dalcanton}, {Seth}, {Weisz},
  {Dolphin}, {Skillman}, {Harris}, {Holtzman}, {Girardi}, {de Jong}, {Olsen},
  {Cole}, {Gallart}, {Gogarten}, {Hidalgo}, {Mateo}, {Rosema}, {Stetson}, \&
  {Quinn}}]{williams09a}
{Williams}, B.~F., {Dalcanton}, J.~J., {Seth}, A.~C., {et~al.} 2009a, \aj, 137,
  419

\bibitem[{{Williams} {et~al.}(2013){Williams}, {Dalcanton}, {Stilp}, {Dolphin},
  {Skillman}, \& {Radburn-Smith}}]{williams13}
{Williams}, B.~F., {Dalcanton}, J.~J., {Stilp}, A., {et~al.} 2013, \apj, 765,
  120

\bibitem[{{Wu} {et~al.}(2014){Wu}, {Tully}, {Rizzi}, {Dolphin}, {Jacobs}, \&
  {Karachentsev}}]{wu14}
{Wu}, P.-F., {Tully}, R.~B., {Rizzi}, L., {et~al.} 2014, \aj, 148, 7

\bibitem[{{Xu} {et~al.}(2022){Xu}, {Cheng}, {Appleton}, {Duc}, {Gao}, {Tang},
  {Yun}, {Dai}, {Huang}, {Lisenfeld}, \& {Renaud}}]{xu22}
{Xu}, C.~K., {Cheng}, C., {Appleton}, P.~N., {et~al.} 2022, \nat, 610, 461

\bibitem[{{Yew} {et~al.}(2018){Yew}, {Filipovi{\'c}}, {Roper}, {Collier},
  {Crawford}, {Jarrett}, {Tothill}, {O'Brien}, {Pavlovi{\'c}}, {Pannuti},
  {Galvin}, {Kapi{\'n}ska}, {Cluver}, {Banfield}, {Schlegel}, {Maxted}, \&
  {Grieve}}]{yew18}
{Yew}, M., {Filipovi{\'c}}, M.~D., {Roper}, Q., {et~al.} 2018, \pasa, 35, e015

\bibitem[{{Zezas} {et~al.}(1999){Zezas}, {Georgantopoulos}, \&
  {Ward}}]{zezas99}
{Zezas}, A.~L., {Georgantopoulos}, I., \& {Ward}, M.~J. 1999, \mnras, 308, 302

\bibitem[{{Zhang} {et~al.}(2021){Zhang}, {Mackey}, \& {Da Costa}}]{zhang21}
{Zhang}, S., {Mackey}, D., \& {Da Costa}, G.~S. 2021, \mnras, 508, 2098

\end{thebibliography}

%

\appendix


\section{Details on sky level and estimates of limiting surface brightness}
\label{app:sky}

\begin{table*}[t!]
\centering
\caption[]{Sky levels and SB limiting magnitudes \mulim\ within 100\,arcsec$^{2}$ areas in empty sky regions} 
\label{tab:sky}
\resizebox{0.7\linewidth}{!}{
\begin{centering}
\begin{tabular}{lcccccc}
\hline
\hline
\noalign{\vskip 3pt}
\multicolumn{1}{c}{Property} & Holmberg\,II & IC\,10 & IC\,342 &  NGC\,2403 & NGC\,6744 & NGC\,6822 \\
\noalign{\vskip 1pt}
\hline
\\ 
\multicolumn{1}{c}{$\mu_{\IE}$ sky } & 22.62 & 22.29 & 22.42 & 22.66 & 22.36 & 21.74\\
\multicolumn{1}{c}{(1) $\mu_{\IE}$ 1$\sigma$ SB limit} & 30.56 & 30.40 & 30.41 & 30.62 & 30.41 & 30.09\\
\multicolumn{1}{c}{(2) $\mu_{\IE}$ 1$\sigma$ SB limit} & 30.67 & 30.69 & 30.59 & 30.76 & 30.57 & 30.19\\
\multicolumn{1}{c}{(3) $\mu_{\IE}$ 1$\sigma$ SB limit} & 30.61 & 30.63 & 30.52 & 30.63 & 30.47 & 30.04\\
& \multicolumn{6}{c}{30.49\ $\pm$\ 0.20$^\mathrm{b}$}\\
\multicolumn{1}{c}{(\magarc)} & \multicolumn{6}{c}{30.3$^\mathrm{c}$}\\ 
\noalign{\vskip 1pt}
\hline
\noalign{\vskip 1pt}
\multicolumn{1}{c}{$\mu_{\YE}$ sky } & 22.64 & 22.28 & 22.40 & 22.52 & 22.22 & 21.60\\
\multicolumn{1}{c}{(1) $\mu_{\YE}$ 1$\sigma$ SB limit} & 29.51 & 29.50 & 29.39 & 29.47 & 29.35 & 28.77\\
\multicolumn{1}{c}{(2) $\mu_{\YE}$ 1$\sigma$ SB limit} & 29.37 & 29.21 & 29.22 & 29.35 & 29.25 & 28.77\\
\multicolumn{1}{c}{(3) $\mu_{\YE}$ 1$\sigma$ SB limit} & 29.16 & 29.06 & 29.09 & 29.17 & 29.05 & 28.53\\
& \multicolumn{6}{c}{29.18\ $\pm$\ 0.26$^\mathrm{b}$}\\
\multicolumn{1}{c}{(\magarc)} & \multicolumn{6}{c}{28.7$^\mathrm{c}$}\\ 
\noalign{\vskip 1pt}
\hline
\noalign{\vskip 1pt}
\multicolumn{1}{c}{$\mu_{\JE}$ sky } & 22.68 & 22.39 & 22.50 & 22.56 & 22.26 & 21.67\\
\multicolumn{1}{c}{(1) $\mu_{\JE}$ 1$\sigma$ SB limit} & 29.71 & 29.32 & 29.59 & 29.66 & 29.51 & 28.97\\
\multicolumn{1}{c}{(2) $\mu_{\JE}$ 1$\sigma$ SB limit} & 29.54 & 29.52 & 29.48 & 29.56 & 29.45 & 29.00\\
\multicolumn{1}{c}{(3) $\mu_{\JE}$ 1$\sigma$ SB limit} & 29.32 & 29.21 & 29.28 & 29.36 & 29.25 & 28.77\\
& \multicolumn{6}{c}{29.36\ $\pm$\ 0.24$^\mathrm{b}$}\\
\multicolumn{1}{c}{(\magarc)} & \multicolumn{6}{c}{28.9$^\mathrm{c}$}\\ 
\noalign{\vskip 1pt}
\hline
\noalign{\vskip 1pt}
\multicolumn{1}{c}{$\mu_{\HE}$ sky } & 22.80 & 22.53 & 22.66 & 22.68 & 22.39 & 21.83\\
\multicolumn{1}{c}{(1) $\mu_{\HE}$ 1$\sigma$ SB limit} & 29.72 & 29.30 & 29.42 & 29.68 & 29.57 & 29.02\\
\multicolumn{1}{c}{(2) $\mu_{\HE}$ 1$\sigma$ SB limit} & 29.57 & 29.19 & 29.51 & 29.60 & 29.48 & 29.08\\
\multicolumn{1}{c}{(3) $\mu_{\HE}$ 1$\sigma$ SB limit} & 29.38 & 29.29 & 29.34 & 29.40 & 29.27 & 28.91\\
& \multicolumn{6}{c}{29.37\ $\pm$\ 0.22$^\mathrm{b}$}\\
\multicolumn{1}{c}{(\magarc)} & \multicolumn{6}{c}{28.9$^\mathrm{c}$}\\ 
\\
\hline
\end{tabular}
\end{centering}
}
\begin{description}
\item
$^\mathrm{a}$~In this and subsequent rows, the numbers in parentheses correspond to the different
approaches: (1) the standard deviation among 100\,arcsec$^2$ tiles following \texttt{gnuastro/noisechisel};
(2) Gaussian fits of the masked sky regions following \citet{roman20}; and
(3) noise in the radial brightness profiles from \texttt{AutoProf} following \citet{stone21}.
\item
$^\mathrm{b}$~Mean and standard deviation of 1$\sigma$ AB magnitude limits averaged over all galaxies and methods.
\item
$^\mathrm{c}$~1$\sigma$ AB magnitude limits taken from \citet{Scaramella-EP1}, after having removed the asinh offset.
\end{description}
\end{table*}
\noindent

\begin{figure*}[h!]
\includegraphics[width=0.48\textwidth]{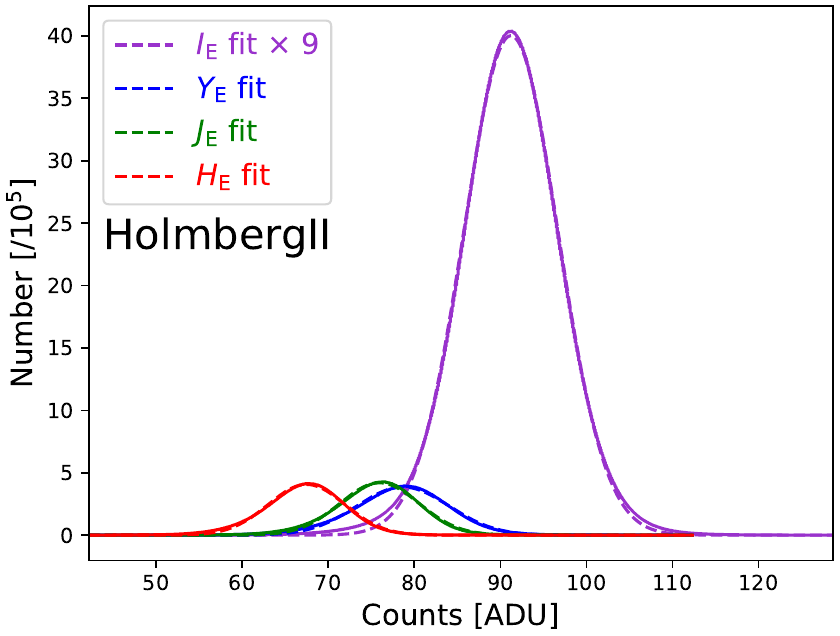}
\hspace{0.04\textwidth}
\includegraphics[width=0.48\textwidth]{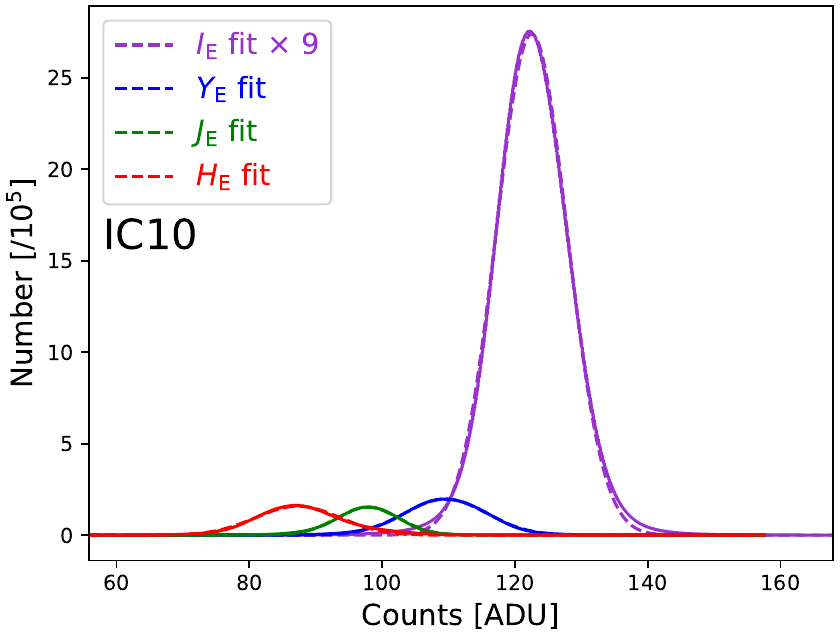} \\
\includegraphics[width=0.48\textwidth]{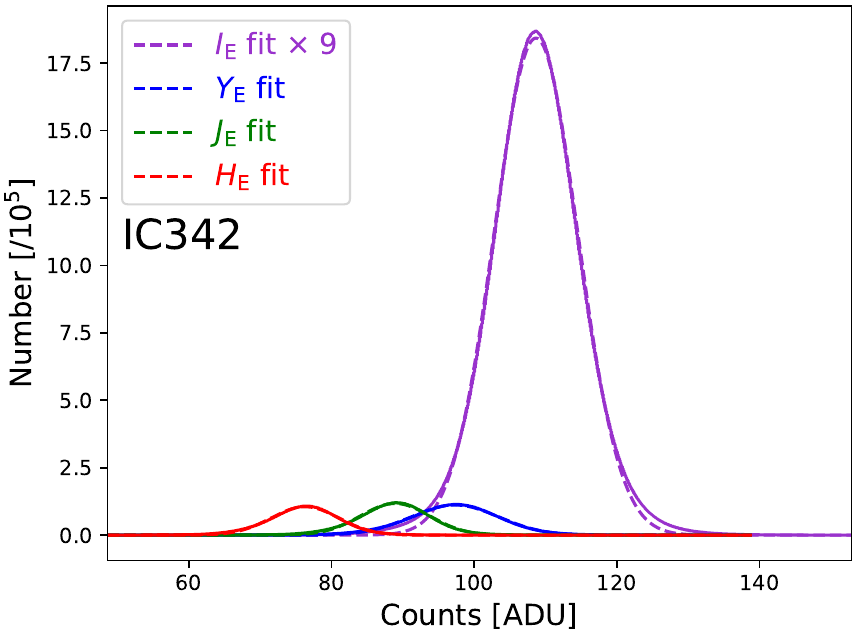}
\hspace{0.04\textwidth}
\includegraphics[width=0.48\textwidth]{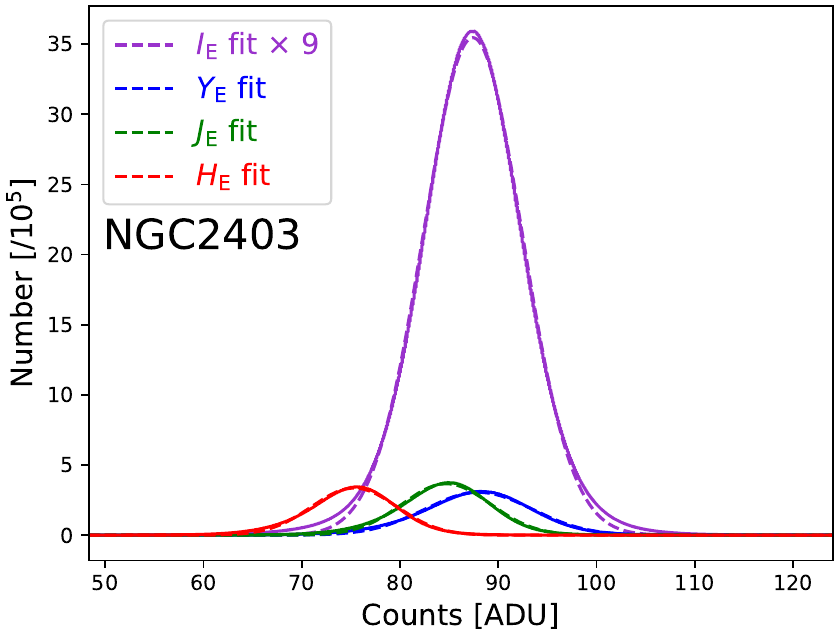}\\
\includegraphics[width=0.48\textwidth]{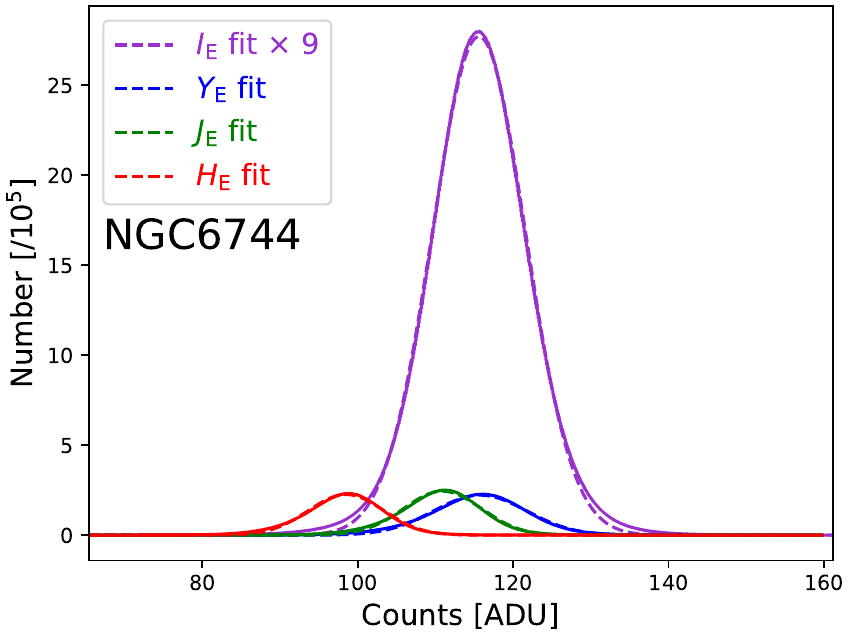} 
\hspace{0.04\textwidth}
\includegraphics[width=0.48\textwidth]{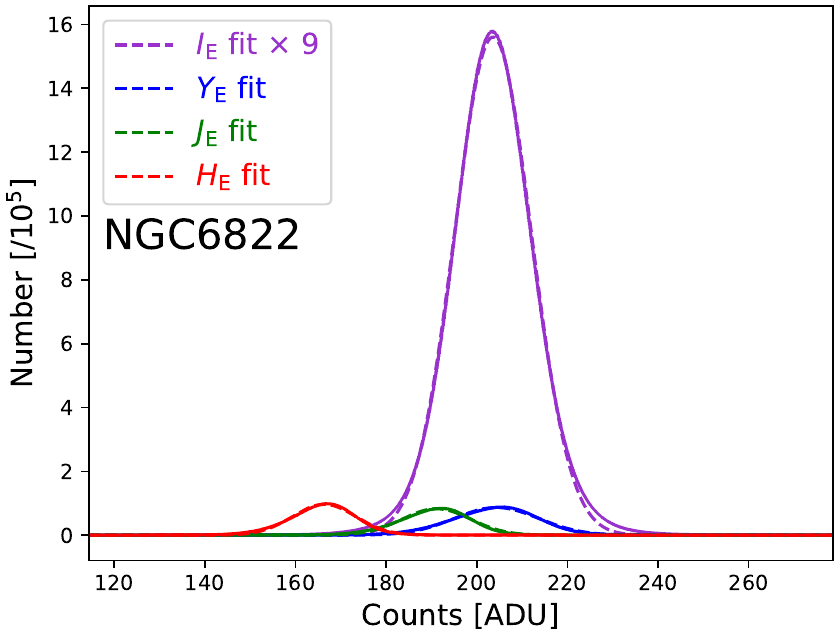} \\
\caption{Distributions of counts (ADU\,pixel$^{-1}$) over the entire masked region of empty sky for NGC\,2403 
and and NGC\,6822 (top panels); 
Holmberg\,II, IC\,10 (middle); and IC\,342, NGC\,6744 (bottom).
The data are shown as solid lines, and the Gaussian best fits as dashed ones.
The four bands are given by purple, blue, green, and red for \IE, \YE, \JE, and \HE, respectively.
The \IE\ VIS band counts have been multiplied by the ratio of the pixel area ($\times 9$) to be
shown together with the NIR bands.
In most cases, 
the data (traced by a solid curve) are so close to a Gaussian as to be indistinguishable from the best fit
(shown by a dashed curve).
}
\label{fig:gaussian_global}
\end{figure*}

We estimate the limiting surface brightness \mulim\ in an equivalent 100\,arcsec$^2$ region 
as a function of the signal standard deviation $\sigma$ using three different approaches. 
\begin{enumerate}[(1)]
\item
\texttt{gnuastro/noisechisel} \citep{akhlaghi15,akhlaghi19a,akhlaghi19b} 
was run on each of the images, with a tile size corresponding to 100\,arcsec$^2$.
\texttt{noisechisel} employs a noise-based approach to detect highly extended and diffuse objects that 
are deeply embedded within a significant background of noise. 
It determines the median $\sigma$ over tiles where there are no detections, rather only empty sky background.
We experimented with various options for \texttt{noisechisel} and found that, 
although there was some variation ($\pm\,0.1$\,mag), on the whole, results were stable.
The median $\sigma$ from \texttt{gnuastro/mkcatalog} was converted to \mulim\
(AB \magarc, Eq.\,\ref{eqn:sblim}), and reported as `(1)' in Table \ref{tab:sky}.
\item
Following \citet{roman20}, we fit a Gaussian to the distribution of the masked signal with sky background only;
the images were masked with the \texttt{noisechisel} detection masks computed as above [approach (1)].
The best-fit Gaussian $\sigma$ for all galaxies are given in Table~\ref{tab:sky} as `(2)', converted to limiting AB magnitudes as above.
Thse fits themselves are shown in Fig.~\ref{fig:gaussian_global}.
The sky levels $\mu_\mathrm{sky}$ in Table~\ref{tab:sky} are calculated from the best-fit central values
from the Gaussian fits for each band.
\item
\texttt{AutoProf} \citep{stone21} was run on each of the images using 5$\sigma$ clipping to better mask bright sources. 
In addition to the surface brightness profiles described in Sect.~\ref{sec:profiles}, \texttt{AutoProf} of necessity
measures the background signal and its uncertainty $\sigma$. These values are reported in Table~\ref{tab:sky} as `(3)',
and converted to \mulim\ in a 100\,arcsec$^2$ region as in Eq. \eqref{eqn:sblim}. 
\end{enumerate}

\begin{figure*}[h!]
\includegraphics[width=0.48\textwidth]{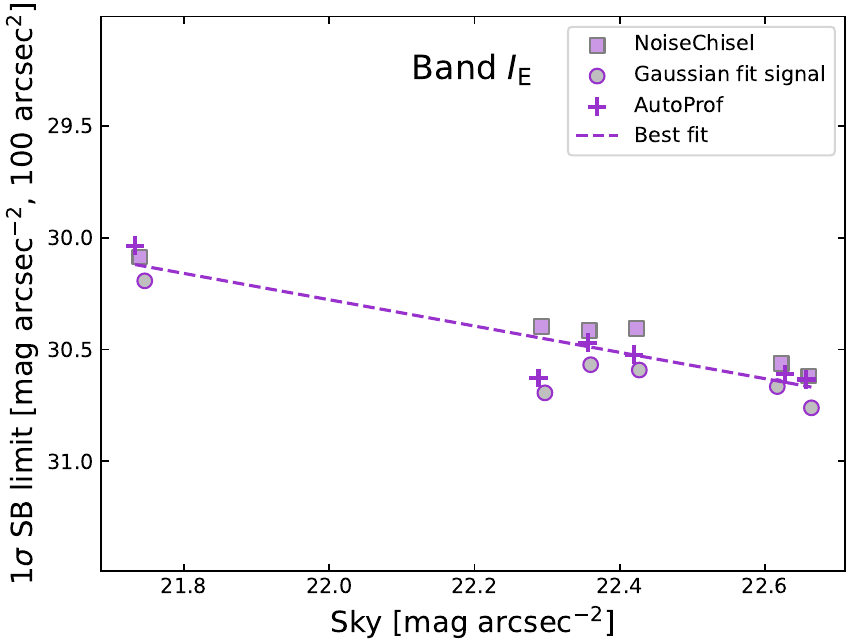}
\hspace{0.04\textwidth}
\includegraphics[width=0.48\textwidth]{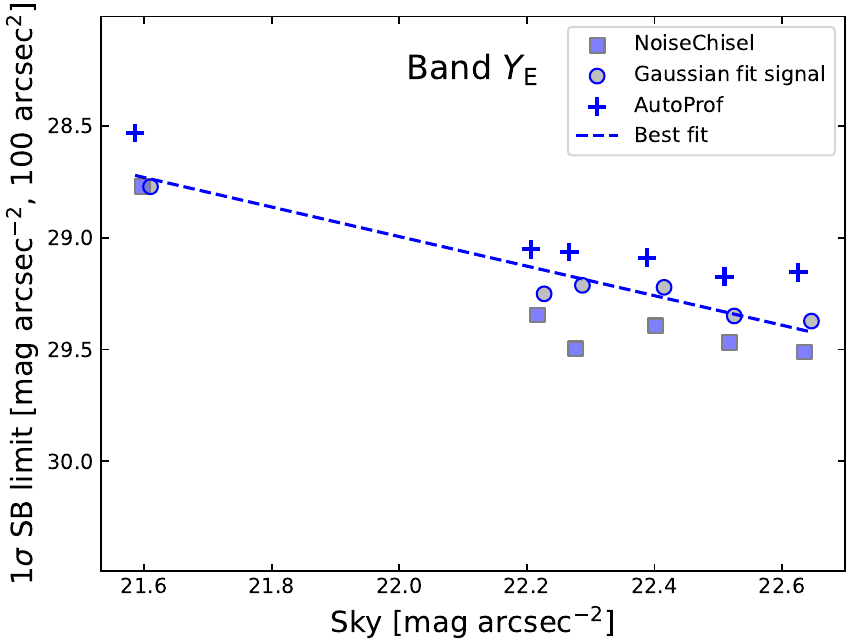} \\
\includegraphics[width=0.48\textwidth]{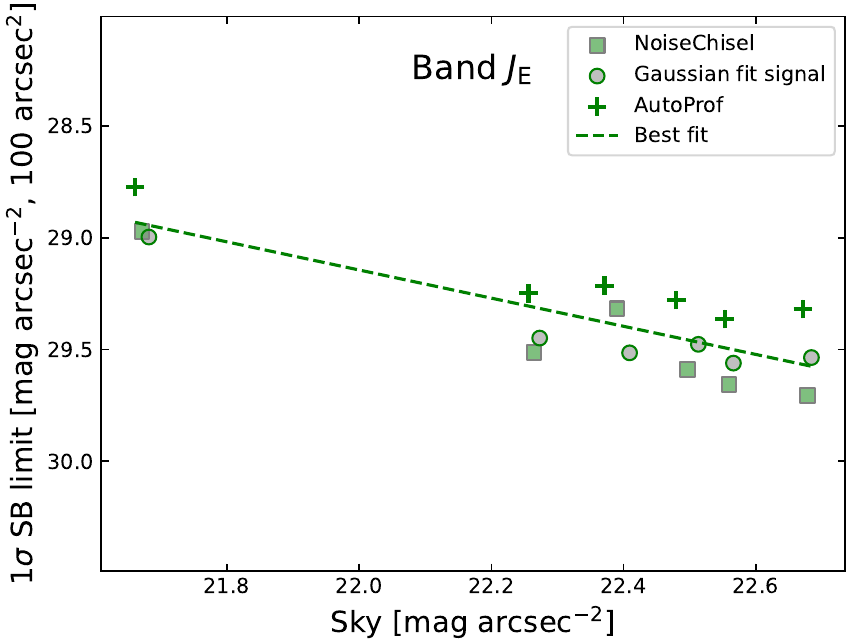}
\hspace{0.04\textwidth}
\includegraphics[width=0.48\textwidth]{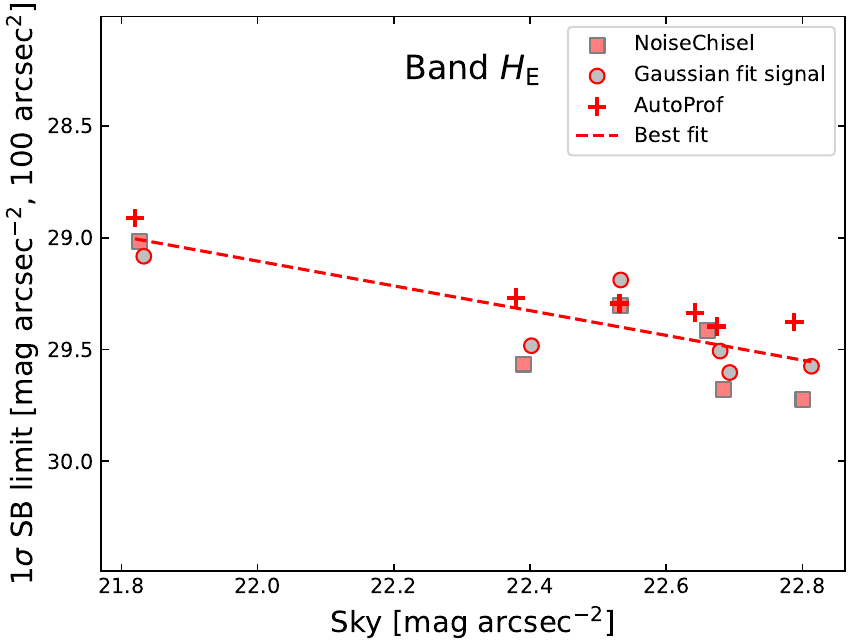} \\
\vspace{-1.5\baselineskip}
\caption{SB limits obtained with the various methodologies plotted against sky brightness for the four \Euclid bands.
For the method that does not directly give sky brightness [Approach (1)], 
we used the mean of the sky values from Approaches (2) and (3).
There is a correlation of the SB limits with sky brightness, with brighter sky having brighter limits,
and a variation with methodology, up to $0.3$--$0.4$\,\magarc. 
}
\label{fig:sblims}
\vspace{-\baselineskip}
\end{figure*}

We measure counts $C$ per pixel in images, with intrinsic uncertainty $\sigma$ per pixel,
and want to determine the uncertainty of surface brightness \mulim\ 
in units of \magarc\ over some spatial scale, $b^2$.
The scaling reported in Eq. \eqref{eqn:sblim} can be explained as follows, considering 
a signal-to-noise of $n$,
a linear pixel size $p$ (arcsec\,pixel$^{-1}$) and a region of area $b^2$\,arcsec$^2$.
\begin{enumerate}[(1)]
\item
Assuming that the noise $\sigma$ inherent in $C$ is uncorrelated from pixel to pixel,
the noise $\sigma_b$ that is obtained in a region of area $b^2$\,arcsec$^2$ adds in quadrature, resulting
in $\sigma_b\,=\,\sigma\,\sqrt{b^2/p^2}\,=\,\sigma\,b/p$, since $b^2/p^2$ is the number of pixels in the region.
We then would need to divide by $b^2$ to convert this value into \magarc, so that
\begin{eqnarray}
\mu_\mathrm{lim}\,=\,\mu(\sigma_b)\,=&\mathrm{ZP} - 2.5\,\log_{10}(n\,\sigma\,b/p) + 2.5\,\log_{10}(b^2)\, \nonumber \\
& \mathrm{mag\,arcsec}^{-2}\ , 
\label{eqn:exp1}
\end{eqnarray}
equivalent to the formulation in Eq.\,(\ref{eqn:sblim}).
This is also the reasoning followed in 
\url{https://www.gnu.org/software/gnuastro/manual/html_node/Surface-brightness-limit-of-image.html}.
\item
Alternatively, 
we can first convert the intrinsic per-pixel uncertainty $\sigma$ to units of \magarc\ considering only the conversion
of pixel size to arcsec$^2$:
\begin{equation}
\mu(\sigma)\,=\,\mathrm{ZP} - 2.5\,\log_{10}(n\,\sigma) + 2.5\,\log_{10}(p^2)\, \quad \mathrm{mag\,arcsec}^{-2}\ .
\label{eqn:exp2}
\end{equation}
However, we seek $\sigma_b$, the noise that is obtained in a region of area $b^2$\,arcsec$^2$,
so consider that $\sigma_b$ is the error of the mean $\sigma$ within
the region of size $b^2$. Thus dividing by the square root of the number of pixels ($b^2/p^2$) within the region gives
\vspace{-0.3\baselineskip}
\begin{equation}
\mu_\mathrm{lim}\,=\,\mu(\sigma_b)\,=\,\mu(\sigma) + 2.5\,\log_{10}(b/p)\, \quad \mathrm{mag\,arcsec}^{-2}\ ,
\label{eqn:exp3}
\end{equation}
again equivalent to the formulation in Eq.\,(\ref{eqn:sblim}).
\end{enumerate}

Results are compared graphically in Fig. \ref{fig:sblims} where the SB limits \mulim\ obtained with different methods
are plotted against sky surface brightness. 
The lack of deviation of the sky value for the individual galaxies, namely the small horizontal scatter, implies 
that the overall sky brightness can be determined quite robustly.

Table \ref{tab:sky} and Fig. \ref{fig:sblims} show that from galaxy to galaxy, 
there can be up to $0.5$\,mag of difference in the derived SB limits, with an excursion that
suggests that the fainter SB limits are associated with fainter sky brightness. 
A large portion of the variation in the measured sky brightness
is almost certainly due to the influence of zodiacal 
light
\citep[see e.g.,][]{EROData}, since there is no trend with foreground extinction (not shown).
For each \Euclid band,
we have fit the data in Fig. \ref{fig:sblims} using a robust-linear method (the python package \texttt{statsmodels}),
shown as a dashed line in each of the panels.
Judging from the fits,
over roughly $1$\,\magarc\ variation in sky surface brightness,
the noise levels can vary up to $0.5$\,\magarc.

Approach (1) calculates the standard deviation $\sigma$ of the sky within 100\,arcsec$^2$ tiles,
while in approaches (2) and (3), $\sigma$ is computed across the entire masked (non-detected) image of the galaxy.
Specifically, approach (3) analyzes the distribution of pixel values in the outer part of the image surrounding the galaxy.
This implies that the first method could measure small-scale variation, 
while the last two could essentially capture large-scale variations across the images.
However, this is not completely borne out by the results in Fig. \ref{fig:sblims}, 
although in the NISP bands, there is a tendency for the latter two methods to give brighter limits.


Here and in Sect. \ref{sec:datareduction} we assess the noise level in the stacked data products used in our analysis,
assuming that the pixels are intrinsically uncorrelated [see Eqs. \eqref{eqn:sblim} and \eqref{eqn:exp3}].
However, the process of stacking images inherently creates covariances between pixels; 
individual pixels in the original frames are spread out over multiple pixels in the final stacked image.
In the simplest bilinear stacking procedures, each pixel is divided into four pixels in the final image, 
although with the more sophisticated stacking in the \IE\ data reduction, a Lanczos kernel is used 
which spreads the pixel over a 7$\times$7 kernel in the final pixel stack.
We can empirically determine the average noise in pixels by taking the standard deviation in dark parts of the image;
then, we can extract the covariance between pixels using dark patches of the frame.
We find that a typical pixel in the \IE\ band has a $10$--$15$\% correlation with its immediate neighboring pixels, and $\la 5$\% 
correlation with pixels two steps away.
For NISP bands (\YE, \JE, and \HE), combined with a bi-linear method, the correlation is 
slightly stronger at $20$--$30$\% for immediate neighbors, and $3$--$8$\% correlation for second neighbors.
These correlations are typical of stacked images and simply allow us to better understand the depth achieved with \Euclid.
Further description of the covariance properties in the stacked images can be found in \citet{EROData},
which describes the process in more detail.



\section{\Euclid imaging capabilities}
\label{app:imaging}

The continuation of Figs. \ref{fig:pretty_holmbergii} and \ref{fig:pretty_ngc6744} (Sect. \ref{sec:integrated}) is shown here
for IC\,10, IC\,342, NGC\,2403, and NGC\,6822.

\begin{figure*}[h!]
\centering
\hbox{
\includegraphics[width=0.40\textwidth]{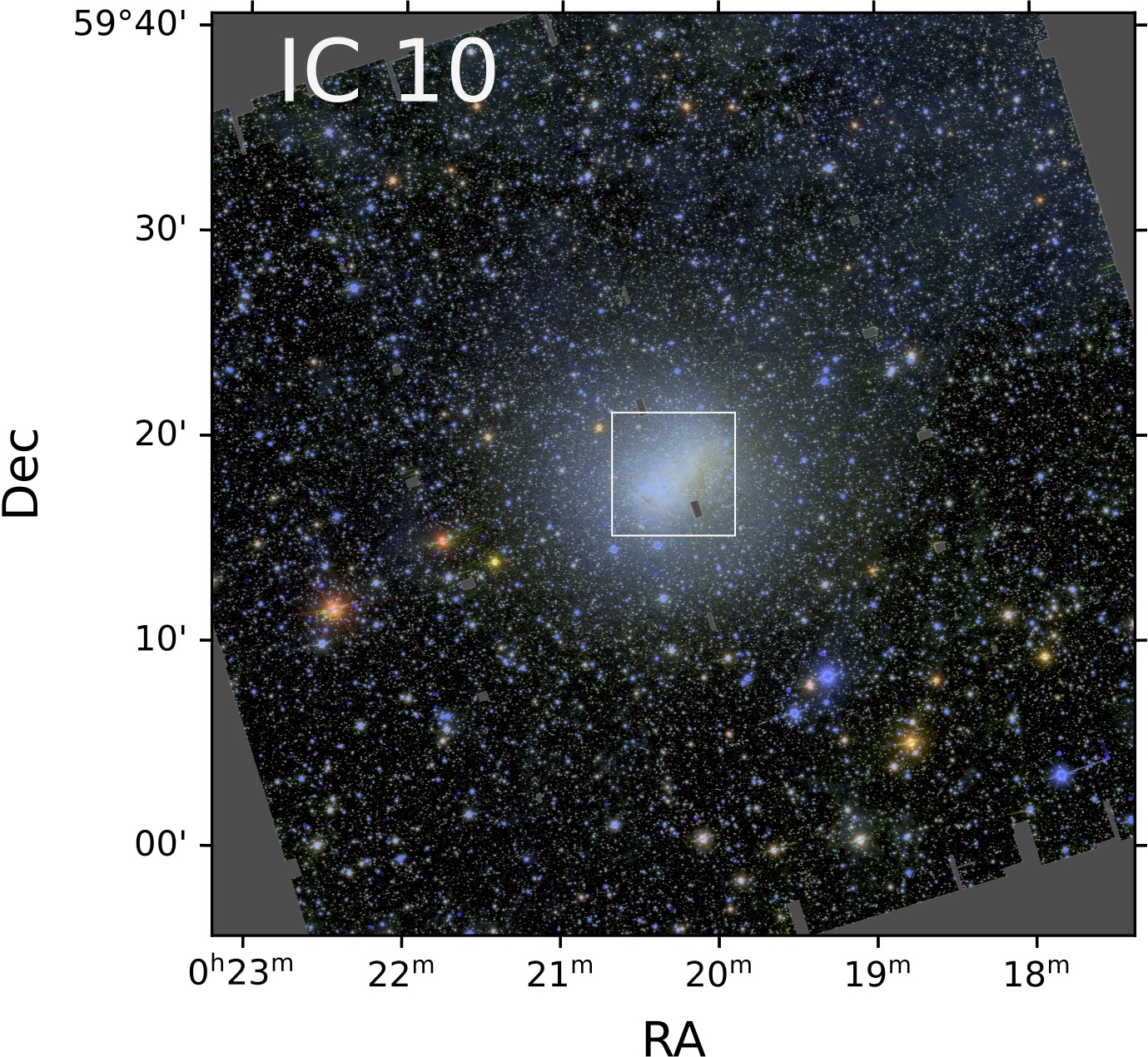}
\hspace{0.11\textwidth}
\includegraphics[width=0.40\textwidth]{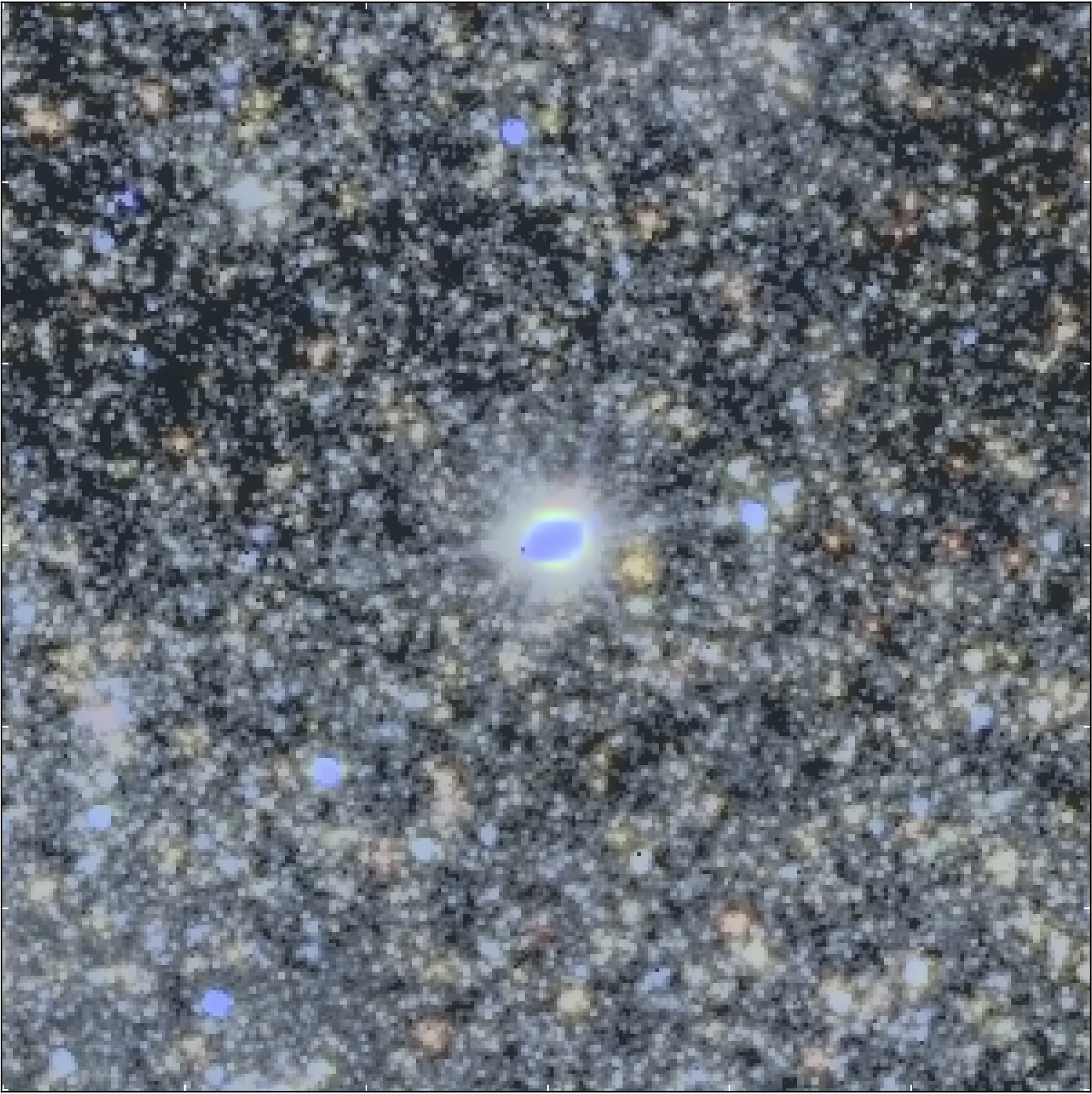}
}
\includegraphics[width=0.85\textwidth]{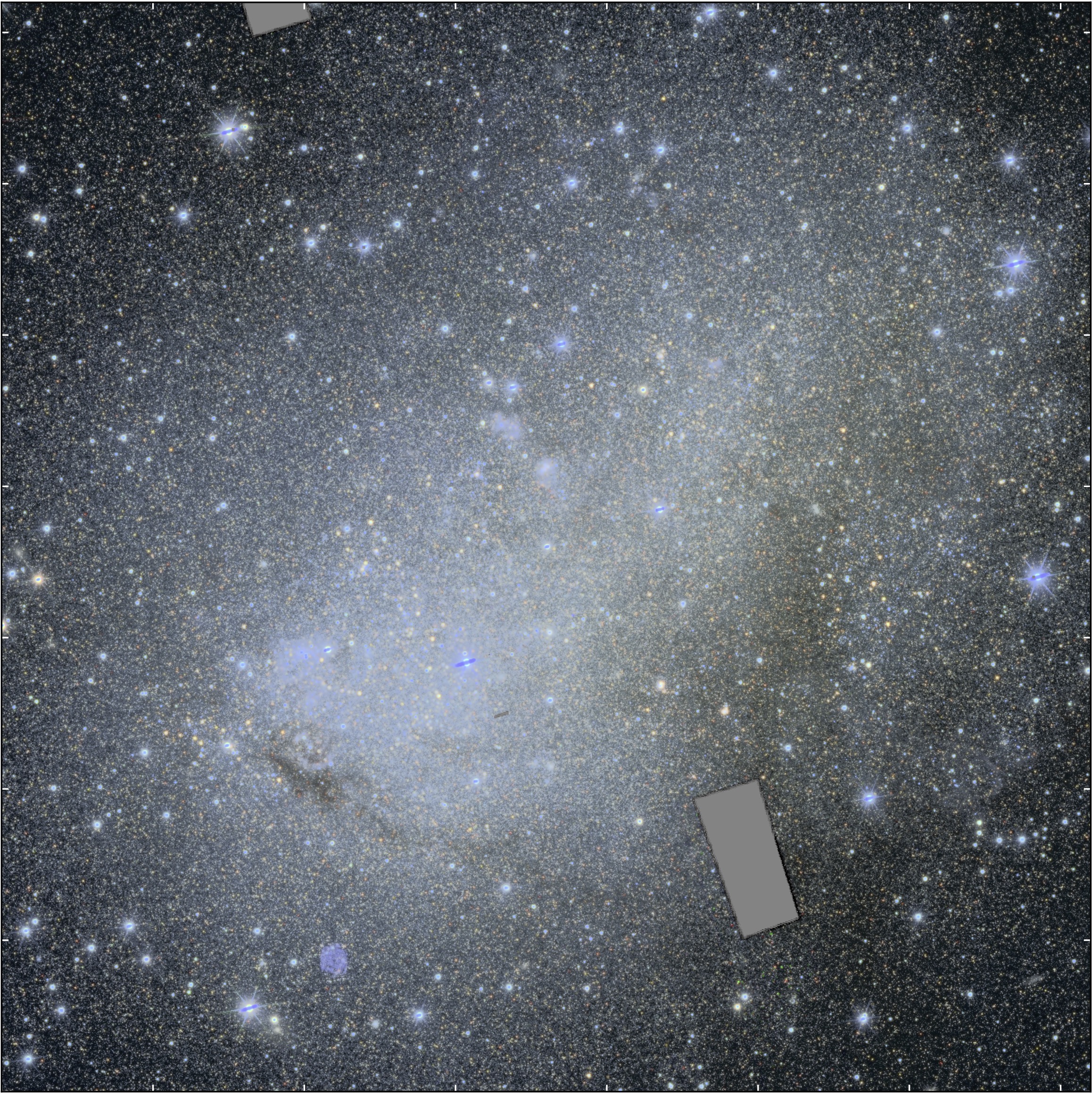}

\caption{Same as for Fig. \ref{fig:pretty_holmbergii}, but for IC\,10, and,
as before,  with \HE\ green, \YE\ red, and \IE\ blue.
Extinction has been corrected and sky subtracted as described in the text (Sects. \ref{sec:sblim} and \ref{sec:extinction}).
In the top left panel, the full FoV of $0\fdg7\,\times\,0\fdg7$ is shown, 
while in the bottom one the inner 6\arcmin\,$\times$\,6\,\arcmin\ region is displayed corresponding to the white box
in the upper panel. 
The top right panel shows the zoomed-in 30\arcsec\,$\times$\,30\arcsec\ RGB image of the blue nucleus also seen in the radial color profiles
(Sect. \ref{sec:profiles}, App. \ref{app:profiles}, and Fig. \ref{fig:profiles_2}).
The dark boxes in IC\,10 result from an incorrect orientation of the instruments relative to the roll angle
of the satellite during observation (this was the first galaxy observed within the ERO Showcase). 
}
\label{fig:pretty_ic10}
\vspace{-2\baselineskip}
\end{figure*}

\begin{figure*}[h!]
\centering
\includegraphics[width=0.40\textwidth]{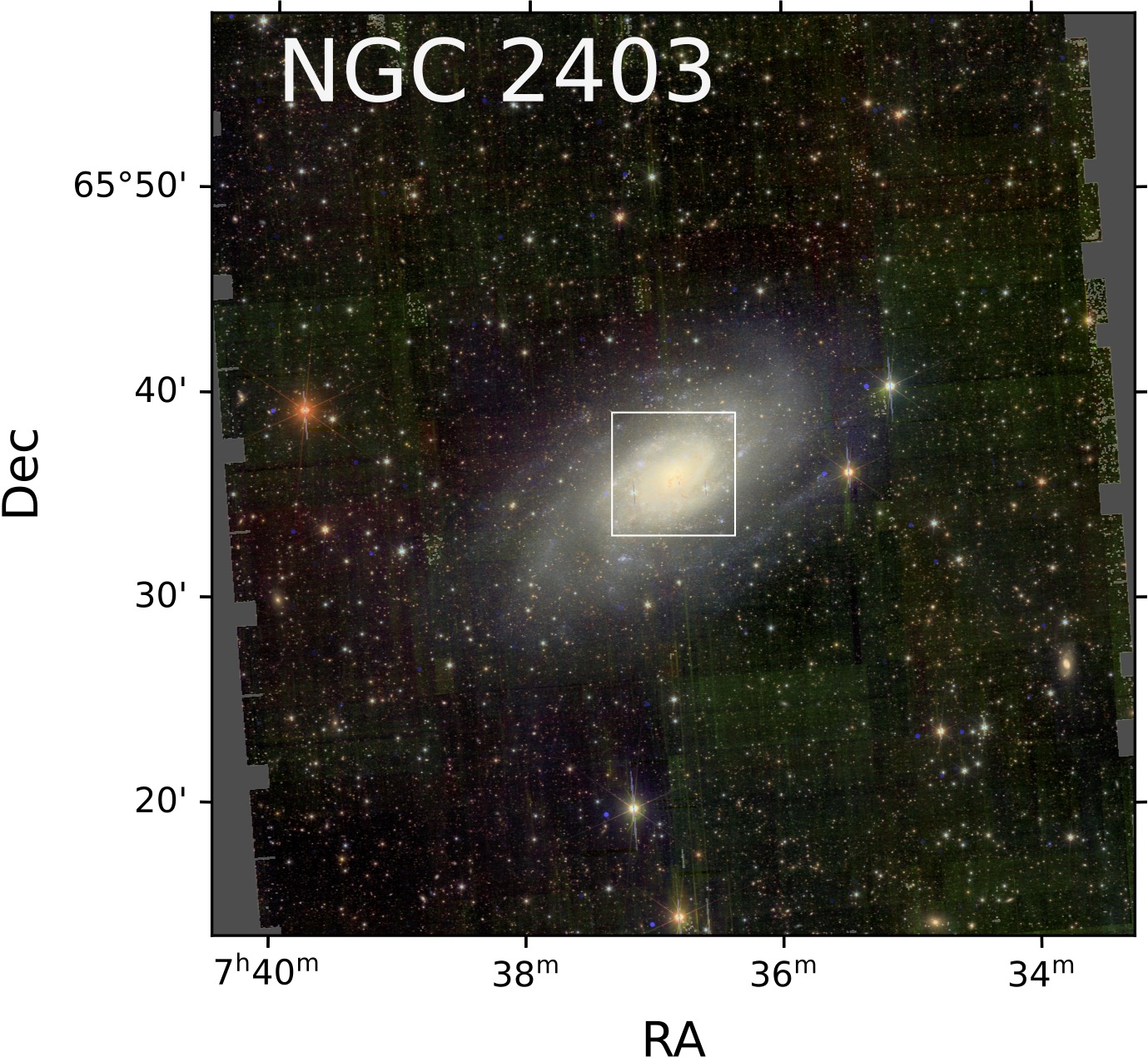}
\includegraphics[width=0.85\textwidth]{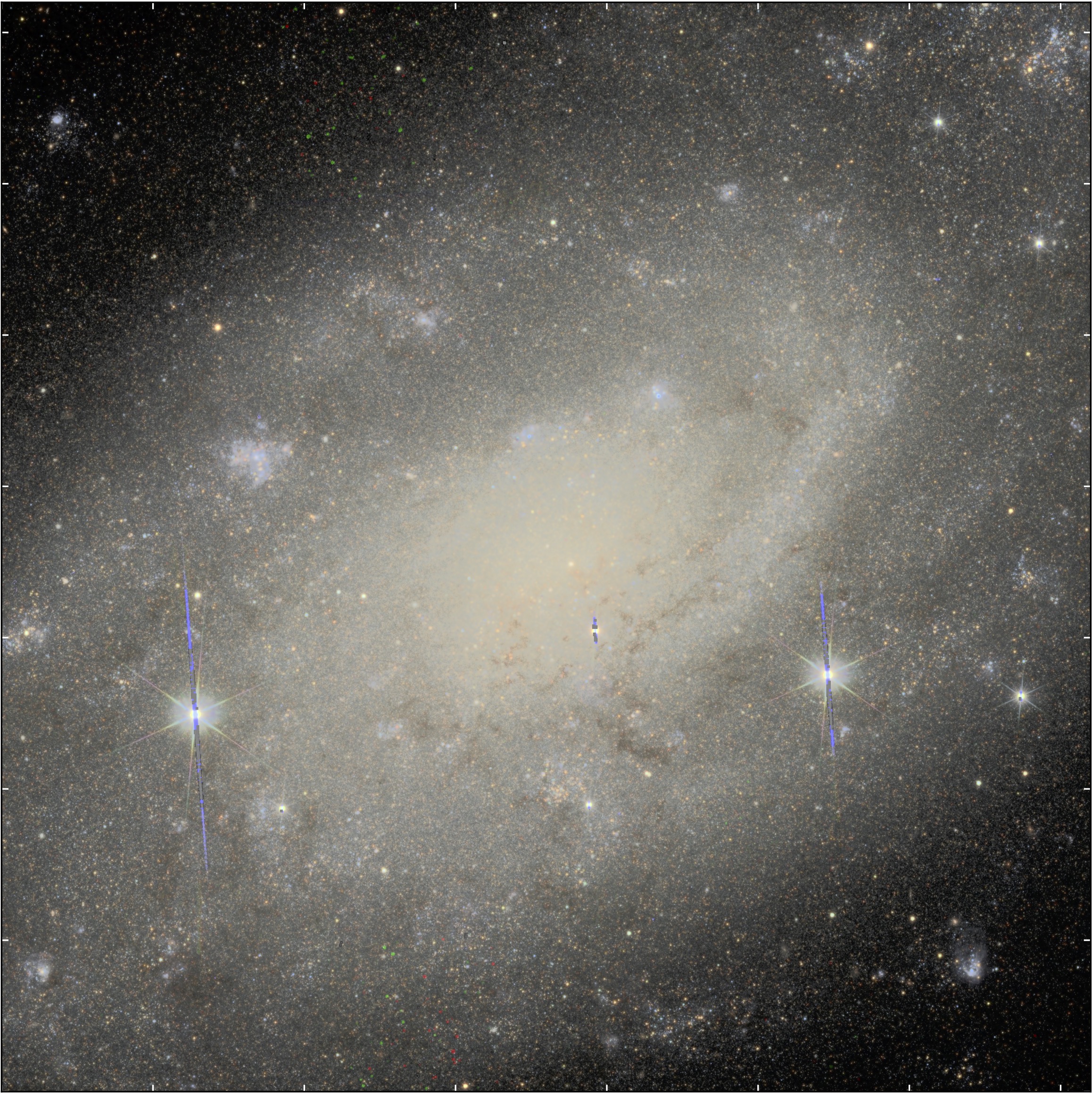}
\caption{Same as for Fig. \ref{fig:pretty_holmbergii}, but for NGC\,2403. 
}
\label{fig:pretty_ngc2403}
\end{figure*}

\begin{figure*}[h!]
\centering
\includegraphics[width=0.40\textwidth]{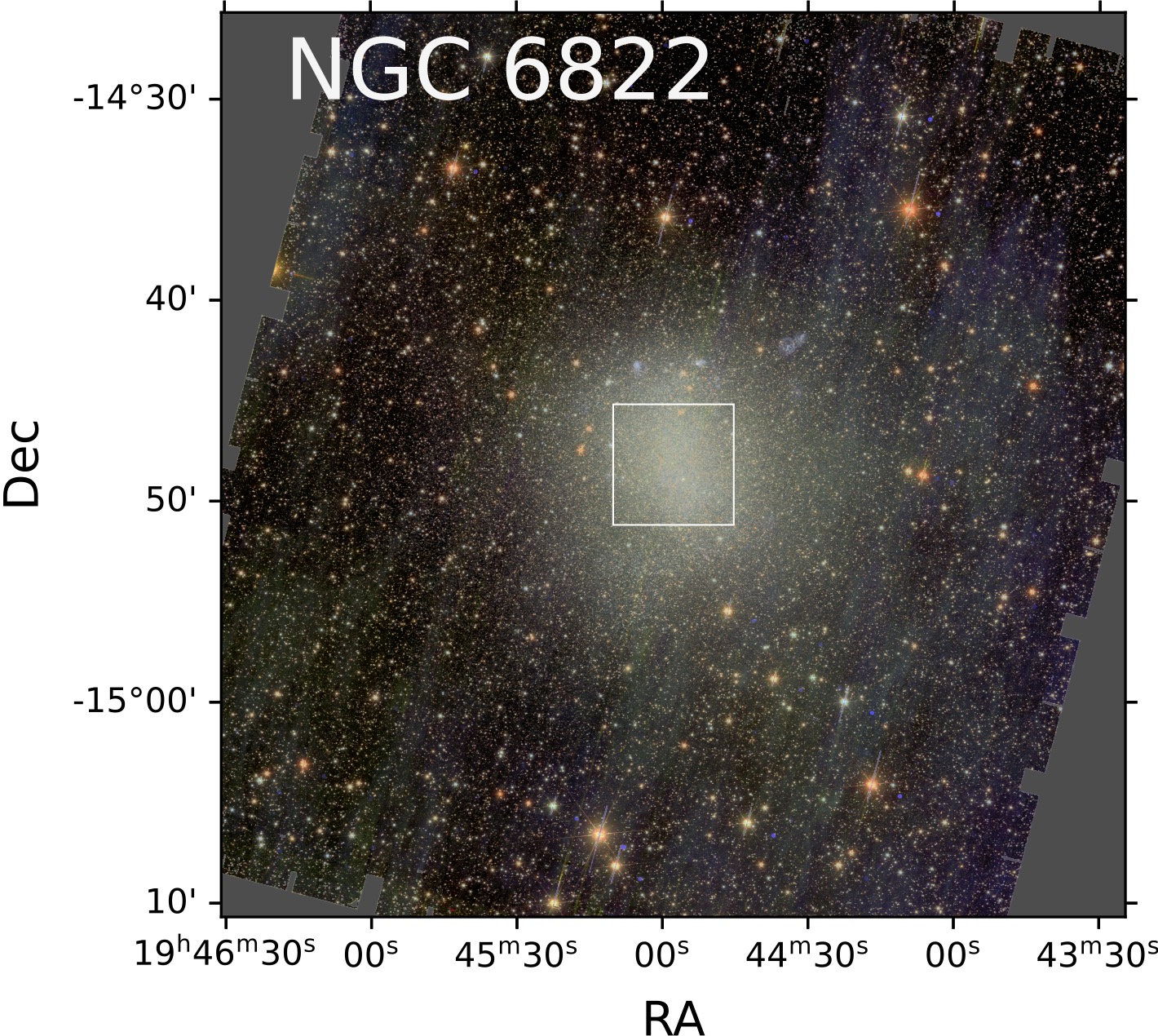}
\includegraphics[width=0.85\textwidth]{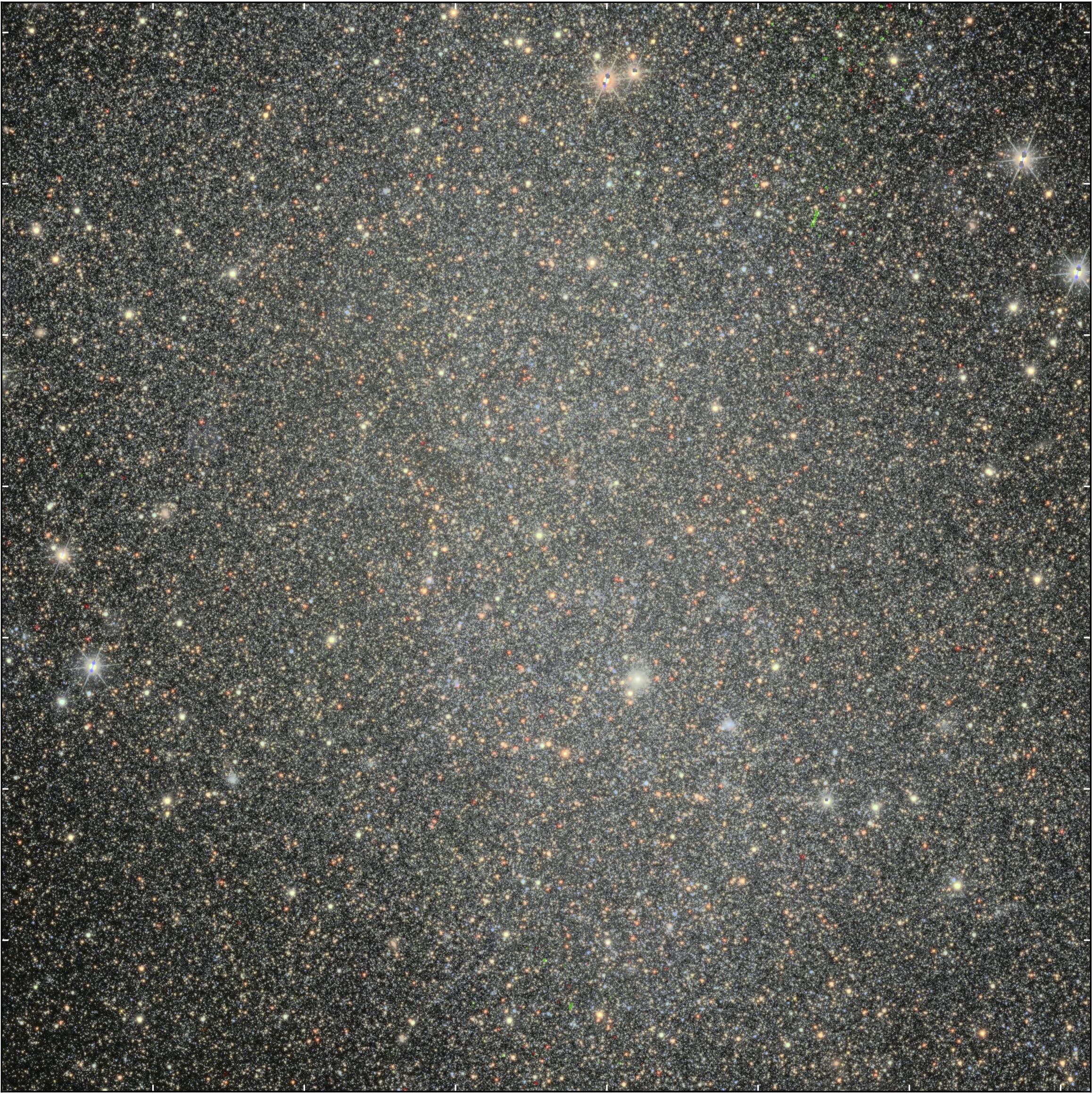}
\caption{Same as for Fig. \ref{fig:pretty_holmbergii}, but for NGC\,6822. 
}
\label{fig:pretty_ngc6822}
\end{figure*}



\section{Surface brightness profiles of four Showcase galaxies}
\label{app:profiles}

Here we show surface-brightness profiles of the remaining four galaxies
that are not given in the main text (Sect. \ref{sec:profiles}).

\begin{figure*}[h!]
\vbox{
\includegraphics[width=0.48\textwidth]{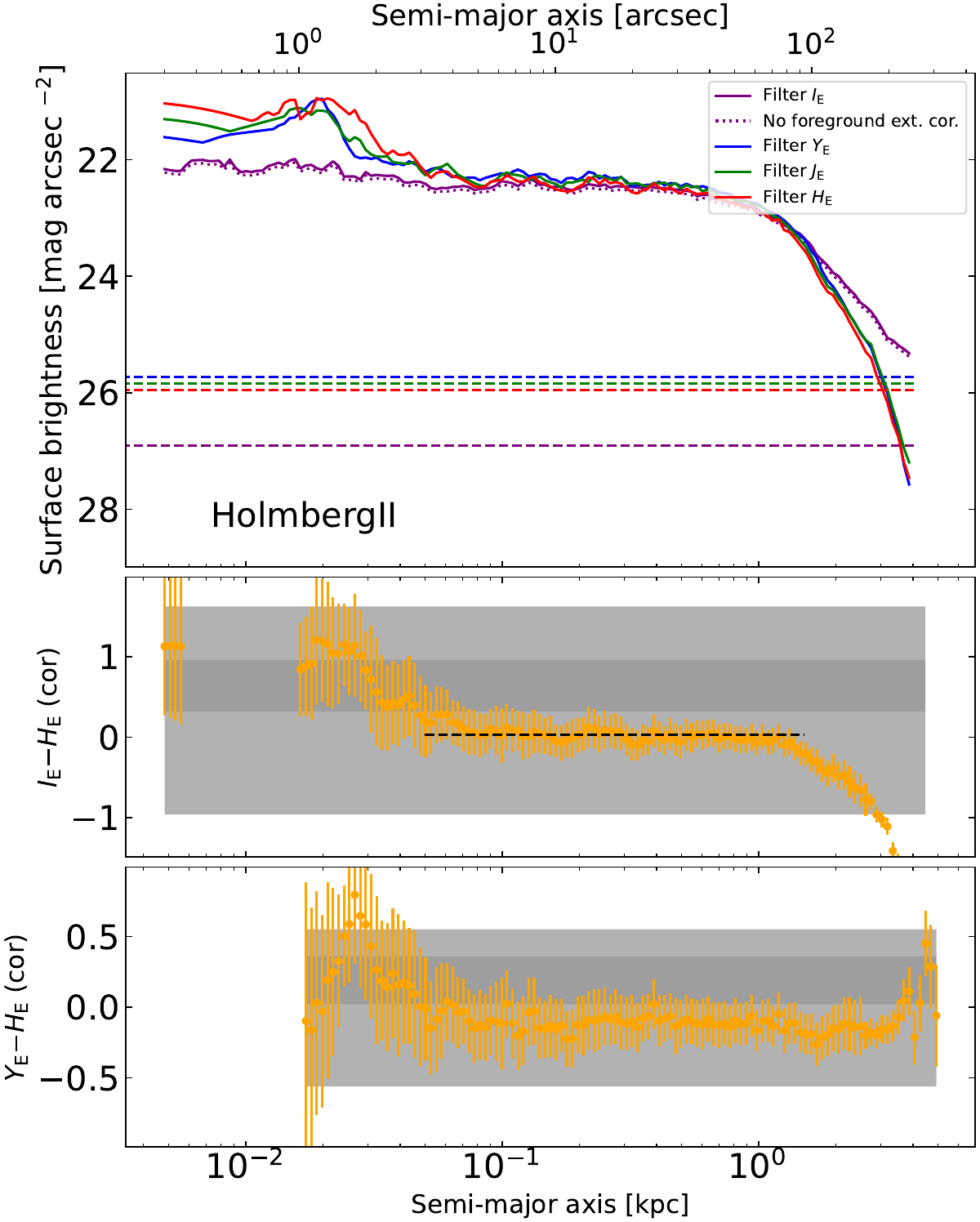}
\hspace{0.04\textwidth}
\includegraphics[width=0.48\textwidth]{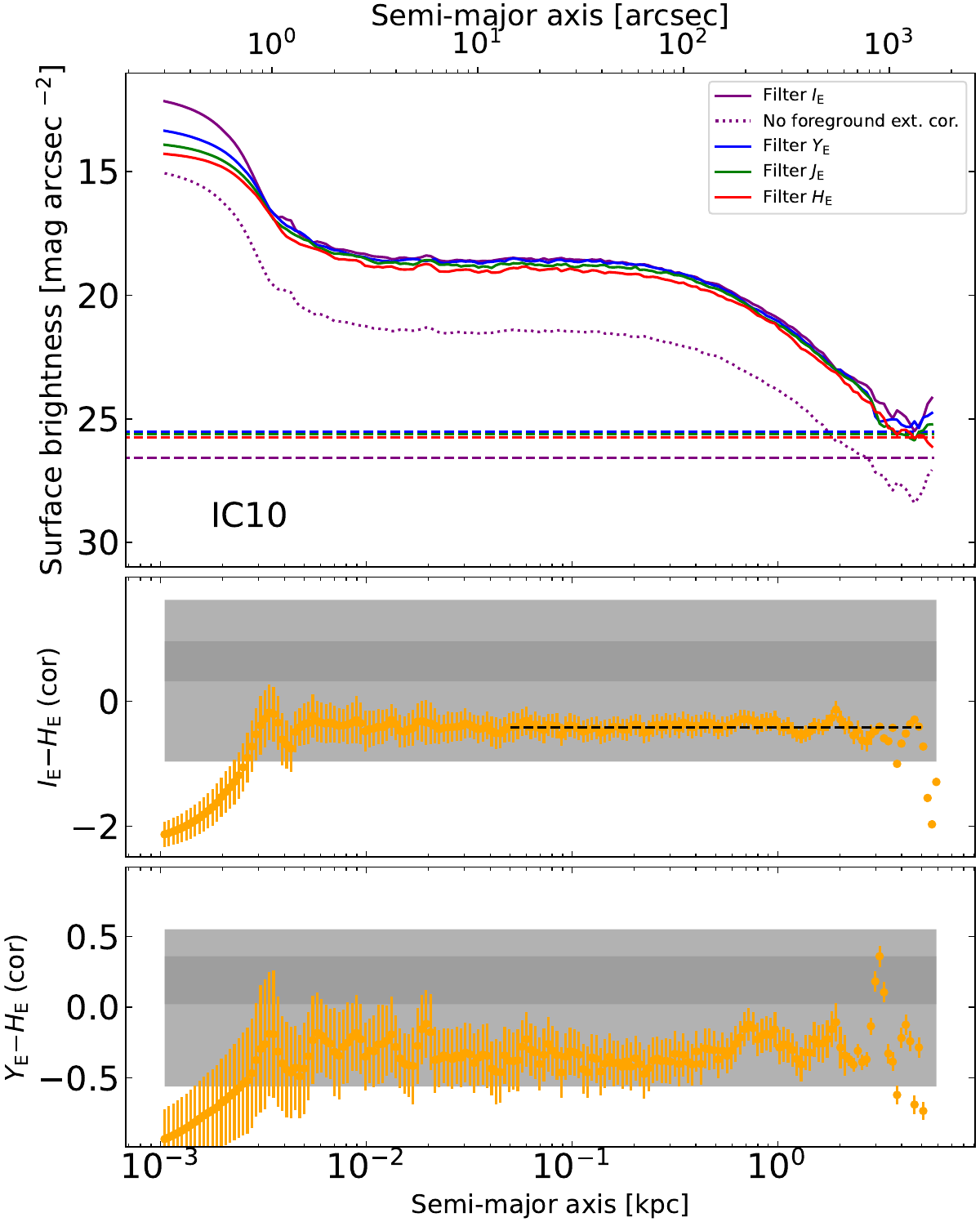} \\
}
\vbox{
\includegraphics[width=0.48\textwidth]{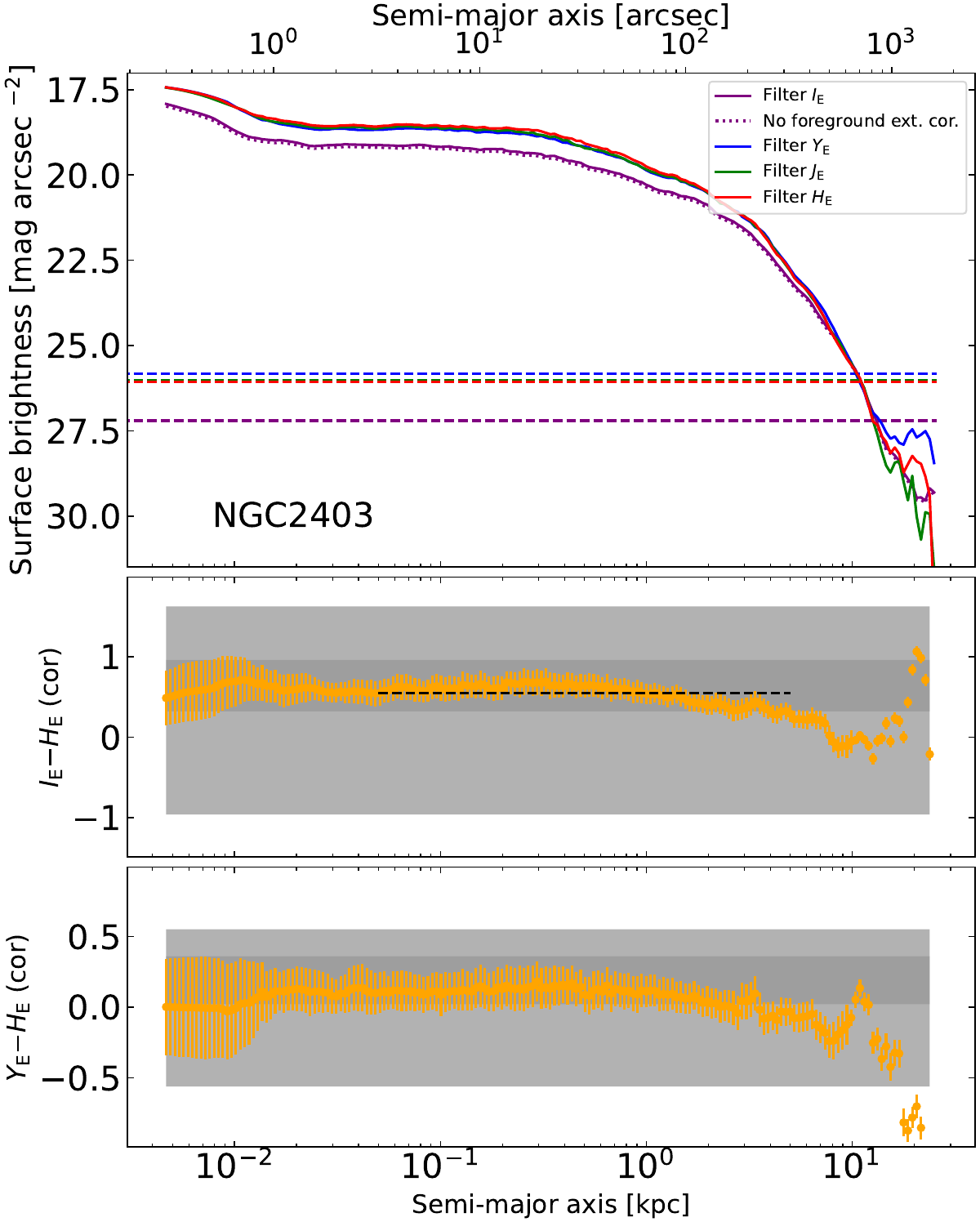}
\hspace{0.04\textwidth}
\includegraphics[width=0.48\textwidth]{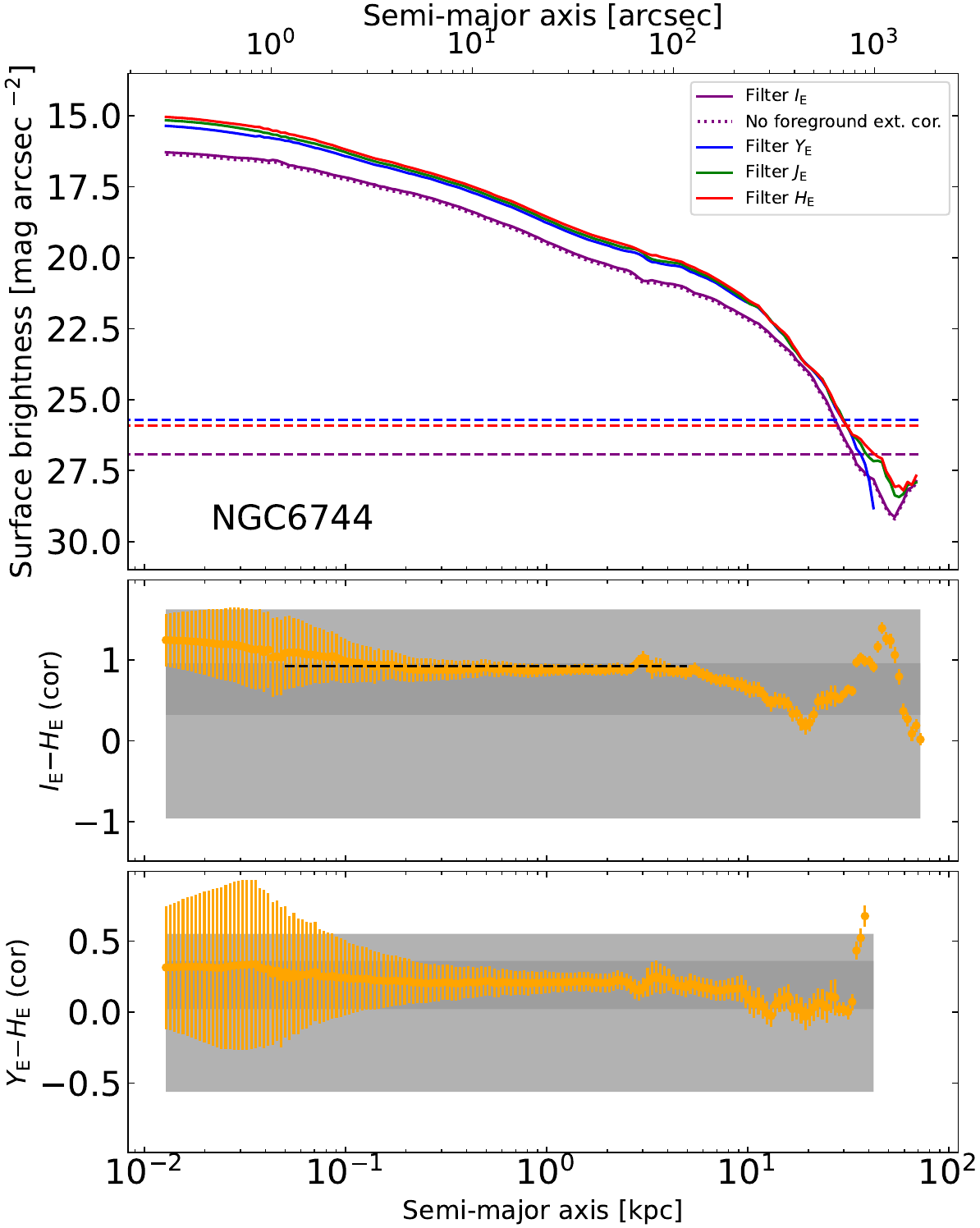} \\
}
\vspace{-\baselineskip}
\caption{Surface brightness profiles extracted by \texttt{AutoProf}, as described
in the text: Holmberg\,II and IC\,10 in the top row; and NGC\,2403 and NGC\,6744 in the bottom.
The four bands are given by purple, blue, green, and red for \IE, \YE, \JE, and \HE, respectively.
The $1\sigma$ SB limits from \texttt{AutoProf} 
in units of \magarc\ 
are shown as dashed horizontal lines, with colors corresponding to the \Euclid bands.
The fluxes have 
been corrected for foreground extinction (Sect. \ref{sec:extinction}) in order to be consistent
with the \IE$-$\HE\ radial color gradient shown in the middle panel, 
and \YE$-$\HE\ shown at the bottom.
The uncorrected \IE\ profile is shown as a dotted (purple) curve in the top panel.
The mean \IE$-$\HE\ color over typically a factor of $100$ in radius in the inner galaxy is shown as a horizontal
dashed line in the middle panel.
}
\label{fig:profiles_2}
\end{figure*}

\vspace{-\baselineskip}



\section{Star-count maps of four Showcase galaxies}
\label{app:starcounts}

Here we present the maps of the star counts of the remaining three galaxies
that are not shown in the main text (Sect. \ref{sec:resolvedstars}).

\begin{figure*}[h!]
\centerline{
\includegraphics[width=0.48\textwidth]{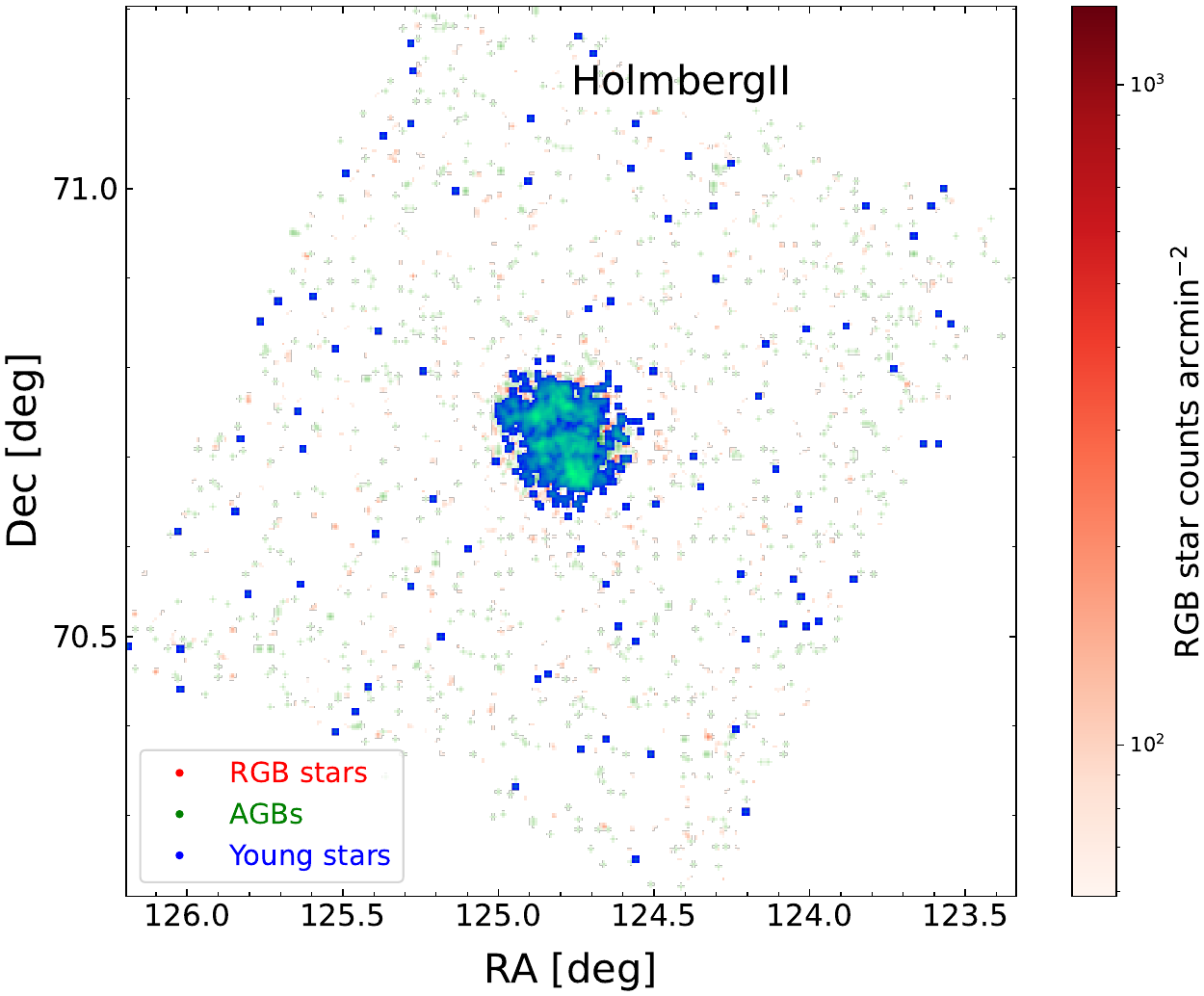}
\hspace{0.04\textwidth}
\includegraphics[width=0.48\textwidth]{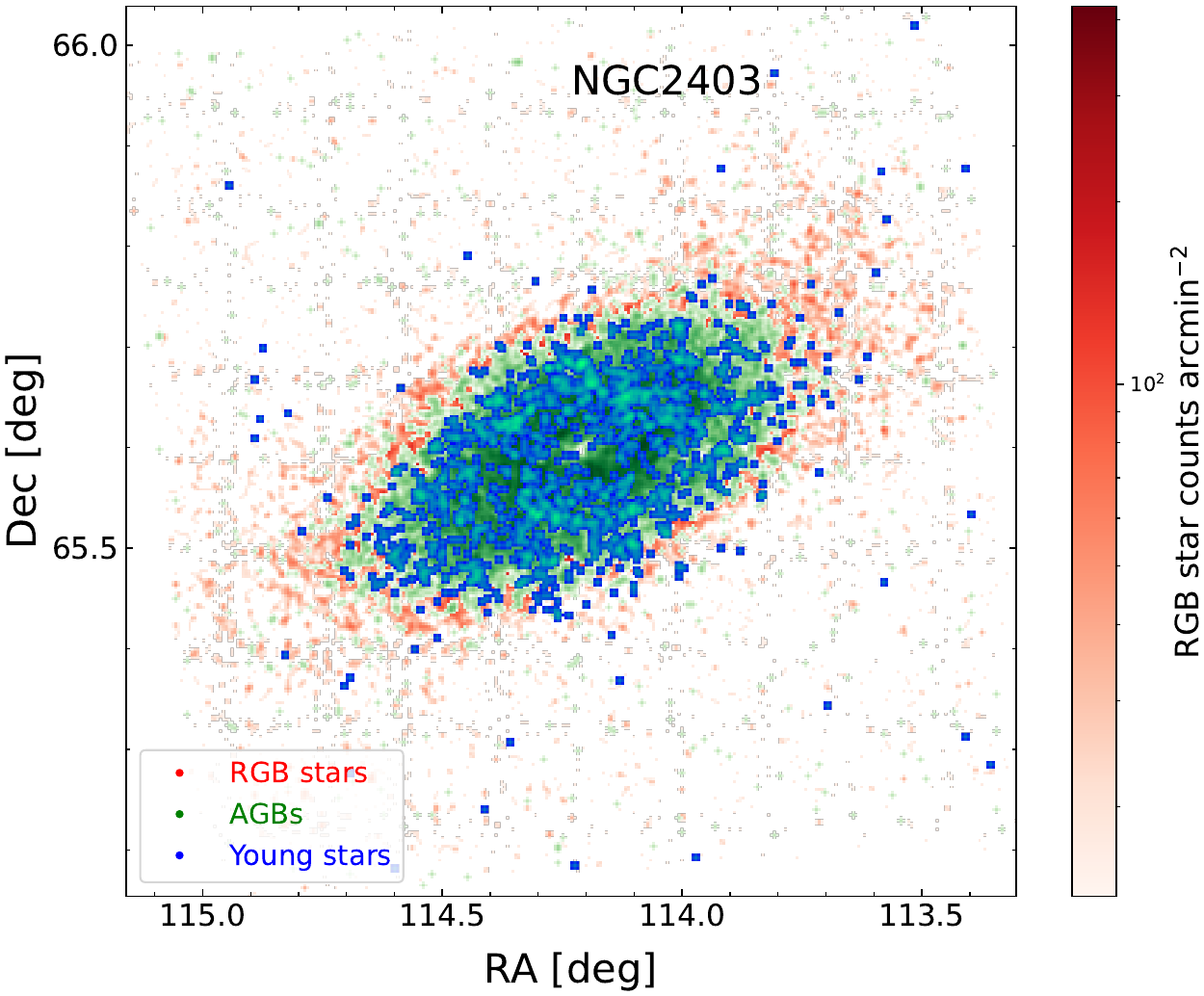}
}
\centerline{
\includegraphics[width=0.48\textwidth]{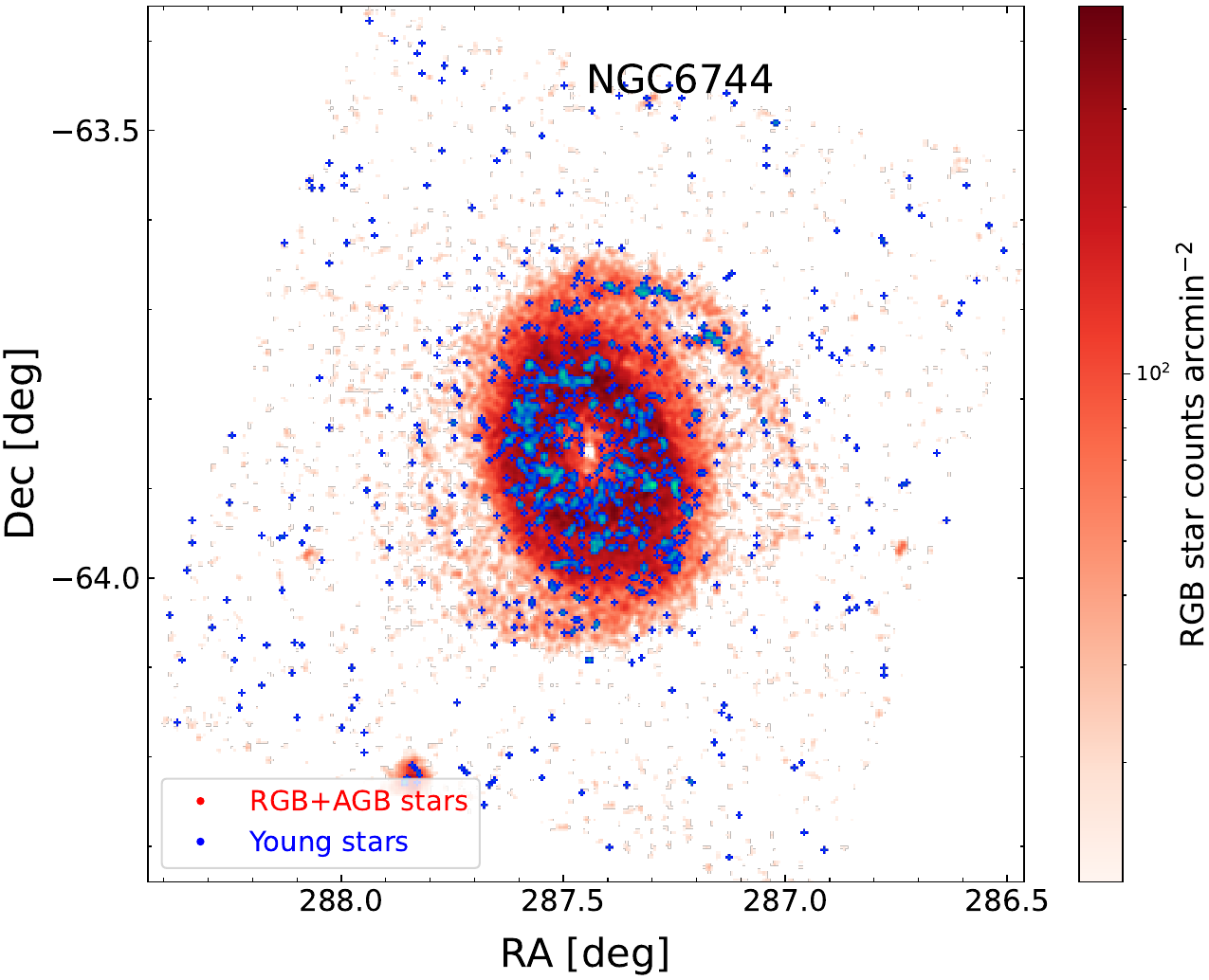}
}
\caption{Star-count maps obtained as described in Sect. \ref{sec:resolvedstars}, but for 
Holmberg\,II (top left), NGC\,2403 (top right), and the most distant Showcase galaxy, NGC\,6744 (bottom).
}
\label{fig:starcounts_3}
\end{figure*}


\end{document}